\documentclass[final,3p,times]{elsarticle}
\biboptions{sort&compress}
\usepackage{wrapfig}

\usepackage{graphics}
\usepackage{graphicx}
\usepackage{epsfig}
\usepackage{amssymb}
\usepackage{amsthm}
\usepackage{amsmath}

\usepackage{caption}
\usepackage[labelfont=bf]{caption}
\usepackage[labelsep=space]{caption}
\usepackage[nodots]{numcompress}

\usepackage{xcolor}
\usepackage{hyperref}
\hypersetup{
  colorlinks,
  citecolor=[violet],
  linkcolor=[red],
  urlcolor=[blue]}

\journal{Coordination Chemistry Reviews}

\begin{document}

\begin{frontmatter}

\title{A comprehensive review on the ferroelectric orthochromates: \\ Synthesis, property, and application}
\tnotetext[]{
\emph{Abbreviations}:
AFM: Antiferromagnetic;
CAFM: Canted AFM;
DM: Dzyaloshinskii Moriya;
FM: Ferromagnetic;
HRS: High resistance state;
IP: Intermediate point;
LRS: Low resistance state;
NPD: Neutron powder diffraction;
NTC: Negative temperature coefficient;
PDF: Pair distribution function;
P-E: Polarization electric;
SHS: Self-propagating high-temperature synthesis;
SR: Spin reorientation;
$T_\textrm{C}$: Ferroelectric transition temperature;
$T_\textrm{N}$: Antiferromagnetic transition temperature;
WFM: Weak FM;
XRPD: X-ray powder diffraction.
}

\author[1]{Yinghao Zhu\fnref{fn1}}
\author[1,3]{Kaitong Sun\fnref{fn1}}
\author[2,1]{Si Wu\fnref{fn1}}
\fntext[fn1]{These authors contributed equally.}
\author[1]{Pengfei Zhou}
\author[1]{Ying Fu}
\author[1]{Junchao Xia}
\author[1]{Hai-Feng Li\corref{cor1}}
\cortext[cor1]{Corresponding author}
\ead{haifengli@um.edu.mo}

\affiliation[1]{organization={Joint Key Laboratory of the Ministry of Education, Institute of Applied Physics and Materials Engineering, University of Macau},
            addressline={Avenida da Universidade, Taipa},
            city={Macao SAR},
            postcode={999078},
            country={China}}
\affiliation[2]{organization={School of Physical Science and Technology, Ningbo University},
            city={Ningbo},
            postcode={315211},
            country={China}}
\affiliation[3]{organization={Guangdong--Hong Kong--Macao Joint Laboratory for Neutron Scattering Science and Technology},
            addressline={No. 1. Zhongziyuan Road, Dalang},
            city={DongGuan},
            postcode={523803},
            country={China}}

\begin{abstract}
Multiferroics represent a class of advanced materials for promising applications and stand at the forefront of modern science for the special feature possessing both charge polar and magnetic order. Previous studies indicate that the family of RECrO$_3$ (RE = rare earth) compounds is likely another rare candidate system holding both ferroelectricity and magnetism. However, many issues remain unsolved, casting hot disputes about whether RECrO$_3$ is multiferroic or not. For example, an incompatibility exists between reported structural models and observed ferroelectric behaviors, and it is not easy to determine the spin canting degree. To address these questions, one key step is to grow single crystals because they can provide more reliable information than other forms of matter do. In this review, the parent and doped ferroelectric YCrO$_3$ compounds are comprehensively reviewed based on scientific and patent literatures from 1954 to 2022. The materials syntheses with different methods, including poly-, nano-, and single-crystalline samples and thin films, are summarized. The structural, magnetic, ferroelectric and dielectric, optical, and chemical-pressure (on Y and Cr sites by doping) dependent chemical and physical properties and the corresponding phase diagrams, are discussed. Diverse (potential) applications, including anti-corrosion, magnetohydrodynamic electrode, catalyst, negative-temperature-coefficient thermistor, magnetic refrigeration, protective coating, and solid oxide fuel cell, are present. To conclude, we summarize general results, reached consensuses, and existing controversies of the past nearly 69 years of intensive studies and highlight future research opportunities and emerging challenges to address existing issues.
\end{abstract}

\begin{highlights}
\item Relationship between materials processing, microscopic and macroscopic properties, and applications.
\item Structural and dielectric properties, incompatibility between structure and ferroelectricity, and origin of ferroelectricity.
\item Nature of weak ferromagnetism in a main antiferromagnetic matrix.
\item Magnetic properties, negative magnetization, and magnetic reversal behavior.
\item Optical properties, local polar domains, leakage current, size of charge displacement, and magnetoelectric effect.
\end{highlights}


\end{frontmatter}


\tableofcontents

\section{Introduction}

In strongly correlated electron systems, there exist cooperative and competitive couplings of charge, orbital, lattice, and spin degrees of freedom, which results in various exotic macroscopic properties, such as superconductivity \cite{matthias1963superconductivity, mcmillan1968transition, maple1976superconductivity, bednorz1986possible, nagamatsu2001superconductivity, tinkham2004introduction, plakida2010high, zimmer1995rate, quebe2000quaternary, kamihara2008iron, chen2008superconductivity, kresin2021superconducting} and multiferroicity \cite{fiebig2002observation, kimura2003magnetic, cheong2007multiferroics, kagawa2010ferroelectricity, valencia2011interface, tokunaga2012electric}. These continue to be exciting fields of research, unravelling the intricate correlations of electrons and obtaining a complete understanding of origins of the interesting phenomena.

The so-called multiferroic materials accommodate manifold orders like magnetic and ferroelectric with potential mutual interactions. This offers multiple parameters for casting the macroscopic functionalities and great challenges in exploring the microscopic origins and coupling mechanisms \cite{li2021understanding, YANG201972, SUN2014124, YOON2003275, Hur2004, Cheong2007, Kagawa2010, Valencia2011, Tokunaga2012, Aken2004, Kenzelmann2005, evans2020domains, lottermoser2020short, davydova2020spin, huang2020manipulating, cano2021multiferroics, wu2021100, zhao2021dzyaloshinskii, schierle2020promising, zhang2021purely, meier2021ferroelectric, choi2020nanoengineering, stein2021combined, zhou2021terahertz, morozovska2021combined, kumar2021recent, wang2021competition, martin2016thin, uchino2018ferroelectric, nicola2005, mishra2021observation, fita2021pressure, sharma2021tuning, otsuka2021effect}. Multiferroic materials provide a dual-mode for information storage \cite{eerenstein2006multiferroic, nan2008multiferroic, dong2015multiferroic, spaldin2019advances} and are particularly important for writing and reading information by promising magnetic and electric devices with multiple functionalities \cite{cheong2007multiferroics, kagawa2010ferroelectricity, valencia2011interface, tokunaga2012electric, fiebig2016evolution, zhao2014near, SEKINE2022214663}. Theoretically, ferroelectricity and ferromagnetism are mutually excluding phenomena in one perovskite \cite{hill2000there}. Many issues need to be addressed for the simultaneous presence of both properties in a single compound from the point of views of experimental observations and theoretical calculations \cite{li2021understanding, YANG201972, SUN2014124, YOON2003275, Hur2004, Cheong2007, Kagawa2010, Valencia2011, Tokunaga2012, Aken2004, Kenzelmann2005, evans2020domains, lottermoser2020short, davydova2020spin, huang2020manipulating, cano2021multiferroics, wu2021100, zhao2021dzyaloshinskii, schierle2020promising, zhang2021purely, meier2021ferroelectric, choi2020nanoengineering, stein2021combined, zhou2021terahertz, morozovska2021combined, kumar2021recent, wang2021competition, martin2016thin, uchino2018ferroelectric, nicola2005, mishra2021observation, fita2021pressure, sharma2021tuning, otsuka2021effect}. Unraveling the factors responsible for these properties is a central topic in the study of multiferroics \cite{cheong2007multiferroics, ederer2011mechanism, balke2012enhanced, SHI2019561}.

Traditionally, multiferroic materials are classified into two catalogs: (i) The first type holds weak interactions between existing ferroelectricity and magnetism owning to their different physical origins. The compound of YMnO$_3$ belongs to such kind of multiferroic materials, in which the ferroelectricity was probably induced by symmetry breaking \cite{Aken2004, gibbs2011high}. (ii) Within the second-type multiferroics, ferroelectricity and magnetism strongly interact mutually, thus a formation of the magnetic ordering could result in ferroelectricity, for example, the compound of TbMnO$_3$ \cite{Kenzelmann2005}. It is evident that magnetization originates from unpaired electrons of the partially-filled \emph{d}- or \emph{f}-shell and is localized in multiferroic materials. Lone electron pairs \cite{hill2000there}, magnetoelectric coupling \cite{mostovoy2006ferroelectricity}, structural geometry \cite{Aken2004}, charge order \cite{choi2008ferroelectricity}, and oxygen vacancy \cite{aschauer2013strain} were previously reported microscopic mechanisms for the appearance of ferroelectricity.

The perovskite-based ABO$_3$ families contribute an important part of multiferroic materials. By substituting the crystallographic A- or B-site elements, chemical pressure is tuned, and different crystalline structures can be realized with potential multiferroics \cite{PhysRevMaterials.2.104414}. Among them, the BiFeO$_3$ compound was considered as a typical representative, which holds a ferroelectric phase transition at $T_\textrm{C}$ = 1103 K and a large value of remanent polarization $P_\textrm{r}$ $\sim$ 55 $\mu$C/cm$^2$ over room temperature. The ordering of lone electron pairs was suggested as the microscopic originn \cite{wang2003epitaxial}. There exist an antiferromagnetic (AFM) phase transition at $T_\textrm{N}$ $\sim$ 76 K and a ferroelectric phase transition at $T_\textrm{C}$ $\sim$ 914 K in the YMnO$_3$ compound \cite{katsufuji2001dielectric}, and the ferroelectric polarization ($\sim$ 6 $\mu$C/cm$^2$) was attributed to geometric variation. In this mechanism, the distortion and the tilting of MnO$_5$ configuration modify the coupling between O and Y ions, consequently, forming the charge dipole of Y ions \cite{Aken2004, fennie2005ferroelectric, gibbs2011high}. In the TbMnO$_3$ compound, there exits a strong magnetoelectric effect, and the charge polarization was due to a specific magnetic order \cite{kimura2003magnetic}.

\begin{figure*} [!t]
\centering \includegraphics[width=0.78\textwidth]{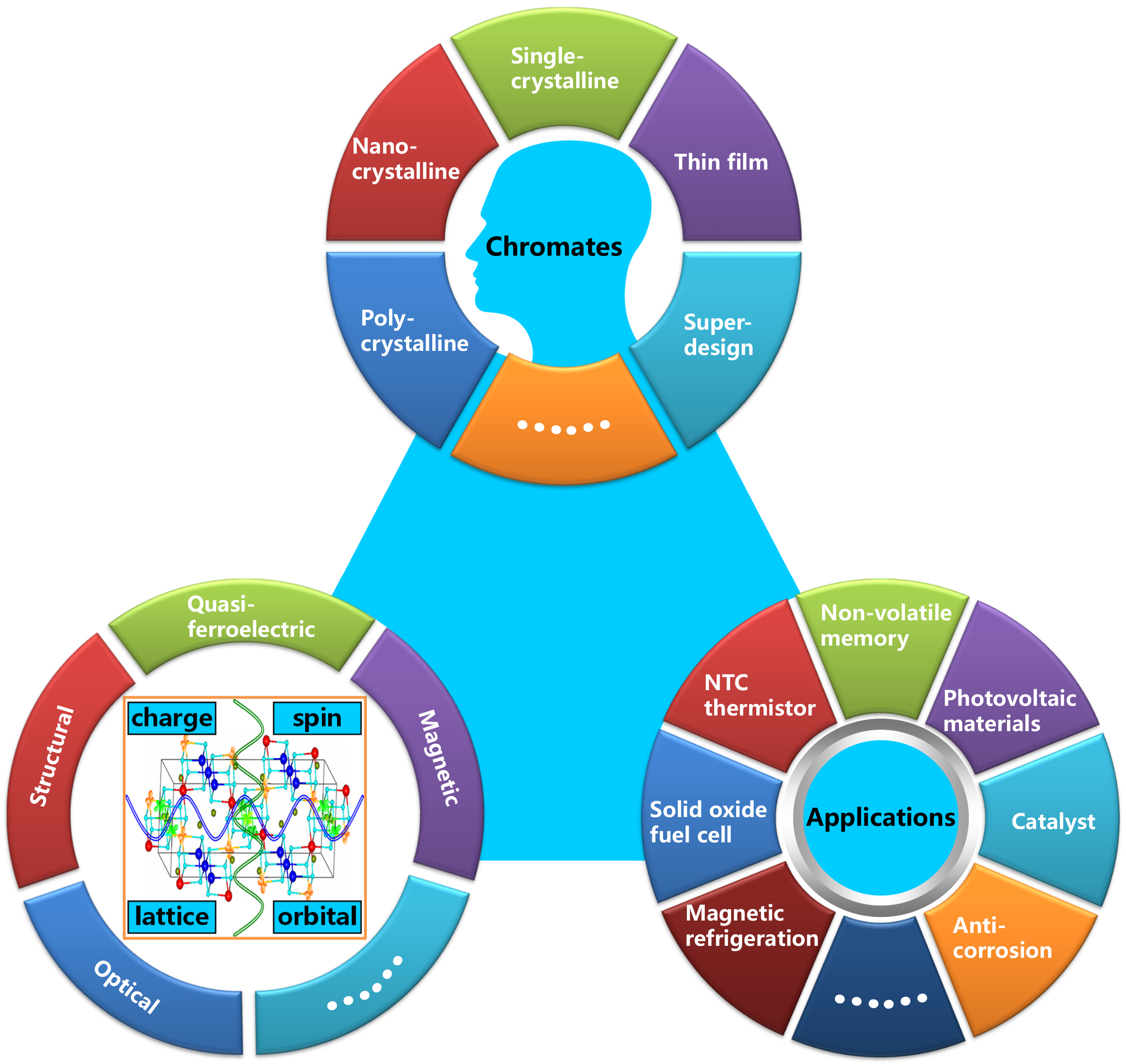}
\caption{
We illustrate the relationship between materials engineering, properties, and applications, taking chromates as one example. (Up corner) Chromates can be synthesized into the forms of polycrystals, nanocrystals, thin films, and single crystals, and one can design theoretically a material satisfying application requirements and try to realize it experimentally. (Left corner) The interesting macroscopic properties, such as ferroelectric, structural, optical, and magnetic, etc. normally originate from microscopic interactions of charge, orbital, spin, and lattice degrees of freedom. (Right corner) Potential applications of the (RE$_{1-x}$A$_x$)(Cr$_{1-y}$B$_y$)O$_{3 \pm {\delta}}$ compounds, where A and B represent doping elements on the RE (rare earth) and Cr sites, respectively, $x$ and $y$ are doping levels, and $\delta$ denotes the non-stoichiometry of oxygen sites.
}
\label{ShortApplications}
\end{figure*}

The system of rare-earth (RE) based RECrO$_3$ materials was proposed as a family of compounds reconciling ferroelectric and weak ferromagnetic (FM) properties in a single compound \cite{Serrao2005, Prado2013}. Among them, the YCrO$_3$ compound is a rare system to hold magnetism and ferroelectricity simultaneously and has generated considerable attention due to its interesting structural, ferroelectric, and magnetic properties \cite{Sardar2011, Zhu2020, Zhu2020-2, Taran2020-1, Mall2020, Shi2020}. It has potentials as a magnetoelectric material and for novel applications in field of thermistor, catalyst, magnetic cooling, non-volatile memory device, and solid-oxide fuel cell, etc., due to the high thermal, electrical, chemical, and structural stabilities (see \textcolor[rgb]{0.00,0.00,1.00}{Fig.}~\ref{ShortApplications}) \cite{Seybolt1966, McCarthy1982, Poplawski2000, Kim2003, Zhang2014, Sharma2014-2, Sharma2014-3, Tiwari2015, Oliveira2016, Goncharov2016, Hussain2017, Malcev2020, Mall2020-1, Tiwari2020, Bhowmik2021, Chakraborty2021, Chakraborty2021-1, Mao2021}.

The history of YCrO$_3$ compound was traced back to 1954 \cite{Looby1954}. The YCrO$_3$ compound adopts the orthorhombic crystalline system with $Pnma$ space group as shown in \textcolor[rgb]{0.00,0.00,1.00}{Figs.}~\ref{Structure-1} and \ref{Structure-2}, and the unit cell contains four formula units. The structure consists of a 3D network of corner-sharing CrO$_6$ octahedra with Cr$^{3+}$ ions located in the octahedral center. The magnetic contribution comes purely from Cr$^{3+}$ ions (3$d^3$, $S = 3/2$) with expected 3D superexchange interactions. Upon cooling, a magnetic phase transition occurs at $T_\textrm{N} =$ 141.5(1) K \cite{Zhu2020-2}, accompanied by the formation of a canted AFM (CAFM) structure with weak ferromagnetism. This was probably caused by the Dzyaloshinskii-Moriya (DM) interaction as schematically illustrated in \textcolor[rgb]{0.00,0.00,1.00}{Fig.}~\ref{Structure-1}(a) \cite{Bertaut1966}. This agrees with the applied-field dependent magnetization measurements, where the measured magnetic moment (unsaturated) is only $\sim$ 0.096 $\mu_\textrm{B}${/}Cr$^{3+}$ at 7 T and 2 K \cite{Zhu2020-2}. A ferroelectric transition was observed at $T_\textrm{C} \sim$ 473 K in the dielectric constant with polarization $\sim$ 2 $\mu$C/cm$^2$ \cite{Serrao2005}. A hysteresis of polarization-electric ($P$-$E$) loop of the polarization curve presents at 300, 325, and 350 K \cite{Serrao2005}. Previous studies of YCrO$_3$ compound focused mainly on polycrystals \cite{Looby1954, Case1984, Kuznetsov1998, Xing2010, Holder1980, Duran2010, Sardar2011, Geller1957-2, Prado-Gonjal2013, Belik2012, Zhang2015}, nanocrystals \cite{Bedekar2007, Duran2012-2, Park2012, Ahmad2016, Apostolov2019, Sinha2021, Chakraborty2021-3}, and thin films \cite{Kagawa1997, Kim2003, Serrao2005, Cheng2010, Seo2013, Seo2015, Seo2016, Araujo2014, Duran2014, Sharma2014-3, Tiznado2014, Duran2016, Arciniega2016, Pal2018, Arciniega2018, Kuang2018, Sharma2020}. The lack of high-quality and large-scale single-crystalline samples has been a long-standing issue for realizing potential applications to some extent and studying their intrinsic physical properties \cite{Remeika1956, Razdan1992, Todorov2011, Yin2017, Sanina2018, Grodkiewics1966, Aoyagi1969, Zhou2002-1}. Consequently, it is still subject of many studies \cite{zvezdin2021multiferroic, otsuka2021effect, bajaj2021magnetoelastic, chakraborty2021structural, taheri2019structural, wang2019structure, ye2020temperature, zhou2020structural, tan2020charge, fita2019spin, yoshii2019dielectric, ahmed2019ab}. For instances, the centrosymmetric orthorhombic structure cannot account for the appearance of ferroelectricity; it is controversial whether ferroelectricity and magnetoelectric coupling really exist in YCrO$_3$; moreover, the magnetic structure and the local distortion modes have not been well understood.

Having checked the relative literatures and patents from 1954-2021, we found that YCrO$_3$ has been the less studied one among the few existing single phase multiferroics mainly because its high dielectric loss has been overshadowing the ferroelectric property \cite{Looby1954, Prado2013}, and there exist few reviews on the YCrO$_3$ compound, which sparks our interest in composing the present comprehensive review.

The review deals with a comprehensive summary of research results, reached consensuses, and existing controversies. The figures and tables are designed in a particular way that they summarize special subjects/properties, trying to combine research results from ploy-, nano-, and single-crystalline and thin-film studies together and include different controversial results, from which prospects are clear.

The outline of the review is as follow:

(i) In Section 2, we show a summary of the syntheses of YCrO$_3$ in four categories: poly-crystalline, nano-crystalline, and single-crystalline samples and thin films. As demonstrated in \textcolor[rgb]{0.00,0.00,1.00}{Figs.}~\ref{Crstate-1}-\ref{Crstate-3}, during the synthesis process, different Cr oxidation states could be formed. In Section 2.3, single crystals can be grown with the flux, molten-salt vaporization, and floating-zone methods; the (dis)advantages of each technique are compared (also see \textcolor[rgb]{0.00,0.00,1.00}{Table}~\ref{Syn-Meths}). For future investigations, we propose a chemical hydrothermal method for the growth of stoichiometric YCrO$_3$ single crystals, for which the reaction kettle must stand a huge pressure. \textcolor[rgb]{0.00,0.00,1.00}{Figures}~\ref{Samples-1} and \ref{Samples-2} show the photos of different forms of YCrO$_3$ material.

(ii) In Section 3, we discuss the structural (see \textcolor[rgb]{0.00,0.00,1.00}{Figs.}~\ref{Structure-1} and \ref{Structure-2}), magnetic (see \textcolor[rgb]{0.00,0.00,1.00}{Figs.}~\ref{mag-1}-\ref{mag-3}), ferroelectric and dielectric (see \textcolor[rgb]{0.00,0.00,1.00}{Figs.}~\ref{FerriPoly-1}-\ref{PEloop-3}), and optical properties.

(iii) In Section 4, we present chemical-pressure dependent properties. The chemical pressures are mainly from chemical dopings on Y and Cr sites. Temperature dependent structural evolution and its effect on ferroelectric and magnetic properties are discussed with \textcolor[rgb]{0.00,0.00,1.00}{Figs.}~\ref{LatticePara-1}-\ref{CrShift-2}.

(iv) In Section 5, we present the phase diagrams of doping dependence of lattice constants (\emph{a}, \emph{b}, and \emph{c}) and unit-cell volume (\emph{V}) (see \textcolor[rgb]{0.00,0.00,1.00}{Fig.}~\ref{LatticePhaseDia}), temperature-dependent magnetic transitions in YCrO$_3$, as well as magnetic structures induced by Mn-doping on the Cr site (see \textcolor[rgb]{0.00,0.00,1.00}{Figs.}~\ref{PhaseDia-1} and~\ref{PhaseDia-2}).

(v) In Section 6, we present possible applications of parent and doped YCrO$_3$, including anti-corrosion, magnetohydrodynamic electrode, catalyst, negative-temperature-coefficient (NTC) thermistor, magnetic refrigeration, protective coating, and solid oxide fuel cell (see \textcolor[rgb]{0.00,0.00,1.00}{Figs.}~\ref{Application-1}-\ref{Application-6}).

(vi) In Section 7, we discuss existing controversies, issues, and potential challenges for capturing the opportunities in practical studies.

(vii) In Section 8, we conclude with perspectives on reached consensuses as well as future opportunities in the field.

\section{Syntheses}

Exploitation and synthesis of new advanced materials with interesting properties and potential macroscopic functionalities are extremely important in condensed matter science. Before preparing a designed material, one should first search for existing related phase diagrams or calculate theoretically the Gibbs free energy ($\Delta{G}$) of the reaction to check whether the expected material is thermodynamically stable or not. When $\Delta{G} <$ 0, the reaction is favorable; otherwise, it is unfavorable ($\Delta{G} >$ 0). The synthesis parameters such as temperature, reaction time, increasing and decreasing temperature speeds, and working gas can usually change the properties of material by tailoring its microscopic structures. The close correlation between synthesis, structure, and properties is often used in scientific discussion \cite{li2008synthesis, li2006correlation, li2007neutron}. Materials can be synthesized into polycrystals \cite{thompson2008scattering}, nanocrystals \cite{CAROLAN2017128, HANUS20131056}, thin films \cite{FU201731, RAMANUJAM2020100619}, and single crystals \cite{ZHOU201487, REVCOLEVSCHI1997321}, as well as alloys \cite{STEMPER2022100873, GALANO2022100831} and composites \cite{GAO2021100813, YANG2021100710}, depending on preparation methods and detailed synthesis parameters. Different forms of materials show different properties \cite{li2008synthesis, li2006correlation, li2007neutron, li2009crystal}, for example, the Nd$_{0.8}$Sr$_{0.2}$NiO$_2$ thin films was discovered to display superconductivity with a critical transition temperature of 9--15 K \cite{li2019superconductivity}, whereas the bulk samples of Nd$_{1-x}$Sr$_x$NiO$_2$ ($x$ = 0, 0.2, 0.4) compounds just show an insulting behavior without the presence of superconductivity, even with an applied pressure up to $\sim $ 50.2 GPa \cite{li2020absence}. The optimization synthesis of material is a labor-intensive and time-consuming process.

An essential prerequisite for a complete chemical reaction is that the starting materials have an average particle size smaller than the diffusion length, that is \cite{li2008synthesis, valenzuela2005magnetic}
\begin{eqnarray}
\sqrt{2Dt} \geq L,
\label{reaction}
\end{eqnarray}
where \emph{D} is the diffusion constant, depending on the materials themselves and reaction temperature, \emph{t} is the reaction time, $\sqrt{2Dt}$ is the diffusion length, and \emph{L} is the particle size. Below the melting temperature of a material, the smaller the particle size, the higher the reaction temperature, and the longer the reaction time are, the higher the reaction rate is \cite{li2008synthesis, valenzuela2005magnetic}. With the equation~(\ref{reaction}), one can determine the preparing parameters of temperature and time based on the estimated particle size of raw materials. Practically, the particle size is not uniform, and there exists a size distribution. To overcome this, one can carry out more times reaction with intermediate ball grinding and mixing of the former resultants \cite{li2008synthesis}.

\subsection{Poly-crystalline}

\begin{figure*} [!t]
\centering \includegraphics[width=0.61\textwidth]{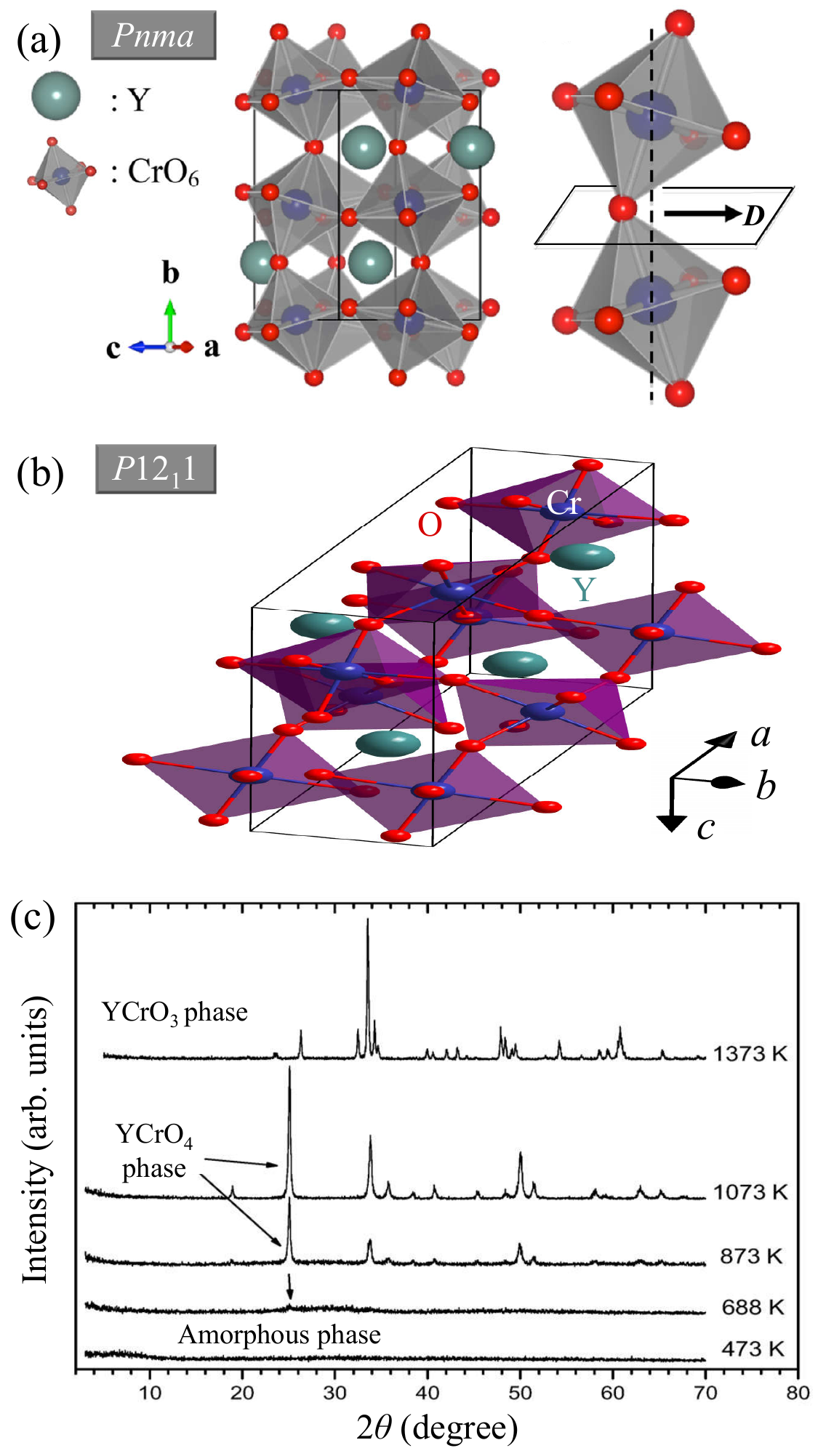}
\caption{
(a) Orthorhombic crystalline structure (\emph{Pnma} space group (No. 62), centrosymmetric) of YCrO$_3$ compound (left), and there is a mirror plane existing between octahedra of CrO$_6$ (right). DM interactions were expressed by the \textbf{\emph{D}} vector whose direction is within the mirror plane as marked by the bold arrow. Adapted with permission from Ref. \cite{Ikeda2015}. Copyright (2015) Springer-Verlag Wien.
(b) Monoclinic crystal structure ($P12_11$ space group (No. 4), noncentrosymmetric) with one unit-cell of YCrO$_3$ at 120 K (solid lines). Reproduced with permission from Ref. \cite{zhao2021insights}. Copyright (2021) Hai-Feng Li.
(c) XRPD patterns of samples with different thermal histories, showing the temperature evolution of YCrO$_3$ phase (orthorhombic) synthesized with amorphous hydroxide precipitates. Reproduced with permission from Ref. \cite{Duran2012-2}. Copyright (2012) Elsevier Ltd.
}
\label{Structure-1}
\end{figure*}

Polycrystals are usually synthesized with traditional solid-state reaction or chemical sol-gel method. The advantages of high-temperature solid-state reaction are the low cost for a bunch of sample preparations and the ready availability of raw materials \cite{li2008synthesis}. On the other hand, the resultants from a high-temperature reaction usually have defects such as cation vacancies and oxygen non-stoichiometry \cite{kroger1977defect, maier1993defect}. In contrast, one need to find suitable solvent for the starting materials with the sol-gel method, but the advantage of this method is with lower reaction temperatures, and one can make resulting materials with finer particle sizes and more stoichiometric.

Looby and Katz reported the earliest synthesis of YCrO$_3$ poly-crystalline samples \cite{Looby1954}. The mixture of raw Y$_2$O$_3$ and Cr$_2$O$_3$ compounds was added into a flux of NaCl, which was then heated up to 900 $^{\circ}\mathrm{C}$ under hydrogen atmosphere. By leaching the flux with water and excess Y$_2$O$_3$ oxide with very dilute hydrochloric acid, the YCrO$_3$ compound was finally synthesized with a mass density of $\sim$ (5.78$\pm$0.05) g/cc \cite{Looby1954}.

We first analyze two important guiding works for the preparation of poly-crystalline YCrO$_3$.

\textbf{(1) Microcracks.} Case and Glinka characterized the microcracks within YCrO$_3$ by elasticity measurements and small-angle neutron scattering study \cite{Case1984}. There exists a phase transition of microcracks: (i) If the sintering temperature $>$ 1100 $^\circ$C, the resulting poly-crystalline YCrO$_3$ specimen has a number of grains ($\sim \mu$m in size) and porosities. (ii) An annealing at temperatures below 1100 $^\circ$C can heal microcracks. (iii) When a subsequent heating temperature is higher than 1100 $^\circ$C, the healed samples will return to their initial microcracked state. Thus normally-prepared YCrO$_3$ polycrystals can stay either a microcracked or a healed state, depending on the detailed heating history. Compared to surface sensitive methods like scanning electron microscopy, neutron scattering can collect microscopic information of a bulk. Thus small-angle neutron scattering can provide decisive information of the crack opening displacement, mean crack radius, volume fraction, and number density of a population of microcracks in poly-crystalline YCrO$_3$ ceramics \cite{Case1984}.

\begin{figure*} [!t]
\centering \includegraphics[width=0.68\textwidth]{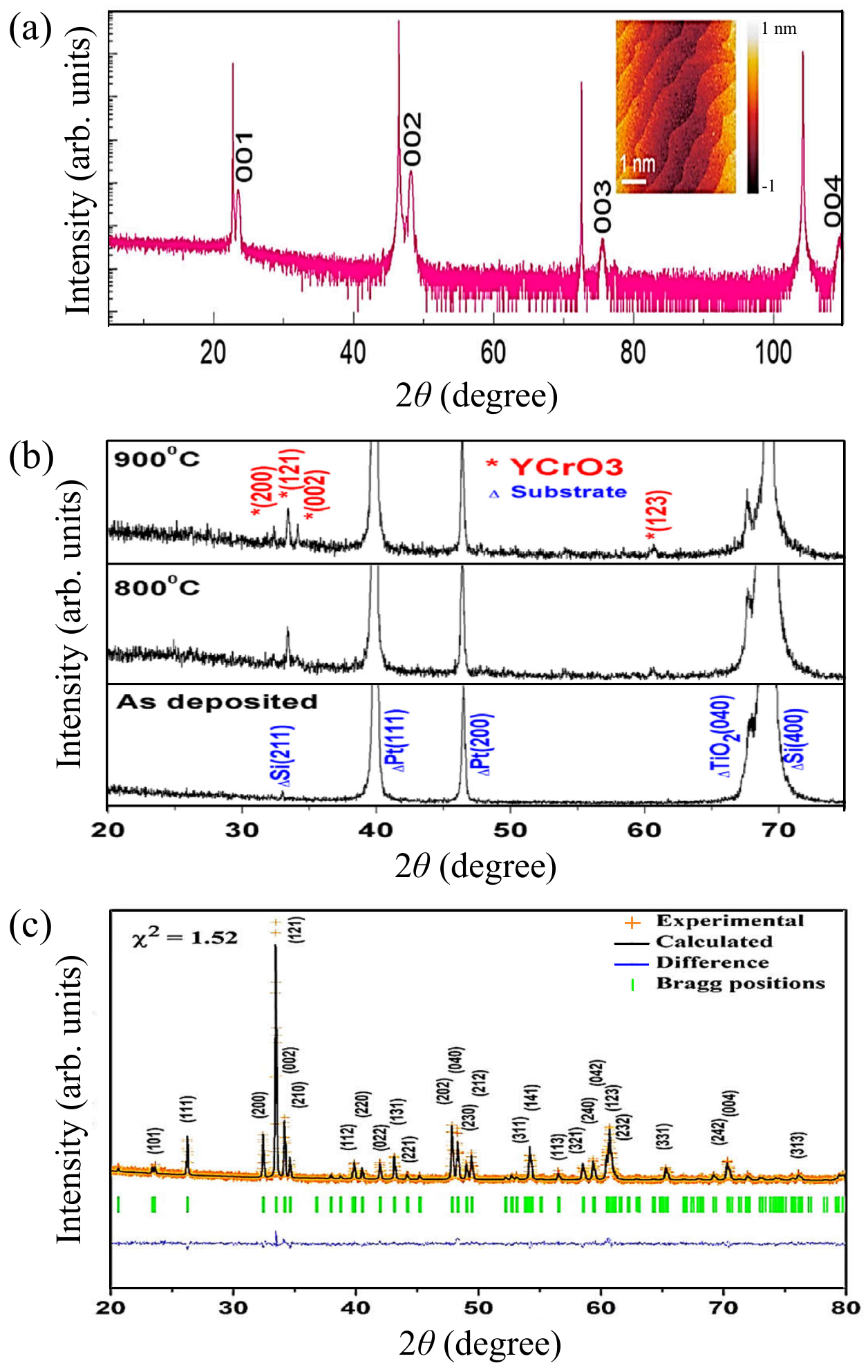}
\caption{
(a) Room-temperature XRPD study of a single-crystal YCrO$_3$ thin film (32 nm). Inset displays its atomic force microscopy image \cite{Sharma2020}.
(b) XRPD patterns of 550 nm YCrO$_3$ thin films as prepared annealed at 800 and 900 $^\circ$C in air. Reproduced with permission from Ref. \cite{Cruz2014}. Copyright (2013) Elsevier B.V.
(c) XRPD pattern and structure refinement of a YCrO$_3$ poly-crystalline pellet with $Pnma$ space group. Reproduced with permission from Ref. \cite{Mall2018-2}. Copyright (2018) Elsevier Ltd.
}
\label{Structure-2}
\end{figure*}

\textbf{(2) CrO$_x$-Y$_2$O$_3$ diagram.} Xing \emph{et al}. studied the phase transformations of the samples (i.e., the CrO$_x$-Y$_2$O$_3$ system synthesized by a deposition precipitation method) by \emph{in situ} measurements under air and N$_2$ atmospheres \cite{Xing2010}. The Raman results revealed that (i) At 100 $^\circ$C, CrO$_3$ reacts with Y$_2$O$_3$, forming YCrO$_4$, i.e., 2CrO$_3$ + Y$_2$O$_3$ $\rightarrow$ 2YCrO$_4$ + $\frac{1}{2}$O$_2 \uparrow$. (ii) At 400 $^\circ$C, YCrO$_4$ decomposes into YCrO$_3$ (i.e., YCrO$_4 \rightarrow$ YCrO$_3$ + $\frac{1}{2}$O$_2 \uparrow$) and Cr$_2$O$_3$ (i.e., 2YCrO$_4 \rightarrow$ Cr$_2$O$_3$ + Y$_2$O$_3$ + O$_2 \uparrow$). These phase transformations begin from the surface region of the sample and then extend to the bulk. The parameters of atmosphere and temperature influence the phase transformation in the surface regime; whereas for the bulk sample, the phase transformation relies only on temperature. If the sample experiencing a high-temperature firing history was exposed to ambient conditions, the Cr$^{3+}$ (in the low oxidation state) ions on the surface regime can be easily reoxidized to Cr$^{5+}$ or Cr$^{6+}$ (in the high oxidation state) \cite{Xing2010}. In addition, it was demonstrated that the YCrO$_3$ compound can form oxide-oxide eutectics with Cr$_2$O$_3$ and Y$_2$O$_3$ materials and oxide-metal eutectics with Mo, W, and Cr metals \cite{Holder1980}.

Second, we present the syntheses of poly-crystalline YCrO$_3$ samples with different methods.

\textbf{(1) Self-propagating high temperature synthesis.} A series of lanthanide orthochromates with general formula RECrO$_3$ (RE = La, Ce, Pr, Nd, Sm, Eu, Gd, Tb, Dy, Ho, Er, Tm, Tb, and Lu), as well as the YCrO$_3$ compound, were prepared by a method so-called self-propagating high temperature synthesis (SHS) \cite{Kuznetsov1998}. There are four different SHS reactions by which RECrO$_3$ forms: (i) Reactions of RE$_2$O$_3$, CrO$_3$, and Cr components with Ba(ClO$_4$)$_2$ or NaClO$_4$ in air. (ii) Reactions of RE$_2$O$_3$, CrO$_3$, and Cr components in oxygen. (iii) Reactions of RECl$_3$, NaO$_2$ (or Na$_2$O$_2$), and CrCl$_3$ in air. (iv) Reactions of RE(NO$_2$)$_2$, melamine, and Cr in air. The study of X-ray powder diffraction (XRPD) indicated that a single phase perovskite with the orthorhombic structure was synthesized for all reactions. Systematic variations in cell volume, pycnometrical densities, and lattice parameters were observed with the change of lanthanide atomic number. The RECrO$_3$ samples synthesized by SHS methods are of weak AFM \cite{Kuznetsov1998}. The YCrO$_3$ compound synthesized via SHS shows lattice constants of $a$ = 5.226 $\textrm{\AA}$, $b$ = 5.534 $\textrm{\AA}$, and $c$ = 7.531 $\textrm{\AA}$, and a unit-cell volume of $V$ = 217.80 $\textrm{\AA}$$^3$ \cite{Kuznetsov1998}.

\begin{figure*} [!t]
\centering \includegraphics[width=0.82\textwidth]{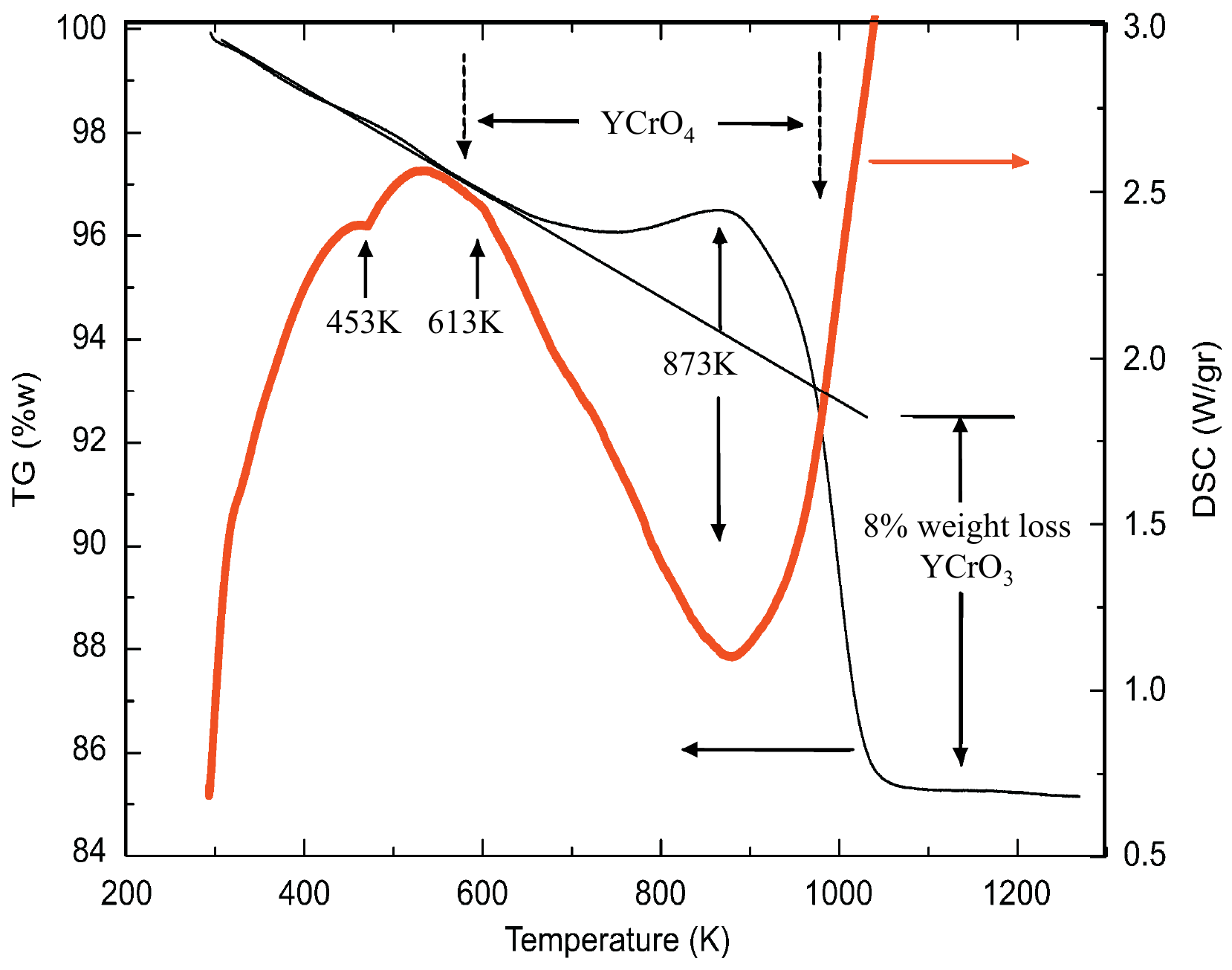}
\caption{
TG curve (left) and DSC curve (right) of the combustion powder. The arrows show the regimes of the physicochemical changes upon heating: that is dehydration (from 85 to 180 $^\circ$C), decomposition of the precursors and releases of N, C, and O in gaseous products (from 180 to 340 $^\circ$C), crystallization of the first stable YCrO$_4$ phase (from 400 to 700 $^\circ$C), and then the decomposition of YCrO$_4$ toward a more stable YCrO$_3$ compound. Reproduced with permission from Ref. \cite{Duran2010}. Copyright (2010) Elsevier Inc.
}
\label{Crstate-1}
\end{figure*}

\textbf{(2) Combustion method and solid-state reaction.} Dur\'{a}n \emph{et al}. synthesized the YCrO$_3$ perovskites by a method of two-step synthesis: (i) combustion method and (2) solid-state reaction \cite{Duran2010}.  Authors summarized the advantages of this procedure and connected the microstructure with thermal, magnetic, and dielectric properties of YCrO$_3$ \cite{Duran2010}. \textcolor[rgb]{0.00,0.00,1.00}{Figure}~\ref{Crstate-1} shows TGA and DSC results of the combustion powder. Upon heating, the first stable phase corresponds to the formation of YCrO$_4$ in a temperature range of 613--873 K. From 973 to 1093 K, YCrO$_4$ undergoes a decomposition into YCrO$_3$.

\textbf{(3) Hydrothermal synthesis.} The rare-earth orthochromates, RECrO$_3$ (RE = La, Pr, Sm, Gd, Dy, Ho, Yb, and Lu) as well as YCrO$_3$ \cite{Sardar2011}, were performed by a hydrothermal synthesis. These materials were synthesized by one step: an amorphous mixed-metal hydroxide was treated hydrothermally above 300 $^{\circ}$C for 24 h. No further post-synthesis annealing was required. The resultant compounds crystallized highly with particle sizes in submicrometer. The XRPD analysis determines that the hydrothermal synthesized YCrO$_3$ adopts an orthorhombic crystalline structure with $Pbmn$ space group, $R_\textrm{A}$ = 1.075 {\AA} (the ionic radius of RE-site ion), and the tolerance factor = 0.867 \cite{Sardar2011, Geller1957-2}.

\textbf{(4) Microwave-assisted method.} Prado-Gonjal \emph{et al}. synthesized orthorhombic ($Pnma$) distorted perovskites, RECrO$_3$ chromates as well as the isostructural YCrO$_3$ compound, by a microwave assisted method \cite{Prado-Gonjal2013}. This route produced good quality materials. Magnetic measurements of the RECrO$_3$ samples indicate that the AFM transition temperatures, $T_\textrm{N}^{\textrm{Cr}}$, depend strongly on the ionic radii of RE$^{3+}$ ions \cite{Belik2012}, displaying a rich of magnetic phase diagrams \cite{Prado-Gonjal2013}. Dielectric inhomogeneity was manifested by at least two relaxation processes in all samples, which was attributed to grain interior and boundary \cite{Prado-Gonjal2013}. Further measurements did not show evidence of potential non-centrosymmetric crystalline structure or concomitant ferroelectricity. It was interesting that a strong connection of dielectric and magnetic features did not occur in the microwave synthesized RECrO$_3$ materials \cite{Prado-Gonjal2013}.

\textbf{(5) High pressure synthesis.} Under high pressure conditions, Zhang \emph{et al}. synthesized perovskites of (1-\emph{x})BiFeO$_3$--(\emph{x})YCrO$_3$ (BFO--YCO) with $x = 0$--0.3 \cite{Zhang2015}. A morphotropic phase transition was observed at $x \approx 0.2$, from rhombohedral ($R3c$ space group) to orthorhombic ($Pbnm$ space group). The combination of BFO and YCO systems improved its magnetic properties and prevented the current leakage. The 0.8BFO–0.2YCO ceramics display a characteristic ferroelectric hysteresis loop with a room-temperature remanent polarization value {=} 14.90 $\mu$C/cm$^2$ and weak ferromagnetism at the structural boundary \cite{Zhang2015}. In the 0.8BFO–0.2YCO ceramics, an anomaly was observed from the temperature-dependent dielectric analysis, i.e., magneto-dielectric effect, indicating a coupling between magnetic and electric orders \cite{Zhang2015}.

\subsection{Nano-crystalline}

Nanomaterials receive attention due to their intrinsic size-dependent properties and resulting various applications \cite{wang2005general, COROT20061471, STOLLE2013160, cozzoli2006synthesis, CHANG2011608}. The nano-scale building blocks, as individual units, are arranged and assembled spatially into functional materials with different techniques \cite{wang2005general, cozzoli2006synthesis}.

A technique to synthesize nano-crystalline YCrO$_3$ powder via a combustion with the corresponding metal nitrates as oxidants and using glycine as the fuel was reported \cite{Bedekar2007}. In order to explore the possibility of preparation of YCrO$_3$ sample with a single phase, authors chose three oxidant-to-fuel ratios, that is fuel-deficient ($1:0.50$, synthesis one), stoichiometric ($1:1.66$, synthesis two), and fuel-excess ($1:2.0$, synthesis three): (i) The synthesis one produced the formation of YCrO$_4$. (ii) The synthesis two resulted in the phase-pure orthorhombic YCrO$_3$ samples. (iii) The synthesis three generated a similar observation as the synthesis two. The phase-pure product calcined at 600 $^{\circ}$C has a crystalline size of 36 nm, which was calculated by the Scherrer's formula \cite{Bedekar2007}. One interesting observation was the formation of onion-like structures comprising of concentric rings in the faceted nanoparticles \cite{Bedekar2007}.

A different method of preparing nanometric YCrO$_3$ samples was used by Dur\'{a}n \emph{et al}. who precipitated a precursor via bubbling gaseous ammonia into a solution of chromium/yttrium salts \cite{Duran2012-2}. This was different from those produced by combustion synthesis. As shown in \textcolor[rgb]{0.00,0.00,1.00}{Fig.}~\ref{Structure-1}(c), unfortunately, the as-made YCrO$_3$ powders are amorphous. Only after a thermal treatment between 1273 and 1373 K, poly-crystalline YCrO$_3$ can be formed with grain sizes around 20 nm. In both amorphous hydroxide and crystalline YCrO$_3$ samples, the chemical environments of Y and Cr ions are uniform \cite{Duran2012-2}. It was noticed that there existed a charge redistribution or a charge transfer between Cr and Y ions of YCrO$_3$ sample, which was indicated by shifts of binding energies of Y 3$d_{5/2}$ and Cr 2$p_{3/2}$ \cite{Duran2012-2}. By this method, the synthesized YCrO$_3$ shows a sharp increase at $\sim$ 473 K in electrical properties. This was attributed to the morphology and grain size of the crystallites \cite{Duran2012-2}. Four types of micrograph were observed from scanning electron microscopy studies \cite{Duran2012-2}. The precipitated hydroxide and the sample thermally-treated at 873 K display no defined forms \cite{Duran2012-2}. The sample synthesized at 1373 K displays a lot of pores and voids, indicating a large amount of nanocrystals formed in the orthorhombic YCrO$_3$ products \cite{Duran2012-2}. The sample sintered at 1623 K shows a formation of botryoidal grains and an agglomeration of particles with size in a range of 300--1500 nm \cite{Duran2012-2}.

The nano-phase of YCrO$_3$ was also prepared by a very high-speed planetary milling process during the preparation of Fe-based oxide dispersion strengthened alloys (84Fe–14Cr–2Y$_2$O$_3$) \cite{Park2012}. The coarse Cr and Fe metal powders with an average particle diameter of $\sim$ 40 $\mu$m and oxide Y$_2$O$_3$ powder ($\sim$ 50 $\mu$m) were mixed and mechanically milled by a high-speed planetary milling process. Based on the XRPD and scanning electron microscopy studies of mechanical alloying materials, Y$_2$O$_3$ raw particles became very finer and homogeniously distributed with a short milling time of 40 min. The consolidated alloys exhibit a nano-phase of YCrO$_3$ particles (10--20 nm) \cite{Park2012}.

Using a citrate precursor method, Ahmad and Lone synthesized single-phase YCrO$_3$ nanoparticles with high crystallinity, an average grain size {=} 22 nm, and a high surface area {=} 344 m$^2$ g$^{-1}$ \cite{Ahmad2016}. Two band-gaps were observed at 1.98 and 2.95 eV, respectively, which was perhaps related to the transition of Cr ions. According to the FM hysteresis (wedge-shaped) measurement, a saturation magnetization {=} 2.48 emu{/}g, a coercive field {=} 12.78 kOe, and a remanent magnetization {=} 0.83 emu{/}g were determined. The ferroelectric loops show a remanent polarization {=} 0.009 $\mu$C cm$^{-2}$, a coercive field {=} -0.66 kV cm$^{-1}$, and a saturation polarization {=} 0.026 $\mu$C cm$^{-2}$ at room temperature. These confirm the multiferroic nature of nanoparticles of YCrO$_3$ \cite{Ahmad2016}.

The YCrO$_3$ nanoparticles were also prepared by a reverse micellar technique via surfactant tergitol after heating the precursors at 800 $^{\circ}$C \cite{Ahmad2018}. The resultant YCrO$_3$ nanoparticles display highly crystalline orthorhombic structure with a grain size {=} 35 nm and a surface area {=} 348 m$^2$ g$^{-1}$. The observed wedge-shaped hysteresis further confirmed the multiferroicity of the nanoparticles \cite{Ahmad2018}.

Apostolov \emph{et al}. studied the dielectric and magnetic properties of GdCrO$_3$ and YCrO$_3$ nanoparticles using a microscopic model and the Green's function method \cite{Apostolov2019}. The magnetization of YCrO$_3$ increases as the particle size decreases. The bulk dielectric constant of YCrO$_3$ is smaller than that in nanoparticles. There exists a strong magnetoelectric coupling in YCrO$_3$ nanoparticles \cite{Apostolov2019}.

\subsection{Single-crystalline}

Single-crystalline materials have translational symmetry in macroscopic lengths. Thus, they can offer more reliable information of intrinsic properties and structures, compared to other forms of material \cite{li2008synthesis}. Single crystals are fundamental important for studying intriguing properties of materials and potential for a wide range of uses. Some studies require rigorously with the quality of single crystals, for example, the modern neutron and synchrotron X-ray scattering methods. The lack of good-quality and large single-crystalline samples of interesting materials has been a long-standing obstacle in condensed matter science \cite{li2008synthesis}.

Many techniques can be used for crystal growths, such as the flux, Czochralski, floating zone, Bridgman, gas-phase growth, and top-seeded solution methods \cite{li2008synthesis}. Different methods are utilized for different specific crystal growths. For example, there is crucible-free, and the seed rod and the feed rod stand freely during the process of single crystal growth with the floating-zone method, therefore, the grown crystals have much higher purity. This method is thus particularly useful and widely used for laboratory crystal growths \cite{li2008synthesis}.

\subsubsection{The flux method}

Small single crystals of YCrO$_3$, as well as other rare-earth orthochromates and orthoferrites, were grown by the flux method \cite{Remeika1956}. For the growth of single-crystalline YCrO$_3$ samples, equivalent molar proportions of the constituent oxides were weighted, and Bi$_2$O$_3$ was taken as the flux with a weight ratio of RECrO$_3 :$ Bi$_2$O$_3 = 1:6$. Within the study \cite{Remeika1956}, PbO was not chosen as the flux because the reaction of PbO and Cr$_2$O$_3$ produces stable lead chromates, preventing the formation of YCrO$_3$ \cite{Remeika1956}. These mixed materials were added into a platinum crucible and then heated at 1300 $^\circ$C for nearly one hour. After that, the temperature was decreased down to 850 $^\circ$C, the solidification temperature, with a uniform manner at $\sim$ 30 $^\circ$C{/}h and then cooled quickly to ambient temperature. The resultant material consists of solidified Bi$_2$O$_3$ and crystals of orthorhombic chromates. After leaching out the flux by a hot dilute solution of HNO$_3$, the crystals were further washed in distilled water and dried. The single crystals of orthochromates were bright green by transmitted light. It was noted that the ratio of sought compound to flux should be not too high because Pt crucible would be damaged to some extent \cite{Remeika1956}.

Razdan \emph{et al}. reported the single-crystal growth of YCrO$_3$ via the flux-melting method using the flux of PbO-PbF$_2$-B$_2$O$_3$ in a Pt crucible and studied elemental compositional changes on the surface regimes of the grown single-crystal samples \cite{Razdan1992}. Various microstructures formed on the host surfaces. The crystal growth procedures are: (i) The Pt crucible was fired at 1260 $^\circ$C for 12 h. (ii) Then it was cooled slowly to 950 $^\circ$C with a cooling rate {=} 1 K/h. (iii) Subsequently, temperature was decreased to ambient temperature by 120 K/h. The crystalline axes can be identified by a Laue camera, however, the crystalline space group cannot be determined \cite{Razdan1992}.

Todorov \emph{et al}. grew the YCrO$_3$ single crystals via a solution-growth method at high temperatures in Pt crucibles for a study of the Raman spectra \cite{Todorov2011}. The complexes PbF$_2 :$ KF$:$ B$_2$O$_3$ with the ratio of $0.75 : 0.23 : 0.02$ were used as solvents. The ratio of the ground material YCrO$_3$ to the solvents varies from $\frac{1}{3}$ to $\frac{1}{4}$. The mixture ($\sim$ 600 g) was semihermetically closed in a Pt crucible and fired up to 1200 $^\circ$C at an increasing rate {=} 50 $^\circ$C/h and hold there for 48 h. Then, the Pt crucible was cooled at a decreasing rate {=} 0.5 $^\circ$C/h down to 920 $^\circ$C, at which the crucible was taken out of oven, and the solvent was poured out. Finally, the resultant YCrO$_3$ single crystals were on the walls and ground of the crucible \cite{Todorov2011}. An optical image of the (010) surface of the grown YCrO$_3$ single crystal with orthorhombic $Pnma$ structure was shown in \textcolor[rgb]{0.00,0.00,1.00}{Fig.}~\ref{Samples-1}(a) \cite{Todorov2011}.

Yin \emph{et al}. reported the growth of single-crystalline YCr$_{0.88}$Fe$_{0.12}$O$_3$ by the flux method \cite{Yin2017}. The raw materials Y$_2$O$_3 :$ Cr$_2$O$_3 :$ Fe$_2$O$_3 :$ PbF$_2 :$ PbO : B$_2$O$_3$ = $1 : 0.5 : 0.5 : 8 : 2 : 1$ (molar ratio) were mixed and placed into a aluminum crucible and fired at 1200 $^\circ$C for 4 days. Subsequently, the mixture was cooled down to 1000 $^\circ$C at a rate {=} 5 $^\circ$C/h. Finally, the furnace was turned off. The grown crystals by this way have a size of 1--2 mm. The chemical composition and crystallographic structure of the grown crystals were confirmed by XRPD and energy dispersive spectroscopy technique, respectively \cite{Yin2017}.

Sanina \emph{et al}. grew single crystals of YCrO$_3$ via the method of spontaneous crystallization in a solution melt (PbO-0.005PbO$_2$ solvent) \cite{Sanina2018}. The thickness of the grown YCrO$_3$ plates is 2--3 mm with an area of 3-5 mm$^2$. The plate plane is perpendicular to either the [110] direction or the \emph{c} axis \cite{Sanina2018}. The growth was at a constant solution-melt temperature of 1350 $^\circ$C \cite{Sanina2018}. During the growth, active oxygen O$^{2-}$ ions recharge Pb$^{2+}$ and Pb$^{4+}$ ions, and the introduction of PbO$_2$ oxide can bond active oxygen and prevent the damage of the used Pt crucible \cite{Sanina2018}.

\subsubsection{Molten-salt vaporization}

\begin{figure*} [!t]
\centering \includegraphics[width=0.68\textwidth]{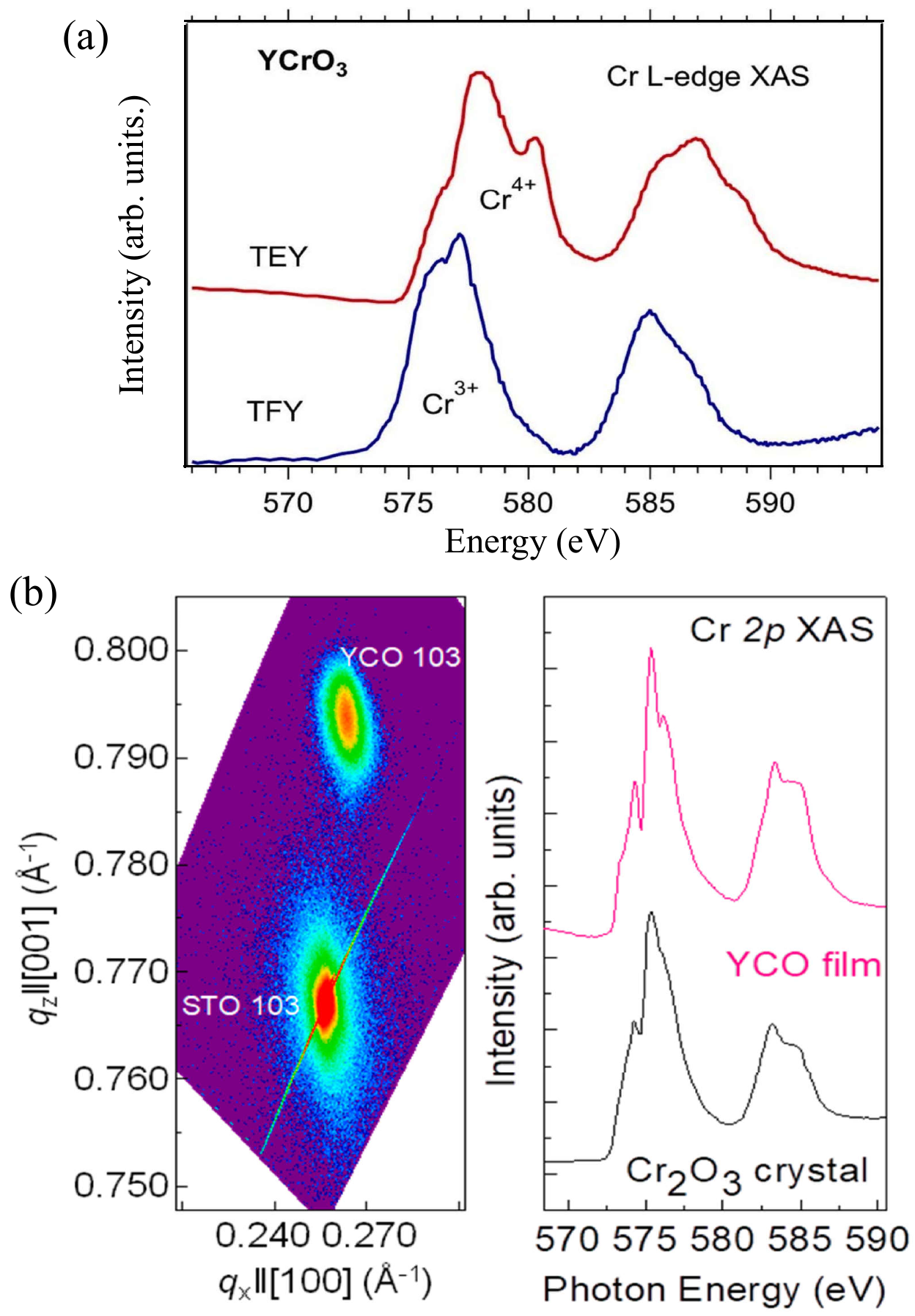}
\caption{
(a) X-ray absorption spectroscopy of amorphous YCrO$_3$ thin films at the Cr \emph{L}-edge \cite{Araujo2014}.
(b) Asymmetric scans in reciprocal space around the Bragg (103) peak of SrTiO$_3$ for the 32 nm YrCrO$_3$ thin film (left panel) and comparison of the X-ray absorption spectroscopy at Cr 2\emph{p} of a YCrO$_3$ thin film to that of Cr$_2$O$_3$ (right panel) \cite{Sharma2020}.
}
\label{Crstate-2}
\end{figure*}

During the earlier single crystal growth of YCrO$_3$ compound, PbO was usually chosen as the flux. However, a large amounts of PbCr$_2$O$_7$ impurity phase forms, which further impedes the crystal growth of the desired compound \cite{Grodkiewics1966}. To solve this issue, a variety of flux materials were tried, according to the principle that the flux would be completely evaporated at high enough temperatures \cite{Grodkiewics1966}. The volatile flux of PbF$_2$ or BiF$_3$ with small amounts of B$_2$O$_3$ was found to be particularly suitable. Using the flux materials with proportions of PbF$_2 :$ B$_2$O$_3 :$ YCrO$_3 = 9:9:1$, dark green rectangular plates (the largest one: 7$\times$4$\times$3 mm$^3$) of YCrO$_3$ crystals were grown by evaporating the flux at $\sim$ 1300 $^\circ$C \cite{Grodkiewics1966}.

Aoyagi \emph{et al}. reported the single crystal growth of YCrO$_3$ by the molten-salt vaporization method at 1300 $^\circ$C \cite{Aoyagi1969}. During the growth process, the compound of lead fluoride was chosen as the molten-salt \cite{Aoyagi1969}.

\subsubsection{The floating-zone method}

To study the thermal conductivity of AFM perovskites, a series of single crystals of ABO$_3$ compounds, including YCrO$_3$, were synthesized by an infrared-heating floating zone furnace \cite{Zhou2002-1}. Unfortunately, for the growth of YCrO$_3$, the massive sublimation prevented a melting of the poly-crystalline YCrO$_3$ rod. Fortunately, large grains of YCrO$_3$ were obtained at the feeding end of the rod, where the temperature was $\sim$ 2000 $^\circ$C, approaching the melting temperature of YCrO$_3$ compound. A disk of YCrO$_3$ with a high density and a large grain size was cut from the end of the rod \cite{Zhou2002-1}.

\begin{figure*} [!t]
\centering \includegraphics[width=0.82\textwidth]{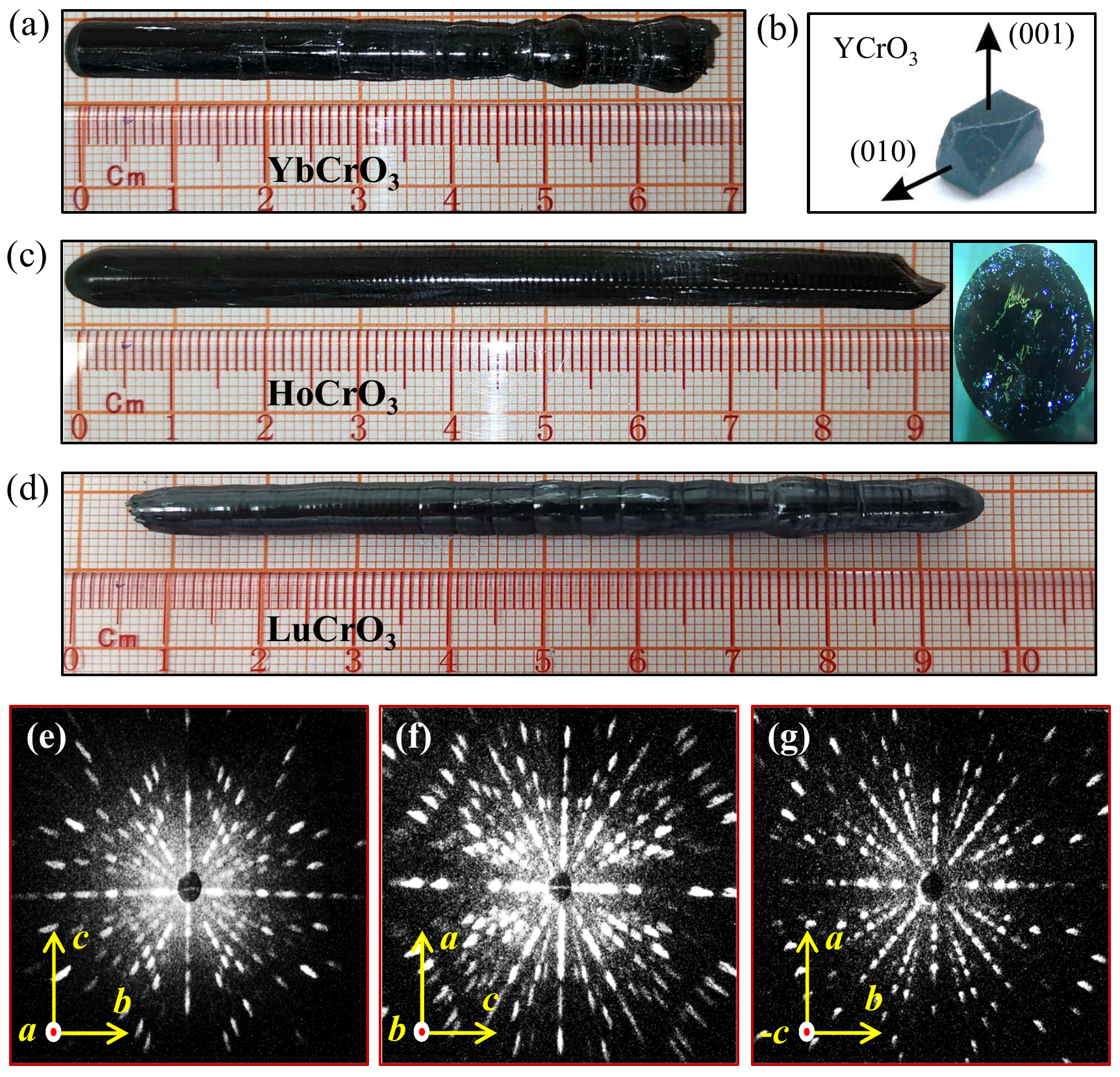}
\caption{
(a-d) Photographs of single crystalline YbCrO$_3$ (a), YCrO$_3$ (b), HoCrO$_3$ (c), and LuCrO$_3$ (d) grown with laser diodes. The right panel of (c) shows the cross section of a HoCrO$_3$ single crystal. Maximum mass $>$ 10 g.
(e-g) Neutron Laue patterns of single crystalline YCrO$_3$. The lattice vectors in real space are illustrated, and the \emph{a} (e), \emph{b} (f), and \emph{c} axes (g) are vertical with respect to the paper. Reproduced with permission from Ref. \cite{Zhu2021, Zhu2021-1}. Copyright (2021) Hai-Feng Li.
}
\label{Samples-2}
\end{figure*}

Zhu \emph{et al}. reported on a good-quality YCrO$_3$ single crystal grown via a laser-diode floating zone technique \cite{Zhu2020}. In the process, poly-crystalline powders of YCrO$_3$ were prepared via traditional solid state reaction, heating at 1000 and 1100 $^{\circ}$C each for 24 h. The powders were pressed into a cylindrical rod ($\sim$ 12 cm) at $\sim$ 70 MPa by hydrostatic press, then the rod was heated to 1300 $^{\circ}$C for 36 h. Before each firing circle, grinding and mixing were carried out by an agate ball with a diameter of $\sim$ 50 mm. Finally, using the poly-crystalline rod with pure phase and additional treatments \cite{Zhu2021}, high-quality YCrO$_3$ single crystal (maximum $>$ 10 g) was grown in a laser-diode floating zone furnace accommodating five infrared lasers (wavelength {=} 975 nm) \cite{Zhu2020}. Representative as-grown single-crystalline RECrO$_3$ (RE = Yb, Y, Tm, and Lu) samples were shown in \textcolor[rgb]{0.00,0.00,1.00}{Figs.}~\ref{Samples-2}(a-d) \cite{Zhu2021, Zhu2021-1}. \textcolor[rgb]{0.00,0.00,1.00}{Figures.}~\ref{Samples-2}(e-g) shows neutron Laue diffraction patterns of YCrO$_3$ \cite{Zhu2021, Zhu2021-1}, implying a good quality of the prepared single crystal.

\subsubsection{Comparison}

Some disadvantages exist with the flux method of growing single-crystal orthochromates: (i) The flux and/or the components of the desired single-crystal materials may react with each other or with the crucible, for example, the Cr$_2$O$_3$ compound would be oxidized into CrO$_3$ below $\sim$ 1000 $^\circ$C, and the resultant compound would react with PbO to form PbCr$_2$O$_7$ \cite{Grodkiewics1966}. (ii) Some elements of the flux materials may occupy the structural positions of rare earths \cite{Grodkiewics1966, Remeika1956}. (iii) The grown single crystals are small (e.g., $\sim$ 55 $\mu$g in Ref.~\cite{Kim2007}) and display different colors due to crystalline defects \cite{Grodkiewics1966, Remeika1956, Aoyagi1969, Todorov2011, Yin2017}. In most cases, the normal floating-zone furnace with four mirrors and four halogen lamps cannot even reach the melting temperatures of rare-earth chromates \cite{Zhou2002-1}. By comparison, with the laser-diode floating-zone method \cite{Wu2020}, larger single crystals (2--10 g) of rare-earth orthochromates can be successfully grown \cite{Zhu2020, Zhu2021, Zhu2020-3}.

\subsection{Thin films}

The YCrO$_3$ compound was deemed as a candidate material for the field of high temperature thermistors \cite{Kagawa1997, Kim2003}. The easy film formation and the densification (inherent to the low sinterbility) are important factors for potential applications \cite{Kagawa1997, QIAO2021100622}.

The spray-pyrolysis combined with an inductively coupled plasma technique at high temperatures was used to form the YCrO$_3$ thin films with spherical particles of 10--50 nm in diameter on substrates of yttrium-stabilized zirconia, single crystalline sapphire, and fused quartz \cite{Kagawa1997}. Different types of substrates led to different patterns of deposited particles. Although the initial solution was Y $: $ Cr $\approx 1 : 1.1$, the final deposits hold a composition of Y $: $ Cr $\approx 1 : 1.3$, indicating that chromium was enriched in the film formation. The crystalline Bragg (121), (002), and (200) peaks could be formed, depending on the type of substrates.

\begin{figure*} [!t]
\centering \includegraphics[width=0.58\textwidth]{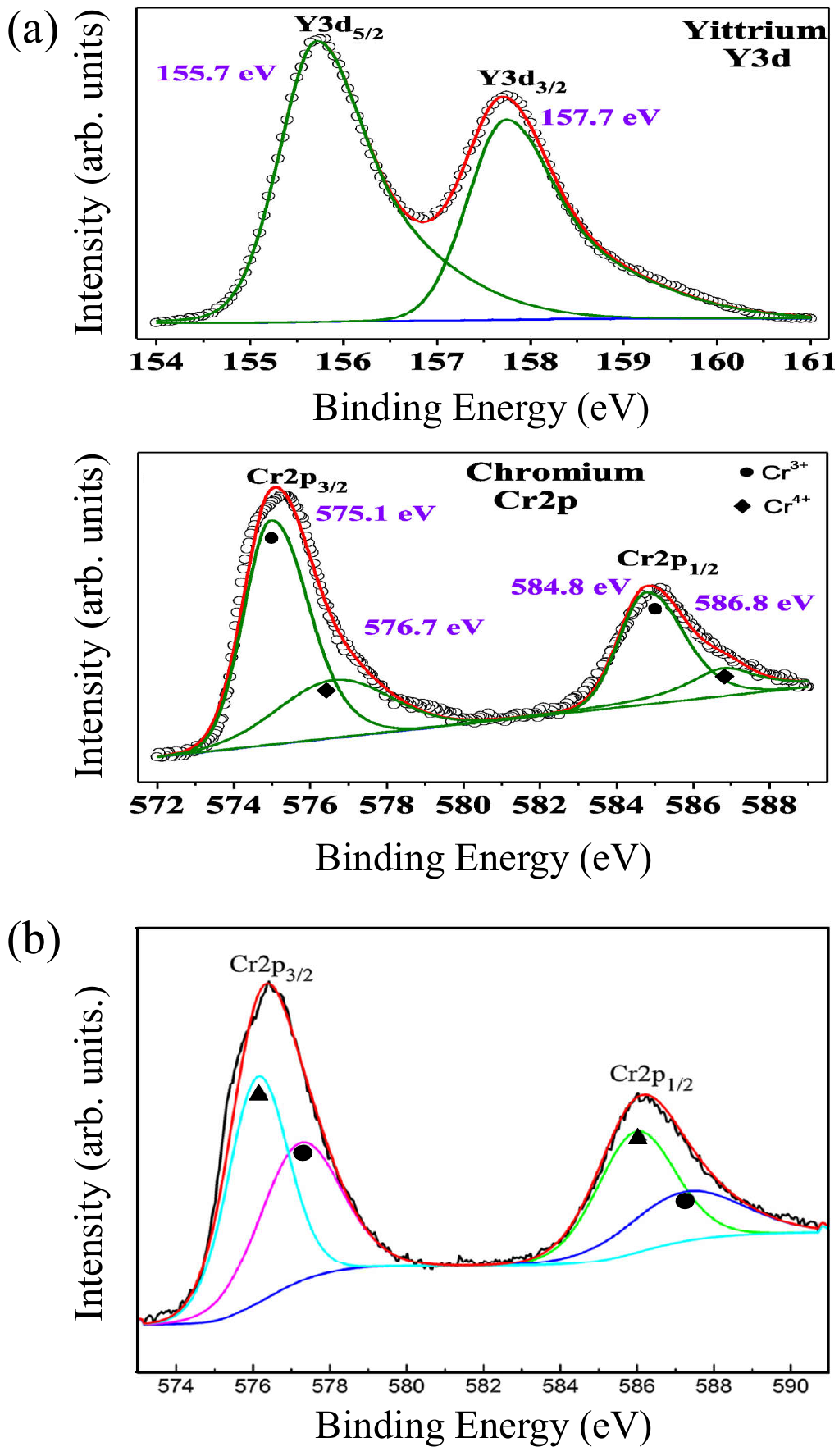}
\caption{
(a) X-ray photoelectron spectroscopy of Y 3\emph{d} (top panel) and Cr 2\emph{p} (bottom panel) regions of a YCrO$_3$ ceramic. Black diamonds and solid circles imply the peaks at the Cr 2\emph{p} spectra that were ascribed to Cr$^{3+}$ and Cr$^{4+}$ ions, respectively. Reproduced with permission from Ref. \cite{Mall2017-1}. Copyright (2017) IOP Publishing Ltd.
(b) X-ray photoelectron spectroscopy around Cr 2\emph{p} region of YCrO$_3$ ceramics. Solid up-triangles and circles imply the peaks at the Cr 2\emph{p} spectra that were attributed to Cr$^{3+}$ and Cr$^{4+}$ ions, respectively. Reproduced with permission from Ref. \cite{Zhang2014}. Copyright (2014) Springer Science{+}Business Media New York.
}
\label{Crstate-3}
\end{figure*}

\begin{figure*} [!t]
\centering \includegraphics[width=0.68\textwidth]{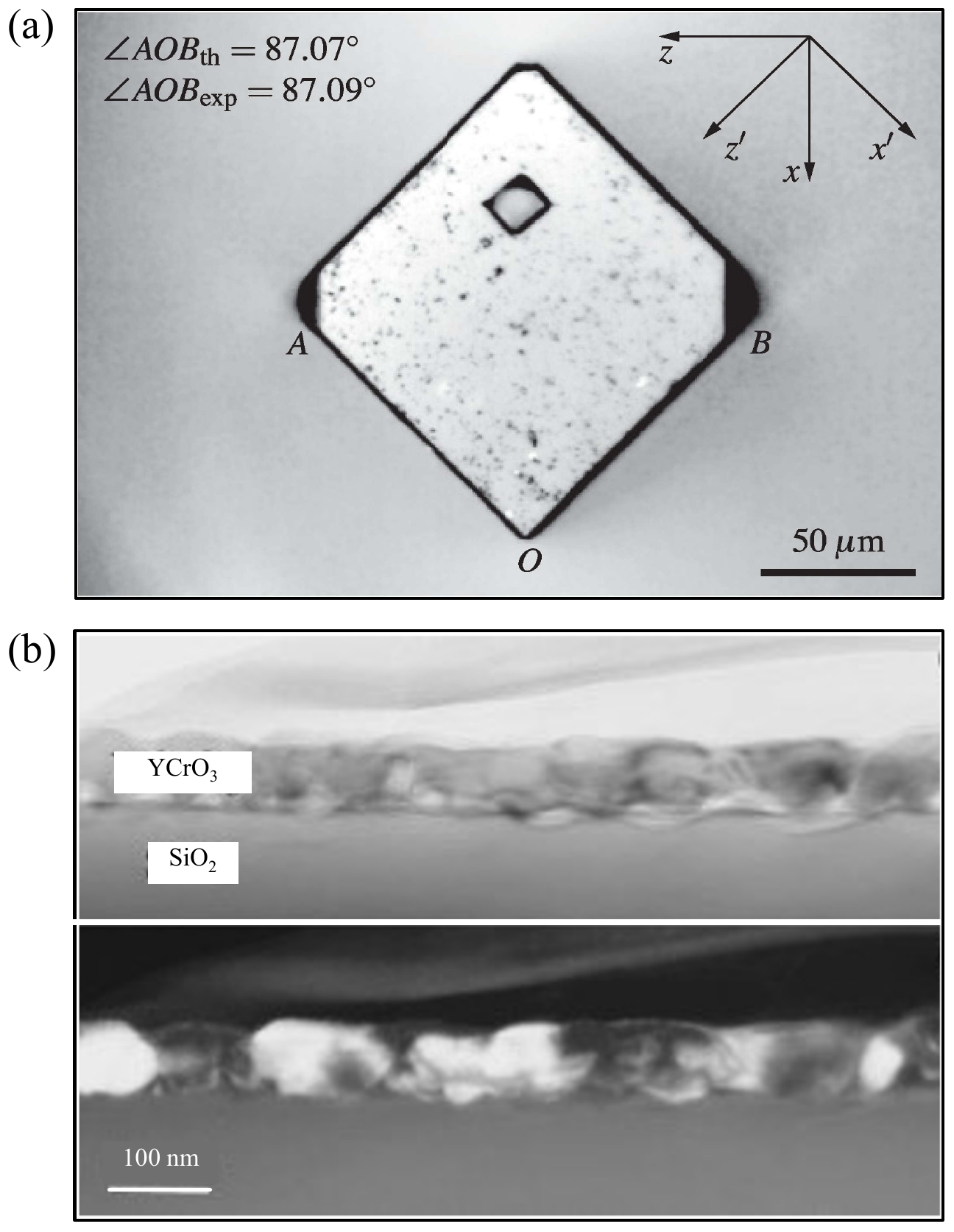}
\caption{
(a) Optical image [(010) surface] of a single-crystal YCrO$_3$ sample (orthorhombic $Pnma$ structure, $\sim$ 120$\times$120 $\mu$m) grown by the flux method \cite{Todorov2011}. The notations are given and used for the crystallographic directions. The calculated angle between vertical ($\overline{1}$01) ($AO$ edge) and (101) ($BO$ edge) surfaces and the measured angle from the image was compared. Reproduced with permission from Ref. \cite{Todorov2011}. Copyright (2011) American Physical Society.
(b) Dark-field (bottom panel) and bright-field (top panel) cross-sectional micrographs of a YCrO$_3$ thin film by transmission electron microscopy. Adapted with permission from Ref. \cite{Kim2003}. Copyright (2003) The Japan Society of Applied Physics.
}
\label{Samples-1}
\end{figure*}

YCrO$_3$ thin films were prepared on SiO$_2$ wafers by the magnetron sputtering method with Ar/O$_2$ as the working gases, and crystallinity and microstructure of the thin films were studied \cite{Kim2003}. First, the YCrO$_3$ target (a disk shape with 2 inch in size) was prepared by a sintering process with the solid-state reaction at 1600 $^\circ$C for 6 h at a heating/cooling rate of 5 $^\circ$C/min in air. Then the thin-film YCrO$_3$ samples were deposited at various depositing conditions. After depositions, the thin films were annealed at different temperatures of 600--800 $^\circ$C for 1 h in an ambient condition. The crystal structure and phase purity were verified by XRPD and transmission electron microscopy techniques. A mixture of the crystalline phase and the amorphous phase existed in the annealed sample at 600 $^\circ$C/1 h; fully crystallized YCrO$_3$ samples formed in annealed samples at $\geq$ 700 $^\circ$C/1 h; the corresponding cross-sectional micrographs of the film were shown in \textcolor[rgb]{0.00,0.00,1.00}{Fig.}~\ref{Samples-1}(b). The YCrO$_3$ thin film was annealed at 800 $^\circ$C/1 h and displayed a linear feature in the log $\sigma T$ versus 1/\emph{T} plot from 300 to 800 $^\circ$C, indicating that the YCrO$_3$ thin film has a potential application as temperature sensors \cite{Kim2003}.

Serrao \emph{et al}. studied the dielectric property of YCrO$_3$ thin films that were deposited via a KrF excimer laser (248 nm, lambda physik) on the $<$111$>$ oriented Pt/TiO$_2$/SiO$_2$/Si substrates \cite{Serrao2005}. The deposition temperature was at 650 $^\circ$C in an oxygen atmosphere (100 mTorr), and the pulse energy was 140 mJ/pulse. Temperature-dependent dielectric constant measurements showed that the grown YCrO$_3$ thin films exhibit a dielectric phase transition at $\sim$ 400 K and 500 Hz (bottom panel of \textcolor[rgb]{0.00,0.00,1.00}{Fig.}~\ref{FerriPoly-1}(a)) \cite{Serrao2005}.

Poly-crystalline thin films of YCrO$_3$ compound were synthesized on Pt/Si substrates by the pulsed laser deposition technique \cite{Cheng2010}. The deposition conditions were at temperatures of 650--730 $^\circ$C in flowing oxygen (100 mtorr) with a cooling rate of 15 $^\circ$C/min. The preferred orientation of the deposited films depends on the deposition temperature, for example, the 300 nm YCrO$_3$ thin film orientates mostly along the $<$001$>$ direction. The YCrO$_3$ compound displayed ambiguous room-temperature ferroelectricity, and no \emph{P}-\emph{E} hysteresis loop was well developed and monitored because of the serious current leakage issue, limiting the ferroelectric performance \cite{Cheng2010}.

\begin{figure*} [!t]
\centering \includegraphics[width=0.82\textwidth]{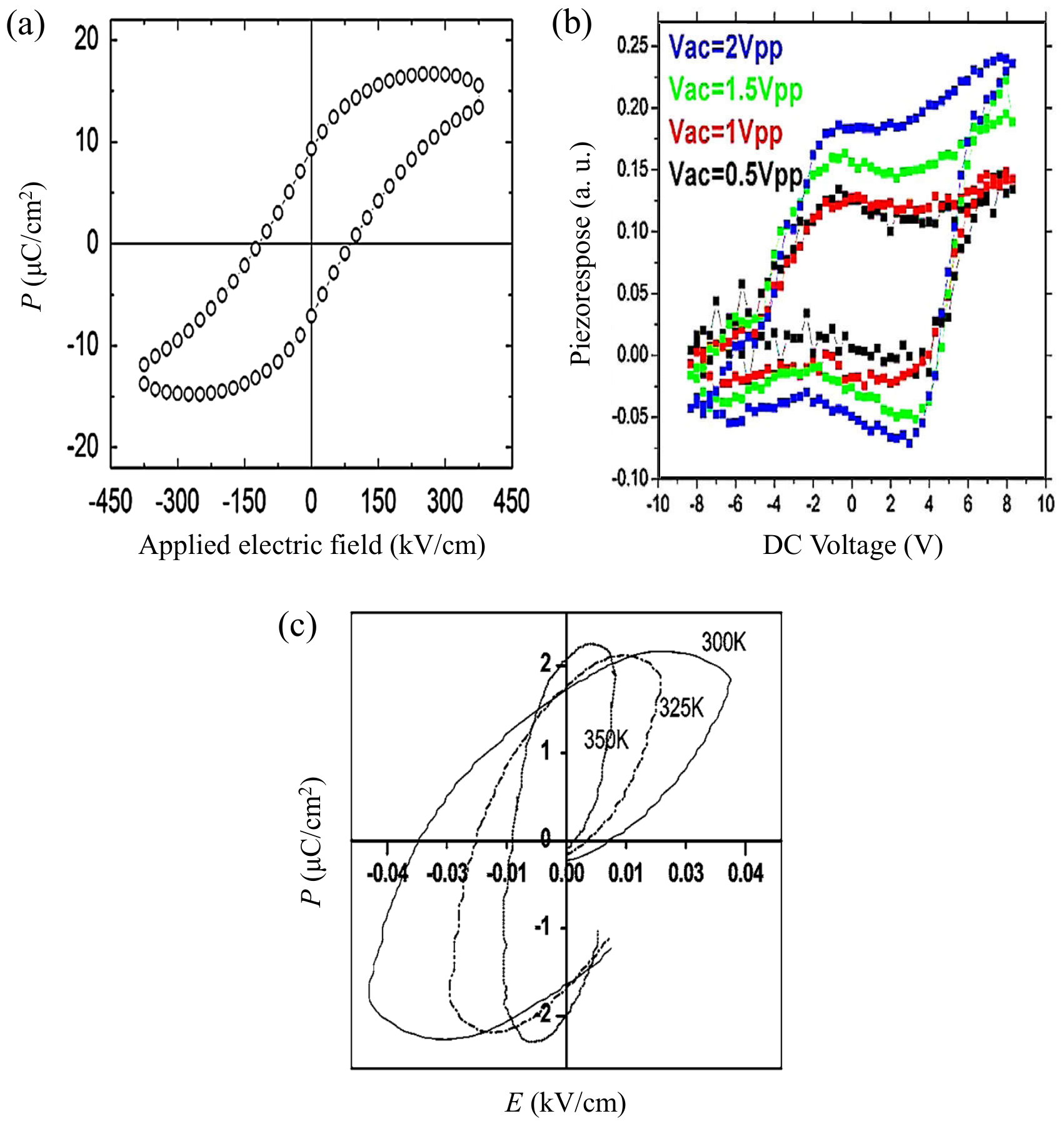}
\caption{
(a) The ferroelectric hysteresis loop of a thin-film YCrO$_3$ sample. The remanent polarization is $\sim$ 9 $\mu$C/cm$^2$ at room temperature. Reproduced with permission from Ref. \cite{Seo2013}. Copyright (2013) Elsevier B.V.
(b) Hysteresis loops in the off dc mode of a 20 nm single-crystal YCrO$_3$ thin film with the (001) plane. Reproduced with permission from Ref. \cite{Arciniega2018}. Copyright (2018) Elsevier B.V.
(c) Dielectric hysteresis of a YCrO$_3$ polycrystal at different temperatures. Reproduced with permission from Ref. \cite{Serrao2005}. Copyright (2005) The American Physical Society.
}
\label{PEloop-1}
\end{figure*}

Seo \emph{et al}. grew thin-film YCrO$_3$ samples on single crystalline Rh substrates (the lattice misfit was $\sim$ 0.7\%) by the pulsed laser deposition \cite{Seo2013}. The multiferroic property was studied, and the ferroelectricity displayed a remanent polarization value {=} 9 $\mu$C/cm$^2$ at room temperature (see \textcolor[rgb]{0.00,0.00,1.00}{Fig.}~\ref{PEloop-1}(a)). A weak ferromagnetism was observed below 150 K in the YCrO$_3$ thin films. The YCrO$_3$ thin films hold a lower domain wall energy, however, they have a larger domain structure than that of the PbTiO$_3$ thin films, indicated by the piezoelectric force microscopy study. This was ascribed to the lower ferroelectric polarizations of YCrO$_3$ \cite{Seo2013}. With the same method, Seo \emph{et al}. also deposited poly-crystalline YCrO$_3$ thin films on the Ta/glass/Pt substrates \cite{Seo2015, Seo2016}. The YCrO$_3$ thin films deposited on glass substrate have a smaller ferroelectric polarization than those on single crystalline Rh substrate, but the ferroelectric switching speed was faster \cite{Seo2015, Seo2016}. Atomic force microscopy indicated that the grain size of YCrO$_3$ thin films deposited on glass substrate was smaller than those on the Rh substrate. Piezoelectric force microscopy indicated that the domain structure of YCrO$_3$ thin films deposited on glass substrate was smaller than those on the Rh substrate, which was due to that the YCrO$_3$ thin films on glass substrate have a smaller domain wall energy than those on the Rh substrate \cite{Seo2015}. The thin-film YCrO$_3$ samples deposited on the (111) Pt/Ta/glass substrates showed a good ferroelectricity, and the remnant polarization value was equal to $\sim$ 5 $\mu$C/cm$^2$. The ferroelectric hysteresis loop and $I$-$V$ curve measurements indicated a large leakage current \cite{Seo2016}.

In amorphous YCrO$_3$ thin films deposited by the pulsed laser, Araujo \emph{et al}. observed a FM order above room temperature \cite{Araujo2014}. The reported order-from-disorder phenomenon was accompanied by a phase transition from an insulating to a metal state, which was attributed to the wide distribution of bond angles (Cr-O-Cr) and the resultant metallization via free carriers \cite{Araujo2014}. X-ray absorption spectroscopy shows the formation of Cr$^{4+}$ oxidation state in the YCrO$_3$ thin film (see \textcolor[rgb]{0.00,0.00,1.00}{Fig.}~\ref{Crstate-2}(a)). Dur\'{a}n \emph{et al}. fabricated the YCrO$_3$/Al$_2$O$_3$ nanocomposites by the thin-film growth technique of atomic layer deposition \cite{Duran2014}. To improve the transport property of YCrO$_3$, Dur\'{a}n \emph{et al}. designed a core-shell architecture, i.e., an amorphous, continuous, and uniform insulating Al$_2$O$_3$ shell with a few nanometer thickness was coated on the YCrO$_3$ core. In-house characterizations with magnetization and alternating-current conductivity measurements showed that the magnetization and the magnetic coercive field depend on the amorphous state and especially the crystallinity of the Al$_2$O$_3$ shell. In addition, the Al$_2$O$_3$ shell acts as a barrier layer and leads to a decrease of the leakage current \cite{Duran2014}. Sharma \emph{et al}. studied the multiferroic properties and photovoltaic effect of a 0.9(BiFeO$_3$) (rhombohedral)–0.1(YCrO$_3$) (orthorhombic) thin-film composite on the Pt/TiO$_2$/SiO$_2$/Si substrate \cite{Sharma2014-3}. Multiferroic properties were measured. A saturation magnetization is equal to $\sim$ 14 emu cm$^{-3}$, larger than the theoretical value. This was caused by the superexchange interactions between Cr and Fe ions. The value of remnant polarization equals $\sim$ 4.5 $\mu$C cm$^{-2}$ \cite{Sharma2014-3}.

Tiznado \emph{et al}. invented a setup, named pulsed-bed atomic layer deposition, for powder coating by changing the configurations of purge and carrier gases \cite{Tiznado2014}. For traditional atomic layer deposition technique, to achieve an uniform coat, one usually uses the method in fluidization conditions which are not easy to reach. Taking advantages of the pulsed-bed mode, Tiznado \emph{et al}. synthesized an multiferroic composite of YCrO$_3$ with both thickness and conformality control of the dielectric Al$_2$O$_3$ shell \cite{Tiznado2014, Duran2016}. In this core–shell Al$_2$O$_3$/YCrO$_3$ structure, the dielectric Al$_2$O$_3$ shell has a function of internal barrier layer, localizing the electrical charges within the core–shell interfacial region, thus improving the electric and magnetic properties of the multiferroics \cite{Tiznado2014, Duran2016}. With increasing the shell thickness, decreases in the dielectric permittivity and in the ac conductivity were observed \cite{Duran2016}.

Different friction coefficients of ferroelectric YCrO$_3$ samples, poly-crystalline, textured, and non-ferroelectric Y-Cr-O films were investigated \cite{Arciniega2016}. The analyses showed that the ferroelectric YCrO$_3$ thin film on Pt(150 nm)/TiO$_2$(30 nm)/SiO$_2$/Si(100) substrate was poly-crystalline and had a lower friction coefficient than that on SrTiO$_3$(110) substrate which was highly textured. As a candidate for ferroelectric memories, poly-crystalline YCrO$_3$ films have more advantages due to the lower friction coefficient \cite{Arciniega2016}.

The heteroepitaxial stabilization of ultra-thin YCrO$_3$ films (5.2$\pm$0.38 nm) was investigated by Pal \emph{et al}. \cite{Pal2018}. The electronic structure was analyzed by atomic multiplet cluster calculations and resonant X-ray absorption spectroscopy. Measurements of polarization dependence at the Cr \emph{L}$_{3,2}$-edges revealed that the presence of X-ray linear dichroism spectrum was uncharacteristic for a 3$d^3$ system with a crystal field of octahedral \cite{Pal2018}. Additionally, the YCrO$_3$ thin films were deposited layer by layer, during which the post-annealing process and the oxygen partial pressure played a crucial role in tuning the phase purity of thin-film YCrO$_3$ samples with suitable Cr$^{3+}$ state and chemical stoichiometry \cite{Pal2018}. Using the magnetron sputtering technique, Gervacio-Arciniega \emph{et al}. deposited locally epitaxial thin-film YCrO$_3$ samples (20 nm, (001) plane) on substrate of SrTiO$_3$(110) at 890 $^\circ$C \cite{Arciniega2018}. The films showed ferroelectricity (see \textcolor[rgb]{0.00,0.00,1.00}{Fig.}~\ref{PEloop-1}(b)). An obvious FM hysteresis loop was obtained at 5 K \cite{Arciniega2018}. By the sol-gel method, Kuang \emph{et al}. prepared thin-film YCrO$_3$/BiFeO$_3$ samples on substrate of quartz by spin coating \cite{Kuang2018}. The bottom YCrO$_3$ layer could favor the grain growth of BiFeO$_3$ films. The YCrO$_3$/BiFeO$_3$ thin films showed lower leakage current density, lower band gap, and enhanced FM and ferroelectric properties, indicating potential applications as ultraviolet and blue-green-driven photo-catalysts \cite{Kuang2018}.

\begin{figure*} [!t]
\centering \includegraphics[width=0.61\textwidth]{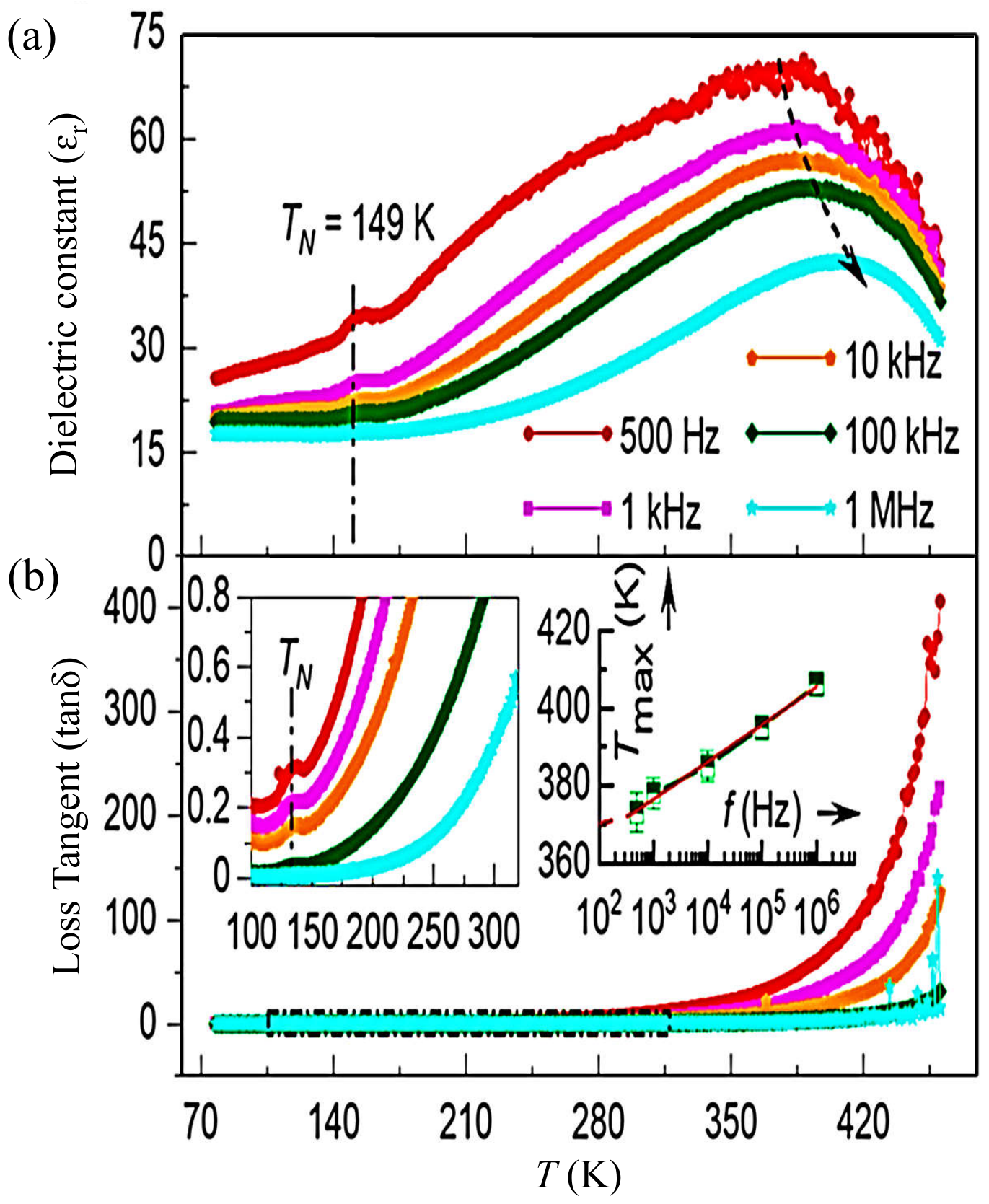}
\caption{
Temperature-dependent dielectric constant $\varepsilon_r$ (a) and loss tangent tan$\delta$ (b) at different frequencies. The dashed line at 149 K indicates a dielectric anomaly close to the magnetic transition. The curved-dash line highlights the relaxor behavior. An enlarged view of the loss data, marked by a narrow-dashed rectangle, is represented in the left inset, showing that the dielectric anomaly can also be observed in loss tangent curves. The right inset shows a Vogel–Fulcher plot with extrapolated freezing temperature ($T_f =$ 370 K, dashed red line), manifesting the relaxor ferroelectricity in an epitaxial single-crystal YCrO$_3$ thin film (32 nm) \cite{Sharma2020}.
}
\label{FerriSingle-1}
\end{figure*}

Sharma \emph{et al}. grew stoichiometric and single-crystal epitaxial thin films of YCrO$_3$ by pulsed laser deposition \cite{Sharma2020}. The (001) SrTiO$_3$ was used as substrates. The grown films were formed with a layer-by-layer fashion and confirmed to be single-crystalline with excellent crystallinity (see \textcolor[rgb]{0.00,0.00,1.00}{Fig.}~\ref{Structure-2}(a)). By comparing the position and shape of the Bragg (103) peak of SrTiO$_3$ substrate and the YCrO$_3$ film (see \textcolor[rgb]{0.00,0.00,1.00}{Fig.}~\ref{Crstate-2}(b)), out-of plane and in-plane strains of the deposited thin film are $-$0.211\% and 0.396\%, respectively, which may influence the magnetic transition temperature \cite{Sharma2020}. Magnetization measurements show a FM-like phase transition at $T_\textrm{N} =$ 144 K. An obvious magnetic hysteresis loop was obtained at 10 K (see \textcolor[rgb]{0.00,0.00,1.00}{Fig.}~\ref{mag-1}(a)). A relaxor ferroelectric feature was observed at $T_\textrm{C} =$ 375--408 K, depending on the frequencies (see \textcolor[rgb]{0.00,0.00,1.00}{Fig.}~\ref{FerriSingle-1}). It is interesting that a close coupling of magnetic and ferroelectric orders was observed in the single-crystal YCrO$_3$ film (32 nm) as evidenced by the appearance of a dielectric anomaly around 149 K (see \textcolor[rgb]{0.00,0.00,1.00}{Fig.}~\ref{FerriSingle-1}) \cite{Sharma2020}.

Finally, we list and compare the synthesis methods for different forms of YCrO$_3$ compound in \textcolor[rgb]{0.00,0.00,1.00}{Table}~\ref{Syn-Meths}.

\section{Properties}

General property of materials corresponds to the responses on outside stimuli such as magnetic field (magnetization), electric field (resistivity), and temperature (thermal property).

\subsection{Structural}

The Y$^{3+}$ ions are smaller than the space between octahedra (CrO$_6$), leading to a mismatch to some extent between Y-O and Cr-O bond lengths. Such kind of distortion could be expressed quantitatively by the tolerance factor $\emph{t}_i$ \cite{Looby1954, Kumar2008, li2008synthesis},
\begin{eqnarray}
t_i = \frac{\textrm{R}_\textrm{Y} + \textrm{R}_\textrm{O}}{\sqrt{2}(\textrm{R}_\textrm{Cr} + \textrm{R}_\textrm{O})},
\label{TFI}
\end{eqnarray}
where $\textrm{R}_\textrm{Y}, \textrm{R}_\textrm{Cr}$, and $\textrm{R}_\textrm{O}$ are ionic radii of Y$^{3+}$ (1.04 {\AA}, coordination number = 6), Cr$^{3+}$ (0.755 {\AA}, coordination number = 6), and O$^{2-}$ (1.24 {\AA}, coordination number = 4) ions \cite{webionicsize}, respectively. Based on the equation~(\ref{TFI}), we calculated $t_i = $ 0.808 for the YCrO$_3$ compound.

A more precise determination of the tolerance factor is based on the averaged bond lengths \cite{li2008synthesis, Platonov2011}, i.e.,
\begin{eqnarray}
t_b = \frac{{<}\textrm{Y}{-}\textrm{O}{>}}{\sqrt{2}{<}\textrm{Cr}{-}\textrm{O}{>}},
\label{TFB}
\end{eqnarray}
where $<$\textrm{Y}-\textrm{O}$>$ and $<$\textrm{Cr}-\textrm{O}$>$ are the averaged lengths of Y-O and Cr-O bonds. Based on the equation~(\ref{TFB}) and previously-reported values of $<$\textrm{Y}-\textrm{O}$>$ = 2.340 {\AA}, and $<$\textrm{Cr}-\textrm{O}$>$ = 1.983 {\AA} \cite{Zhu2020, Zhu2020-2}, we calculated $t_b =$ 0.834 at 300 K in agreement with the value of $t_i$ \cite{Looby1954, Katz1955, Sardar2011}.

The tolerance factor normally covers a range of 0.75--1.00 \cite{Kumar2008} and has been widely used to discuss the perovskite structural distortion. When $t = 1$, the perovskite holds a perfect cubic structure. When $t < 1$, the octahedra (e.g., CrO$_6$) will cooperatively rotate and/or tilt to relieve the internal chemical pressure. As a consequence, the structural symmetry will be decreased from cubic to a lower symmetric one such as orthorhombic, monoclinic, or even triclinic, depending on the detailed value of \emph{t}.

\begin{figure*} [!t]
\centering \includegraphics[width=0.82\textwidth]{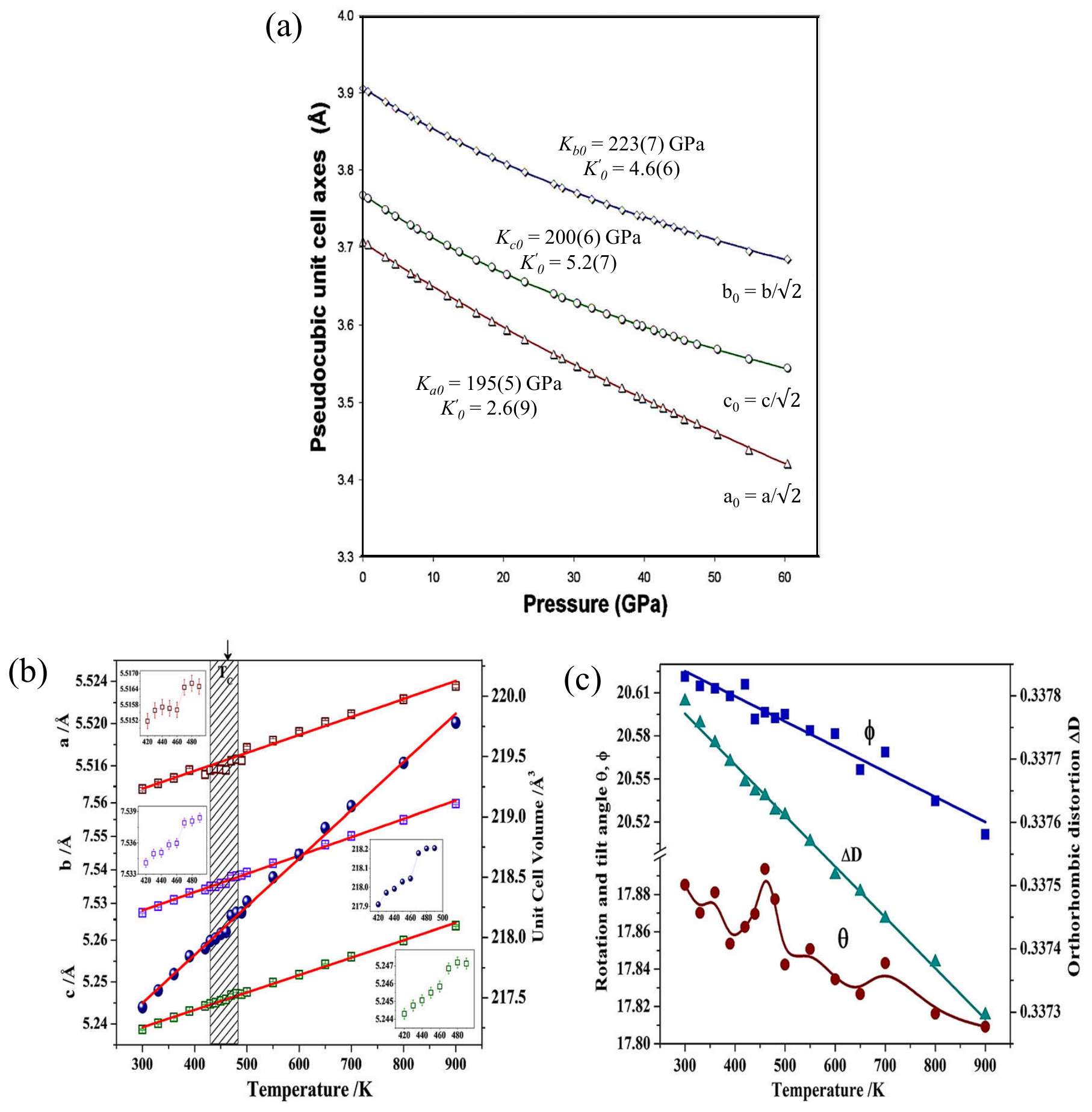}
\caption{
(a) Change in the unit-cell axes of YCrO$_3$ compound between room and 60.4 GPa pressures. Reproduced with permission from Ref. \cite{Ardit2010}. Copyright (2010) The American Physical Society.
(b) Variation of lattice parameters with temperature of poly-crystalline YCrO$_3$ ceramics. The solid line is a linear fit. The region around $T_{\textrm{c}} \sim$ 460 K was shaded. Inset shows the zoomed view of the shaded regime across $T_{\textrm{c}}$.
(c) Change of the CrO$_6$ octahedral tilting angle (circle and square) and the orthorhombic crystalline distortion (triangle) of poly-crystalline YCrO$_3$ ceramics as a function of temperature obtained from refining XRPD patterns. Reproduced with permission from Ref. \cite{Mall2019}. Copyright (2019) 2019 John Wiley \& Sons, Ltd.
}
\label{LatticePara-1}
\end{figure*}

\textbf{(1) Orthorhombic.} Using the YCrO$_3$ single crystal (without impurity phase) grown by Remeika \cite{Remeika1956}, Geller and Wood performed a XRPD study and determined that YCrO$_3$ belongs to an orthorhombic crystalline system ($Pbnm$ space group) with lattice constants of $a$ = 5.238 {\AA}, $b$ = 5.518 {\AA}, and $c$ = 7.54 {\AA} \cite{Geller1956}. In the true crystallographic cell of this structure, there has four distorted perovskite units. Fabian \emph{et al}. prepared YCrO$_{3+\gamma}$ ($\gamma =$ 0 and 1) compounds by the co-precipitation technique and performed studies with X-ray absorption near edge structure spectra and \emph{in situ} XRPD \cite{Fabian2017}. The study indicates that (i) The onset of the crystallization process of the YCrO$_4$ phase (tetragonal, $I41/amd$ space group) begins at $\sim$ 325 $^\circ$C. (ii) At higher temperatures ($680 \leq T \leq 720 ^\circ$C), the phase of YCrO$_3$ (orthorhombic, $Pbnm$ space group) gradually forms. The Cr$^{5+}$ and Cr$^{3+}$ oxidation states depend on the preparation temperature. The temperature-dependent magnetic susceptibilities indicate that the zircon-like YCrO$_4$ phase undergoes a ferromagnetic phase transition at 9 K, and YCrO$_3$ displays an AFM transition at $\sim$ 120 K \cite{Fabian2017}. Poly-crystalline samples of YCrO$_3$ were prepared, and Mall \emph{et al}. measured temperature dependence of XRPD from 300 to 900 K \cite{Mall2019}. By Rietveld refinements, it was found that all XRPD patterns could be well fit by an orthorhombic system with $Pnma$ space group. The obtained lattice constants (\emph{a}, \emph{b}, and \emph{c}) and unit-cell volume (\emph{V}) show a linear increase with temperature with a small anomaly appearing around $T_\textrm{C} \sim$ 460 K (see \textcolor[rgb]{0.00,0.00,1.00}{Fig.}~\ref{LatticePara-1}(b)), indicating that there exist very small changes in lattice constants and unit-cell volume across the ferroelectric to paraelectric phase transition. Therefore, no structural phase transition happens for YCrO$_3$ within the investigated temperature range. As shown in \textcolor[rgb]{0.00,0.00,1.00}{Fig.}~\ref{LatticePara-1}(c) \cite{Mall2019}, the orthorhombic distortion $\Delta$\emph{D} was calculated, displaying almost a linear decrease upon warming and no response to the ferroelectric phase transition; in contrast, the refined CrO$_6$ octahedral tile angles, $\theta$ and $\phi$, also display a monotonic decrease but with a clear anomaly across $T_\textrm{C}$, which can induce a Y displacement.

\textbf{(2) Monoclinic.} Looby and Katz indexed the XRPD data of YCrO$_3$ compound (with Cr$_2$O$_3$ impurity) based on a monoclinic crystal structure ($P2_1/n$ space group) with lattice constants $a$ ($\approx$ $c$) = 7.61$\pm$0.01 {\AA}, $b$ = 7.54$\pm$0.01 {\AA}, and $\beta$ = 92$^{\circ}$56$'$$\pm$6$'$ \cite{Looby1954, Katz1955}. Compared to the fundamental perovskite unit, the monoclinic cell has each of its axes doubled and thus corresponds to eight molecules \cite{Looby1954}. The atomic positions had not yet been determined. First-principles calculations were performed on a basis of spin-dependent density functional theory. The study shows that the lowest energy structure of YCrO$_3$ belongs to the monoclinic structure with a noncentrosymmetric $P12_11$ space group (No. 4, noncentrosymmetric) \cite{Serrao2005} rather than the previously-reported centrosymmetric structure $P2_1/n$ (No. 14) \cite{Looby1954, Katz1955}, reconciling the existence of ferroelectricity in YCrO$_3$. It was also found that there exist tiny distortions, indicating a further lowering of the structure. The specific heat was studied for single-crystalline YCrO$_3$ by Kim \emph{et al}., trying to explore its crystalline structure \cite{Kim2007}. Since the mass of the grown YCrO$_3$ single crystal was too small ($\sim$ 55 $\mu$g) \cite{Razdan1992, Kim2007}, it was difficult to determine its crystalline structure by conventional X-ray diffractometer. To overcome this difficulty, Kim \emph{et al}. invented a new microcalorimeter. It holds a high sensitivity of $\sim$ 1 $\mu$J/K for specific heat measurements \cite{Kim2007}. The obtained heat capacity displays a curve shape of $\lambda$ type at $T_\textrm{N} \approx$ 140 K. Authors subtracted the phonon specific heat obtained from an analogue of non-magnetic LaGaO$_3$ (orthorhombic), and it was found that the left entropy has a much higher value than the theoretical magnetic entropy of Cr$^{3+}$ ($S = \frac{3}{2}$) spins \cite{Kim2007}. The additional phonon entropy was ascribed to a potential monoclinic structure of YCrO$_3$ ($P2_1/n$ space group, No. 14, centrosymmetric) \cite{Looby1954, Katz1955, Kim2007}. It was also suspected that the excessive Debye-phonon modes in the remaining heat capacity could be correlated to the $P12_11$ space group (No. 4, noncentrosymmetric) \cite{Serrao2005}.

\textbf{(3) High-pressure synchrotron diffraction study.} A synchrotron XRPD study of YCrO$_3$ with a high pressure up to 60.4 GPa was performed by Ardit \emph{et al}. \cite{Ardit2010}. The lattice constants, atomic coordinates, and isotropic atomic displacements were extracted at ambient pressure, 4.6 and 42.6 GPa \cite{Ardit2010}. The values of these parameters at room pressure are in agreement with the ones reported previously \cite{Ramesha2007}. The study shows anisotropic elastic modulus along the three crystallographic axes, that is, the \emph{b} axis [$K_{b0}$ = 223(7) GPa] is clearly less compressible than the \emph{a} and \emph{c} axes [$K_{a0}$ = 195(5) GPa and $K_{c0}$ = 200(6) GPa] (see \textcolor[rgb]{0.00,0.00,1.00}{Fig.}~\ref{LatticePara-1}(a)). Although the YCrO$_3$ structure gets more distorted as pressure increases, the space group remains to be $Pbnm$ \cite{Ardit2010}. The YCrO$_3$ nanocrystals ($\sim$ 26 nm) were synthesized by the Pechini method, and Jana \emph{et al}. performed a study with pressure up to $\sim$ 17 GPa \cite{Jana2018-1, Jana2018-2}. Synchrotron XRPD shows that all profiles can be indexed to the orthorhombic structure ($Pnma$ space group), and there exists no structure transition over the whole range of pressure. However, there indeed exist anomalies in structure parameters of the unit cell at $\sim$ 4.3 GPa, where a maximum shift in position occurs to the Y atoms. The lattice constants change with the same decreasing rate up to 4.3 GPa, after which the lattice constant \emph{a} gets less compressible than the \emph{b} and \emph{c} axes indicative of a more distorted structure \cite{Jana2018-1, Jana2018-2}.

\textbf{(4) Neutron diffraction study.} A neutron powder diffraction (NPD) study on 3 g YCrO$_3$ from 15 to 550 K was reported \cite{Ramesha2007}. This high-resolution NPD study revealed that below (in the ferroelectric state) and above (in the paraelectric state) the dielectric anomaly occurring at $\sim$ 473 K (see \textcolor[rgb]{0.00,0.00,1.00}{Fig.}~\ref{FerriPoly-1}(a), top panel) \cite{Serrao2005}, the average crystal structure is orthorhombic (centrosymmetric) with space group of $Pnma$ rather than the monoclinic as shown in \textcolor[rgb]{0.00,0.00,1.00}{Fig.}~\ref{CrShift-1}(a) \cite{Looby1954, Katz1955, Serrao2005}. The lattice parameters and atomic positions as well as isotropic thermal parameters were obtained \cite{Ramesha2007}. Local non-centrosymmetry was reported as the origin of the ferroelectric transition in YCrO$_3$ \cite{Serrao2005}. Zhu \emph{et al}. pulverized a YCrO$_3$ single crystal \cite{Zhu2021} grown by the super-necking technique with a laser-diode floating zone furnace \cite{Wu2020} and carried out a time-of-flight NPD study from 321 to 1200 K (see \textcolor[rgb]{0.00,0.00,1.00}{Fig.}~\ref{CrShift-1}(b)) \cite{Zhu2020}. Within the present experimental resolution, the collected data was indexed sufficiently with the $Pmnb$ space group in the entire temperature range. The structural parameters at 321 K, including lattice parameters ($a$ = 7.5332(3) {\AA}, $b$ = 5.5213(2) {\AA}, $c$ = 5.2418(2) {\AA}, and unit-cell volume $V = $ 218.02(1) {\AA}$^3$), thermal parameters, atomic positions, local distortion parameter, bond lengths, bond angles, bond valence states, and local distortion modes were obtained by structural refinements. The temperature-dependent thermal expansions (along with the \emph{a}, \emph{b}, and \emph{c} axes) and the corresponding cell volume were theoretically modelled using the Gr\"{u}neisen function, and the extracted Debye temperature is $\Theta_\textrm{D}$ = 580 K. The thermal expansions exhibit an anisotropic characteristic along the \emph{a}, \emph{b}, and \emph{c} axes with an onset of anomaly around 900 K (see \textcolor[rgb]{0.00,0.00,1.00}{Fig.}~\ref{LatticePara-2}(a)) \cite{Zhu2020}. For the series of compounds isostructural to GdFeO$_3$ (orthorhombic, space group: $Pbnm$), some of them can transform to rhombohedral ($R$\={3}$m$) structure at high temperatures, for example, SmAlO$_3$ and LaGaO$_3$ transform to the rhombohedral structure at 800 $^\circ$C and 875 $^\circ$C, respectively \cite{Geller1957-1}. In contrast, the YCrO$_3$ compound keeps no structural phase transition within $12 \leq T \leq 1200$ K (see \textcolor[rgb]{0.00,0.00,1.00}{Figs.}~\ref{LatticePara-2}(a) and \ref{LatticePara-2}(b)) \cite{Zhu2020, Zhu2020-2}. As shown in \textcolor[rgb]{0.00,0.00,1.00}{Fig.}~\ref{LatticePara-2}(b), the lattice parameters (\emph{a}, \emph{b}, and \emph{c}) deviate largely from the theoretical calculations below $T_\textrm{N}$, showing an anisotropic magnetostriction effect as well as a magnetoelastic effect. Upon cooling, the sample contraction is increased below $T_\textrm{N}$ \cite{Zhu2020-2}.

\textbf{(5) Neutron pair distribution study.} By neutron pair distribution function (PDF) method, one can extract the information of local structures or short-range from the diffuse scattering, it is thus a powerful technique for detecting the local crystalline structures. Such a study on YCrO$_3$ \cite{Ramesha2007} showed that the data at 550 K ($>$ 440 K) was well indexed with the $Pnma$ space group over the range of 1.6--6 {\AA}; for a comparison, the low-temperature data (collected in the ferroelectric state) can be better fitted with the $P12_11$ model (No. 4, noncentrosymmetric) (see \textcolor[rgb]{0.00,0.00,1.00}{Fig.}~\ref{CrShift-2}(a)), resulting in an off-centring displacement of Cr ions ($\sim$ 0.01 {\AA}) in the $z$ direction (see \textcolor[rgb]{0.00,0.00,1.00}{Fig.}~\ref{CrShift-2}(b)) \cite{Ramesha2007}.

\textbf{(6) Theoretical expectation.}
The band structure of YCrO$_3$ was examined by first-principles calculations \cite{Weber2019}. A band inversion near the Fermi energy was calculated in the centrosymmetric $P6_3/mmc$ space group. There could be a phase transition from $P6_3/mmc$ to noncentrosymmetric $P6_3cm$, taking into account soft $K_3$ and $\Gamma_2^-$ phonons. The calculation predicted possible topological {``}drumhead{''} surface states, and that synthesizing YCrO$_3$ with the hexagonal structure could provide an interesting platform for investigating the interplay between multiferroicity and topology \cite{Weber2019}.

\subsection{Magnetic}

The study of weak ferromagnetism appearing in main AFM matrix has been an interesting topic \cite{fu2021gapless, pan2021giant, kim2021spin, davidson2021pressure, ning2021antisite, flynn2021two, li2021antiferromagnetic, zeng2021smoothing, das2021anisotropic, zhou2020weak}. A full understanding of the intriguing magnetism necessitates distinguishing the origin of the weak ferromagnetism. In orthorhombic chromates two mechanisms are possible: The first one is related to the single ion anisotropy, and the second one is connected with antisymmetric superexchanges. YCrO$_3$ is such a fascinating compound that orders antiferromagnetically with weak ferromagnetism below $T_\textrm{N}$ $\approx$ 141 K \cite{Judin1966}.

\textbf{(1) Magnetization.} Temperature- and applied-magnetic-field-dependent magnetization along various crystallographic \emph{a}, \emph{b}, and \emph{c} axes of an orthorhombic ($Pbnm$) single crystalline YCrO$_3$ sample was measured \cite{Judin1966}. Magnetic-field-dependent measurements show spontaneous magnetization only along the \emph{c} axis at 77 K from 0 to 21 KOe, and the magnetic susceptibility $\chi_c > \chi_a > \chi_b$. Temperature-dependent magnetic susceptibilities show $\chi_a > \chi_c > \chi_b$ in a certain temperature range below $T_\textrm{N}$, observing the Brillouin curve of Cr$^{3+}$ ions ($S = \frac{3}{2}$). Among them, $\chi_a$ displays ferromagnetically, and $\chi_c$ shows a sharp maximum around $T_\textrm{N}$, displaying an AFM characteristic \cite{Judin1966}. Therefore, there exit strong competitations between spin exchange, anisotropy, and Dzyaloshinsky fields \cite{Judin1966, HFLi2016}. Based on these in-house characterizations, Judin \emph{et al}. inferred that anisotropic superexchange plays an important role in producing the weak ferromagnetism in YCrO$_3$ \cite{Judin1966}.

\textbf{(2) High-temperature measurements.} The paramagnetic susceptibility of a YCrO$_3$ single crystal was measured with temperature up to 800 K \cite{Tsushima1970}. The Curie-Weiss temperature was fit as $\theta =$ -230$\pm$5 K. Based on the following molecular-field relations \cite{Tsushima1970},

\begin{table*}[!b]
\small
\caption{\newline Summary of the synthesis methods for parent and doped YCrO$_3$ compounds. ICP = Inductively coupled plasma.}
\label{Syn-Meths}
\setlength{\tabcolsep}{3.9mm}{}
\renewcommand{\arraystretch}{1.2}
\begin{tabular}{l|lll}
\hline
\hline
                               & Methods                      & Remark                                                           & Refs.                                                \\ [2pt]
\hline
                               & Combustion technique         &36 nm (600 $^{\circ}$C), onion-like layers                        &\cite{Bedekar2007}                                    \\
                               & Sol-gel method               &38--125 nm (600 $^{\circ}$C, 2 h), non-linear optical property    &\cite{Krishnan2011}                                   \\
                               & High-speed planetary milling &Finer and homogeneous (10--20 nm)                                 &\cite{Park2012}                                       \\
                               & Microwave-assisted synthesis &Amorphous phase (10 min irradiation);                             &\cite{Prado2013}                                      \\
Nano-crystal                   &                              &Poly-crystalline YCrO$_4$ (500 $^{\circ}$C, 2 h);                 &\cite{Prado2013}                                      \\
                               &                              &Formation of YCrO$_3$ (800 $^{\circ}$C, 2 h).                     &\cite{Prado2013}                                      \\
                               & Citrate precursor method     &High crystallinity (22 nm)                                        &\cite{Ahmad2016}                                      \\
                               & Reverse micellar method      &Highly crystalline (35 nm)                                        &\cite{Ahmad2018}                                      \\ [1pt]
\hline
                               & Pulse laser deposition       &(001) orientation @300 nm                                         &\cite{Cheng2010}                                      \\
                               & Atomic layer deposition      &Conformality and thickness control                                &\cite{Tiznado2014}                                    \\
Thin film                      & Spray pyrolysis plus ICP     &Ultrafine particles, impurity (Cr$_2$O$_3$ \& Y$_2$O$_3$)         &\cite{Kagawa1997}                                     \\
                               & Spin coating                 &Low current leakage                                               &\cite{Kuang2018}                                      \\
                               & Magnetron sputtering         &20 nm, highly oriented (001)                                      &\cite{Arciniega2018}                                  \\ [1pt]
\hline
                               & Flux method                  &Impurity                                                          &\cite{Looby1954}                                      \\
                               & Sol-gel method               &Low-temperature, defects, oxygen vacancies                        &\cite{Carini1991-1}                                   \\
                               & SHS                          &Self-propagating, 800 $^{\circ}$C, 2 h                            &\cite{Kuznetsov1998}                                  \\
Poly-crystal                   & Hydrazine method             &Low-temperature, YCrO$_4$                                         &\cite{Tachiwaki2001}                                  \\
                               & Solid state reaction         &High-temperature, defects, non-stoichiometric                     &\cite{Cheng2010, Sahu2008}                            \\
                               & Hydrothermal synthesis       &Low-temperature, highly crystalline                               &\cite{Sardar2011}                                     \\ [1pt]
\hline
                               & Flux method                  &Slowly cooling, bright green, small, impurity                     &\cite{Razdan1992, Todorov2011, Yin2017, Sanina2018}   \\
                               & Molten-salt vaporization     &Keeping high temperature, dark green, 7$\times$4$\times$3 mm$^3$  &\cite{Grodkiewics1966}                                \\
Single-crystal                 &                              &Keeping high temperature, small                                   &\cite{Aoyagi1969}                                     \\
                               & Floating-zone method         &Large ($> 10$ g), black, cracks, non-stoichiometric               &\cite{Zhu2021}                                        \\
                               & Hydrothermal synthesis       &High pressure, stoichiometric, hope 100--200 mg                   &Propose                                               \\ [1pt]
\hline
\hline
\end{tabular}
\end{table*}

\begin{eqnarray}
\theta = \frac{2S(S + 1)(6J_1 + 12J_2)}{3k}, \textrm{and}
\label{CWtheta}
\end{eqnarray}
\begin{eqnarray}
T_\textrm{N} = \frac{2S(S + 1)(-6J_1 + 12J_2)}{3k},
\label{AFMTem}
\end{eqnarray}
where $J_1$ is the averaged value of the two exchange interactions for the 6 nearest neighbor Cr$^{3+}$ ions along the (\emph{a}$\pm$\emph{b}) and \emph{c} axes, $J_2$ is the averaged value of the 3 exchange interactions for the 12 next-nearest Cr$^{3+}$ ions, $T_\textrm{N} =$ 141 K, $S = \frac{3}{2}$, and $k$ = 1.38 $\times$ 10$^{-23}$ J/K is the Boltzmann constant. Based on these values and the equations~(\ref{CWtheta}) and (\ref{AFMTem}), the calculated $J_1$ = -8.7 cm$^{-1}$ and $J_2$ = -1.0 cm$^{-1}$ \cite{Tsushima1970}.

Zhu \emph{et al}. studied on the high-temperature magnetism of YCrO$_3$ from 300 to 1200 K \cite{Zhu2020}. Using the Curie-Weiss law:
\begin{eqnarray}
M = \frac{m}{(T-\theta_\textrm{CW})},
\label{1}
\end{eqnarray}
the $M$-$T$ data was first fit, extracting a paramagnetic Curie-Weiss temperature $\theta_\textrm{CW}$ = $-$264.0(1) K. For a more accurate fitting, the temperature-dependent inverse magnetic susceptibility $\chi^{-1}$ was fit by:
\begin{eqnarray}
\chi^{-1} (T) = \frac{3k_B(T-\theta_\textrm{CW})}{N_A\mu^2_\textrm{eff}},
\label{2}
\end{eqnarray}
in five temperature regimes: 300--400, 400--540, 540--640, 640--750, and 750--980 K. As temperature increases from 300--400 K to 750--980 K, the extracted effective paramagnetic moment $\mu_{\textrm{meas}}$ decreases from 4.09(1) to 3.47(1) $\mu_\textrm{B}$. The frustration parameter can be expressed as:
\begin{eqnarray}
f = \frac{|\theta_\textrm{CW}|}{T_\textrm{N}},
\label{3}
\end{eqnarray}
and $f$ decreases from 2.34(1) to 0.46(1) upon warming \cite{Zhu2020}. These could be the ramifications of formation of magnetic polarons and strong magnetic frustrations, relying on the complex low-temperature magnetic structure of YCrO$_3$ compound.

\begin{figure*} [!t]
\centering \includegraphics[width=0.82\textwidth]{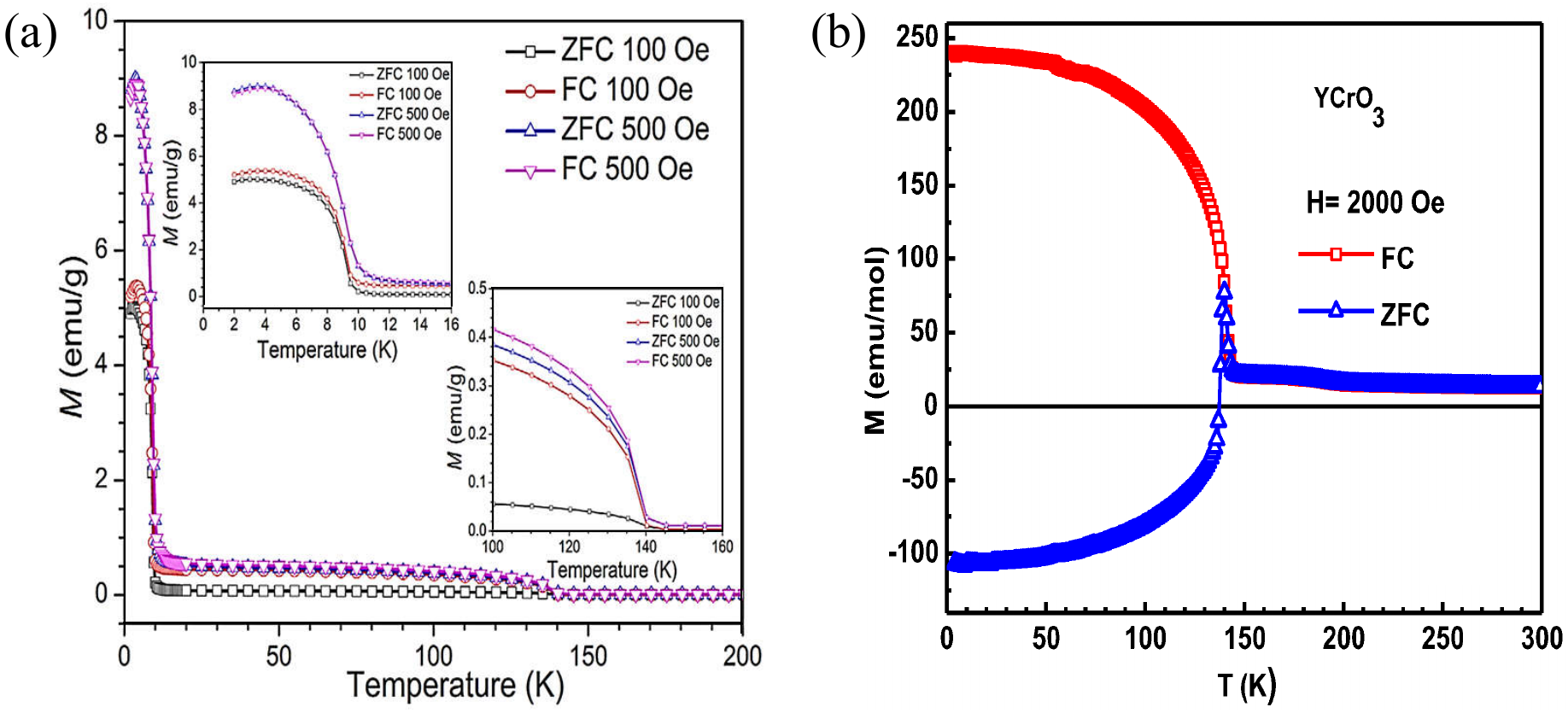}
\caption{
(a) Variation of magnetization (\emph{M}) as a function of temperature in a nano-crystalline YCrO$_3$ sample synthesized at 700$^\circ$C. The two insets show the zoomed curves near 10 and 140 K, respectivly. Below 140 K, there is a sudden change in the behavior of \emph{M}. Reproduced with permission from Ref. \cite{Singh2013}. Copyright (2018) AIP Publishing LLC.
(b) Magnetization (\emph{M}) of a poly-crystalline YCrO$_3$ sample as a function of temperature measured at 0.2 T with protocols of field cooling (FC) and zero-field cooling (ZFC). Reproduced with permission from Ref. \cite{Kumar2017}. Copyright (2017) Published by Elsevier B.V.
}
\label{mag-2}
\end{figure*}

\textbf{(3) Grain size effect.} Grain size effect on magnetic properties of YCrO$_3$ was studied by Singh \emph{et al}. \cite{Singh2013}. In the study, YCrO$_3$ particles were prepared by the confinement with miniemulsion droplets in water-in-oil systems. It was noticed that if the calcination temperature was higher than 900 $^\circ$C, pure YCrO$_3$ ceramics decompose into Y$_2$O$_3$ and Cr$_2$O$_3$. Temperature-dependent magnetization measurements show an interesting feature that the magnetization displays a sharp increase before plateauing around 5 K as shown in \textcolor[rgb]{0.00,0.00,1.00}{Fig.}~\ref{mag-2}(a), indicating a magnetic instability in the YCrO$_3$ nanoparticles ($<$ 20 nm) below 10 K \cite{Singh2013}. This anomalous behavior could be understood by involving the idea of mesocrystals or elongated grain. There are two contributions to the magnetic hysteresis loop measured below 10 K: One is similar to FM, and the other is from normal AFM system \cite{Singh2013}.

\textbf{(4) Magnetic structure.} Orthorhombic ($Pbnm$) RECrO$_3$ (RE = rare earth) perovskites were investigated \cite{Bertaut1966}. The results show that spins of the Cr sublattice order in a \emph{G}-type mode. As RE sites change from element of Lu to La, the corresponding $T_\textrm{N}$ increases from 112 to 282 K. There exists a coupling between spins of RE- and Cr-sites in other RECrO$_3$ compounds, except for TbCrO$_3$ where Tb$^{3+}$ and Cr$^{3+}$ ions differ in their respective magnetic unit cells. Moreover, the spin ordering of RE sites was observed for Er, Ho, Nd, Pr, and Tm above 4.2 K; below 4.2 K, spins of Tb$^{3+}$ and Dy$^{3+}$ ions order magnetically. For the case of YCrO$_3$, Y$^{3+}$ ions ($4d^0$$5s^0$) is nonmagnetic, and a $G_x$-type spin ordering of Cr$^{3+}$ sublattice forms below $T_\textrm{N}$ $\approx$ 141 K \cite{Bertaut1966}. In this \emph{G}-type magnetic structure, all six nearest spins are antiparallel. Additionally, it is pointed out that in the pseudo-tetragonal structure, the spin direction cannot be specified in the $xy$-plane since the \emph{G}-mode does not have a lower index \cite{Bertaut1966}.

\begin{figure*} [!t]
\centering \includegraphics[width=0.82\textwidth]{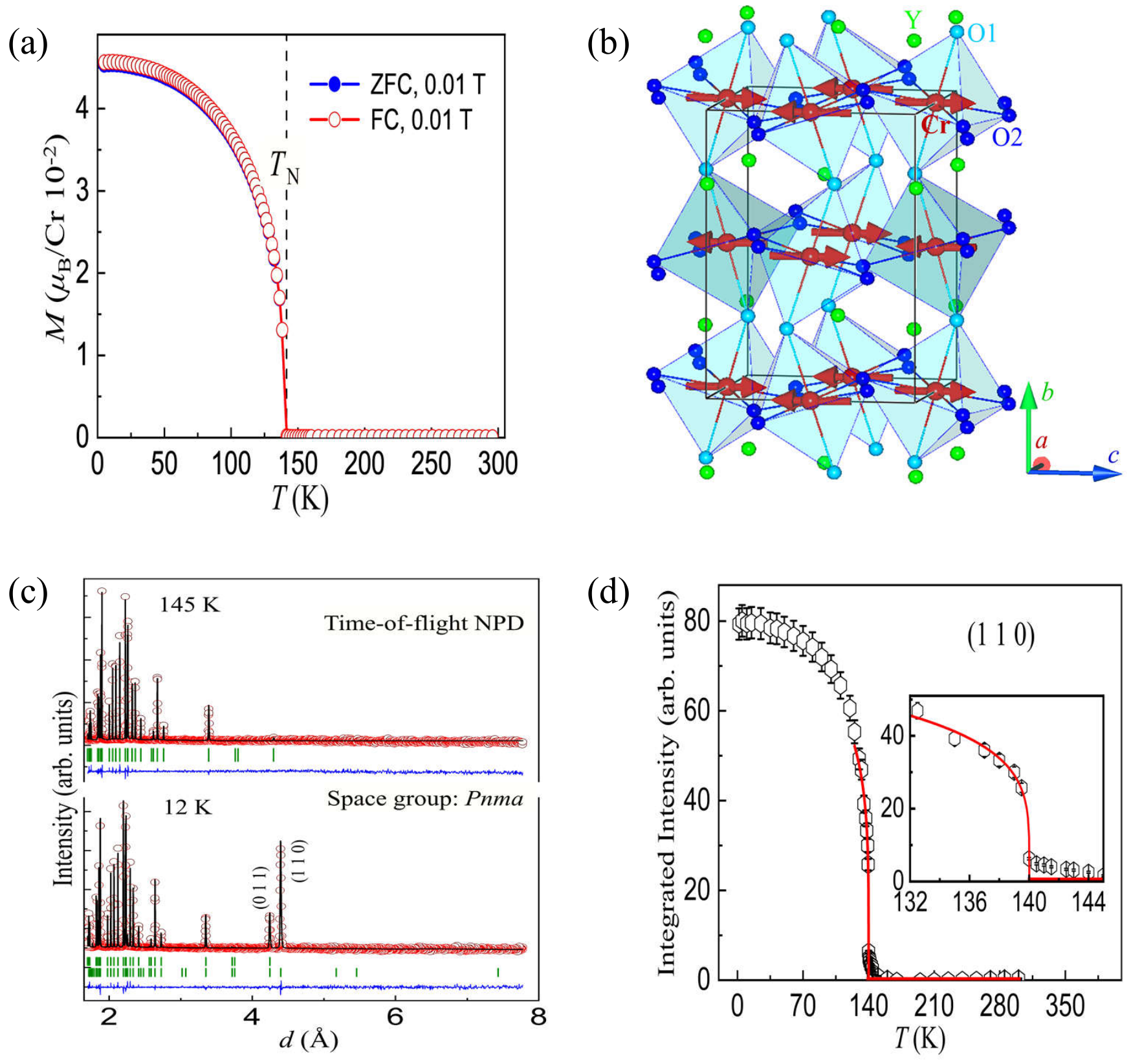}
\caption{
(a) Magnetization (\emph{M}) of chromium ions in single crystal YCrO$_3$ versus temperature, measured with zero-field cooling (ZFC) and field cooling (FC) modes at 0.01 T.
(b) Orthorhombic crystalline structure ($Pnma$ space group) and the AFM magnetic structure (propagation vector \emph{k} = (110)) of single-crystalline YCrO$_3$ below $T_{\textrm{N}}$. The arrows sitting on Cr ions are the chromium spins.
(c) Calculated and observed NPD patterns of single-crystal YCrO$_3$ from a time-of-flight study at 145 K (top panel) and 12 K (bottom panel). The lower curves are the difference between calculated and observed patterns. The vertical bars represent the locations of magnetic and nuclear Bragg peaks.
(d) Integrated magnetic intensity from the Bragg (110) reflection (void pentagons). Inset is an enlarged figure in the temperature range of 132--145 K (around $T_\textrm{N}$). The data was fit to a power law (solid line). The error bars were obtained by propagation calculations. Figures (a-d) were reproduced with permission from Ref. \cite{Zhu2020-2}. Copyright (2020) American Physical Society.
}
\label{mag-3}
\end{figure*}

\textbf{(5) Magnetic phase transition.} The magnetic properties of single-crystal YCrO$_3$ were studied with in-house magnetic and heat capacity measurements \cite{Zhu2020-2}. According to temperature-dependent magnetization measurements and the corresponding inverse magnetic susceptibility (see \textcolor[rgb]{0.00,0.00,1.00}{Fig.}~\ref{mag-3}(a)), YCrO$_3$ displays a sharp magnetic transition with the transition temperature of $T_\textrm{N}$ = 141.5(1) K. The corresponding frustration factor was calculated as 3.061(5) ($f=|\theta_\textrm{CW}|/T_\textrm{N}$). The study of magnetic hysteresis loop carried out at 2 K indicates that YCrO$_3$ is a soft ferromagnet. The characteristic feature was also observed in the heat capacity data of canted antiferromagnet, i.e., a shift of the $\lambda$-shape peak from $\sim$ 138.8 K (at 0 T) to $\sim$ 139.9 K (at 5 T).

\begin{figure*} [!t]
\centering \includegraphics[width=0.82\textwidth]{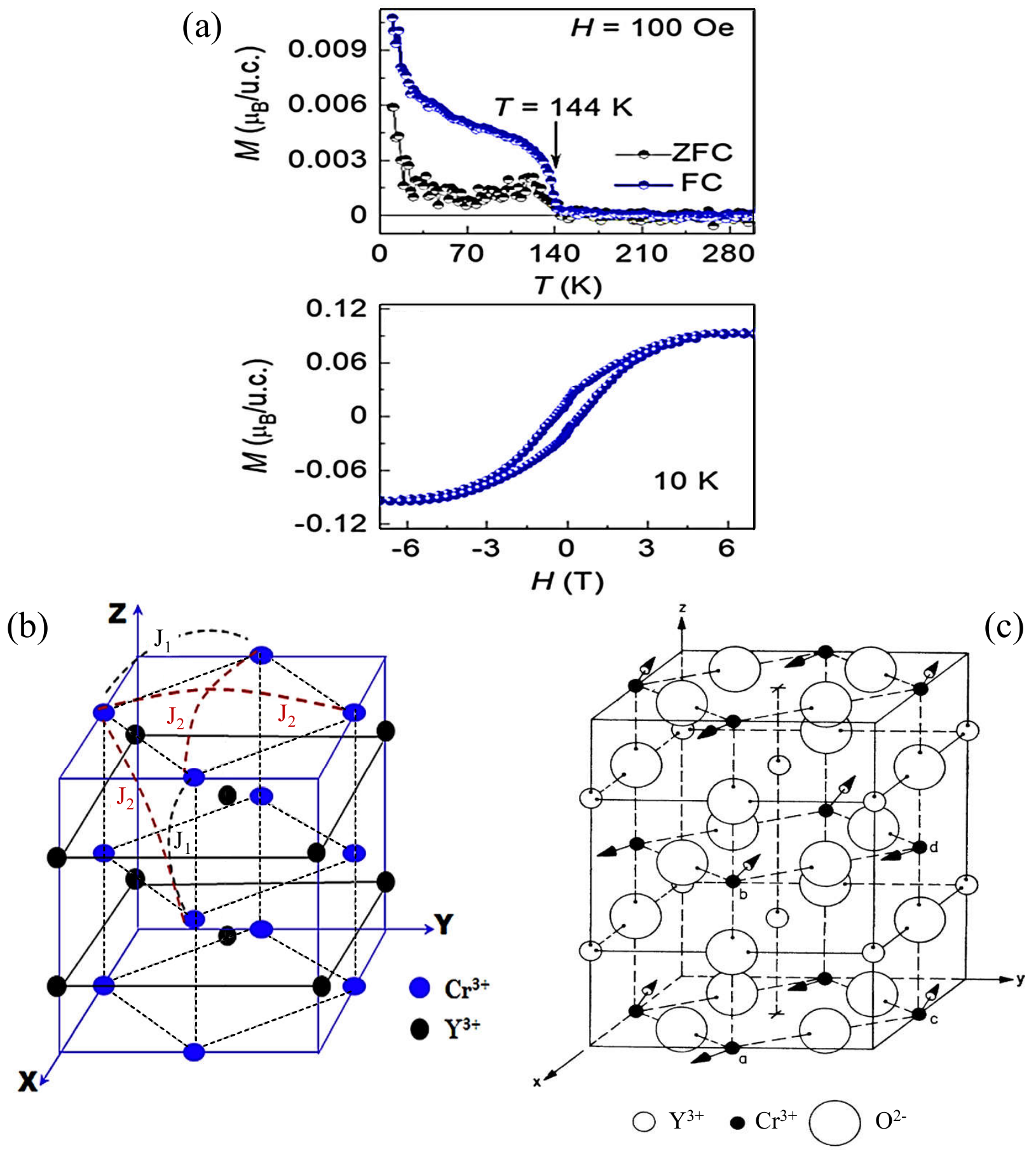}
\caption{
(a) Temperature dependence of magnetization (\emph{M}-curves) in the protocols of zero-field cooling (ZFC) and field-cooling, displaying a magnetic phase transition at 144 K (top panel), and the in-plane magnetic field dependent \emph{M} loop of a single-crystal YCrO$_3$ thin film (32 nm) at 10 K (bottom panel) \cite{Sharma2020}.
(b) Illustration of the two exchange interactions ($J_1$ and $J_2$) in the unit cell of YCrO$_3$. Reproduced with permission from Ref. \cite{Sharma2014-1}. Copyright (2014) AIP Publishing LLC.
(c) Unit cell of YCrO$_3$ shows the directions of the Cr$^{3+}$ spins at low temperatures. Reproduced with permission from Ref. \cite{Ziel1969-3}. Copyright (1969) The American Physical Society.
}
\label{mag-1}
\end{figure*}

The magnetothermal property of YCrO$_3$ polycrystals produced by the solid state reaction was studied \cite{Duran2010}. Magnetization measurements indicate that the polycrystal shows a spin CAFM ordering of Cr$^{3+}$ ions below $T_\textrm{N} \approx$ 140 K. Around $T_\textrm{N}$, an anomaly ($\lambda$ type) appears in the specific heat data. Detailed analysis reveals the existence of spin fluctuation when $T > T_\textrm{N}$ and that a spin-reorientation (SR) phase transition occurs within 140--60 K as decreasing temperature \cite{Duran2010}. It was speculated that the AFM easy \emph{a}-axis rotates smoothly within the \emph{a}-\emph{c} plane of space group $Pbnm$ around 60 K \cite{Duran2010}.

\textbf{(6) Spin flop transition.} Jacobs \emph{et al}. induced SRs in the YCrO$_3$ single crystal with applied high dc magnetic fields along the AFM easy \emph{a} axis \cite{Jacobs1971}. As an applied magnetic field increases, the AFM easy axis continuously rotates from crystallographic \emph{a} to \emph{c} axis within the $a$-$c$ plane, arriving at the \emph{c} axis at the critical field of $H_\textrm{cr} =$ 40(1) kOe at 4.2 K. This phenomenon agrees with the theoretically predicated second-order type spin-flop transition \cite{HFLi2016}. The detailed analysis of the magnetization data based on a two-sublattice approximation shows that there exists only the uniaxial anisotropy in YCrO$_3$, without the quartic anisotropy in the $a$-$c$ plane. The appearance of the weak ferromagnetism at 4.2 K was attributed to the antisymmetric exchange ($H_\textrm{D}$) and single-ion anisotropy ($H_\textrm{A}$) with $|\frac{H_\textrm{A}}{H_\textrm{D}}| < 0.05$. The applied magnetic-field induced SR in YCrO$_3$ was confirmed by Sugano \emph{et al}. \cite{Sugano1971-1, Sugano1971-2} via observing the $t_2^3$ $^4A_2$ $\rightarrow$ $t_2^3$ $^2E$ single-ion excitations \cite{Sugano1971-1}. In this study, a magnetic field was applied along the crystallograhpic \emph{a} axis up to 50 kOe. With high magnetic fields ($>$ 40 kOe), the spectrum resembles the selection rule of \emph{G}$_z$-type AFM state, which indicates that there exists a magnetic-field-induced SR transition from crystallographic \emph{a}-axis to \emph{c}-axis \cite{Sugano1971-1, Sugano1971-2}.

\textbf{(7) Neutron diffraction study.} A NPD study \cite{Zhu2020-2} revealed that below $T_\textrm{N}$, YCrO$_3$ forms an AFM structure (see \textcolor[rgb]{0.00,0.00,1.00}{Fig.}~\ref{mag-3}(b)), and the propagation vector was determined as $\textbf{k}$ = (1 1 0) as displayed in \textcolor[rgb]{0.00,0.00,1.00}{Fig.}~\ref{mag-3}(c). A neutron diffraction study was performed on a YCrO$_3$ single crystal with $(H K 0)$ as the scattering plane \textcolor[rgb]{0.00,0.00,1.00}{Fig.}~\ref{mag-3}(d) \cite{Zhu2020-2}. The magnetic Bragg (110) reflection was observed at 2 K, implying an appearance of the AFM structure. The power law can be expressed by the following equation:
\begin{eqnarray}
I(T) = I_0{\left(\frac{|T-T_\textrm{N}|}{T_\textrm{N}}\right)}^{2\beta}.
\label{4}
\end{eqnarray}
This was used to fit the integrated intensity of magnetic (110) Bragg peak as shown in \textcolor[rgb]{0.00,0.00,1.00}{Fig.}~\ref{mag-3}(d). The critical exponent was extracted, that is $2\beta = 0.215(6)$, indicative of a phase transition with the second-order type. Thus, the spin interactions within the (110) scattering plane could probably be a two-dimensional Ising type \cite{Zhu2020-2}. The moment size was refined as 2.45(6) $\mu_\textrm{B}$ at 12 K for Cr$^{3+}$ ions, which is about 82$\%$ of the corresponding theoretical value (3 $\mu_\textrm{B}$). This is in agreement with that a magnetic frustration exists in YCrO$_3$ compound \cite{Zhu2020-2}.

\textbf{(8) High-pressure study.} The temperature-dependent ac susceptibility of YCrO$_3$ ceramic samples synthesized via traditional solid state reactions was measured by Zhou and Goodenough \cite{Zhou2002-2}. The measurements were under pressure of 0--20 kbar, and the values of $T_\textrm{N}$ were tracked. The curve of $T_\textrm{N}$ versus pressure observes the Bloch rule for localized-electron antiferromagnetism. Pressure does not change the AFM ordering in YCrO$_3$, but linearly increases $T_\textrm{N}$ over the entire pressure range \cite{Zhou2002-2}.

\textbf{(9) Theoretical study.} The connection between structural instabilities and magnetic ordering of YCrO$_3$ compound was studied theoretically by Ray and Waghmare, using first-principles calculations based on density functional theory, trying to correlate the nonmagnetic, FM, and AFM states with structural features \cite{Ray2008}. The calculated relative energies of nonmagnetic, FM, and AFM states are 0 eV, -2.02 eV, and -2.4 eV, respectively. Therefore, the \emph{G}-type AFM order is most stable. The main factors that lead to structural instabilities are angle values of Cr-O-Cr bonds and rotations of octahedra (CrO$_6$). These are coupled strongly with the formation of magnetic ordering. The calculations show that the local noncentrosymmetry and the small ferroelectric polarization are probably due to weak ferroelectric (WFM) instabilities as well as their competition with various magnetic and structural instabilities \cite{Ray2008}.

Nair \emph{et al}. studied the magnetic and electronic properties of YCrO$_3$ compound by first-principles calculations \cite{Nair2013}. The calculations were based on generalized gradient approximation by ultra-soft pseudo-potential with plane wave basis. Structural optimization revealed that the relative total energies of different types of magnetic structure are: 0 meV/f.u. (FM), -23.5 meV/f.u. (\emph{A}-type AFM), -23.4 meV/f.u. (\emph{C}-type AFM), and -39.9 meV/f.u. (\emph{G}-type AFM). Therefore, the most stable magnetic structure is the \emph{G}-type AFM. The insulating YCrO$_3$ compound has a 1.43 eV band gap. For all types of magnetic structure, the calculated spin moment size equals $\sim$ 3 $\mu_\textrm{B}$ for the Cr site. In order to consider correlated electron interactions, the introduction of Hubbard \emph{U} parameter (\emph{U} = 1 eV) results in an increased band gap from 1.43 to 1.64 eV \cite{Nair2013}. Based on the Heisenberg model taking into account the nearest neighbor spin couplings only, the \emph{G}-type magnetic ordering temperature was simulated to be 137 K for YCrO$_3$ by mean field theory \cite{Nair2013}. This temperature is quite close to the experimentally determined one ($\sim$ 140 K).

The re-normalized phonon energy of RECrO$_3$ (RE = Sm, Er, Dy, Gd, Pr, and Y) multiferroic compounds versus magnetic field, temperature, and RE-ionic radius was calculated using a Green's function with a microscopic model by Apostolov \emph{et al}. \cite{Apostolov2017}. The calculation showed that when the RE-site is magnetic, spin-phonon coupling is essential for the SR transition on lattice vibrations; when RE-site is Y, a nonmagnetic element, a new microscopic model was proposed for YCrO$_3$, indicating that the induced DM interaction leads to the spontaneous polarization and temperature induced SR phase transition \cite{Apostolov2017}.

Magnetic properties of YCrO$_3$ bulk perovskite were studied by calculations with Monte Carlo simulations and density functional theory by Ahmed \emph{et al}., showing that YCrO$_3$ is a 3D Heisenberg antiferromagnet with two-sublattice model \cite{Ahmed2019}. First, the volume parameter in a cubic $Pm3m$ structure was optimized. The refined lattice parameter \emph{a} = 3.78 {\AA}. The calculations show that the crystallographic $<$110$>$ is the easy magnetic direction with minimum energy. Taking into account the 6 nearest and 12 next-nearest neighbours, the exchange coupling constants were calculated as $J_1 =$ 13.9 meV, and $J_2 =$ 6.74 meV. There exists a low value of the single-ion anisotropic parameter. The values of orbital and spin moments, i.e., $-$0.094 $\mu_\textrm{B}$ and 2.38 $\mu_\textrm{B}$, respectively, were also calculated, indicating that the orbital moment has a small contribution to the total magnetic moment \cite{Ahmed2019}.

\textbf{(10) Other measurements.} The fluorescence spectrum of a YCrO$_3$ single crystal (space group: $Pbnm$) grown by the flux method was measured at 4.2 K by Ziel and Uitert \cite{Ziel1969-1, Ziel1969-3}, from which values of the spin exchange interactions were extracted with the magnon sidebands that are from the emission of an optical photon and the creation of a magnon at a neighboring lattice site. The intrinsic $^4A_2$-to-$^3E$ emission was not observed in the fluorescence spectrum of YCrO$_3$. At slightly lower energies, the sharp emission lines were attributed to transitions of $^2E$-to-$^4A_2$ of impurity-perturbed chromium ions. A spin-wave model was constructed, taking into account the six isotropic exchange interactions around one Cr$^{3+}$ ion. In this case, $J_1$ and $J_2$ are the exchange energies along the crystallographic \emph{c} axis and in the \emph{ab} plane (see \textcolor[rgb]{0.00,0.00,1.00}{Fig.}~\ref{mag-1}(b)), respectively, neglecting the higher-order neighbour exchanges and the net FM-moment component along the crystallographic \emph{c} axis (see \textcolor[rgb]{0.00,0.00,1.00}{Fig.}~\ref{mag-1}(c)). By fitting the strongest chromium emission peak and the two magnon sideband peaks, the best results were obtained with considering the nearest-neighbor exchanges only (Hamiltonian $H = +J_{ij}$S$_i{\cdot}$S$_j$), resulting in $J_1$ = 24.8 cm$^{-1}$, $J_2$ = 20.8 cm$^{-1}$, and zero anisotropy \cite{Ziel1969-1, Ziel1969-3}.

The AFM resonance of YCrO$_3$ single crystals was measured along the \emph{a}-, \emph{b}-, and \emph{c} axes using pulse magnetic fields with $\lambda$ = 1--6 mm by Sanina \emph{et al}. \cite{Sanina1970}. By fitting the data with a two-sublattice model, the effective field of anisotropic exchange interactions was exacted as $H_\textrm{D} =$ 74(3) kOe, that of the single-ion anisotropic interactions $H_\textrm{A} =$ 3.8(5) kOe, and the components of the \emph{g}-factor $g_{xx} \approx g_{zz} = g =$ 2.07(2) \cite{Sanina1970}.

The magnetic ordering of YCrO$_3$ poly-crystalline samples was studied by microwave power absorption measurements \cite{Alvarez2010}. The electron paramagnetic resonance study was performed in a temperature range from 85 to 300 K. Within the entire temperature range, the spectra display a broad and symmetric Lorentzian single line that was attributed to Cr$^{3+}$ ions, indicating that at $\sim$ 139 K, a magnetic transition occurs, entering an AFM state from the paramagnetic one. A weak ferromagnetism was observed below 139 K, which was ascribed to a canting of magnetic Cr$^{3+}$ sublattices in the AFM matrix. The temperature-dependent \emph{g}-factor shows a value of $\sim$ 1.86 which is smaller than 2.0023 for a free electron \cite{Alvarez2010}.

The magnetic anisotropy of YCrO$_3$ powders was studied by measuring high-field electron spin resonances \cite{Ikeda2015}. The \emph{g}-factor value was extracted as 1.983 at 265 K for the paramagnetic state. This was used to calculate the magnitude of \textbf{\emph{D}} vector (as illustrated in \textcolor[rgb]{0.00,0.00,1.00}{Fig.}~\ref{Structure-1}(a)) of DM interactions from Moriya's formula. As listed in \textcolor[rgb]{0.00,0.00,1.00}{Table}~\ref{SM-parameters}, the calculated \emph{D}-value = 0.36 K, and a magnetic field for the spin-flop transition was estimated as 3.5 T \cite{Ikeda2015, HFLi2016}. The magnetic structure of YCrO$_3$ is not consistent with a two-sublattice antiferromagnet, suggesting an unconventional AFM ordering \cite{Ikeda2015}.

The electron paramagnetic resonance was measured on high-quality poly-crystalline YCrO$_3$ samples in a temperature range from 120 to 298 K while warming the samples \cite{Mall2017-1}. The spectra parameters consist of line width, integrated intensity, and \emph{g}-factor. The variation of \emph{g}-factor with temperature indicates that a new magnetic phase occurs at $\sim$ 230 K (above $T_\textrm{N} \sim$ 140 K) in the paramagnetic state. This was ascribed to the appearance of short range CAFM correlations. The anomaly observed at $\sim$ 230 K also appears in the temperature-dependent dielectric measurements (see \textcolor[rgb]{0.00,0.00,1.00}{Fig.}~\ref{FerriPoly-1}(b)), indicating the presence of a magneto-dielectric coupling in YCrO$_3$ \cite{Mall2017-1}. Furthermore, the study also implies the existence of Cr$^{4+}$ ions (see \textcolor[rgb]{0.00,0.00,1.00}{Fig.}~\ref{Crstate-3}(a)) and a possible spin-orbital coupling in YCrO$_3$ \cite{Mall2017-1}.

The optical spectroscopy and electron paramagnetic resonance of YCrO$_3$ nanoparticles (roughly spherical shape, size: 100--200 nm) were monitored by Jara \emph{et al} \cite{Jara2018}. The reflectance spectrum was diffuse and displayed well-defined bands corresponding to the $^4A_2$ $\rightarrow$ $^4T_1$, $^4A_2$ $\rightarrow$ $^4T_2$, $^4A_2$ $\rightarrow$ $^2E$, and $^4A_2$ $\rightarrow$ $^2T_1$ transitions. These are characteristic features of Cr$^{3+}$ ions ($^4F_{3/2}$, 3$d^3$) in a crystal field of octahedral symmetry. The paramagnetic Cr$^{3+}$ ions have an effective \emph{g}-values of 1.97--1.98. This value is typical for Cr$^{3+}$ ions in an orthorhombic ligand. The study showed that the AFM transition at $\thicksim$ 140 K was ascribed to the exchange interactions of Cr$^{3+}$-Cr$^{3+}$ ions with \emph{J} = $-$41.5 cm$^{-1}$ \cite{Jara2018}.

In the review, we summarize the structural and magnetic properties of different forms of YCrO$_3$ compound in \textcolor[rgb]{0.00,0.00,1.00}{Table}~\ref{SM-parameters}.

\subsection{Ferroelectric and dielectric}

Until now, there has been a long debate on whether YCrO$_3$ is multiferroic or not \cite{Saha2014}.

\begin{figure*} [!t]
\centering \includegraphics[width=0.61\textwidth]{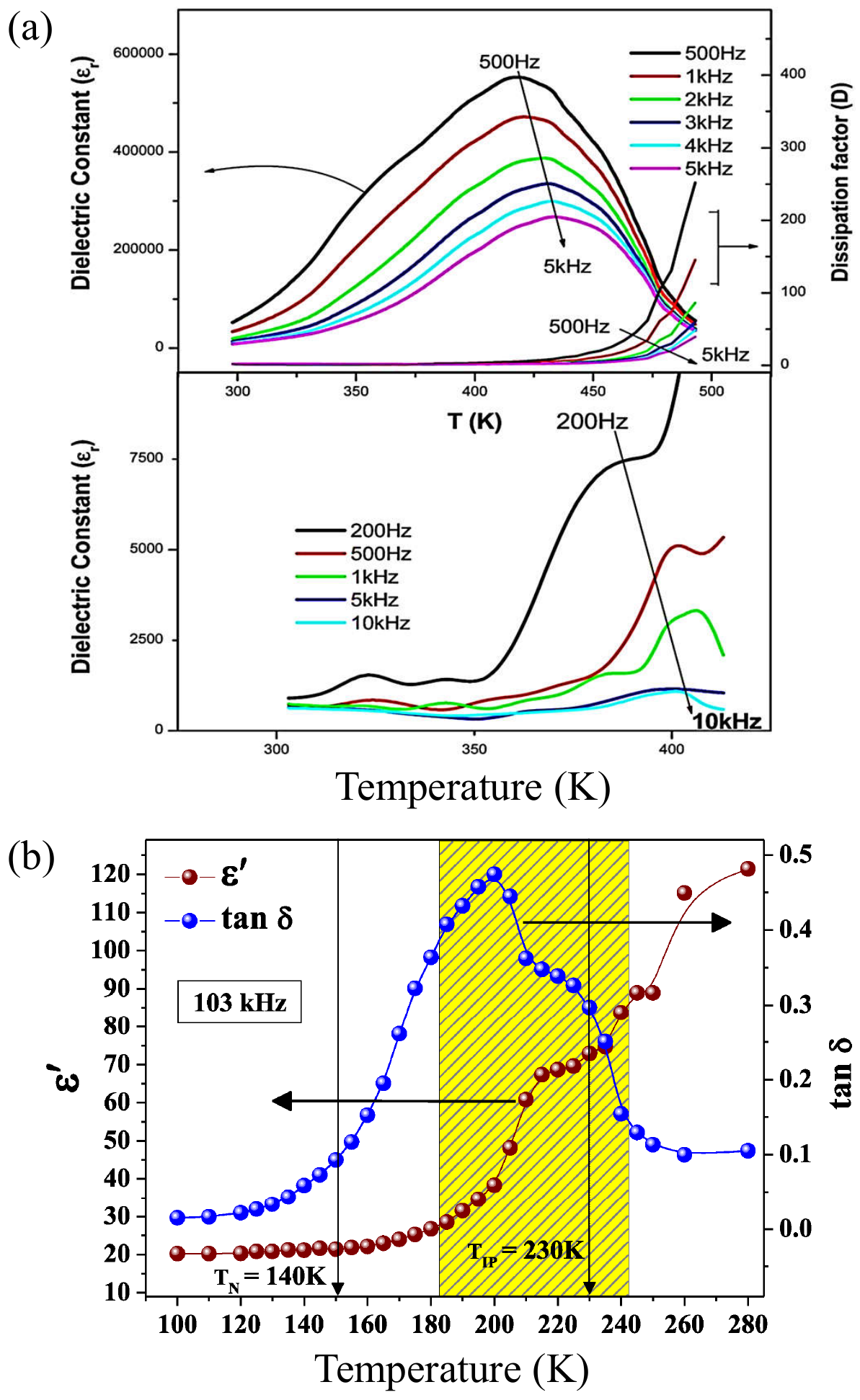}
\caption{
(a) Temperature dependent dielectric constant of poly-crystalline (top panel) and thin-film (bottom panel) YCrO$_3$ samples at different frequencies. Reproduced with permission from Ref. \cite{Serrao2005}. Copyright (2005) The American Physical Society.
(b) Real part ($\varepsilon'$) of the dielectric constant (left) and tangent loss (tan$\delta$) (right) of a YCrO$_3$ polycrystal versus temperature across $T_\textrm{N}$ and $T_{\texttt{IP}}$ (an intermediate point of temperature at which a new magnetic phase appears in the paramagnetic state) at 103 kHz. Reproduced with permission from Ref. \cite{Mall2017-1}. Copyright (2017) IOP Publishing Ltd.
}
\label{FerriPoly-1}
\end{figure*}

\begin{figure*} [!t]
\centering \includegraphics[width=0.82\textwidth]{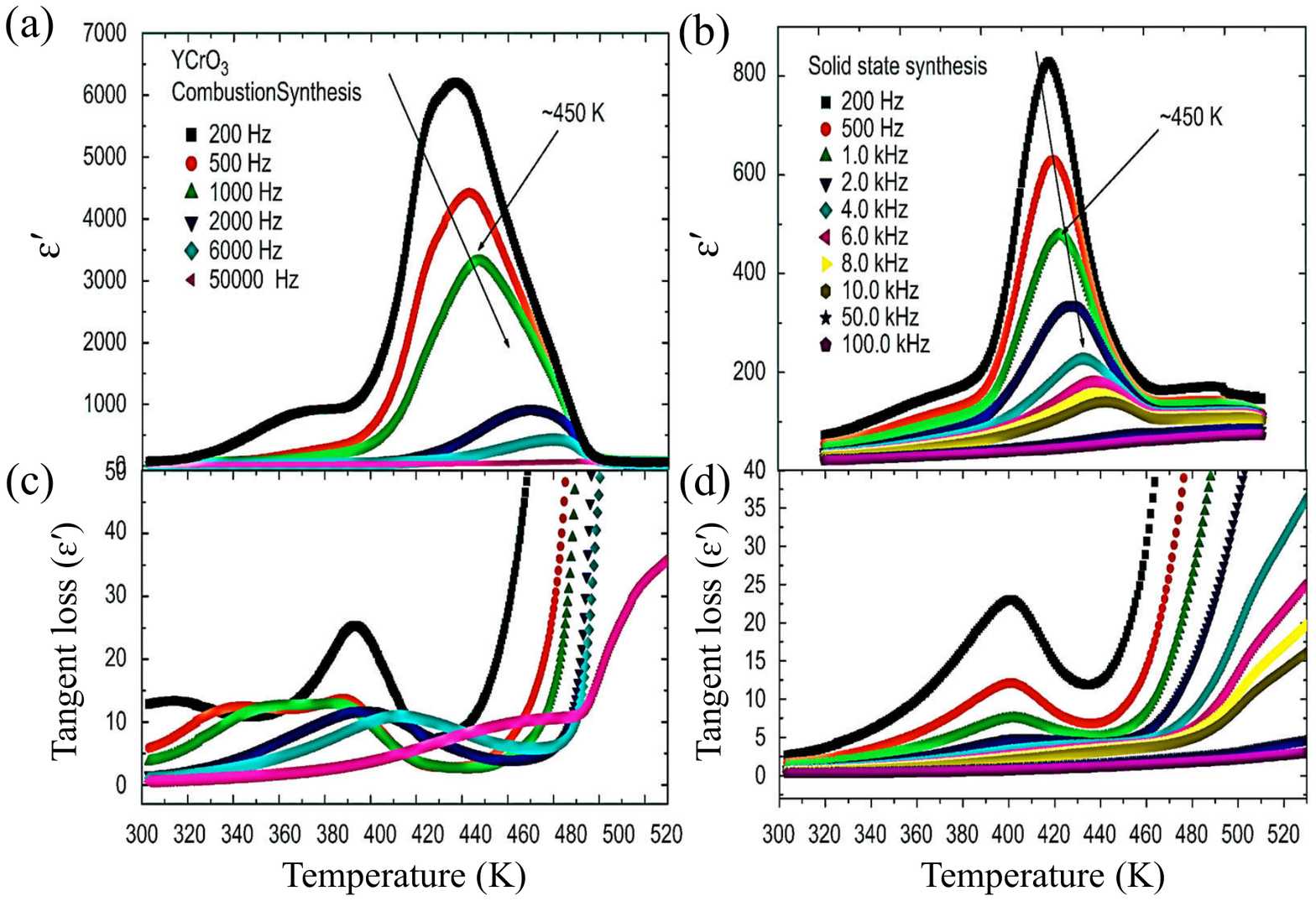}
\caption{
Dielectric constants (a, b) and the corresponding dielectric losses (c, d) versus temperature at different frequencies of the combustion YCrO$_3$ sample (a, c), as well as the YCrO$_3$ sample prepared by solid state reaction (b, d). Reproduced with permission from Ref. \cite{Duran2010}. Copyright (2010) Elsevier Inc.
}
\label{FerriPoly-2}
\end{figure*}

\begin{figure*} [!t]
\centering \includegraphics[width=0.82\textwidth]{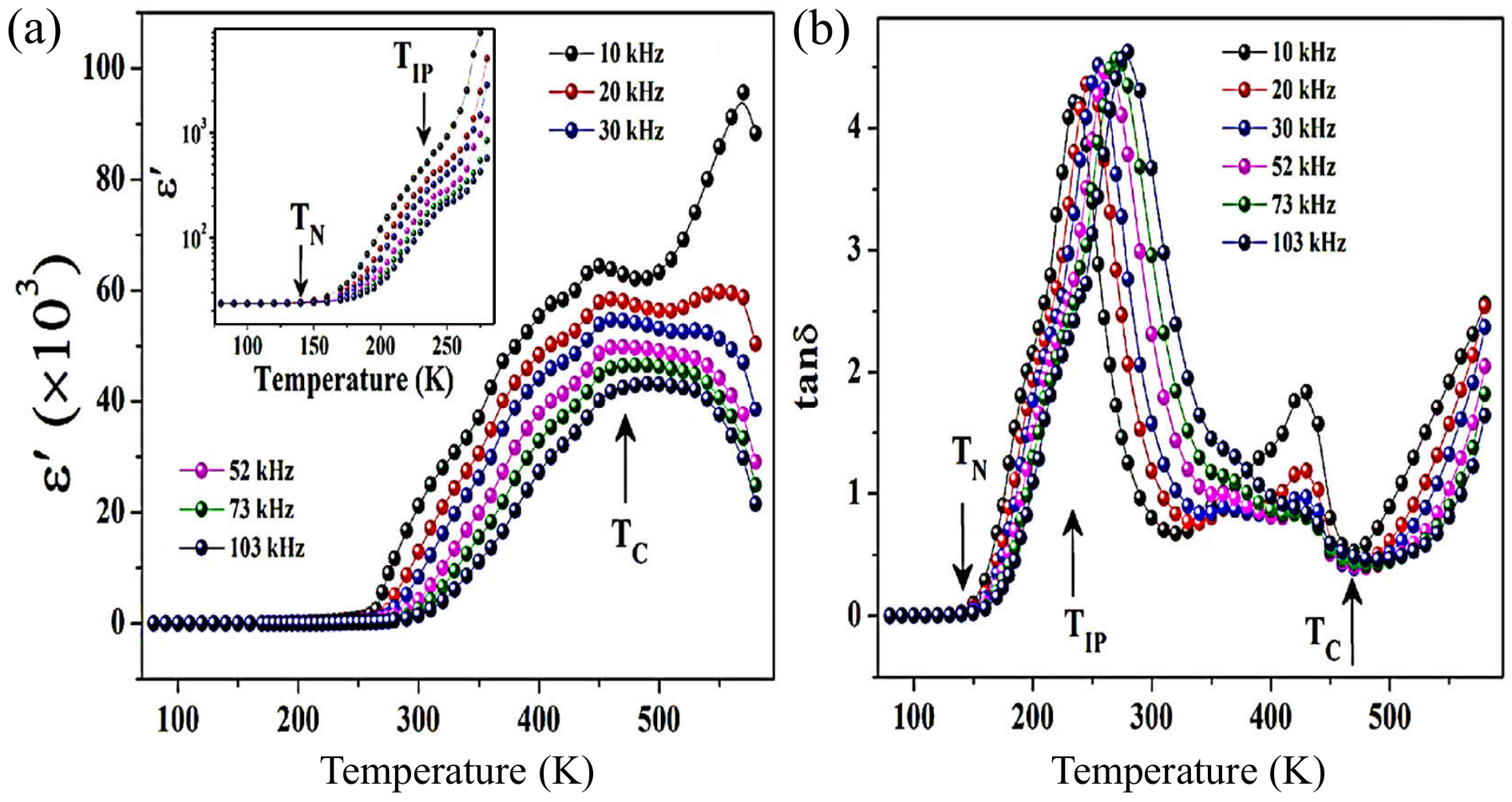}
\caption{
Frequency dependent variation of real dielectric constant versus temperature plot (a) and tangent loss versus temperature plot (b) of a poly-crystalline YCrO$_3$ sample. Reproduced with permission from Ref. \cite{Mall2018-2}. Copyright (2018) Elsevier Ltd.
}
\label{FerriPoly-3}
\end{figure*}

\begin{figure*} [!t]
\centering \includegraphics[width=0.78\textwidth]{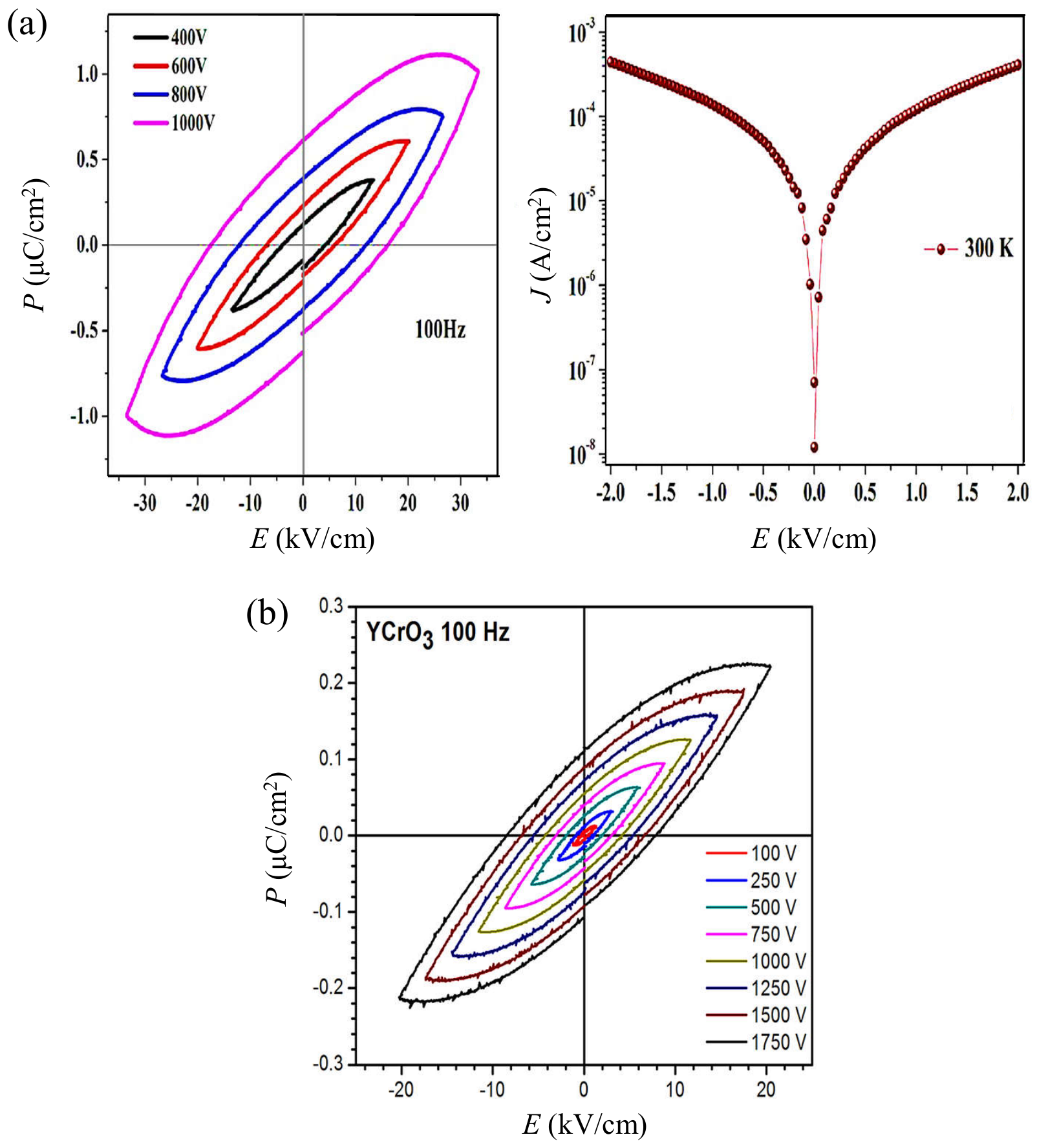}
\caption{
(a) Electric field-dependent ferroelectric hysteresis loops (left panel) and current leakage versus electric field (i.e., \emph{J}-\emph{E} plot) (right panel), measured at room temperature and 100 Hz for poly-crystalline YCrO$_3$ ceramics. Reproduced with permission from Ref. \cite{Mall2018-2}. Copyright (2018) Elsevier Ltd.
(b) Evolution of the room-temperature $P$-$E$ hysteresis loops of nano-crystalline YCrO$_3$ compound at 100 Hz and various applied voltages. The loops display similar features as those shown in a relaxor ferroelectric. Reproduced with permission from Ref. \cite{Singh2013}. Copyright (2018) AIP Publishing LLC.
}
\label{PEloop-2}
\end{figure*}

\textbf{(1) Polycrystals.} Rao \emph{et al}. suspected that some poly-crystalline rare-earth-based orthochromates like YCrO$_3$ probably show the biferroic property, which is not compatible with the centrosymmetric $Pbnm$ space group \cite{Rao1968}. Serrao \emph{et al}. studied the dielectric properties of pressed pellets of YCrO$_3$ (see \textcolor[rgb]{0.00,0.00,1.00}{Fig.}~\ref{FerriPoly-1}(a)) \cite{Serrao2005}. The dielectric transition occurs at $T_\textrm{C} =$ 473 K in a pellet over a temperature range from 410 to 440 K. The maximum occurs at 418 K and 500 Hz, which is frequency dependent. The dielectric-constant distribution displays a huge dispersion below $T_\textrm{C}$, and it is frequency independent above $T_\textrm{C}$, indicating that YCrO$_3$ is a relaxor ferroelectric \cite{Serrao2005}. The effects of oxygen vacancies and local polar regions on these observed relaxor-like phenomena were not ruled out \cite{Serrao2005}. Dur\'{a}n \emph{et al}. measured the dielectric permittivity of YCrO$_3$ powders with temperature, where a broad peak was centered at 40 K in the real part (see \textcolor[rgb]{0.00,0.00,1.00}{Fig.}~\ref{FerriPoly-2}) \cite{Duran2010}. The peak intensity decreases, and the peak position shifts to elevated temperatures as frequency increases. There exists a broad ferroelectric transition around $\sim$ 450 K, characteristic of a relaxor whose dipoles were attributed to the local non-centric symmetry \cite{Duran2010}. Mall \emph{et al}. studied dielectric properties of YCrO$_3$ ceramics \cite{Mall2016}. The dielectric constant, ac conductivity, and tangent loss as a function of frequency increase with temperature. A relaxor-like transition happens at $\sim$ 460 K in the dielectric measurements \cite{Mall2016}. AC conductivity measurements show an 0.25 eV activation energy. It was suggested that vacancies of oxygen ions play a crucial role in the conduction process albeit with the polaron hopping at elevated temperatures \cite{Mall2016}. Mall \emph{et al}. studied the temperature-dependent loss tangent (tan$\delta$) and real dielectric constant ($\varepsilon'$) of poly-crystalline YCrO$_3$ as shown in \textcolor[rgb]{0.00,0.00,1.00}{Figs.}~\ref{FerriPoly-1}-\ref{FerriPoly-3} \cite{Mall2017-1}. Upon warming, there is a sudden increase in $\varepsilon'$ above $T_\textrm{N}$, and an anomaly appears in $\varepsilon'(T)$ curve around $T_\textrm{IP} \sim$ 230 K, suggesting a new phase formed in the paramagnetic state probably due to an onset of short-range CAFM correlations. The observed dielectric anomaly around $T_\textrm{IP}$ may indicate a coupling between lattice and spin orders of degree mediated by magneto-dielectric correlations \cite{Mall2017-1}. Mall \emph{et al}. performed a study of detailed temperature- and frequency-dependent dielectric behaviours of poly-crystalline YCrO$_3$ ceramics \cite{Mall2018-2}. The XRPD study as well as the corresponding structural refinement demonstrated a single phase with space group of $Pnma$ (orthorhombic) as displayed in \textcolor[rgb]{0.00,0.00,1.00}{Fig.}~\ref{Structure-2}(c). As shown in \textcolor[rgb]{0.00,0.00,1.00}{Fig.}~\ref{FerriPoly-3}, the temperature-dependent real part of dielectric permittivity measurements revealed a dielectric relaxation peak around $T_{\textrm{IP}} \sim$ 230 K (the intermediate point of temperature). This was related to the ferroelectric ordering in YCrO$_3$ \cite{Mall2018-2}. Moreover, there exists a broad peak with maximum located at $T_{\textrm{C}} \sim$ 450 K. The strength of the maximum peak weakens, and its position moves to elevated temperatures with increasing frequency, resembling the behaviors of a YCrO$_3$ relaxor \cite{Mall2018-2}. Additionally, one peak sitting around $\sim$ 570 K was also observed \cite{Mall2018-2}. The dc conductivity also shows abrupt changes in the temperature-dependent activation energy around 230 and 450 K, and its nature is of Arrhenius type. The activation-energy study indicated that the hopping mechanism of polarons gets stable at low temperatures; the process at elevated temperatures is associated with the diffusion of double ionized oxygen vacancies \cite{Mall2018-2}. In contrast, the ac conductivity study suggested that below 300 K, the overlapping of tunnelling conduction mechanism from large polarons induces the ac conduction; above 300 K, it is coherent with the mechanism of coupled barrier hopping conduction \cite{Mall2018-2}. The \emph{P}-\emph{E} loop was also measured at applied voltages of 400--1000 V and a frequency of 100 Hz as shown in \textcolor[rgb]{0.00,0.00,1.00}{Fig.}~\ref{PEloop-2}(a) \cite{Mall2018-2}. Unfortunately, the loops seem to be unsaturated, indicating strong electrical leakage effect. The leakage current was determined as $\sim$ 10$^3$ A/cm$^2$ at 2 kV/cm (the maximum field) as shown in the right panel of \textcolor[rgb]{0.00,0.00,1.00}{Fig.}~\ref{PEloop-2}(a). The room-temperature remnant polarization 2$P_\textrm{r} \approx$ 1.1 $\mu$C/cm$^2$; at 1000 V, the coercive-field 2$E_\textrm{C} \approx$ 33.8 kV/cm \cite{Mall2018-2}. Similar $P$-$E$ hysteresis loops were also observed in nano-crystalline YCrO$_3$ samples at room temperature (see \textcolor[rgb]{0.00,0.00,1.00}{Fig.}~\ref{PEloop-2}(b)) \cite{Singh2013}.

\textbf{(2) Nanocrystals.} Apostolov \emph{et al}. investigated theoretically the dielectric property of YCrO$_3$ and GdCrO$_3$ nanoparticles \cite{Apostolov2019}, which was compared to the corresponding bulk properties. The dielectric constant $\epsilon'$ of bulk YCrO$_3$ has a smaller value than that of nanoparticles, however, it was observed an opposite behavior in GdCrO$_3$ compound. The magnetic field influences $\epsilon'$ in YCrO$_3$ nanoparticles, indicating a possible existence of the strong magnetoelectric coupling \cite{Apostolov2019}.

\textbf{(3) Thin films.} The dielectric phase transition was observed in thin-film YCrO$_3$ around 400 K at 500 HZ as displayed in of \textcolor[rgb]{0.00,0.00,1.00}{Fig.}~\ref{FerriPoly-1}(a) (the bottom panel) \cite{Serrao2005}. During these measurements, there exists the dc current leakage. The \emph{P}-\emph{E} loop was measured, displaying a presence of hysteresis (see \textcolor[rgb]{0.00,0.00,1.00}{Fig.}~\ref{PEloop-1}(c)). For YCrO$_3$ pellets, the polarization has a maximum value of $\sim$ 2 $\mu$C/cm$^2$ near room temperature; in contrast, for thin films, the maximum polarization is $\sim$ 3 $\mu$C/cm$^2$ at 178 K \cite{Serrao2005}. Cruz \emph{et al}. investigated the thickness-dependent piezoelectric and ferroelectric properties of thin-film YCrO$_3$ samples \cite{Cruz2014}. The 20--180 nm YCrO$_3$ thin films were synthesized by r.f. sputtering on Pt/TiO$_2$/Si substrates at room temperature. The thin-film samples were subsequently annealed in air at 900 $^\circ$C. As displayed in \textcolor[rgb]{0.00,0.00,1.00}{Fig.}~\ref{Structure-2}(b), the 550 nm film as synthesized at room temperature is amorphous. After annealed at 800 and 900$^{\circ}$C in air, poly-crystalline YCrO$_3$ phase was present as evidenced by the sharpness of the Bragg (200) and (002) reflections \cite{Cruz2014}. In the thinner films, there were better decrease of charge accumulation at grain boundaries and grain coalescence, resulting in an improved piezoelectric property. For example, for the 20 nm film, the coercive voltage is $\sim$ 12.5 V ($V_\textrm{c}$), and the piezoelectric coefficient is $\sim$ 6.4 pm/V ($d_{33}$). Moreover, the thin film (20 nm) displays a magnetic hysteresis loop, characteristic of a ferromagnet. Therefore, it was demonstrated that when the thickness of thin films was decreased down to 20 nm, YCrO$_3$ maintains multiferroicity \cite{Cruz2014, Arciniega2018}.

\begin{figure*} [!t]
\centering \includegraphics[width=0.78\textwidth]{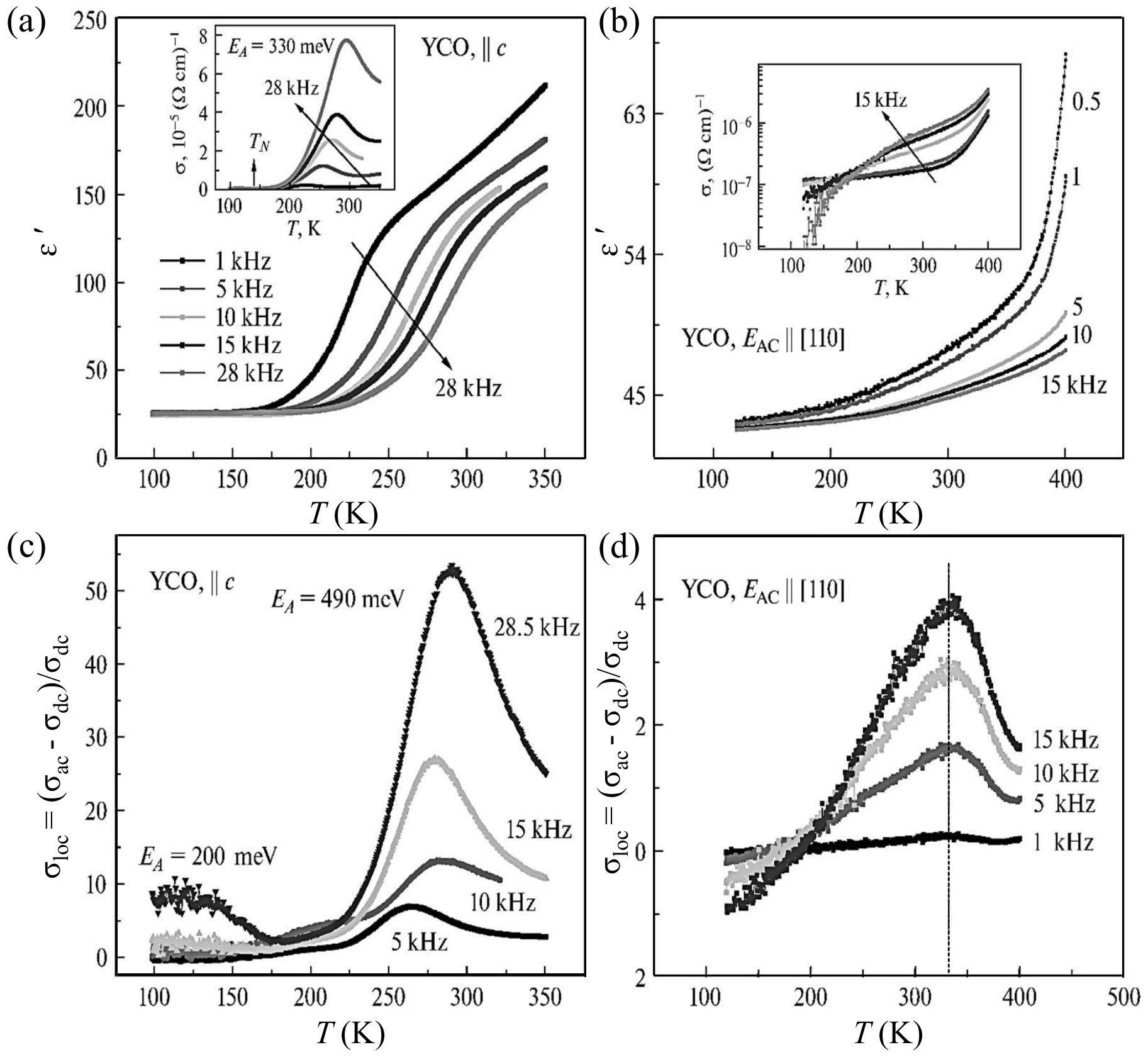}
\caption{
Temperature dependences at various frequencies of the dielectric permittivity $\varepsilon'$ along the \emph{c}-axis (a) and in the transverse (110) plane (b) and local conductivity $\sigma_{\textrm{loc}}$ along the \emph{c}-axis (c) and in the transverse (110) plane (d) of a YCrO$_3$ single crystal (space group: $Pbnm$) grown by the flux method. Reproduced with permission from Ref. \cite{Sanina2018}. Copyright (2018) Pleiades Publishing, Ltd.
}
\label{FerriSingle-2}
\end{figure*}

\begin{figure*} [!t]
\centering \includegraphics[width=0.82\textwidth]{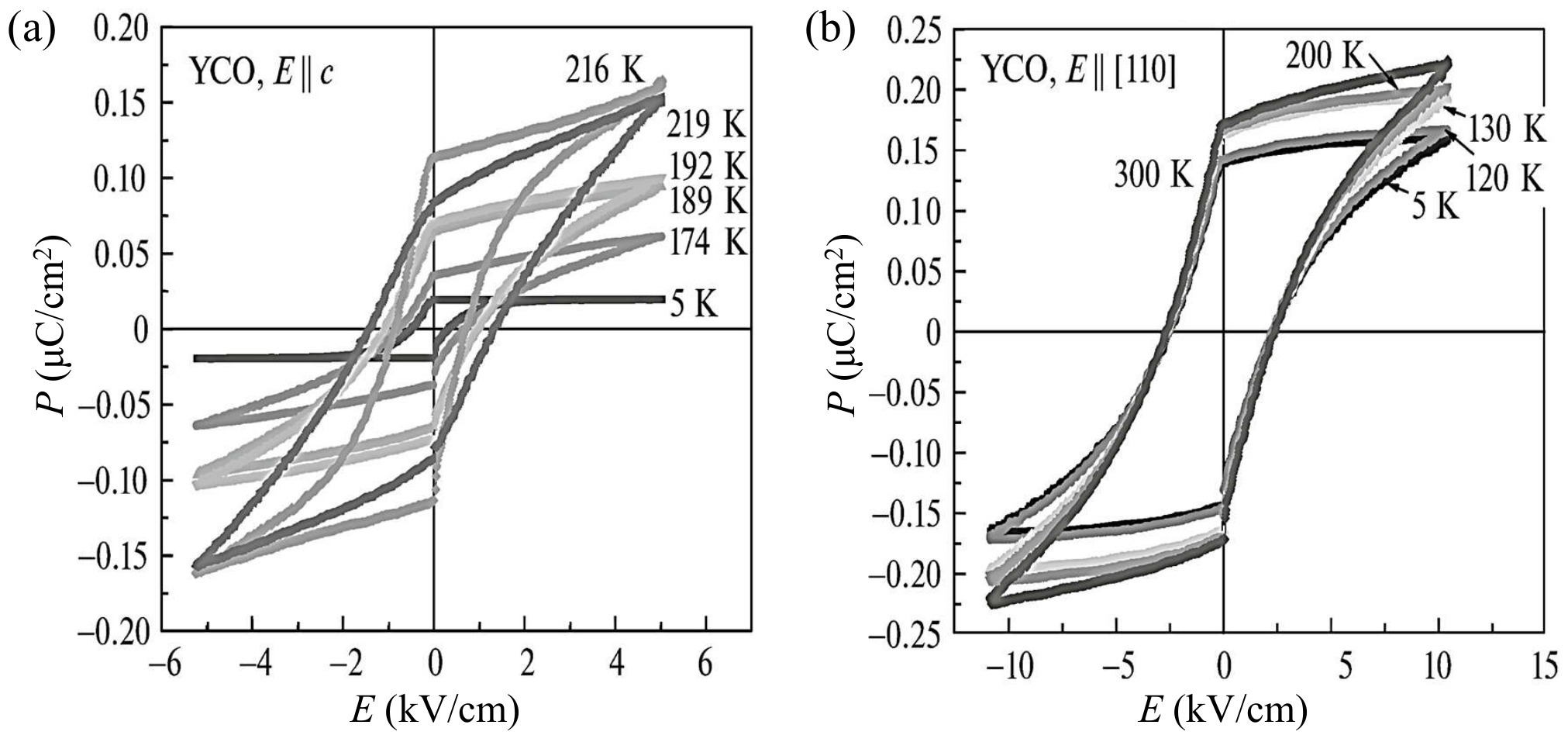}
\caption{
Electric polarization $P$-$E$ loops at marked temperatures along the crystallographic \emph{c} axis (a) and in the transverse (110) plane (b) of a YCrO$_3$ single crystal (space group: $Pbnm$) grown by the flux method. Reproduced with permission from Ref. \cite{Sanina2018}. Copyright (2018) Pleiades Publishing, Ltd.
}
\label{PEloop-3}
\end{figure*}

\textbf{(4) Single crystals.} Sanina \emph{et al}. for the first time studied the electric polarization in YCrO$_3$ single crystals, trying to figure out the polar configuration as well as its nature \cite{Sanina2018}. The study found that the local electric polarization in single-crystal YCrO$_3$ is along with the [110] direction and the crystallographic \emph{c} axis of $Pbnm$ symmetry, and the corresponding hysteresis loops with remanent polarization were shown in \textcolor[rgb]{0.00,0.00,1.00}{Fig.}~\ref{PEloop-3}. The appearance of the local electric polarization relies on the direction of applied-electric field relative to the crystallographic axes, and it seems that there is no coupling between local electric polarization and antiferromagnetism (see \textcolor[rgb]{0.00,0.00,1.00}{Fig.}~\ref{FerriSingle-2}). It was argued that Cr$^{2+}$ ions appear in the grown single crystal. In the measured temperature dependence of $\epsilon'$ (dielectric constant) and local conductivity $\sigma$ of single crystalline YCrO$_3$ at different frequencies, no free dispersion maximum $\epsilon'$ and $\sigma$ were observed over the whole temperature range in the transverse plane (110) and along with the crystallographic \emph{c} axis (see \textcolor[rgb]{0.00,0.00,1.00}{Fig.}~\ref{FerriSingle-2}), which indicates that no ferroelectric phase transitions exist in the single crystalline YCrO$_3$ sample up to 350--400 K. Therefore, no homogeneous and intrinsic ferroelectricity was found in the single crystalline YCrO$_3$ samples from 5 to 350 K in the study \cite{Sanina2018}.

\textbf{(5) Amorphous.} Liz\'{a}rraga \emph{et al}. predicted theoretically the structure of amorphous YCrO$_3$ compound by first-principles calculations \cite{Lizarraga2012}. The study shows that the Cr$^{3+}$ ions sit in the slightly-distorted oxygen octahedra within the amorphous structure. The distribution of these octahedra is disordered. It was suggested that the amorphous YCrO$_3$ phase may display ferroelectric properties resulting from the presence of the same Cr$^{3+}$ local environments \cite{Lizarraga2012}.

\subsection{Optical}

Infrared transmittance spectra of YCrO$_3$ ceramics were measured by Patil \emph{et al}. \cite{Patil1988}. In YCrO$_3$ compound, the stretching vibration band of Cr-O bonds was located at 610 cm$^{-1}$, and the Y-O stretching vibration occurs at $\sim$ 490 cm$^{-1}$. Therefore, the observed absorption bands were correlated to the basic lattice vibrations of the oxide ions rather than the cations, consistent with the covalency of Cr-O and Y-O bonds. These vibration bands were compared to that of Cr$_2$O$_3$, Y$_2$O$_3$, and Sr-doped YCrO$_3$ compounds \cite{Patil1988}.

\begin{figure*} [!t]
\centering \includegraphics[width=0.78\textwidth]{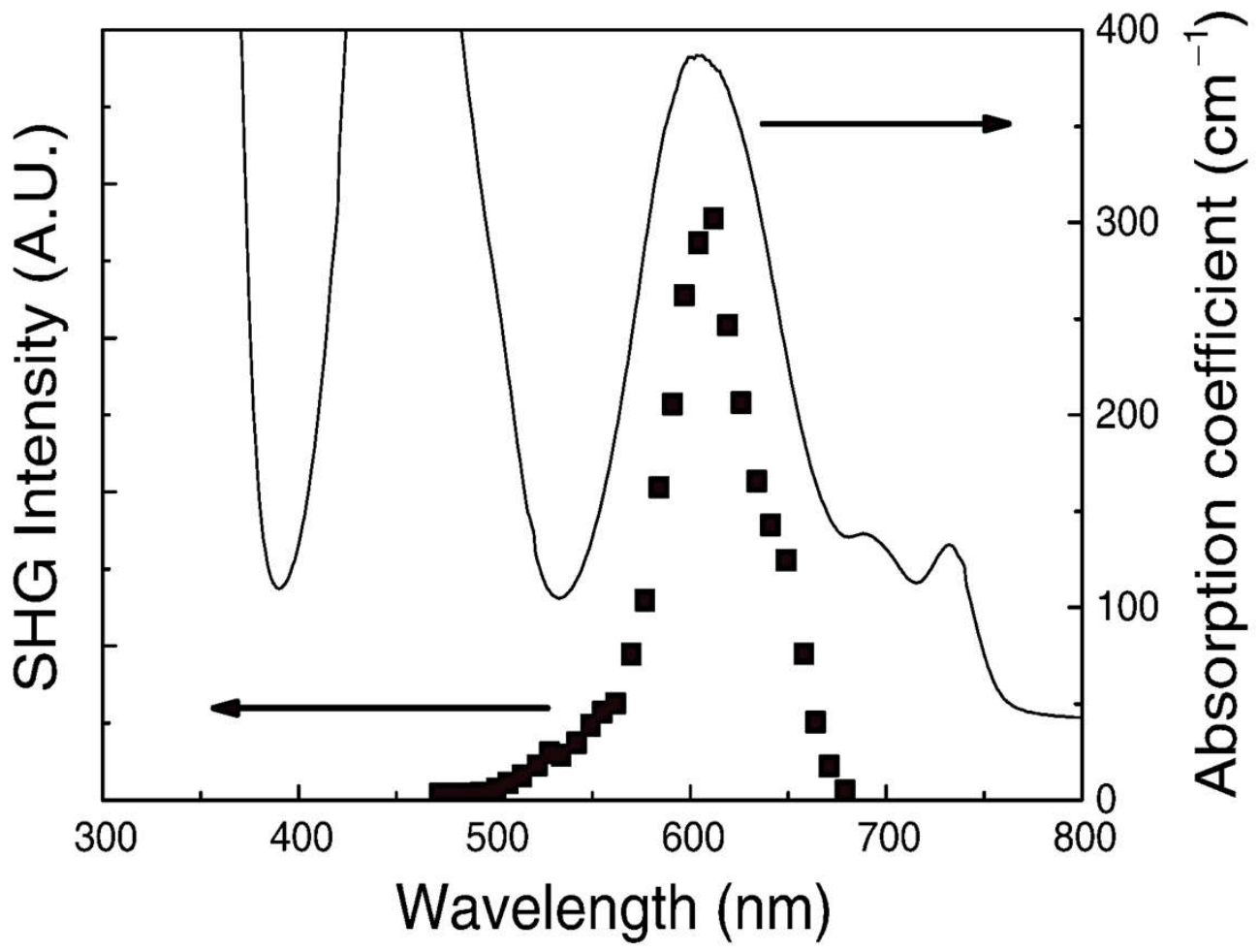}
\caption{
Absorption and the second harmonic generation spectra of YCrO$_3$ observed at room temperature. Reproduced with permission from Ref. \cite{Eguchi2005-1}. Copyright (2005) The Physical Society of Japan.
}
\label{Application-1}
\end{figure*}

The spectrum of second harmonic generation of YCrO$_3$ single crystals was measured by Eguchi \emph{et al}., and based on its polarization and spectral features, it was correlated with a magnetic dipole transition \cite{Eguchi2005-1}. The second-order optical nonlinearity, i.e., the second harmonic generation, was believed to be the most powerful method for detecting the detail of a magnetic structure that cannot be understood by magneto optical and conventional magnetic measuring methods \cite{Eguchi2005-1}. The method sensitivity strongly depends on the symmetry of the crystal and magnetic structures. During measurements, irradiation was from an infrared light of pulsed lasers from 450 to 700 nm; the temperature of YCrO$_3$ started from 300 K and decreased by 20 K{/}15 min until 20 K; and then the temperature was increased back to 300 K with the same rate. The intensity of the second harmonic generation displays a resonant peak at 600 nm corresponding to the transition related to a magnetic dipole around $T_\textrm{N}$, which indicates that magnetic order and related fluctuations have strong effect on the optical nonlinearity (see \textcolor[rgb]{0.00,0.00,1.00}{Fig.}~\ref{Application-1}) \cite{Eguchi2005-1}.

The nonlinear absorption behavior of biferroic YCrO$_3$ nanoparticles was studied by Krishnan \emph{et al}., employing the open aperture with Z-scan measurements using 5 ns and 532 nm laser pulses \cite{Krishnan2011, Krishnan2013}. This study revealed that YCrO$_3$ displays strong nonlinear optical properties, indicating a potential for optical limiting applications \cite{Krishnan2011}. The optical response of YCrO$_3$ has a size dependence \cite{Krishnan2014}. The absorption coefficient of three photons increases with the particle size. The nonlinearity originates from a mechanism of three-photon-like absorption. It was suggested that the nonlinearity was attributed to two photon absorption, after which the absorption of excited states happens \cite{Krishnan2012, Krishnan2014}. These studies indicate that the YCrO$_3$ compound could open up several new possibilities for device applications exploiting the features of ferroelectricity, ferromagnetism, and nonlinearity \cite{Krishnan2011, Krishnan2013, Krishnan2014, Krishnan2012}.

Todorov \emph{et al}. carried out a polarized Raman investigation on YMnO$_3$ and YCrO$_3$ single crystals \cite{Todorov2011}, from which authors determined the symmetry of the polarized spectra lines. The lines were ascribed to definite atomic vibrations on the basis of lattice dynamical calculations; two new lines appearing in YCrO$_3$, $B_{\textrm{1g}}$ (556 cm$^{-1}$) and $B_{\textrm{2g}}$ (611 cm$^{-1}$), were observed. The observed lines of YCrO$_3$ and YMnO$_3$ compounds have close frequencies, but differ significantly in intensity, indicating a larger Jahn-Teller distortion for the YMnO$_3$ compound \cite{Todorov2011}.

Raman spectra of orthorhombic \emph{RE}CrO$_3$ (\emph{RE} = Y, Pr, La, Gd, Sm, Ho, Dy, Lu, and Yb) perovskites were measured \cite{Weber2012}. The poly-crystalline powder samples were prepared by a direct hydrothermal synthesis method. Based on the room-temperature data as well as the associated phonon mode assignment, crystal structural information of a series of \emph{RE}CrO$_3$ orthochromates was studied, and phonons were calculated by \emph{ab}-\emph{initio}. Two $A_\textrm{g}$ modes were assigned as soft modes of octahedral rotation because the scale of their positions with respect to the tilt angle of CrO$_6$ octahedra was linear. For YCrO$_3$, it was obtained that $R_\textrm{A}$ = 1.019 {\AA} (ionic radius of Y$^{3+}$ ions) and \emph{t} = 0.852 (tolerance factor) \cite{Weber2012}.

The optical properties of YCrO$_3$ and HoCrO$_3$ compounds were measured for a controlled study \cite{Tiwari2013}. Both compounds display similar structural parameters but differ hugely in lower wavenumber (50--300 cm$^{-1}$) Raman active phonon modes, which was ascribed to the different Y$^{3+}$ and Ho$^{3+}$ ionic motions associated with atomic masses. The diffuse reflectance spectra were measured in a range of 350--850 nm to understand the effect of Ho$^{3+}$ and Y$^{3+}$ ions on the \emph{d}–\emph{d} transition of Cr$^{3+}$ ions (3\emph{d}$^3$, {$e^0_\textrm{g}$}{$t^3_{2\textrm{g}}$}, $^4F_{3/2}$). Using the Kubelka–Munk function, the diffuse reflectance spectra of YCrO$_3$ and HoCrO$_3$ were transferred into absorption spectra. Compared to YCrO$_3$, HoCrO$_3$ displayed extra optical transitions. This was ascribed to the asphericity of Ho ions and/or the local magnetic field \cite{Tiwari2013}.

The temperature-dependent Raman spectra of poly-crystalline YCrO$_3$ pellets were studied from 90 to 300 K \cite{Mall2014}. This study revealed that a spin-phonon coupling may exist around the magnetic transition temperature $T_\textrm{N}$ $\sim$ 140 K of YCrO$_3$ \cite{Mall2014}. The stretching of the Cr-O bonds in CrO$_6$ octahedra and the Y$^{3+}$ cation displacement lead to that the Raman peak position of all modes decreases monotonically to lower frequency with increasing temperature \cite{Mall2016}.

A temperature-dependent Raman study was carried out on distorted (orthorhombic) YCrO$_3$ perovskites in a temperature range of 20--300 K by Sharma \emph{et al}. \cite{Sharma2014-1}. The parameters of Raman-line shape of some modes were connected with the rotation of octahedral CrO$_6$ and Y$^{3+}$-site displacement within the unit cell. The temperature dependence of these parameters revealed an abnormal shift of phonons in YCrO$_3$ near $T_\textrm{N}$. Moveover, an anomalous phonon was observed at $\sim$ 60 K ($T^\ast$). This was ascribed to the modification of spin dynamics. From the magnetization isotherms, spin exchange constants ($J_1$ and $J_2$, as marked in \textcolor[rgb]{0.00,0.00,1.00}{Fig.}~\ref{mag-1}(b)) were estimated by mean-field approximation. In Raman frequencies, positive and negative shifts were observed between $T^\ast$ and $T_\textrm{N}$. This study indicates that there exists a competition between the weak ferromagnetism and antiferromagnetism below $T_\textrm{N}$ in YCrO$_3$. There is a close coincidence between the square of sublattice magnetization and the temperature dependencies of phonon frequency of $B_{3\textrm{g}}$ (3)-octahedral rotation mode between $T^\ast$ and $T_\textrm{N}$, indicating a presence of the coupling between spin and phonon in YCrO$_3$ compound \cite{Sharma2014-1}.

The local structural inhomogeneity in YCrO$_3$ was studied by Mannepalli \emph{et al}. \cite{Mannepalli2016}. One of the Raman mode $B_{3\textrm{g}}$ (3) was observed softening and disappearing at temperatures higher than the dielectric transition, which was attributed to the CrO$_6$ octahedral tilting mode \cite{Mannepalli2016, Mannepalli2017-3}, indicating that the local structural inhomogeneity may be the reason for the ferroelectricity of YCrO$_3$. By the way, the substitution of Bi for Y ions in YCrO$_3$ also resulted in the local structural inhomogeneity. In addition, above the dielectric transition temperature, calculations based on the density functional theory reveal that a room-temperature orthorhombic structure with Y-cation distortion transfers into an orthorhombic one without any distortion \cite{Mannepalli2017-3}.

Anomalous phonon modes were observed from temperature-dependent Raman measurements \cite{Mall2018-1}. It was proposed that there are two competing mechanisms working in YCrO$_3$ compound, and both determine its ferroelectricity \cite{Mall2018-1}. One is the rotation of CrO$_6$ octahedra, which favors a lower symmetry state, inducing the ferroelectricity; the other is the displacement of Y atoms, which tends to suppress the ferroelectricity, leaving it to exhibit only an incipient ferroelectric state \cite{Mall2018-1}.

We collect and list some of the results from dielectric and optical studies of different forms of YCrO$_3$ compound in \textcolor[rgb]{0.00,0.00,1.00}{Table}~\ref{DO-parameters}.

\section{Chemical-pressure dependence}

In this section, we will give an overview on the doping-induced chemical pressure as well as its effect on the ferroelectric, structure, magnetic, and optical properties of the parent and chemically-doped YCrO$_3$ compounds. We divided the doping into two parts: One is on the Y$^{3+}$ site, and the other relates to the Cr$^{3+}$ site.

\subsection{Y-sites doping}

\textbf{(1) Y$_{1-x}$M$_x$CrO$_3$ (M = Ca, Mg, Ba, and Sr).} The electrical and thermal transport properties of Y$_{1-x}$M$_x$CrO$_3$ (M = Ca, Mg, Ba, and Sr, $x = 0$--0.15) compounds were measured by Weber \emph{et al}. in order to study the effect of Y substitution with divalent metal ions in YCrO$_3$ \cite{Weber1986}. The measured electrical conductivity agrees with the hoping-type conducting mechanism of a concentration (independent on temperature) of small polarons. The seebeck coefficient depends almost linearly on temperature. Both measurements are strongly dependent on dopant size, i.e., ionic radius. Carrier concentration has the highest value when Ca was the dopant due to the similarity of ionic radii of Y$^{3+}$ and Ca$^{2+}$ ions. Moreover, the thermal conductivity displayed a slight decrease with dopant concentration and temperature \cite{Weber1986}.

\textbf{(2) Y$_{1-x}$Sr$_x$CrO$_3$.} Patil \emph{et al}. sintered Y$_{1-x}$Sr$_x$CrO$_3$ $(x = 0.0, 0.1$, and 0.2) compounds and measured infrared spectra to make their electronic properties and oxidation chemistry clear \cite{Patil1988}. Sintered pellets were obtained with 85 to 90$\%$ densification. The pure YCrO$_3$ sample displays a color of green, while the doped YCrO$_3$ samples are black \cite{Patil1988}. It was noted that no distortion occurred in the crystalline structure of Sr-doped YCrO$_3$ compounds, i.e., no deformation of CrO$_6$ octohedra \cite{Patil1988}. Tiwari \emph{et al}. synthesized poly-crystalline Y$_{1-x}$Sr$_x$CrO$_3$ $(x = 0.0$ and 0.1) samples via the solid state reaction method \cite{Tiwari2019}. The materials crystallize into the orthorhombic system with $Pbnm$ space group. The doping with Sr element was believed to be homogenous at Y sites and decreased the dielectric property \cite{Tiwari2019}.

\textbf{(3) Y$_{1-x}$Ca$_x$CrO$_3$.} Using a modified liquid mix method, Carini II \emph{et al}. fabricated Y$_{1-x}$Ca$_x$CrO$_3$ ($x$ = 0, 0.05, 0.10, 0.15, and 0.20) samples with a single phase and a grain size of $\sim$1 $\mu$m. These powders display a reversible oxidation-reduction behavior and a high stability \cite{Carini1991-1}. The dependence of the defect structure on temperature and oxygen activity was studied. The electrical conductivity agrees well with the previously-developed model. This indicates a great effect of temperature, oxygen activity, and Ca content with a general trend of the compensation shift (from electronic to ionic state) to higher oxygen activity as the Ca-doping level or temperature increases \cite{Carini1991-1}. The measurements of the Seebeck, electrical conductivity, and mobility versus temperature and oxygen activity indicate that the conduction of Y$_{1-x}$Ca$_x$CrO$_3$ samples processes by the small polaron hopping mechanism \cite{Carini1991-2}. Dur\'{a}n \emph{et al}. prepared Y$_{1-x}$Ca$_x$CrO$_3$ ($x$ = 0, 0.025, 0.050, 0.075, 0.10, and 0.15) powders and studied the effects of Ca substitution on the magnetism, structure, and electrical properties \cite{Duran2012-1}. As a consequence of Ca doping (charge compensation), the oxidation state of Cr atoms changes from Cr$^{3+}$ to Cr$^{4+}$, leading to a decrease in the cell volume (see \textcolor[rgb]{0.00,0.00,1.00}{Fig.}~\ref{LatticePhaseDia}(d)). Whereas, the AFM transition temperature was kept as $\sim$ 140 K, and there was no obvious change. In contrast, the Ca-doping indeed induced strong changes in the electrical transport properties. This is attributed to small polarons \cite{li2012-1} that are localized mainly at the Cr$^{3+}$-Cr$^{4+}$ sites through the lattice distortion of O-Cr-O bonds. The local non-centrosymmetric distortion in undoped parent compounds and the charge compensation of Ca$^{2+}$ in Y$^{3+}$ sites of doped samples result in the formation of small polarons \cite{Duran2012-1}. Escamilla \emph{et al}. synthesized Y$_{1-x}$Ca$_x$CrO$_3$ ($x = 0.000, 0.025, 0.050$, and 0.100) poly-crystalline ceramics \cite{Escamilla2017}. The average Cr-O bond length decreases as the Ca composition level increases, indicating that the charge compensation is via the oxidation modification from Cr$^{3+}$ to Cr$^{4+}$ rather than oxygen vacancies. The study of X-ray photoelectron spectroscopy confirmed the above hypothesis, i.e., core-levels of Cr$^{3+}$ (located at 576.22 eV) and Cr$^{4+}$ (located at 577.57 eV) ions were clearly observed \cite{Escamilla2017}.

\textbf{(4) Y$_x$Cr$_{0.8}$Ga$_{0.2}$O$_3$.} Pure and gallium-doped YCrO$_3$ samples with the formula Y$_x$Cr$_{0.8}$Ga$_{0.2}$O$_3$ $(x = 0.8, 0.9$, and 1.0) were synthesized by Westphal \emph{et al}. \cite{Westphal2000}. The effects of Y-site cation deficiency on structure, conductivity, and electrode behaviour in a special modified galvanic cell were studied. The Y-site deficiency induces an appearance of the second-phase Y$_3$Ga$_5$O$_{12}$ adopting the garnet structure (the pure YCrO$_3$ compound belongs to orthorhombic structure). Electrical conductivity was studied on pressed and sintered (1273 K, 10 h) samples. Electrodes were deposited on both sides of the sintered pellets by platinum paint. The conductivity was measured by impedance spectroscopy with a frequency response analyser over a frequency range of 100--100000 Hz. It was strongly influenced by the formation of the second phase, i.e., a higher conductivity was observed in the Y-site deficient samples \cite{Westphal2000}.

\textbf{(5) Y$_{1-y}$Ca$_y$Cr$_{1-x-y}$Al$_x$Ti$_y$O$_3$.} Y$_{1-y}$Ca$_y$Cr$_{1-x-y}$Al$_x$Ti$_y$O$_3$ perovskite pigments with $0 < x < 0.5$ and $0 < y < 0.2$ were synthesized by Ardit \cite{Ardit2009}. Except for the sample of Y$_{1.0}$Cr$_{0.5}$Al$_{0.5}$O$_3$, the changes of bond distances and cell parameters are nearly linear with the compositional variation while the crystal symmetry keeps the same. The modifications of stoichiometry and perovskite crystalline structure affected optical properties, e.g., the green colour of the samples turned into gray-brown with the change of oxidation states from Cr$^{3+}$ to Cr$^{4+}$. The replacement of Y$^{3+}$ by Ca$^{2+}$ ions and the heating in oxidizing atmosphere promote the appearance of Cr$^{4+}$ ions \cite{Ardit2009}.

\textbf{(6) Y$_{1-x}$Ho$_x$CrO$_3$.} Poly-crystalline samples of Y$_{1-x}$Ho$_x$CrO$_3$ (0 $\leq$ x $\leq$ 0.1) were synthesized by Mall \emph{et al}. via solid state reaction \cite{Mall2015}. The increase of holmium content in Y$_{1-x}$Ho$_x$CrO$_3$ leads to a quasi-continuous tuning of the lattice without magnetic interference effects of Ho sites. As Ho content increases, Raman mode frequencies decrease, depending on the average ionic radius of Y/Ho site \cite{Mall2015}.

\textbf{(7) Y$_{1-x}$Bi$_x$CrO$_3$.} Mannepalli and Ramadurai prepared Y$_{1-x}$Bi$_x$CrO$_3$ ($x = 0, 0.05, 0.1$, and 0.15) samples by the sol-gel route in order to study the local-structure inhomogeneity \cite{Mannepalli2016}. The study indicated that the unit-cell volume expansion by substituting Bi$^{3+}$ in Y-sites could soften the B$_{3g}$ (3) Raman mode associated with the tilting of CrO$_6$ octahedra around the structural phase transition. This was believed to induce the ferroelectric behavior in YCrO$_3$ \cite{Mannepalli2016}. The conductivity was measured using dielectric spectroscopy. This measurement indicated that the conductivity increases with increasing the composition. Dielectric loss indicated that chromates display the characteristic of charge carrier to itinerary of oxygen vacancy with increasing temperature and the Bi content \cite{Mannepalli2017-1}. Moreover, the grain contribution dominated the conductivity in orthochromates \cite{Mannepalli2017-2}. An increase of grain size in Bi-doped samples largely decreases the conductivity. This was attributed to inherent Cr$^{3+}$ electronic configuration or charge-carrier hopping resulting from intrinsic defects of Cr$^{4+}$ in Bi-doped YCrO$_3$ \cite{Mannepalli2017-2}.

\textbf{(8) Y$_{1-x}$La$_x$CrO$_3$.} Sinha \emph{et al}. synthesized and characterized La-doped Y$_{1-x}$La$_x$CrO$_3$ ($x = 0, 0.01, 0.05$, and 0.10) nano-powders by the sol-gel technique \cite{Sinha2016}. With increasing the La-doping level, the samples have a decreased optical band gap. In the visible range ($\sim$ 630 nm), red light emission was clearly observed. Dielectric permittivity in the temperature regime of 298--523 K follows a power law. The samples observe a semiconductor behavior. According to the analyses of dc and ac conductivities, the activation energy decreased. The conductivity increased with improving the doping level. This observation indicates that the compounds of chromates can take the role of interconnect materials utilized in solid-oxide fuel cells \cite{Sinha2016}.

\textbf{(9) Y$_{1-x}$Pr$_x$CrO$_3$.} Poly-crystalline samples of YCrO$_3$ and Y$_{0.95}$Pr$_{0.05}$CrO$_3$ were prepared by Kumar \emph{et al}. via the traditional solid-state reactions, and the effect of low Pr-doping level on magnetic properties was investigated \cite{Kumar2017}. XRPD results confirmed that both compounds adopt a distorted structure of orthorhombic system with $Pnma$ space group. The zero-field cooling magnetization measurement on YCrO$_3$ displays a negative magnetization under an applied-magnetic filed of 2000 Oe. This phenomenon disappears in the field-cooling measurement (see \textcolor[rgb]{0.00,0.00,1.00}{Fig.}~\ref{mag-2}(b)). In contrast, the Pr-doped Y$_{0.95}$Pr$_{0.05}$CrO$_3$ compound always exhibits positive magnetization in both filed- and zero-field-cooling modes. For diamagnetic materials, it is possible to display negative magnetization behavior. Here the observation could be from a competition between FM and AFM interactions \cite{Tiwari2013}. It is stressed that to perform such measurements, one should first rule out the trapped field effect from magnetometer itself by using oscillatory mode for decreasing applied magnetic field \cite{Tiwari2013, Kumar2010}. Furthermore, the Pr-doped sample displays a magnetic phase transition of SR behavior, which does not occur in the pristine YCrO$_3$ sample. Moreover, the magnetic hysteresis curves exhibit a possible negative exchange bias effect \cite{Kumar2017}. Dur\'{a}n \emph{et al}. synthesized Y$_{1-x}$Pr$_x$CrO$_3$ ($x = 0.0, 0.025, 0.050, 0.075, 0.200$, and 0.300) poly-crystalline samples by the SHS (combustion) method and studied the thermal property, magnetic property, and crystal structure of the samples \cite{Duran2018-3}. With increasing the Pr-doping level, the AFM transition temperature ($T_\textrm{N}$) increases, and notable magnetic behaviors were observed. The magnetic property depends strongly on the structure parameters. The effects of magnetization reversal, SR, and exchange bias were observed in a small regime of Pr compositions with temperature. The SR transition temperature has a range of 35--149 K, corresponding to the compositions of 0.025 $\leq$ x $\leq$ 0.1, respectively. Moreover, the Cr magnetic sublattices rotate in a second-order mode from SR temperature to a new spin structure at lower temperatures \cite{Duran2018-3}. Furthermore, cooling magnetic field strongly affects the magnetic phase of Y$_{0.9}$Pr$_{0.1}$CrO$_3$ compound \cite{Deng2015}: zero filed cooling produces $G_zF_x$ spin structure, whereas a higher field ($\geq$ 100 Oe) cooling leads to a magnetic order of $G_xF_z$.

\textbf{(10) Y$_{1-x}$Gd$_x$CrO$_3$.} Gd-doped Y$_{1-x}$Gd$_x$CrO$_3$ ($x = 0.0, 0.01, 0.05$, and 0.10) nano-materials were prepared by Sinha \emph{et al}. following the sol-gel method \cite{Sinha2018}. The transport property was investigated in a temperature regime of 298--523 K. The free-charge conductivity and the space-charge conductivity increase as a function of temperature. The time for relaxation decreases as temperature increases. The dc and the ac activation energies increase as the Gd-doping concentration increases \cite{Sinha2018}.

\textbf{(11) Y$_{1-x}$Nd$_x$CrO$_3$.} Nd-doped Y$_{1-x}$Nd$_x$CrO$_3$ $(x = 0.0, 0.01, 0.05$, and 0.10) nanomaterials were prepared by Sinha \emph{et al}. with the sol-gel method, and the electrical, relaxor ferroelectric, and FM properties were studied \cite{Sinha2019}. The samples display a semiconductor behavior, and the conductivity increases upon Nd-doping, accompanied by the decrease of the dc activation energy. The conductivity dependences of free-charge and space-charge carriers increase as the doping content increases. Moreover, the relaxer ferroelectric behavior is enhanced with increasing the Nd-doping level in YCrO$_3$ \cite{Sinha2019}.

\subsection{Cr-sites doping}

\textbf{(1) YCr$_{1-x}$Mg$_x$O$_3$.} Single-phase perovskite compounds of YCr$_{1-x}$Mg$_x$O$_3$ with $x =$ 0, 0.025, 0.05, 0.075, 0.10, 0.12, and 0.15, Cr-sites substituted by Mg$^{2+}$ ions, were synthesized by Tachiwaki \emph{et al}. \cite{Tachiwaki2001}. Thermo-gravimetry measurements and differential thermal analysis were carried out with a heating rate 10 $^\circ$C/min in air. The results indicate that Y(Cr$_{1-x}$Mg$_x$)O$_4$ crystallizes firstly from amorphous materials at $\sim$ 600 $^\circ$C by the hydrazine technique; then submicrometer-sized Y(Cr$_{1-x}$Mg$_x$)O$_3$ particles form at elevated temperatures ($\sim$ 850 $^\circ$C); finally, densed Y(Cr$_{1-x}$Mg$_x$)O$_3$ ceramics were obtained by sintering at the higher temperature of 1800 $^\circ$C with atmospheric pressure. With increasing the Mg$^{2+}$-content, the grain sizes, electrical conductivity, and relative densities increase \cite{Tachiwaki2001}.

\textbf{(2) YCr$_{1-x}$M$_x$O$_3$ (M = Fe and Ni).} Phase-pure YCr$_{1-x}$M$_x$O$_3$ poly-crystalline ceramics (M = Fe and Ni, and $x = 0$ and 0.1) were prepared by Mall \emph{et al}. via the solid state reactions, and modifications of the structure, magnetism, and Raman spectra were studied \cite{Mall2017-3}. Upon doping, there is no change in the crystalline structure of orthorhombic with $Pnma$ space group. The Raman scattering shows the certain broadening Raman modes, B$_{3g}$(5) and B$_{1g}$(3), because of the doping-induced lattice disorder. The samples are AFM. The corresponding transition temperature $T_\textrm{N}$ increases with the Fe doping and decreases for the Ni doping, which was attributed to the modification of exchange interactions of the (next) nearest neighbors \cite{Mall2017-3}.

\textbf{(3) YCr$_{1-x}$Ti$_x$O$_3$.} Ti-doped YCr$_{1-x}$Ti$_x$O$_3$ ($x = 0.0, 0.01, 0.03, 0.05$, and 0.10) poly-crystalline samples were prepared by Dur\'{a}n \emph{et al}. with conventional solid-state reaction method, and the dielectric properties and crystal structure were investigated \cite{Duran2018-2}. The Ti-doping leads to noticeable changes in the ferroelectric and dielectric properties. A paraelectric to ferroelectric transition was observed at $\sim$ 240 $^\circ$C upon cooling. The permittivity peaks are broad, and their frequency dependence indicate a relaxor-like ferroelectric characteristic. The Ti-doping results in decreases of the mobility of free-charge carriers and dielectric loss, improving the ferroelectric and dielectric properties of the parent YCrO$_3$ compound \cite{Duran2018-2}.

\textbf{(4) YCr$_{1-x}$Mn$_x$O$_3$.} The YCr$_{1-x}$Mn$_x$O$_3$ ($x = 0.0, 0.1, 0.2$, and 0.3) poly-crystalline compounds were synthesized by Sahu \emph{et al}. with the high-temperature ceramic method. The Mn-substitution effect on the multiferroic and magnetic properties was studied \cite{Sahu2008}. The samples possess an orthorhombic system with $Pnma$ space group and display a CAFM state below $T_\textrm{N}$. With increasing the Mn-doping level, the AFM transition temperature ($T_\textrm{N}$) decreases \cite{Sahu2008}. For the measurement of dielectric properties with a precision Agilent impedance analyser (4294 A), the samples were firstly polished and then coated in an Ar gas with gold. To make sure good electrical contacts, the coated samples were heated with a heating rate 1 $^\circ$C/min up to 300 $^\circ$C for 2 h. The transition temperature of the ferroelectricity decreases with an increase of the Mn-doping level \cite{Sahu2008}. For the YCr$_{0.7}$Mn$_{0.3}$O$_3$ compound, it has a low AFM transition temperature (106 K) and high values of the remnant magnetization (784 emu/mol at 10 K) and coercive field (14770 Oe at 10 K) along with a ferroelectric phase transition in a temperature regime of 418--425 K \cite{Sahu2008}. Kamlo \emph{et al}. reported on the preparation of YCr$_{1-x}$Mn$_x$O$_3$ ($x = 0.0, 0.2, 0.4, 0.6$, and 0.8) ceramics which, in the final step, were sintered at 1500 and 1600 $^\circ$C under a flowing of N$_2$ gas for 5 h \cite{Kamlo2011}. By substituting Mn for Cr element, when $x \leq 0.6$, YCr$_{1-x}$Mn$_x$O$_3$ holds an orthorhombic perovskite structure; for $x \geq 0.8$, the second phase forms with a similar structure to the hexagonal phase of YMnO$_3$. Substituting Cr with Mn element leads to a decreased porosity. As the Mn-doping level increases, the material constant characterizing the thermal sensitivity decreases when $x \leq 0.6$, so does the resistivity (10$^4$--10$^8$ $\Omega$ cm at 25 $^\circ$C, with an activation energy range of 0.28--0.99 eV at low and high temperatures) \cite{Kamlo2011}. Sinha \emph{et al}. prepared Mn-doped YCr$_{1-x}$Mn$_x$O$_3$ ($x = 0.0, 0.01, 0.05$, and 0.10) nano-materials by the sol-gel method and studied the electrical and dielectric properties of the compounds \cite{Sinha2015}. The samples' optical band-gap increases with increasing the Mn-doping concentration, estimated by analysis of the UV-vis absorption spectroscopy. The samples' dc resistivity decreases as temperature increases. The ac conductivity changes as a function of frequency, consistent with the hoping-conduction mechanism of correlated barriers. The activation energy induced by dielectric response and dc resistance increases with increasing the Mn-doping content \cite{Sinha2015}. El hachimi \emph{et al}. studied the electronic structure and ferromagnetism of YCrO$_3$ perovskites with Mn doping using the coherent-potential approximation based on first-principles within the KKR technique \cite{hachimi2017}. The YCrO$_3$ with Mn doping holds a larger magnetic moment value than the parent compound YCrO$_3$, showing a linear increase with the Mn concentration, which was attributed to exchanges of the half-metallic semiconductor mediated by carriers. The origin of the magnetic moments is mainly due to transition Cr and Mn metals, with a small contribution from oxygen ions because of the hybridization between oxygen-2\emph{p} and 3\emph{d} electrons \cite{hachimi2017}. When $0.05 \leq x \leq 0.15$, YCr$_{1-x}$Mn$_x$O$_3$ accommodates a more favorable FM state. A ferrimagnetic ground state is stable at a higher Mn-doping level of $x = 0.2$ but with weakened total magnetic moment. The electronic structure implies that the YCrO$_3$ compounds with Mn doping display a semi-metallic characteristic that can exist even at room temperature. These theoretical calculations are important in guiding the development of spintronics device. Li \emph{et al}. synthesized YCr$_{1-x}$Mn$_x$O$_3$ ($x = 0.0, 0.1, 0.15, 0.20, 0.30, 0.40$, and 0.45) poly-crystalline ceramics via a modified pechini technique and studied their magnetic and structural properties \cite{Li2017}. This study reported an interesting phenomenon of negative magnetization, resulting from the competing multiple-exchange interactions, i.e., the AFM interactions between WFM moments of Cr$^{3+}$-Cr$^{3+}$ and Mn$^{3+}$-Mn$^{3+}$ with a coupling of Cr$^{3+}$-Mn$^{3+}$ ions. Below the compensation temperature, the absolute value of negative magnetization increases, followed by a decrease, with increasing the doping level \emph{x}. This was attributed to the enhanced Mn$^{3+}$-Mn$^{3+}$ and decreased Cr$^{3+}$-Cr$^{3+}$ interactions. An increase in magnetization with increasing \emph{x} was observed in the measurement of hysteresis loops, which was probably caused by the FM exchange interactions between Cr$^{3+}$ and Mn$^{3+}$ ions \cite{Li2017}. Panwar \emph{et al}. synthesized the YCr$_{0.85}$Mn$_{0.15}$O$_3$ poly-crystalline compound following the conventional solid-state reactions and investigated the low-temperature magnetocaloric and magnetic properties of the compound \cite{Panwar2018}. Compared to pristine YCrO$_3$, YCr$_{0.85}$Mn$_{0.15}$O$_3$ has a larger lattice volume and a lower $T_\textrm{N}$ ($\sim$ 132 K). The magnetization reversal behavior was observed at low applied magnetic fields whereas it disappears under higher applied magnetic fields. The magnetocaloric effect was measured around 36 K. The resultant change in magnetic entropy was $\sim$ 0.186 J kg$^{-1}$ K$^{-1}$ ($-\Delta$\emph{S}), corresponding to a relative cooling power of $\sim$ 6.65 J kg$^{-1}$, when the applied magnetic field was equal to 5 T \cite{Panwar2018}.

\begin{table*}[!t]
\small
\caption{\newline Structural and magnetic parameters of YCrO$_3$ compound, including theoretical quantum numbers of spin (\emph{S}), orbital (\emph{L}), total angular momentum (\emph{J}), Land$\acute{\textrm{e}}$ factor ($g_J$), and the ground-state term ($^{2S+1}L_J$), theoretical (theo.) effective (eff) ($\mu_{\textrm{eff{\_}theo}}$) and saturation (sat) ($\mu_{\textrm{sat{\_}theo.}}$) chromium moments. Here, Pars = Parameters; Nano = Nanocrystal; PC = Polycrystal; SC = Single crystal; Str = Structure; Orth. = Orthorhombic; MON. = Monoclinic; SGr = Space group; The Orth. ($Pbnm$ and $Pnma$) and MON. ($P2_1/n$) crystal systems are centro-symmetric; NCS = Noncentrosymmetric; $T_\textrm{SR}$ represents the spin-reorientation temperature. $T_{\texttt{IP}}$: an intermediate point of temperature at which a new short-ranged magnetic phase appears within the paramagnetic (PM) state. $\Theta_{\textrm{CW}}$ = PM Curie-Weiss temperature; SFF = Spin-flop field; \textbf{\emph{D}} is the vector of DM interactions.}
\label{SM-parameters}
\setlength{\tabcolsep}{1.1mm}{}
\renewcommand{\arraystretch}{1.0}
\begin{tabular}{l|lllllllllllllll}
\hline
\hline
\multicolumn {16}{c}{YCrO$_3$}                                                                                                                                                                                                                             \\[3pt]
\multicolumn {16}{c}{($3d^3$, $S$ = $\frac{3}{2}$, $L$ = 3, $J$ = $\frac{3}{2}$, $g_J$ = 2, $^4F_{\frac{3}{2}}$, $\mu_{\textrm{eff{\_}theo}}$ = 3.873 $\mu_\textrm{B}$, $\mu_{\textrm{sat{\_}theo.}}$ = 3 $\mu_\textrm{B}$)}                               \\[3pt]
\hline
Pars    &Str   &SGr     &$a$     &$b$     &$c$     &$T_\textrm{SR}$ &$T_\textrm{N}$ &$T_\textrm{IP}$ &SFF  &$\Theta_{\textrm{CW}}$ &$\mu_{\textrm{eff}}$ &$\mu_{\textrm{sat}}$ &\emph{\textbf{D}}     &Remark                        &Refs.                \\[2pt]
(unit)  &      &        &(\AA)   &(\AA)   &(\AA)   &(K)             &(K)            &(K)             &(T)  &(K)                    &($\mu_\textrm{B}$)   &($\mu_\textrm{B}$)   &(K)                   &                              &                     \\[2pt]
\hline
        &Orth. &$Pnma$  &5.5248  &7.5408  &5.2425  &                &140            &                &     &                       &                     &                     &                      &$T_{\textrm{FM}} \sim$ 110 K  &\cite{Kim2007}       \\
        &Orth. &$Pnma$  &5.52    &7.53    &5.23    &10              &140            &                &     &                       &                     &                     &                      &$<$ 20 nm                     &\cite{Singh2013}     \\
        &Orth. &$Pnma$  &5.497   &7.512   &5.224   &                &               &                &     &                       &                     &                     &                      &38 nm                         &\cite{Krishnan2013}  \\
Nano    &Orth. &$Pnma$  &5.509   &7.520   &5.233   &                &               &                &     &                       &                     &                     &                      &80 nm                         &\cite{Krishnan2013}  \\
        &Orth. &$Pnma$  &5.514   &7.525   &5.239   &                &               &                &     &                       &                     &                     &                      &100 nm                        &\cite{Krishnan2013}  \\
        &Orth. &$Pnma$  &5.5181  &7.5293  &5.2389  &                &140            &                &     &$-$448                 &4.5                  &                     &                      &100--200 nm                   &\cite{Jara2018}      \\[1pt]
\hline
        &      &        &        &        &        &                &150            &                &     &                       &                     &                     &                      &100 nm, PC                    &\cite{Seo2013}       \\
Film    &Orth. &$Pnma$  &5.523   &7.534   &5.242   &                &               &                &     &                       &                     &                     &                      &20 nm, SC                     &\cite{Arciniega2018} \\
        &Orth. &$Pbnm$  &        &        &        &                &144            &                &     &                       &                     &                     &                      &32 nm, SC                     &\cite{Sharma2020}    \\[1pt]
\hline
        &MON.  &$P2_1/n$&7.61    &7.54    &7.61    &                &               &                &     &                       &                     &                     &                      &$\beta = 92^\circ56(6)'$      &\cite{Katz1955}      \\
        &Orth. &$Pbnm$  &        &        &        &                &141            &                &     &                       &                     &2.96                 &                      &$G_x$, 4.2 K                  &\cite{Bertaut1966}   \\
        &Orth. &$Pbnm$  &        &        &        &                &141            &                &     &                       &                     &2.56                 &                      &$G_x$, 80 K                   &\cite{Bertaut1966}   \\
        &      &        &        &        &        &                &               &                &4.0  &                       &                     &                     &                      &4.2 K                         &\cite{Jacobs1971}    \\
        &MON.  &$P12_11$&        &        &        &                &140            &                &     &                       &                     &                     &                      &NCS, theoretical              &\cite{Serrao2005}    \\
        &Orth. &$Pnma$  &5.52    &7.53    &5.24    &                &141            &                &     &$-$360                 &                     &                     &                      &                              &\cite{Sahu2008}      \\
        &Orth. &$Pnma$  &5.5239  &7.5363  &5.2438  &60              &140            &                &     &$-$325                 &3.77                 &                     &                      &                              &\cite{Duran2010}     \\
        &MON.  &$P2_1/n$&        &        &        &                &139            &220             &     &                       &                     &                     &                      &                              &\cite{Alvarez2010}   \\
        &Orth. &$Pbnm$  &5.2434  &5.5242  &7.5356  &                &               &                &     &                       &                     &                     &                      &0 GPa                         &\cite{Ardit2010}     \\
        &Orth. &$Pbnm$  &4.9403  &5.2767  &7.1995  &                &               &                &     &                       &                     &                     &                      &42.6 GPa                      &\cite{Ardit2010}     \\
        &      &        &        &        &        &                &151            &                &     &$-$239                 &3.92                 &                     &                      &                              &\cite{Cheng2010}     \\
PC      &Orth. &$Pbnm$  &5.2547  &5.5204  &7.5360  &                &140            &                &     &$-$260                 &3.84                 &                     &                      &                              &\cite{Sardar2011}    \\
        &Orth. &$Pbnm$  &5.2437  &5.5235  &7.5360  &                &140            &                &     &$-$401                 &4.24                 &                     &                      &                              &\cite{Duran2012-1}   \\
        &Orth. &$Pnma$  &5.519   &7.534   &5.244   &                &140.8          &                &     &$-$449                 &4.7                  &                     &                      &                              &\cite{Tiwari2013}    \\
        &Orth. &$Pnma$  &5.5122  &7.5349  &5.2461  &60              &142            &                &     &$-$440                 &3.76                 &                     &                      &                              &\cite{Sharma2014-1}  \\
        &Orth. &$Pnma$  &5.5150  &7.5218  &5.2346  &                &140            &                &3.5  &$-$325                 &                     &                     &0.36                  &                              &\cite{Ikeda2015}     \\
        &Orth. &$Pnma$  &5.5147  &7.5358  &5.2465  &                &               &                &     &$-$309                 &5.80                 &                     &                      &                              &\cite{Kumar2017}     \\
        &Orth. &$Pnma$  &        &        &        &                &140            &230             &     &$-$431                 &                     &                     &                      &                              &\cite{Mall2017-1}    \\
        &Orth. &$Pnma$  &5.5182  &7.5358  &5.2456  &                &               &                &     &                       &                     &                     &                      &                              &\cite{Mall2018-2}    \\
        &Orth. &$Pnma$  &5.437   &7.678   &5.432   &                &140            &                &     &                       &                     &                     &                      &                              &\cite{Taran2020}     \\
        &Orth. &$Pbnm$  &5.2781  &5.5520  &7.5829  &                &               &                &     &                       &                     &                     &                      &                              &\cite{Saxena2020}    \\
        &Orth. &$Pnma$  &5.5146  &7.5349  &5.2458  &140.6           &150            &                &     &$-$234.53              &3.88                 &                     &                      &                              &\cite{Shi2020}       \\
        &Orth. &$Pnma$  &5.5146  &7.5349  &5.2458  &                &141            &                &     &$-$459                 &                     &                     &                      &$^{89}$YCrO$_3$               &\cite{Mall2021-2}    \\[1pt]
\hline
        &      &        &        &        &        &                &141            &                &     &$-$230                 &                     &                     &                      &$J_1$ = --8.7 cm$^{-1}$,      &\cite{Tsushima1970}  \\
        &      &        &        &        &        &                &               &                &     &                       &                     &                     &                      &$J_2$ = --1.0 cm$^{-1}$.      &\cite{Tsushima1970}  \\
        &      &        &        &        &        &                &140            &                &     &$-$342                 &3.8                  &                     &                      &                              &\cite{Kim2007}       \\
SC      &Orth. &$Pbnm$  &5.243   &5.524   &7.536   &                &142            &                &     &                       &                     &                     &                      &                              &\cite{Sanina2018}    \\
        &Orth. &$Pnma$  &5.5198  &7.5297  &5.2392  &                &141.5          &                &     &$-$433.2               &3.95                 &2.45                 &                      &0.01 T                        &\cite{Zhu2020-2}     \\
        &      &        &        &        &        &                &144.5          &                &     &                       &                     &                     &                      &5 T                           &\cite{Zhu2020-2}     \\[1pt]
\hline
\hline
\end{tabular}
\end{table*}

\begin{figure*} [!t]
\centering \includegraphics[width=0.82\textwidth]{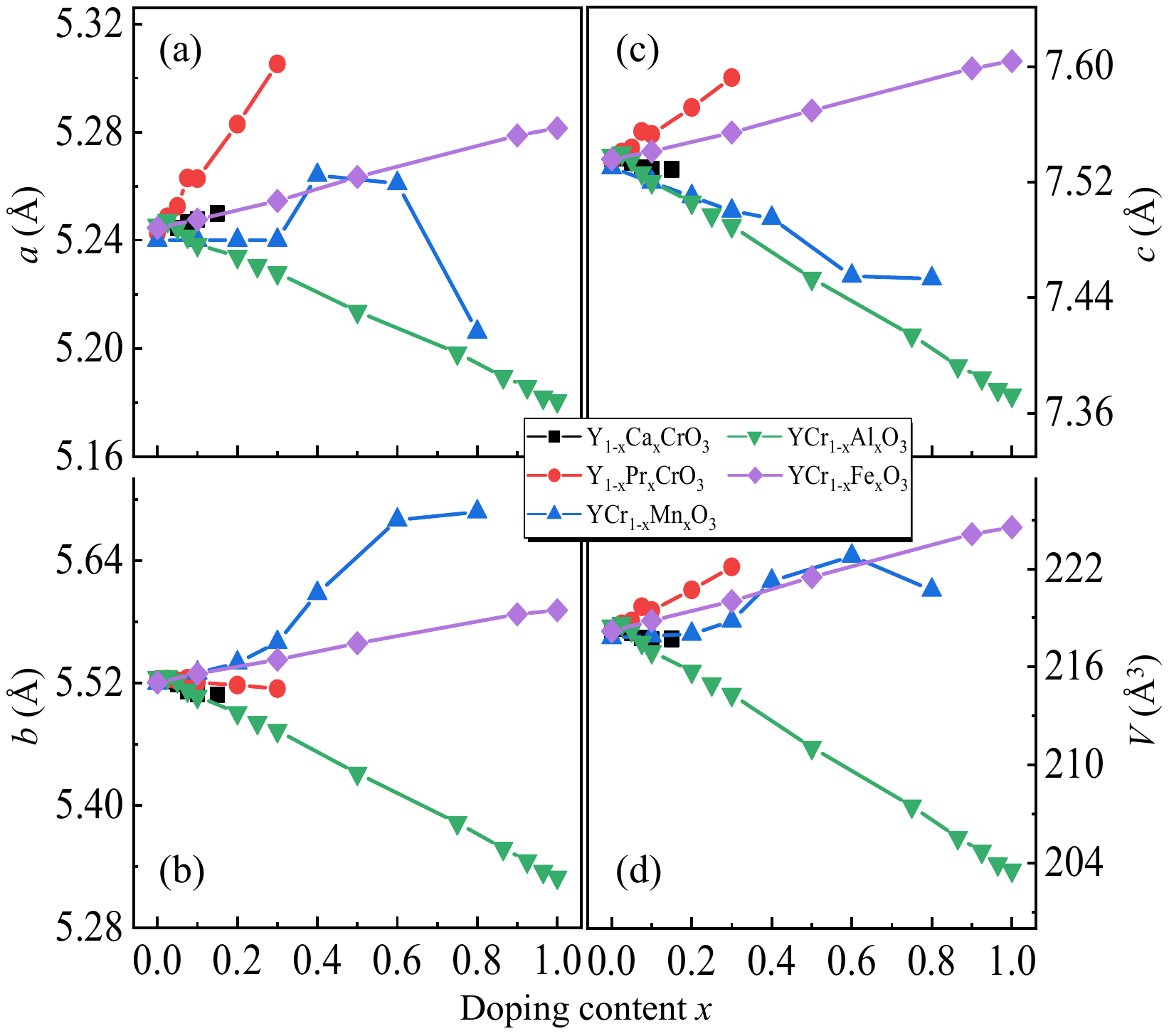}
\caption{
Lattice parameters of \emph{a} (a), \emph{b} (b), and \emph{c} (c) and the unit-cell volume \emph{V} (d) for the Y-site doped Y$_{1-x}$Ca$_x$CrO$_3$ \cite{Duran2012-1} and Y$_{1-x}$Pr$_x$CrO$_3$ \cite{Duran2018-3} and the Cr-site doped YCr$_{1-x}$Mn$_x$O$_3$ \cite{Sahu2008, Kamlo2011}, YCr$_{1-x}$Al$_x$O$_3$ \cite{Cruciani2009, Duran2018-1}, and YCr$_{1-x}$Fe$_x$O$_3$ \cite{Fabian2016} systems.
}
\label{LatticePhaseDia}
\end{figure*}

\textbf{(5) YAl$_{1-x}$Cr$_x$O$_3$.} Poly-crystalline YAl$_{1-x}$Cr$_x$O$_3$ perovskites ($x = 0.000, 0.035, 0.075, 0.135, 0.250, 0.500, 0.750$, and 1.000) were prepared by Cruciani \emph{et al}. with the sol gel combustion method together with a polymeric gel-action with the fast-combustion reaction, and the structural relaxation was studied by electron absorption spectra \cite{Cruciani2009}. The lattice constants $(a, b,$ and $c)$ (see \textcolor[rgb]{0.00,0.00,1.00}{Fig.}~\ref{LatticePhaseDia}) and the averaged lengths of Y-O and (Cr/Al)-O bonds increase linearly as the Cr content increases. As the Cr composition decreases within the range of $x \leq 0.4$, the strength of crystal field (10$Dq$) decreases, and the optical parameters of polarizability ($B_{55}$) and covalency ($B_{35}$) increase; but some excess 10\emph{Dq} exists with a nonlinear trend. The local Cr-O bond lengths decrease from 1.98 {\AA} ($x = 1.000)$ to 1.95 {\AA} ($x = 0.035)$, which leads to the lowest relaxation coefficient ($\varepsilon =$ 0.54) of the compound in contrast to corundum (0.58), spinel (0.68), and garnet (0.74) \cite{Cruciani2009}. The Cr$^{3+}$-O-Cr$^{3+}$ bond has a characteristic of covalent bond, which influences the flexibility of the perovskite polyhedral network. Dur\'{a}n \emph{et al}. synthesized YAl$_{1-x}$Cr$_x$O$_3$ ($x = 0.500, 0.700, 0.800, 0.900, 0.925, 0.950, 0.975$, and 1.000) powder samples by the combustion method \cite{Duran2018-1}. The structure, magnetic, and dielectric properties were studied. As the value of \emph{x} increases, the lattice constants (\emph{a}, \emph{b}, and \emph{c}) and the unit-cell volume (\emph{V}) almost increase linearly (see \textcolor[rgb]{0.00,0.00,1.00}{Fig.}~\ref{LatticePhaseDia}), while keeping the oxidation state of Cr ions as 3+. Two mechanisms were provided for understanding the magnetic and electronic properties. Firstly, the deformation of local structure is expressed by the octahedral inclination (corresponding to the bond angles of O-Al/Cr-O) and the octahedral distortion (corresponding to the changes in Al/Cr-O bond lengths), which strongly influences the AFM transition temperature ($T_\textrm{N}$) and the strength of coercive field at high values of \emph{x}. Secondly, with decreasing Cr$^{3+}$ concentration, the Cr 3\emph{d} states have a decreased contribution at the Fermi level, thus decreasing the values of remanent magnetization and coercive field strength \cite{Duran2018-1}.

\textbf{(6) YCr$_{1-x}$Fe$_x$O$_3$.} Wold and Croft prepared LnCr$_{1-x}$Fe$_x$O$_3$ (Ln = La, Nd, Sm, and Y) materials by heating mixtures of the rare earth oxide and the corresponding proportions of iron (III) oxide and chromium (III) oxide in air for 96 h \cite{Wold1959}. The space group of parent and doped YCrO$_3$ compounds is $Pbnm$. Each unit cell consists of four formula weight. The YCrO$_3$ compound has lattice constants 5.245 {\AA} ($a$), 5.518 {\AA} ($b$), and 7.540 {\AA} ($c$), and its unit-cell volume ($V$) is 218.22 {\AA}$^3$. The increasing substitution of Cr$^{3+}$ for Fe$^{3+}$ results in a reducing of the unit-cell volume \cite{Wold1959}. Kovachev \emph{et al}. prepared YCr$_{1-x}$Fe$_x$O$_3$ ($x = 0.000, 0.125, 0.250, 0.330, 0.500, 0.670, 0.750, 0.875$, and 1.000) materials by an original route implementing a modified solution combustion method, followed by thermal treatment at 800 $^\circ$C in air \cite{Kovachev2010}. The magnetic and structure properties were studied \cite{Kovachev2010}. When $x =$ 0, the compound of YCrO$_3$ that was heated at 500 $^\circ$C for 1 h remains amorphous; after a heating at 600 $^\circ$C for 1 h, the phase of YCrO$_4$ was identified from the initial material; after heated at 700 $^\circ$C for 1 h, the original materials transfer into the YCrO$_3$ perovskite-type phase \cite{Kovachev2010}. These are consistent with the previously-reported results obtained on YCrO$_3$ samples prepared by the sol-gel technique \cite{Tachiwaki2001-1}: (i) The preparation temperature range for the YCrO$_4$ phase is 490--530 $^\circ$C. (ii) The phase transition from YCrO$_4$ to YCrO$_3$ happens within the temperature regime of 750--830 $^\circ$C. For a comparison, the synthesis of YCr$_{1-x}$Fe$_x$O$_3$ compounds proceeds without the formation of the intermediate YCrO$_4$ phase \cite{Kovachev2010}. XRPD results confirmed the formation of YCr$_{1-x}$Fe$_x$O$_3$ solid solutions within the whole concentration interval $(0.000 \leq x \leq 1.000)$. All samples crystallize in the orthorhombic system with $Pnma$ space group. Some of the samples display an AFM order at room temperature, which was studied by neutron diffraction measurements. The obtained structural results for YCr$_{1-x}$Fe$_x$O$_3$ and the specific thermomagnetic behavior of the system give strong indication that the replacement of Cr with Fe leads to frustrated magnetic interactions \cite{Kovachev2010}. MalagaReddy \emph{et al}. prepared YCr$_{1-x}$Fe$_x$O$_3$ ($x = 0.00, 0.25, 0.33, 0.50, 0.67, 0.75$, and 1.00) samples by the simple mechanochemical synthesis route and studied their magnetic properties \cite{MalagaReddy2012}. With decreasing the Fe-doping level \emph{x}, the lattice constants (\emph{a}, \emph{b}, and \emph{c}) display a linear reduction. Hysteresis loops were observed at 5 K in magnetization measurements. The substitution of Fe for Cr weakens the magnetization for $x \geq 0.67$. For lower doping levels, $0 \leq x \leq 0.67$, the system displays magnetic frustration. At $x = 0$, the parent YCrO$_3$ compound shows a week ferromagnetism below $T_\textrm{N}$ ($\sim$ 140 K). A simple method of synthesizing pure multiferroic compounds was provided with the chloride-salt-based mechanochemical route \cite{MalagaReddy2012}. Fabian \emph{et al}. prepared a series of YCr$_{1-x}$Fe$_x$O$_3$ ($x = 0.0, 0.10, 0.30, 0.50, 0.90$, and 1.0) samples (with grains of micrometer in size) by the co-precipitation method and studied the behavior of reversal magnetization with the Fe and Cr oxidation states \cite{Fabian2016}. The samples have an orthorhombic $Pnma$ structure. With increasing the Fe concentration, the unit-cell volume has an expansion (see \textcolor[rgb]{0.00,0.00,1.00}{Fig.}~\ref{LatticePhaseDia}(d)), and the magnetic transition temperature shows a continuous increase. Magnetic measurements display the reversal magnetization effect around $x =$ 0.5, which was ascribed to the DM interactions. The X-ray absorption near-edge spectroscopy study at the Cr and Fe \emph{K}-edges indicates that both ions of Fe and Cr show mixed-valence states, probably there exist magnetic interactions between Cr$^{2+}$, Cr$^{3+}$, Fe$^{2+}$, and Fe$^{3+}$ ions. The difference between observed magnetization and calculations by the Curie-Weiss law and the observation of a WFM state were thus ascribed to the strong magnetic interactions of Cr and Fe ions with the mixed oxidation states \cite{Fabian2016}. Yin \emph{et al}. grew a single crystal of YCr$_{0.88}$Fe$_{0.12}$O$_3$ by the flux technique and studied its magnetic, dielectric, and structural properties \cite{Yin2017}. The single crystalline sample is AFM and has a weak ferromagnetism, undergoing three magnetic phase transitions at $T_{\textrm{N1}}$ $\sim$ 127.3 K, $T_{\textrm{N2}}$ $\sim$ 115.2 K, and $T_{\textrm{SR}}$ $\sim$ 99.0 K. Moreover, below the SR transition temperature $T_{\textrm{SR}}$, an applied magnetic field can also arouse spins to re-orientate. The dielectric relaxation observed around $T_{\textrm{N1}}$ was attributed to electron hopping \cite{Yin2017}.

Some of the magnetic and dielectric properties as well as doping dependent structural parameters are compiled into the following phase diagrams (Section 5.1 and 5.2).

\section{Phase diagram}

For more elaborate studies of interesting materials, one should first get to know how microscopic parameters change with modifications of macroscopic parameters, such as temperature, pressure, or chemical doping with addition of a certain amount of different substances. Phase diagrams are guiding tool for effective assessment of structural and magnetic characteristics of material in a large range of parameters. Different phase diagrams \cite{schiffer1995low, hemberger2002structural} compiled based on previous intensive studies can provide different information on the system such as: (i) What is the magnetic state of material at a certain temperature? (ii) What is the crystalline structure of material at given temperature and chemical pressure? (iii) How many different crystal and magnetic structures can a particular material form? (iv) Can the substance that we add to the material soluble in that material and up to what concentration?

In this section, we will summarize the phase diagrams of doped YCrO$_3$: (i) lattice constants (\emph{a}, \emph{b}, and \emph{c}) and unit-cell volume (\emph{V}) versus doping levels with different elements on Y and Cr sites. (ii) Magnetic phase diagrams of YCrO$_3$ versus temperature and YCr$_{1-x}$Mn$_x$O$_3$ versus Mn-doping level at different temperatures.

\begin{table*}[!t]
\small
\caption{\newline Summarized results from dielectric and optical studies of YCrO$_3$ compound. Here, $T_{\textrm{abno}}$ represents the temperature points where anomaly appears in dielectric measurements; $T_{\textrm{C}}$: ferroelectric transition temperature; dcon = dielectric constant; RT = room temperature; $P_\textrm{r}$: remanent polarization; $d_{33}$ = piezoelectric coefficient; $\sigma$: electrical conductivity.}
\label{DO-parameters}
\setlength{\tabcolsep}{1.8mm}{}
\renewcommand{\arraystretch}{1.1}
\begin{tabular}{l|lllllllll}
\hline
\hline
Pars                &$T_{\textrm{abno}}$  &$T_{\textrm{C}}$  &dcon       &$P_\textrm{r}$    &$d_{33}$    &bandgap &$\sigma$                &Remark                                         &Refs.               \\ [3pt]
(unit)              &(K)                  &(K)               &           &({$\mu$}C/cm$^2$) &(pm/V)      &(eV)    &($\Omega^{-1}$m$^{-1}$) &                                               &                    \\ [3pt]
\hline
                    &                     &450               &           &                  &            &        &                        &10 kHz; relaxor-like                           &\cite{Duran2012-2}  \\
Nano                &                     &                  &           &                  &            &3.39    &9.35$\times10^{-6}$     &320 \textrm{K}, 35--45 nm                      &\cite{Sinha2016}    \\
                    &                     &                  &           &                  &            &3.72    &                        &100--200 nm                                    &\cite{Jara2018}     \\ [1pt]
\hline
                    &                     &400               &           &3                 &            &        &                        &178 K                                          &\cite{Serrao2005}   \\
                    &                     &                  &           &$\sim$ 9          &            &        &                        &100 nm, PC; $P_{\textrm{max}} \sim$ 15         &\cite{Seo2013}      \\
Film                &                     &                  &           &                  &6.4         &        &                        &20 nm                                          &\cite{Cruz2014}     \\
                    &149                  &375--408          &           &                  &            &3.75    &                        &32 nm, SC                                      &\cite{Sharma2020}   \\ [1pt]
\hline
                    &                     &473               &8000       &2                 &            &        &                        &300 K, 5 kHz; weak ferroelectric               &\cite{Serrao2005}   \\
                    &                     &                  &           &3                 &            &        &                        &Along the $a$-axis, theoretical                &\cite{Serrao2005}   \\
PC                  &230                  &450               &           &                  &            &        &                        &Spin-charge coupling; relaxor-like             &\cite{Mall2018-2}   \\
                    &                     &                  &           &0.421             &            &        &                        &                                               &\cite{Taran2020}    \\
                    &373                  &453               &7700       &5.62              &            &2.31    &                        &$P_{\textrm{max}} \sim$ 5.62                   &\cite{Saxena2020}   \\ [1pt]
\hline
SC                  &                     &                  &           &$\sim$ 0.22       &            &        &                        &300 K, $\parallel$[110]; local polar domains   &\cite{Sanina2018}   \\ [1pt]
\hline
\hline
\end{tabular}
\end{table*}

\begin{figure*} [!t]
\centering \includegraphics[width=0.82\textwidth]{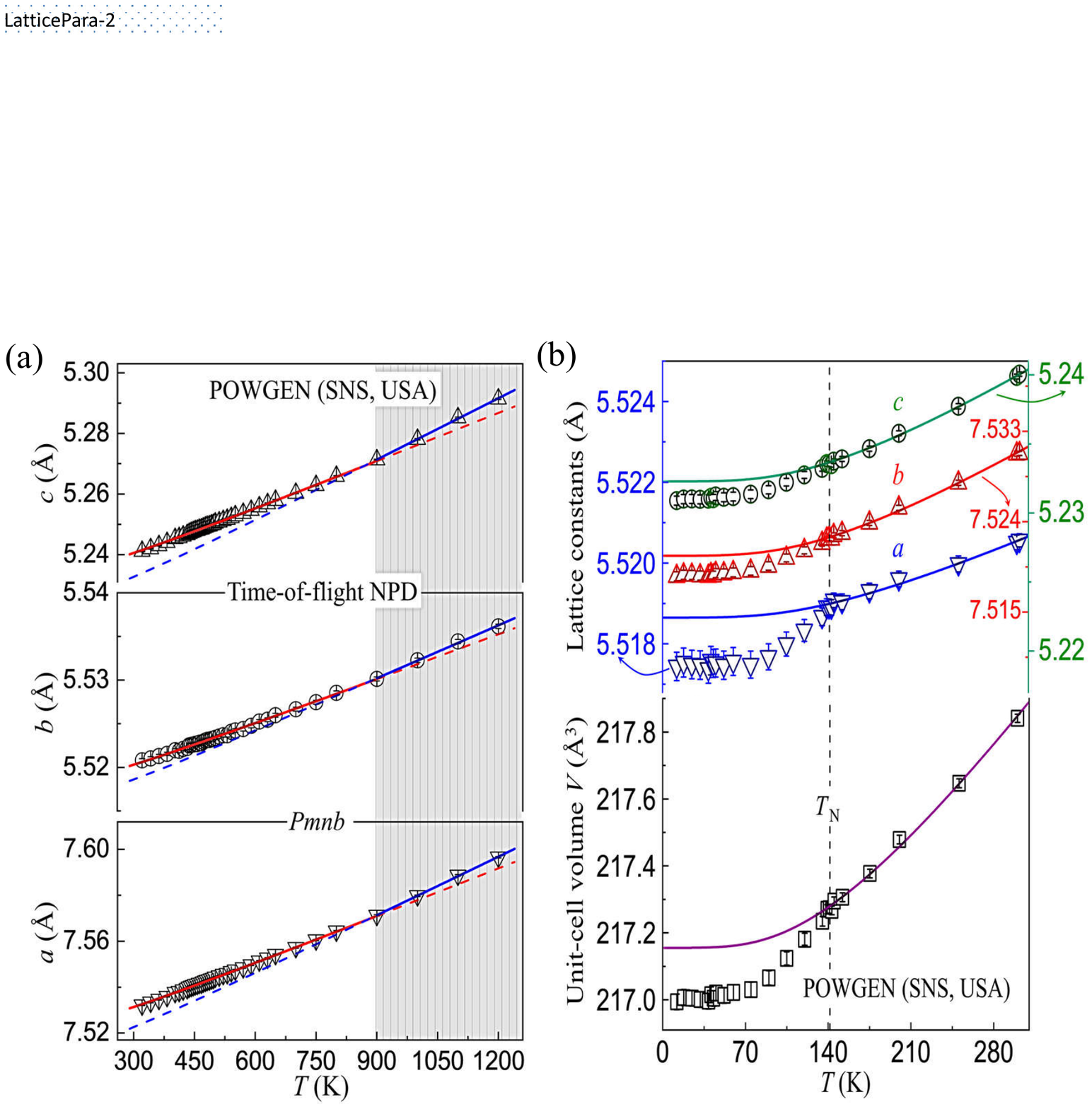}
\caption{
(a) Temperature-dependent lattice constants of \emph{a}, \emph{b}, and \emph{c} ($Pmnb$) of single-crystalline YCrO$_3$ (void symbols) from 321 to 1200 K from a time-of-flight NPD study. The solid lines represent theoretical calculations of the structure parameters. These lines were extrapolated to the entire temperature regime (dashed lines). Error bar is standard deviation. Reproduced with permission from Refs. \cite{Zhu2020, Zhu2020-2}. Copyright (2020) American Physical Society.
(b) Temperature dependence of lattice constants of \emph{a}, \emph{b}, and \emph{c} ($Pnma$) for single-crystal YCrO$_3$ in a temperature range of 12--302 K. Solid lines are the corresponding theoretical calculations. $T_\textrm{N}$ = 141.5(1) K is the AFM transition temperature. Error bars are the standard deviations.
}
\label{LatticePara-2}
\end{figure*}

\begin{figure*} [!t]
\centering \includegraphics[width=0.72\textwidth]{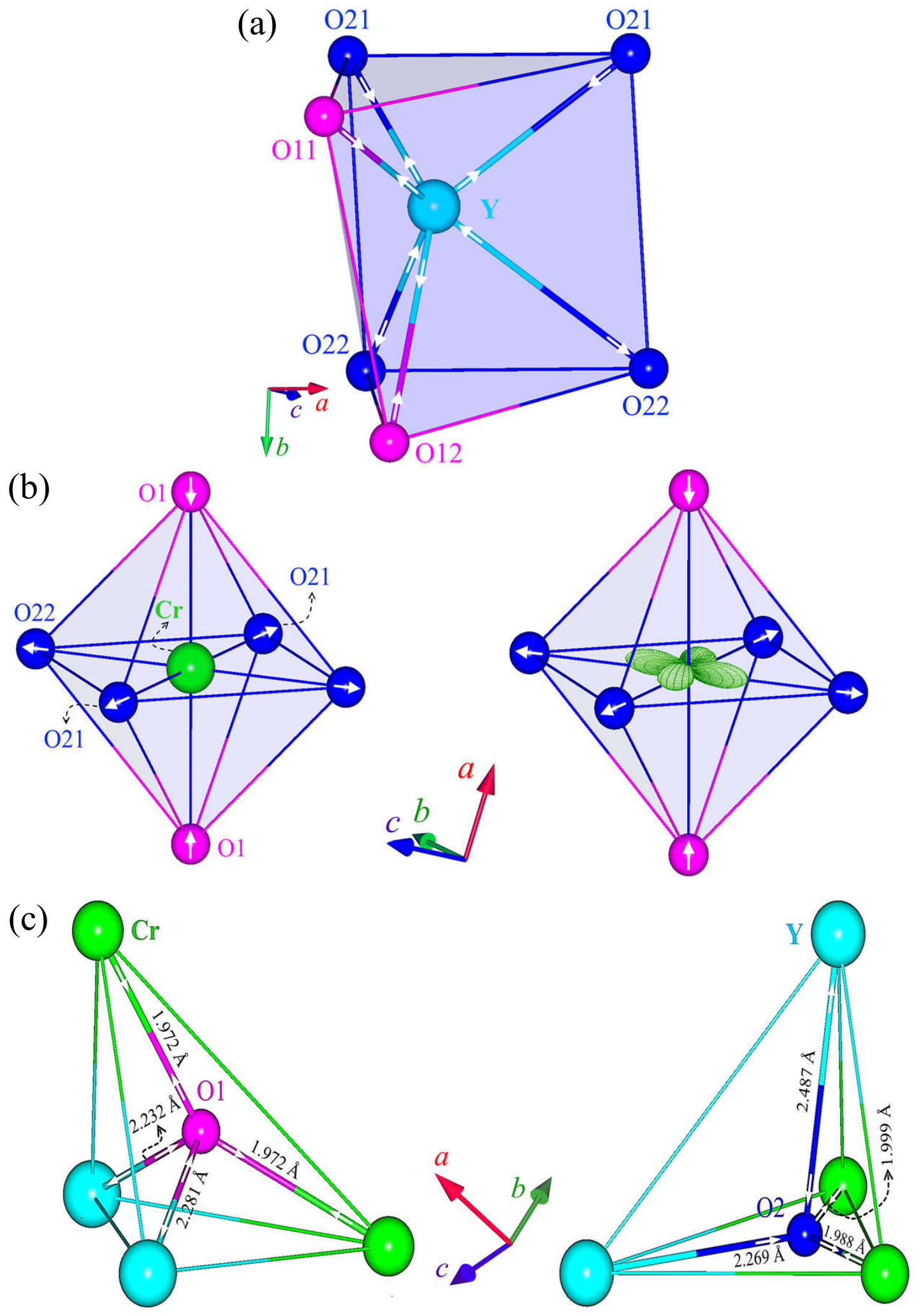}
\caption{
Local pentahedron environment of Y ion (a), octahedral environment of Cr ion (b), and tetrahedral environments of O1 and O2 ions (c) in a single-crystal YCrO$_3$, The Y, Cr, O11, O12, O21, and O22 ions are
labeled as displayed. The arrows sitting on the Y-O, O-Y, and O-Cr bonds and drawn through the oxygen ions schematically show the deduced pentahedron, octahedral, and tetrahedral distortion modes, respectively. In the octahedral geometry (right panel of (b)), the approximate 3$d_{\textrm{yz}}$ orbital shape in real space was schematically drown. Reproduced with permission from Refs. \cite{Zhu2020}. Copyright (2020) American Physical Society.
}
\label{LatticePara-3}
\end{figure*}

\begin{figure*} [!t]
\centering \includegraphics[width=0.64\textwidth]{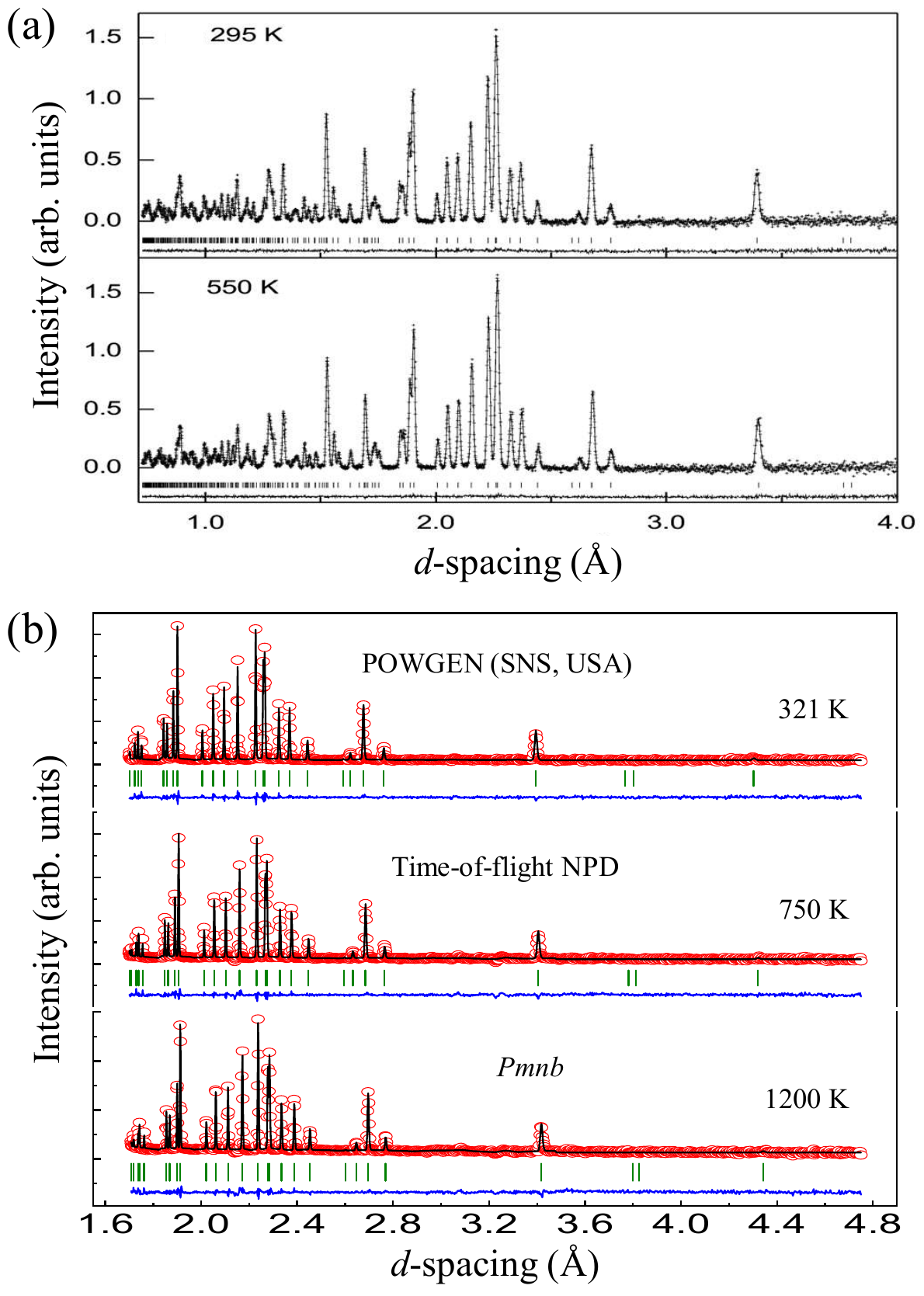}
\caption{
(a) NPD patterns of YCrO$_3$ in the ferroelectric-like (top panel, 295 K) and paraelectric (down panel, 550 K) states, measured at the NPDF instrument (located at the Los Alamos Neutron Scattering Center, USA). Tick marks point out Bragg peak positions. The bottom line represents the difference between calculated and observed patterns. Reproduced with permission from Ref. \cite{Ramesha2007}. Copyright (2007) IOP Publishing Ltd.
(b) Calculated (solid lines) and observed (circles) NPD patterns of the YCrO$_3$ single crystal, measured on the SNS POWGEN diffractometer (USA) at temperatures of 321, 750, and 1200 K. The vertical bars locate the nuclear Bragg peak positions ($Pmnb$ space group). The lower curves depict the difference between calculated and observed patterns. Reproduced with permission from Ref. \cite{Zhu2020}. Copyright (2020) American Physical Society.
}
\label{CrShift-1}
\end{figure*}

\begin{figure*} [!t]
\centering \includegraphics[width=0.55\textwidth]{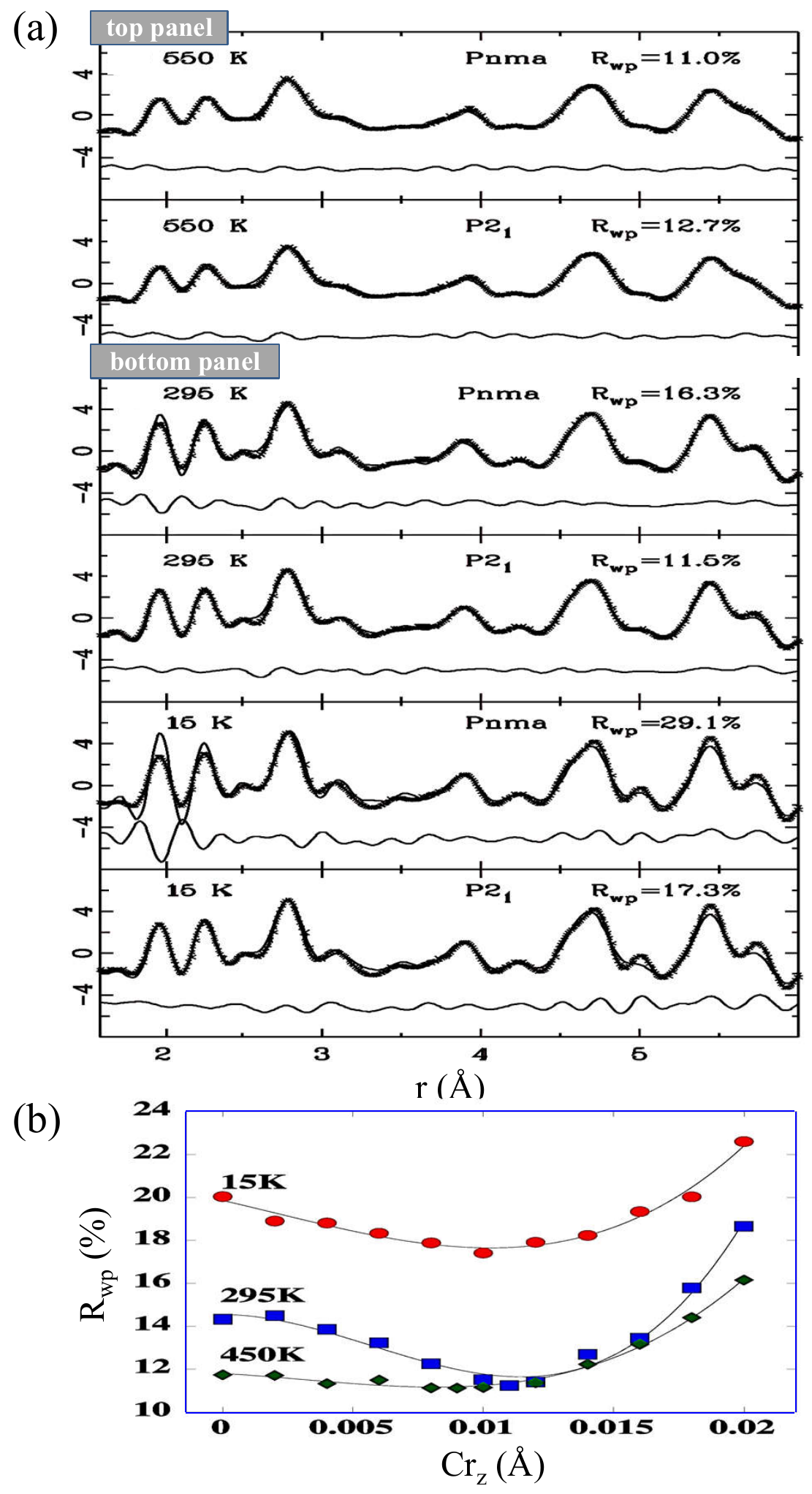}
\caption{
(a) Neutron PDF refinements over 1.6--6 {\AA} for a YCrO$_3$ compound in the paraelectric state (top panel, 550 K) and in the ferroelectric state (bottom panel, 15 and 295 K). The study shows that the $Pnma$ space group is in better agreement with collected data in the paraelectric regime, in contrast to the $P12_11$ model (No. 4, noncentrosymmetric). The $P12_11$ space group displays better agreement in the ferroelectric state. The space groups used and the corresponding value of goodness of fit $R_{\textrm{wp}}$ are marked. The differences between calculated and observed data are displayed below each graph.
(b) Values of $R_{\texttt{wp}}$ as a function of Cr displacement, extracted from the PDF measurement with $r_{\texttt{max}}$ = 6 {\AA} along the \emph{z} direction in the ferroelectric regime of YCrO$_3$ compound.
Reproduced with permission from Ref. \cite{Ramesha2007}. Copyright (2007) IOP Publishing Ltd.
}
\label{CrShift-2}
\end{figure*}

\begin{figure} [!t]
\centering \includegraphics[width=0.75\textwidth]{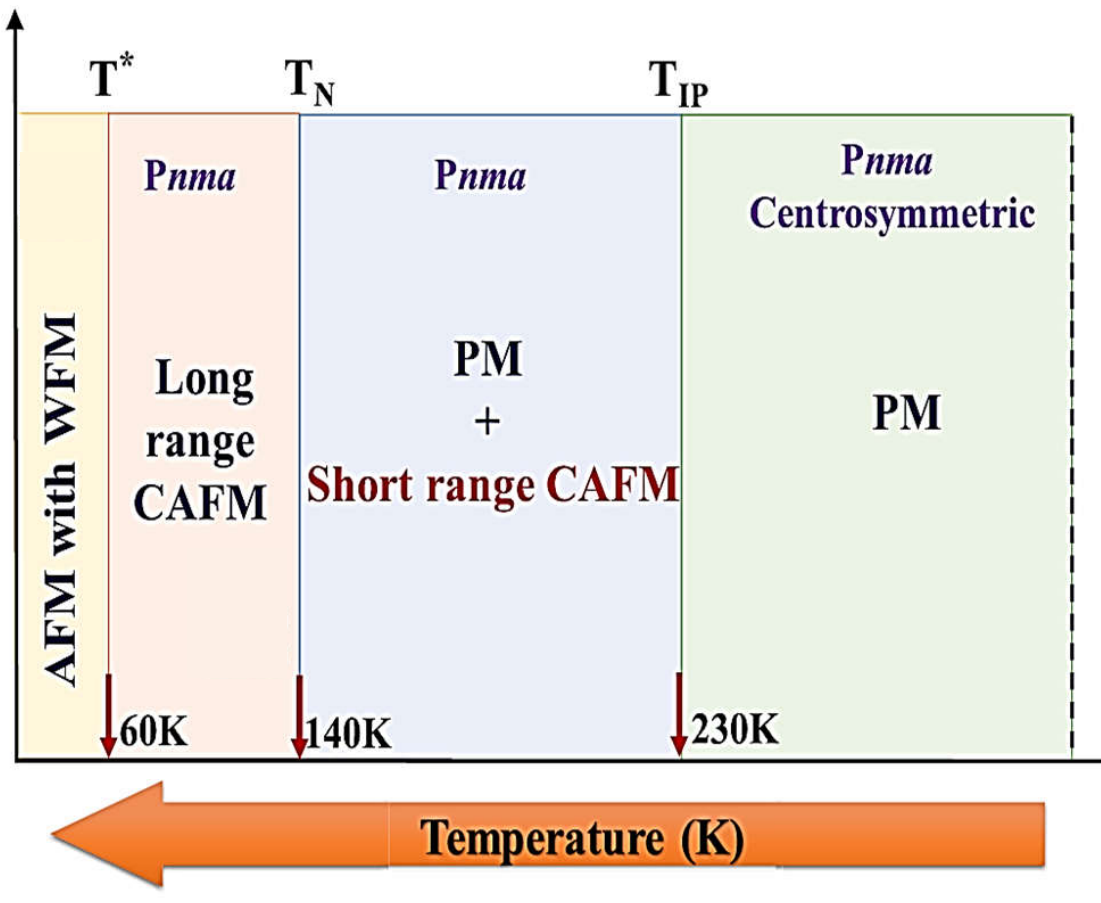}
\caption{
Reported magnetic phase diagram for YCrO$_3$, implying an formation of a new phase with short-range AFM correlations below $\sim$ 230 K. Here PM = Paramagnetic; IP = Intermediate point; AFM = Antiferromagnetic; CAFM = Canted antiferromagnetic; WFM = Weak ferromagnetic. To see detailed discussion in the text. Reproduced with permission from Ref. \cite{Mall2017-1}. Copyright (2017) IOP Publishing Ltd.
}
\label{PhaseDia-1}
\end{figure}

\begin{figure} [!t]
\centering \includegraphics[width=0.75\textwidth]{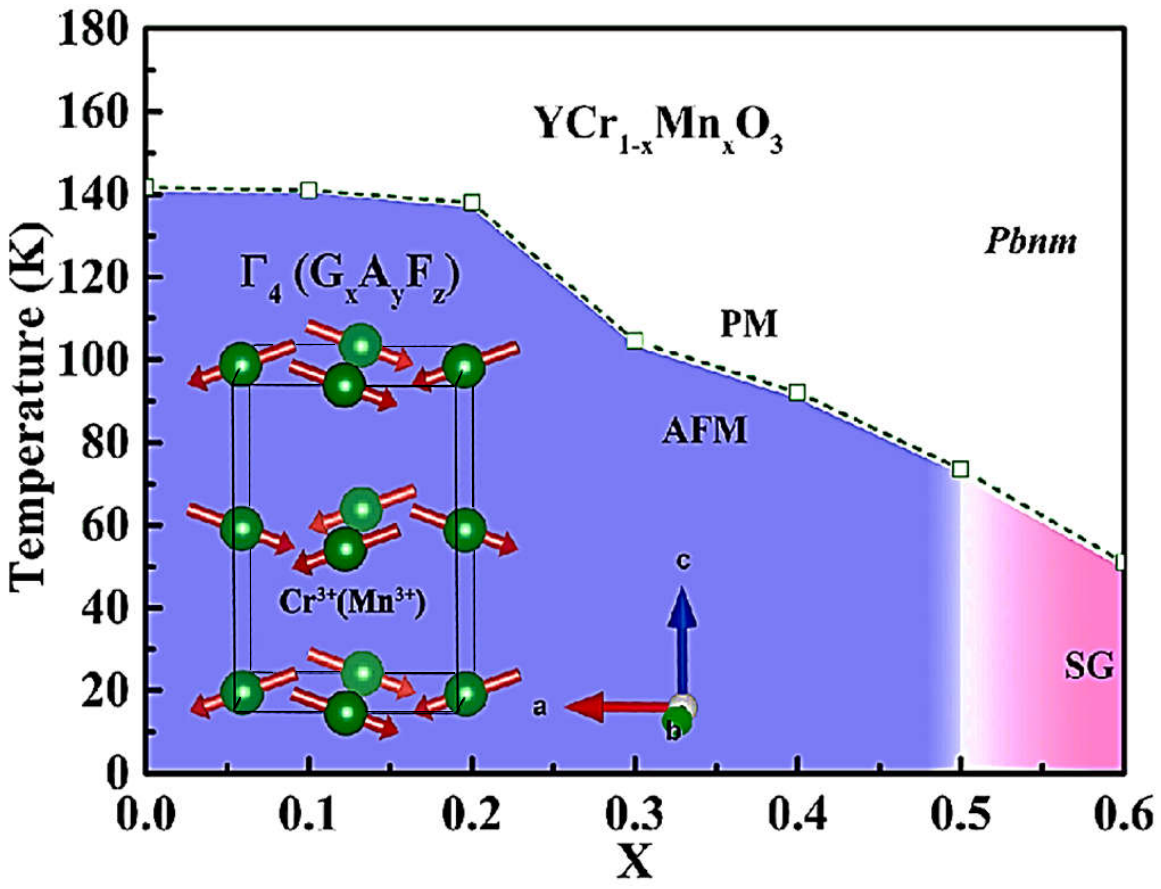}
\caption{
Temperature- and composition-dependent magnetic phase diagram of the system of YCr$_{1-x}$Mn$_x$O$_3$ compounds. Inset displays a schematic configuration of the $\Gamma$4 Cr$^{3+}$ (Mn$^{3+}$) spin structure. Here PM = Paramagnetic; AFM = Antiferromagnetic; SG = Spin glass. To see detailed discussion in the text.
Reproduced from Ref. \cite{Wang2021}. Copyright (2021) Elsevier B.V.
}
\label{PhaseDia-2}
\end{figure}

\begin{figure*} [!t]
\centering \includegraphics[width=0.78\textwidth]{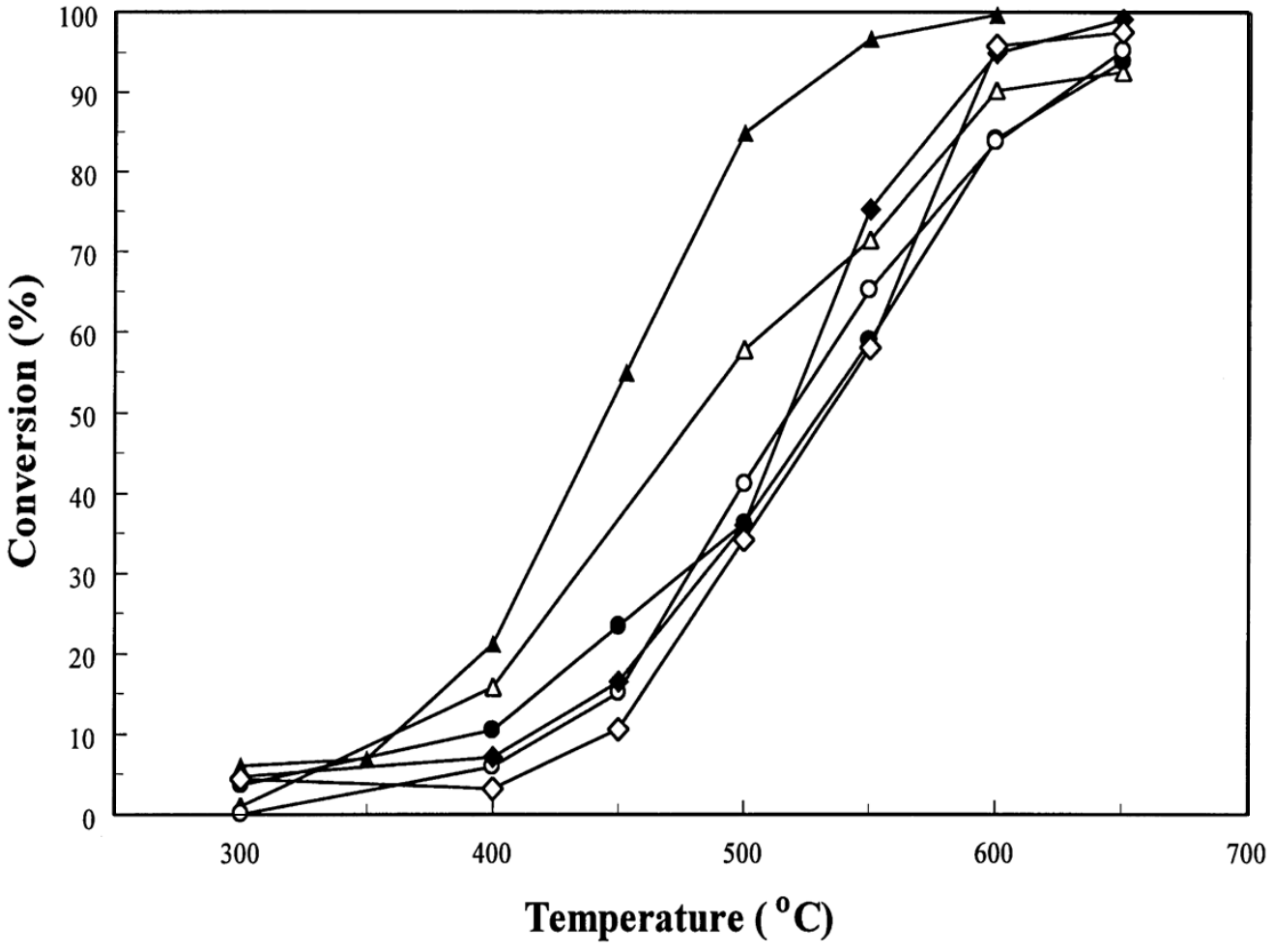}
\caption{
Steady-state \emph{o}-DCB conversion as a function of temperature for different catalysts tested: ({\large{$\blacktriangle$}}) YCrO$_3$, ({\large\large\large{$\triangle$}}) LaCrO$_3$, ({\large{\large{\large{$\bullet$}}}}) YMnO$_3$, ($\bigcirc$) LaMnO$_3$, ($\blacklozenge$) YFeO$_3$, and ({\large\large\large{$\diamondsuit$}}) LaFeO$_3$ (600 ppm \emph{o}-DCB, 10\% O$_2$, 1\% H$_2$O). Reproduced with permission from Ref. \cite{Poplawski2000}. Copyright (2000) Elsevier Science B.V.
}
\label{Application-2}
\end{figure*}

\begin{figure*} [!t]
\centering \includegraphics[width=0.75\textwidth]{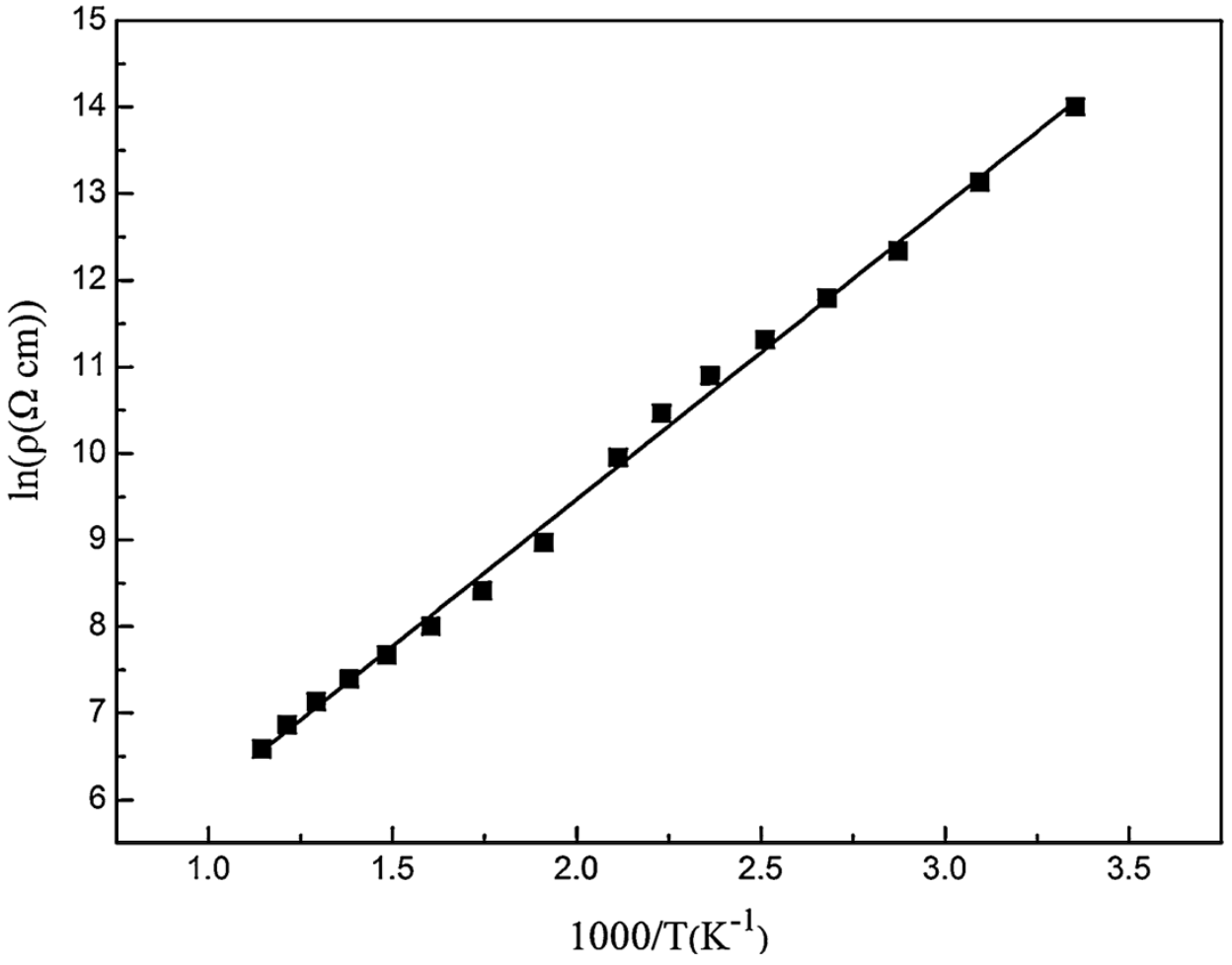}
\caption{
Relationship between ln$\rho$ and 1,000/\emph{T} for YCrO$_3$ ceramics, indicating a potential for NTC thermistors. Reproduced with permission from Ref. \cite{Zhang2014}. Copyright (2014) Springer Science+Business Media New York.
}
\label{Application-3}
\end{figure*}

\begin{figure*} [!t]
\centering \includegraphics[width=0.82\textwidth]{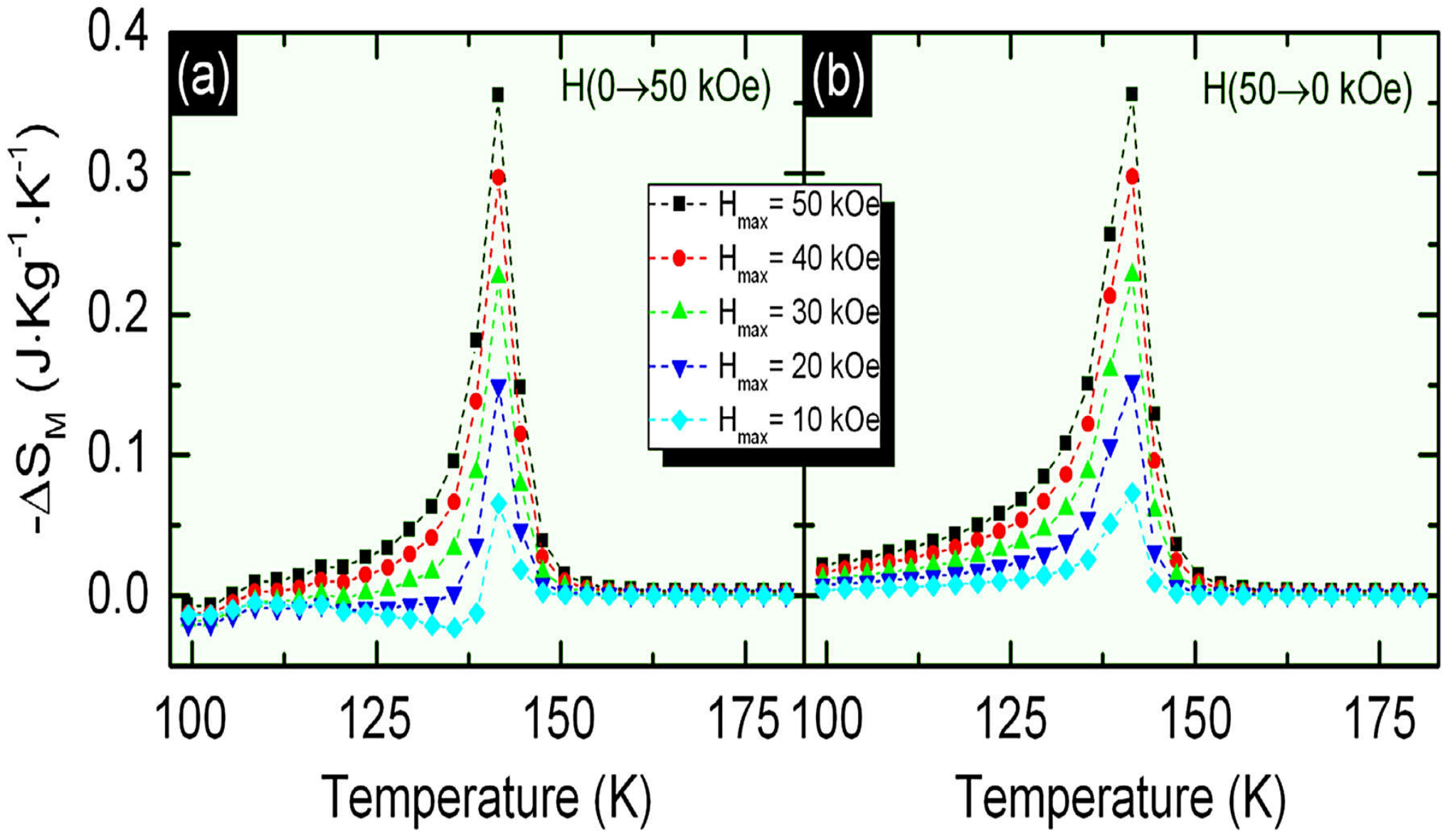}
\caption{
Magnetic entropy change with temperature at various field changes for YCrO$_3$ compound: (a) 0 to 50 kOe; (b) 50 to 0 kOe. Reproduced with permission from Ref. \cite{Oliveira2016}. Copyright (2015) Elsevier Ltd.
}
\label{Application-4}
\end{figure*}

\begin{figure*} [!t]
\centering \includegraphics[width=0.75\textwidth]{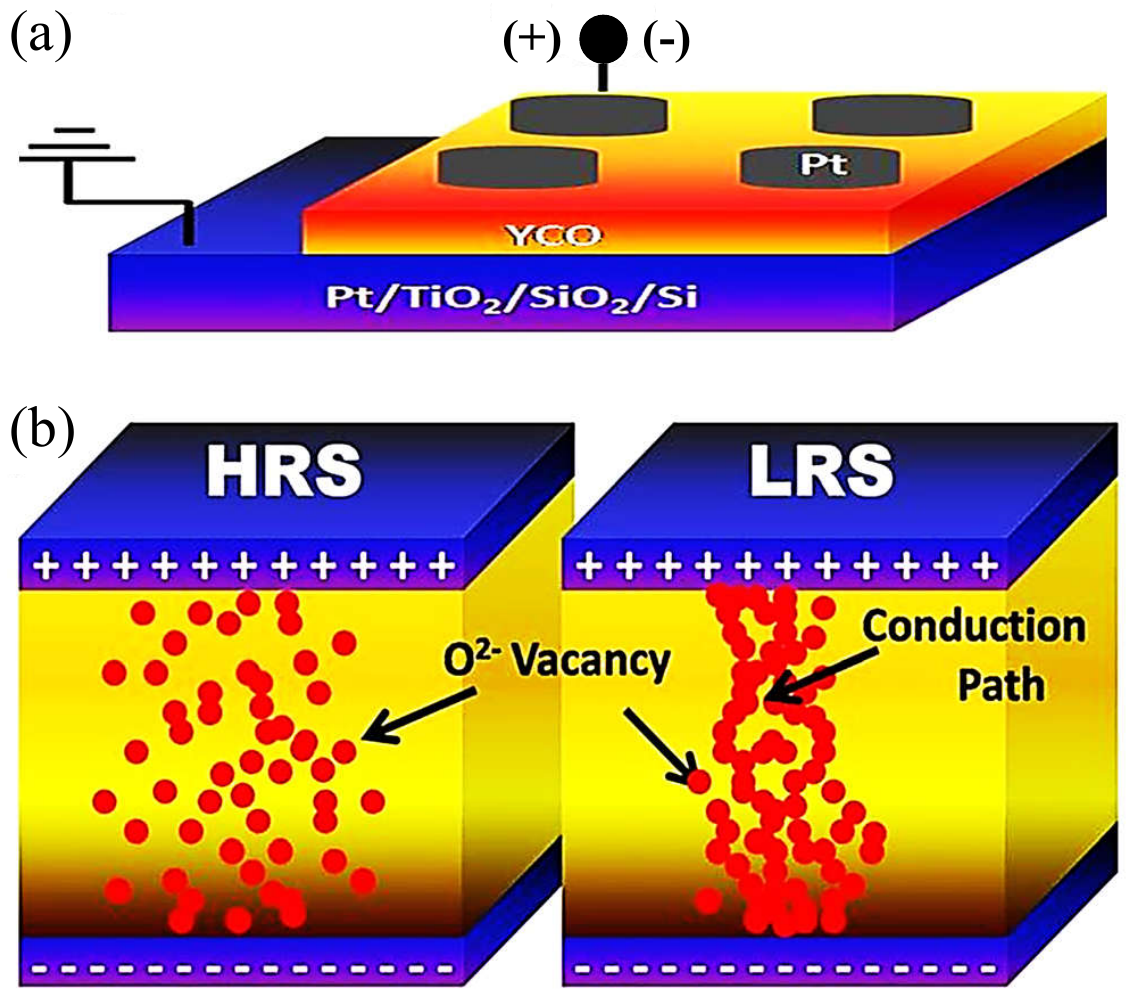}
\caption{
(a) The illustration diagrams of Pt/YCrO$_3$/Pt device. (b) Configurations of oxygen vacancies in the high-resistance state (HRS) and the low-resistance state (LRS), respectively. Reproduced with permission from Ref. \cite{Sharma2014-2}. Copyright (2014) AIP Publishing LLC.
}
\label{Application-5}
\end{figure*}

\begin{figure*} [!t]
\centering \includegraphics[width=0.82\textwidth]{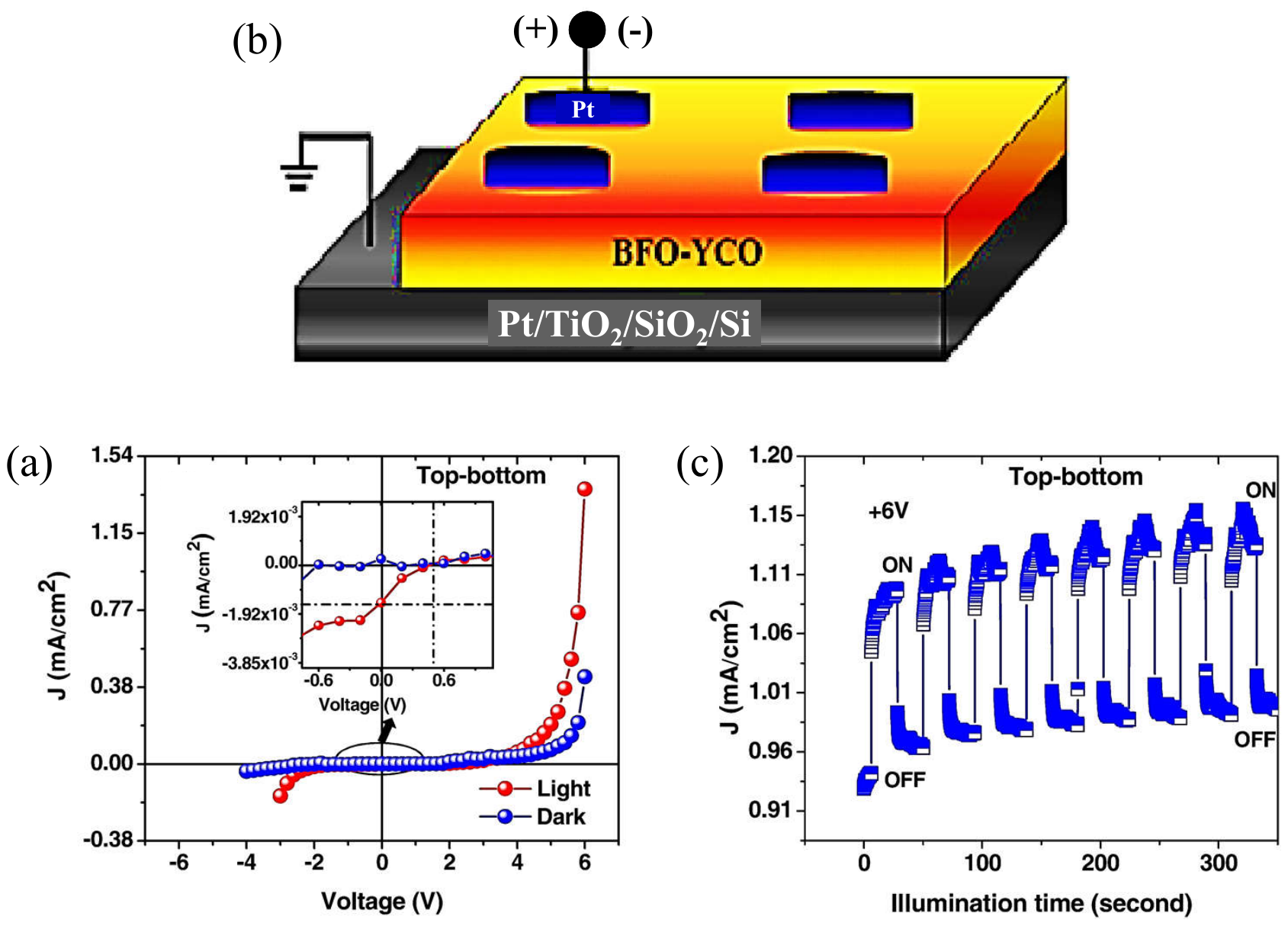}
\caption{
The dark current density and the illuminated current density as a function of voltage (i.e., \emph{J}–\emph{V} curves) (a) in the top-bottom electrode structure (b). Inset of (a) shows a zoomed view around the origin. (c) Change of current versus time upon turning the light on and off for the composite BiFeO$_3$–YCrO$_3$ thin film with the top–bottom electrode configuration. Reproduced with permission from Ref. \cite{Sharma2014-3}. Copyright (2014) IOP Publishing Ltd.
}
\label{Application-6}
\end{figure*}

\subsection{Lattice parameters versus doping level}

As shown in \textcolor[rgb]{0.00,0.00,1.00}{Fig.}~\ref{LatticePhaseDia}, there is no obvious change in lattice constants (\emph{a}, \emph{b}, and \emph{c}) and unit-cell volume (\emph{V}) of Y$_{1-x}$Ca$_x$CrO$_3$ system \cite{Duran2012-1}. As the Pr-doping level in Y$_{1-x}$Pr$_x$CrO$_3$ system \cite{Duran2018-3} increases, lattice constant \emph{b} almost remains unchanged, whereas, the linear increases of \emph{a} and \emph{c} jointly lead to a linear increase of \emph{V}. For the YCr$_{1-x}$Al$_x$O$_3$ system \cite{Cruciani2009, Duran2018-1}, lattice parameters \emph{a}, \emph{b}, \emph{c}, and \emph{V} all decrease linearly with increasing the value of \emph{x}. In contrast, all lattice parameters increase linearly with doping level $x$ in the YCr$_{1-x}$Fe$_x$O$_3$ system \cite{Fabian2016}. For the YCr$_{1-x}$Mn$_x$O$_3$ system \cite{Sahu2008, Kamlo2011}, it seems that there display domes in lattice parameters of \emph{a}, \emph{b}, and \emph{V}, while the lattice constant \emph{c} keeps a linear decrease. It is clear that there is no regular rule of changes in lattice parameters with chemical doping on the Y or Cr sites.

\subsection{Magnetic structure versus temperature}

The magnetic phase diagram of YCrO$_3$ compound was proposed by Mall \emph{et al}. as shown in \textcolor[rgb]{0.00,0.00,1.00}{Fig.}~\ref{PhaseDia-1} \cite{Mall2017-1}. It was divided into four regimes: (i) When $T > T_{\textrm{IP}} \sim$ 230 K (IP = intermediate point, where the lowest \emph{g}-factor appears, extracted from the electron paramagnetic resonance study \cite{Mall2017-1}), YCrO$_3$ stays in a pure paramagnetic state, and the Cr moments orient randomly as usual. (ii) With decreasing temperature from $T_{\textrm{IP}}$ down to $T_\textrm{N} \sim$ 140 K, spin correlations of short-range CAFM form gradually with the paramagnetic background \cite{Mall2017-1}. (iii) At $T_\textrm{N} \sim$ 140 K, YCrO$_3$ undergoes a magnetic phase transition from the short-range CAFM to a long-range ordered CAFM \cite{Mall2017-1, Oliveira2016, Mall2017-3, Sharma2014-1}. This magnetic phase persists down to 60 K. It is pointed out that Sharma \emph{et al}. observed a clear loop opening in magnetization isotherms at 125 K and no saturation of magnetization up to 20 kOe indicative of a coexistence of WFM and AFM phases at this temperature. (iv) Below 60 K, YCrO$_3$ transfers from the long-range CAFM state to an AFM phase with WFM component \cite{Mall2017-1, Sharma2014-1}. It is pointed out that some phenomena displayed in \textcolor[rgb]{0.00,0.00,1.00}{Fig.}~\ref{PhaseDia-1} have not been demonstrated by modern neutron scattering that is a powerful and unique probe for monitoring magnetism-related behaviors.

Wang \emph{et al}. synthesized a set of powder samples of orthorhombic YCr$_{1-x}$Mn$_x$O$_3$ ($0 \leq x \leq 0.6$) and carried out XRPD, magnetization, and NPD studies \cite{Wang2021}. As shown in \textcolor[rgb]{0.00,0.00,1.00}{Fig.}~\ref{PhaseDia-2}, when $0.2 \leq x \leq 0.5$, the Cr$^{3+}${/}Mn$^{3+}$ sublattice has a $\Gamma$4 $(G_x, A_y, F_z$) AFM structure, accompanied by a canted FM one. When x = 0.1 and 0.2, the compounds show a behavior of negative magnetization. An applied-magnetic field is able to suppress and reverse the phenomenon. With increasing the doping level to 0.6, the sample shows a clear spin-glass feature below the temperature of spin freezing. The results were associated with competing interactions of the weakened AFM Cr$^{3+}$-Cr$^{3+}$ ions and the enhanced FM Cr$^{3+}$-Mn$^{3+}$ ions while doping with Mn$^{3+}$ element \cite{Wang2021}. As the doping level (x) increases from 0.1 to 0.6, the AFM transition temperature decreases, and the cell distortion becomes larger and larger.

\section{Applications}

In this section, we summarize some (potential) applications of parent and doped YCrO$_3$ compound based on journal and patent literatures (Table~\ref{application}). The compound of YCrO$_3$ crystallizes with a perovskite-type structure and exhibits P-type behaviors of semiconductors \cite{Jin2004}. The melting temperature of refractory YCrO$_3$ is (2290$\pm$30) $^{\circ}$C \cite{Jin2004}, and it holds high stabilities of electrical, thermal, structural, and chemical properties, therefore, (potential) applications are numerous. The YCrO$_3$ can be used solely in the form of nanocrystal, polycrystal, or thin film. It can also be hybridized as a component of alloys or composites when more functionalities are required. The selection depends on a detailed application as reviewed below.

\subsection{Anti-corrosion}

The YCrO$_3$ compound can be formed at the scale-alloy (chromium plus 5 volume percent Y$_2$O$_3$) interface during the long-period oxidation process in air. The formation of YCrO$_3$ layer between Cr$_2$O$_3$ and chromium alloy may constitute an increasingly effective blocking layer to continued chromium ion migration into the grown Cr$_2$O$_3$ layer through the scale, indicating an improved oxidation resistance of Fe-Cr alloys or pure chromium by an addition of yttrium \cite{Seybolt1966}. Therefore, YCrO$_3$ displays anti-corrosion effect in alloys \cite{Seybolt1966, Zenzo2009} or stainless steels \cite{Tatsuaki2013}.

\subsection{Magnetohydrodynamic electrode}

Doped lanthanum chromates with Mg, Ca, or Sr display good electrical and electronic conductivity as well as high melting temperature ($>$ 2500 K). However, above $\sim$ 1850 K, Cr evaporates easily, and a hygroscopic La$_2$O$_3$ phase forms, leading to a large change in volume and the loss of mechanical integrity \cite{McCarthy1982}. In contrast, yttrium-chromates have similar properties to La-chromates, but do not hydrate. The electrochemical corrosion of Mg-doped La and Y chromates was investigated, suggesting that Y-chromates could be a better material for viable magnetohydrodynamic electrodes than La-chromates, and the Y-chromate with an addition of 5 mol{\%} MgO shows a greater resistance to electrochemical and chemical corrosion than its analogous La-chromates in molten coal-slags \cite{McCarthy1982}.

\subsection{Catalyst}

Perovskite ABO$_3$ compounds (A = rare earth, or alkali earth ion, B = $3d, 4d$, or $5d$ transition metal ions) are interesting materials for oxidation catalysts due to their properties for catalytic applications such as stability of valence states, defect chemistry, and mobile oxygen ions \cite{Poplawski2000}. The catalytic property of perovskite ABO$_3$ oxides, where A = Y, La, Gd, and Nd, B = Mn, Fe, Co, and Cr, was investigated by Poplawski \emph{et al}. for the oxidation of 1,2-dichlorobenzene (i.e, \emph{o}-DCB), and this is a typical compound for the highly-toxic polychlorinated dibenzodioxins \cite{Poplawski2000}. The study demonstrates that ACrO$_3$ perovskites are more active than the compounds with other transition metals. The replacement of La with other rare-earth trivalent cations at the A-sites leads to a stable surface area and, to some extent, an improved catalytic activity. Most important, the catalyst of YCrO$_3$ compound is highly more active among the various compounds studied (see \textcolor[rgb]{0.00,0.00,1.00}{Fig.}~\ref{Application-2}) \cite{Poplawski2000}. Moveover, YCrO$_3$ does not display any loss of the initial activity even after several hours on stream; in contrast, other studied compounds loss 10--20{\%} within the first 5--10 h \cite{Poplawski2000}.

\subsection{Negative-temperature-coefficient thermistor}

Thermistors, as a tool for accurate temperature measurements, are very important for industry development and our daily life. Compared to traditional positive-temperature-coefficient-type thermistors, NTC thermistors show a decreasing trend in resistance upon warming. The broadband high-temperature thermistors could be made of thin films of oxide semiconductors like YCrO$_3$ (a P-type semiconductor) and doped YCrO$_3$ that have been believed to be a potential candidate for thermistor applications due to its thermoelectric property \cite{Zhang2014, Mori1996, Kuzuoka2005, Takeuchi2021, iwaya1997ceramic, sorg2001yttrium, taira2020temperature, iwaki2011temperature, Taira2020, Takeuchi2021-1, Murata1998, Yoshihara2021}. The YCrO$_3$ ceramics (nanoparticles with size $\sim$ 72 nm) were synthesized by Zhang \emph{et al}. with the field-assisted sintering technique. During the sintering process, a high current flow leads to the simultaneous applications of high temperature and axial pressure, resulting in densed YCrO$_3$ with a relative density of 97.6{\%}. The electrical properties of the samples as sintered were studied. The as-sintered YCrO$_3$ ceramics have the values of $\rho_{25} =$ 1.21 $\times$ 10$^6$ $\Omega$ cm, $\rho_{600} =$ 7.26 $\times$ 10$^2$ $\Omega$ cm, $B_{25/600} =$ 3,358 K, and an activation energy = 0.293 eV. As depicted in \textcolor[rgb]{0.00,0.00,1.00}{Fig.}~\ref{Application-3}, the results show that YCrO$_3$ ceramics are potential materials for designing NTC automotive thermistor used in a wide temperature regime of 25--600 $^\circ$C \cite{Zhang2014, Kim2003, Jin2004}.

A method for manufacturing a NTC-type thermistor with YCrO$_3$ thin films deposited on a silicon wafer was provided to reduce the power consumption using a high-frequency magnetron sputtering \cite{Jin2004}.

The powder of (Mn$\cdot$Cr)O$_4$ is one of the high-temperature thermistor materials, displaying a higher resistivity and a higher temperature coefficient of resistance than those of YCrO$_3$ \cite{Kuzuoka1999}. By varying the mixing ratio of (Mn$\cdot$Cr)O$_4$ and YCrO$_3$ powders and firing them at temperatures from 1400--1700 $^{\circ}$C, components of the mixture will react with each other and produce (Mn$_x$$\cdot$Cr$_y$)O$_4$ spinel, where $x >$ 0, $y \leq$ 2 and $x + y =$ 3, and Y(Cr$\cdot$Mn)O$_3$ perovskite. Such resultant has stable thermistor properties for a high-temperature thermistor \cite{Kuzuoka1999}.

The NTC thermistor made of YCrO$_3$ can be used as one layer of a multi-layer sensor integrated into an object, monitoring harsh working environments such as high temperature, pressure, or mechanical strain within the object \cite{Hardwicke2006}.

\subsection{Magnetic refrigeration}

The magnetocaloric effect of poly-crystalline YCrO$_3$ samples was investigated by Oliveira \emph{et al}. for potential applications in magnetic refrigeration \cite{Oliveira2016}. The $M(H)$ isotherms curves were studied at applied-magnetic fields from 0 to 50 kOe in the temperature regime of 98--182 K. The extracted magnetic entropy changes at different magnetic-field changes show a peak at the CAFM transition temperature $T_\textrm{N} \sim$ 140 K. As shown in \textcolor[rgb]{0.00,0.00,1.00}{Fig.}~\ref{Application-4}, as the magnetic-field change increases, the maximum increases and attains a value of $-\Delta$\emph{S}$_\textrm{M}$ $\sim$ 0.36 J kg$^{-1}$K$^{-1}$ at a field change of 50 kOe. This is much larger than that of LaCrO$_3$ ($T_\textrm{N}$ = 285 K), $\sim$ 0.067 J kg$^{-1}$K$^{-1}$, at the same field variation \cite{Tiwari2015}. The extracted refrigerant capacity for YCrO$_3$: $R_\textrm{C} =$ 7.1 J kg$^{-1}$, demonstrating that YCrO$_3$ can act as a candidate compound for magnetic refrigeration \cite{Oliveira2016, Tiwari2020, Bhowmik2021}.

\subsection{Protective coating}

Goncharov \emph{et al}. carried out an investigation on the diffusion permeability of different kinds of Y-Cr-O materials to enhance the quality of those heat-resistant coatings \cite{Goncharov2016}. The substrates were made of the chromium-based alloy, i.e., Cr-0.7V-0.17La. The protective coatings on substrates display a good barrier effect on diffusion fluxes in the substrate-coating-environment system. Among the studied protective coatings, the system of (Y + YCr + YCrO$_3$) alloys shows the best protection of the substrates against gas saturation \cite{Goncharov2016}.

The YCrO$_3$ compound can be used as a component of a multi-layer coating for protecting alloys and metals against high-temperature oxidation. The coating could be utilized in devices at high temperatures, especially for coating interconnect material in solid-oxide electrolytic devices, such as electrolysis cells and solid-oxide fuel cells \cite{Hendrickson2011}.

\subsection{Solid oxide fuel cell}

The doped YCrO$_{3-\delta}$ ceramics were investigated as promising anode oxides for low-temperature solid-oxide fuel cells (400--650 $^\circ$C) \cite{Hussain2017, jabbar2019chromate, barnett2003fuel}. One of the studied materials shows an electrical conductivity of 1.7 and 3.5 S/cm at 600 $^\circ$C, which was measured under reducing gases and air, respectively. The conductivity of the doped YCrO$_{3-\delta}$ ceramics is higher than that of traditional LaCrO$_3$-based anode materials. To achieve catalytic properties for H$_2$ oxidation, Ni-GDC nanoparticles were introduced on the surface of the doped-YCrO$_{3-\delta}$ ceramic scaffold. The electrochemical properties of the newly-developed material were determined that a peak power density of 200 mW/cm$^2$ at 0.4 A/cm$^2$ was obtained at 650 $^\circ$C in H$_2$/3{\%} H$_2$O using an electrolyte supported solid oxide fuel cell \cite{Hussain2017}.

The YCrO$_3$ based material Y$_{1-x}$A$_x$Cr$_{1-y}$B$_y$O$_3$, where A is an alkaline metal element Sr or Ca, $x =$ 0--0.3, B is Fe or cobalt, and $y =$ 0--0.3, can be used as the cathode material of a sulfur-oxygen fuel cell \cite{Chu2013}. The working temperature of such a battery is in the range of 500--800 $^{\circ}$C. When the working temperature is at 800 $^{\circ}$C, the voltage is up to 0.74 V \cite{Chu2013}. The cathode materials can be synthesized by solid-state reactions, a chemical coprecipitation, or the sol-gel technique.

The LaCrO$_3$ and YCrO$_3$ compounds were believed to be good connecting material utilized in solid oxide fuel cells \cite{Chen2017}. For the LaCrO$_3$ compound, below 1300 $^{\circ}$C, it shows good chemical stability with the yttria stabilized zirconia electrolyte; above 1400 $^{\circ}$C, it can react with zirconium oxide (ZrO$_2$) to form the La$_2$Zr$_2$O$_7$ phase that has a high resistance, which restricts the application of LaCrO$_3$ in solid-oxide fuel cells. In contrast, YCrO$_3$ shows a higher chemical stability and doesn't react with zirconium oxide at the preparation temperature of solid-oxide fuel cells \cite{Chen2017}. Therefore, YCrO$_3$ is a better connection material for solid-oxide cells.

\subsection{Non-volatile memory}

The resistance switching effect can be utilized in designing the resistive random access memory \cite{Sharma2014-2}. Such kind of device has some advantages such as simple structure as illustrated in \textcolor[rgb]{0.00,0.00,1.00}{Fig.}~\ref{Application-5}, high switching speed, low power consumption, and high storage density, therefore, it is of great interesting \cite{Sharma2014-2}. To study the resistive switching phenomenon, Sharma \emph{et al}. fabricated amorphous YCrO$_3$ films on the Pt/TiO$_2$/SiO$_2$/Si substrate via the pulsed laser deposition technique \cite{Sharma2014-2}. The consequent Pt/YCrO$_3$/Pt device exhibits stable unipolar resistive switching behavior with excellent retention and endurance characteristics. The resistance ratio between LRS and HRS is $\sim$ 10$^5$. The observed resistive switching behavior was attributed to oxygen vacancies as illustrated in \textcolor[rgb]{0.00,0.00,1.00}{Fig.}~\ref{Application-5}. The study indicates that YCrO$_3$ could act as a possible candidate for non-volatile memory application \cite{Sharma2014-2}.

\subsection{Photovoltaic material}

Ferroelectric photovoltaic ceramics are promising materials for exploiting solar energy due to potential large open circuit photovoltages \cite{Sharma2014-3}. The photovoltaic effect of a composite 0.9(BiFeO$_3$)–0.1(YCrO$_3$) thin film was studied \cite{Sharma2014-3}, and in the top-bottom electrode structure (see \textcolor[rgb]{0.00,0.00,1.00}{Fig.}~\ref{Application-6}), the short-circuit current-density is 1.48 $\mu$A cm$^{-2}$ ($J_{\textrm{SC}}$), and the open circuit voltage ($V_{\textrm{OC}}$) is 0.51 V, while in the lateral electrode configuration, $V_{\textrm{OC}}$ = 0.32 V, and $J_{\textrm{SC}}$ = 0.44 $\mu$A cm$^{-2}$, indicating that the composite thin film has an excellent photovoltaic effect \cite{Sharma2014-3}.

\begin{table*}[!t]
\small
\caption{\newline Summary of the applications of parent and doped YCrO$_3$ compounds.}
\label{application}
\begin{tabular}{p{2.8cm}|p{3.1cm}p{6.9cm}l}
\hline
\hline
Applications                  &  Materials                               &  Properties and Advantages                                                                                                                                                & Refs.                                                      \\ [2pt]
\hline
P-type semiconductor          &  YCrO$_3$                                &  Refractory, high stabilities of electrical, thermal, structural, and chemical properties. Nanocrystal, polycrystal, thin film, component of alloys or composites.        & \cite{Jin2004}                                             \\
\hline
Anti-corrosion                &  YCrO$_3$ blocking layer                 &  YCrO$_3$ formed at the interface of scale-alloys, improving oxidation resistance.                                                                                        & \cite{Seybolt1966,Zenzo2009,Tatsuaki2013}                  \\
\hline
Magnetohydrodynamic electrode &  Mg-doped Y-chromates                    &  Good electrical and electronic conductivity, high melting temperature, not hydrate, great resistance to electrochemical and chemical corrosion.                          & \cite{McCarthy1982}                                        \\
\hline
Catalyst                      &  YCrO$_3$ perovskites                    &  Stability of valence states, defect chemistry, mobile oxygen ions, high catalytic activity, no loss of catalytic activity, good catalyst for the oxidation of $o$-DCB.   & \cite{Poplawski2000}                                       \\
\hline
                              &  YCrO$_3$ ceramics                       &  Negative temperature-resistance dependance, wide working temperature regime of 25-600 $^{\circ}$C, used for NTC automotive thermistor.                                   & \cite{Zhang2014,Kim2003,Jin2004}                           \\
                              &  YCrO$_3$ thin films                     &  Deposited on silicon wafers, reducing power consumption.                                                                                                                 & \cite{Jin2004}                                             \\
NTC thermistor                &  Y(Cr$\cdot$Mn)O$_3$ perovskite          &  High resistivity, high temperature coefficient of resistance, stable thermistor properties for high temperature thermistor.                                              & \cite{Kuzuoka1999}                                         \\
                              &  Integrated into a multi-layer sensor    &  Monitoring harsh working environments such as high temperature, pressure, or mechanical strain.                                                                          & \cite{Hardwicke2006}                                       \\
\hline
Magnetic refrigeration        &  YCrO$_3$ polycrystal                    &  Magnetocaloric effect, large magnetic entropy changes, large refrigerant capacity.                                                                                       & \cite{Oliveira2016,Tiwari2020,Bhowmik2021}                 \\
\hline
                              &  (Y+YCr+YCrO$_3$) alloys                 &  Barrier effect on diffusion fluxes, protecting substrate against gas saturation.                                                                                         & \cite{Goncharov2016}                                       \\
Protective coating            &  YCrO$_3$ on multi-layer coating         &  Protecting alloy and metals against high-temperature oxidation, utilized in electrolytic devices.                                                                        & \cite{Hendrickson2011}                                     \\
\hline
                              &  YCrO$_{3-\delta}$ ceramics              &  High conductivity, promising anode oxides for solid-oxide fuel cells.                                                                                                    & \cite{Hussain2017,jabbar2019chromate,barnett2003fuel}      \\
Solid oxide fuel cell         &  Y$_{1-x}$A$_x$Cr$_{1-y}$B$_y$O$_3$      &  Cathode material for sulfur-oxygen fuel cell, working temperature range: 500-800 $^{\circ}$C.                                                                            & \cite{Hussain2017}                                         \\
                              &  YCrO$_3$                                &  High chemical stability, resistant to ZrO$_2$, connecting materials for solid-oxide fuel cells.                                                                          & \cite{Chu2013}                                             \\
\hline
Non-volatile memory           &  YCrO$_3$ amorphous films                &  Resistance switching effect, high switching speed, low powder consumption, high storage density, simple structure, utilized in resistive random access memory            & \cite{Chen2017}                                            \\
\hline
Photovoltaic materials        &  0.9(BiFe$_3$)-0.1(YCrO$_3$) thin film   &  Large open circuit photovoltage, excellent photovoltaic effect.                                                                                                          & \cite{Sharma2014-2}                                        \\
\hline
\hline
\end{tabular}
\end{table*}

\begin{figure*} [!t]
\centering \includegraphics[width=0.68\textwidth]{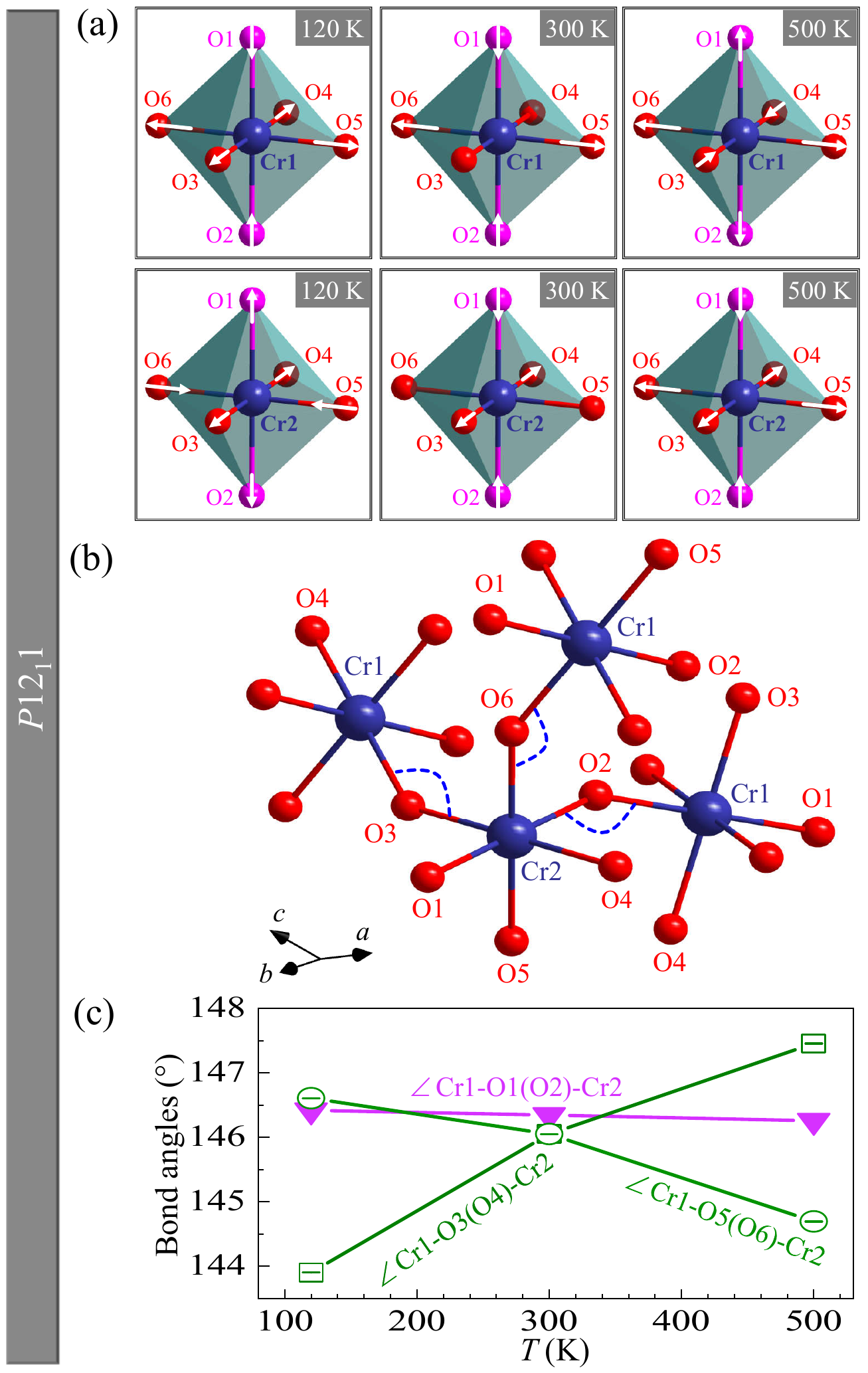}
\caption{
(a) Local octahedral environments for Cr1(Cr2) ions in $P12_11$ space group (No. 4, noncentrosymmetric) at 120, 300, and 500 K.
(b) With the $P12_11$ space group, schematic figure of the Cr1-O1(O2)-Cr2, Cr1-O3(O4)-Cr2, and Cr1-O5(O6)-Cr2 bond angles.
(c) Temperature dependence of Cr1-O1(O2)-Cr2, Cr1-O3(O4)-Cr2, and Cr1-O5(O6)-Cr2 bond angles.
Figures (a-c) were reproduced with permission from Ref. \cite{zhao2021insights}. Copyright (2021) Hai-Feng Li.
}
\label{Mon4}
\end{figure*}

\begin{table*}[!t]
\small
\caption{\newline Extracted structure parameters of single-crystal YCrO$_3$ at 120, 300, and 500 K with $P12_11$ (No. 4, noncentrosymmetric) and \emph{Pmnb} (No. 62, centrosymmetric) space groups, including $\beta$ angle, lattice constants, atomic positions, unit-cell volume, and goodness of fit. We listed the Wyckoff site for each ion. The numbers in parenthesis are the estimated standard deviations of the (next) last significant digit. Reproduced with permission from Ref. \cite{zhao2021insights}. Copyright (2021) Hai-Feng Li.}
\label{RefinParas1}
\setlength{\tabcolsep}{1.9mm}{}
\renewcommand{\arraystretch}{0.9}
\begin{tabular}{llllllllll}
\hline
\multicolumn {10}{c}{Controlled structure refinement for a YCrO$_3$ single crystal}                                                                                                     \\[2pt]
\hline
Space group                                   &&\multicolumn{3}{c}{$P12_11$}                         &&\multicolumn{4}{c}{$Pmnb$}                                                       \\[2pt]
$T$ (K)                                       &&120             &300                &500             &&                             &120             &300              &500             \\[1pt]
\hline
$a$ ({\AA})                                   &&7.52424(1)      &7.53407(1)         &7.54756(1)      &&                             &7.52377(1)      &7.53350(1)       &7.54703(1)      \\
$b$ ({\AA})                                   &&5.52091(1)      &5.52315(1)         &5.52623(1)      &&                             &5.52045(1)      &5.52272(1)       &5.52583(1)      \\
$c$ ({\AA})                                   &&9.16631(2)      &9.17854(2)         &9.19557(2)      &&                             &5.23462(1)      &5.24183(1)       &5.25227(1)      \\
$\alpha (= \gamma)$ $(^\circ)$                &&90              &90                 &90              &&                             &90              &90               &90              \\
$\beta$ $(^\circ)$                            &&145.17181(8)    &145.17030(7)       &145.16501(6)    &&                             &90              &90               &90              \\
$V$ ({\AA}$^3$)                               &&217.467(1)      &218.138(1)         &219.086(1)      &&                             &217.418(1)      &218.088(1)       &219.039(1)      \\
\hline
Y1(2\emph{a}) \emph{x}                        &&0.2669(1)       &0.2670(1)          &0.2675(1)       &&Y1(4\emph{c}) \emph{x}       &0.25            &0.25             &0.25            \\
Y1(2\emph{a}) \emph{y}                        &&0.0673(1)       &0.0669(1)          &0.0661(1)       &&Y1(4\emph{c}) \emph{y}       &0.56757(3)      &0.56689(2)       &0.56608(3)      \\
Y1(2\emph{a}) \emph{z}                        &&0.2674(1)       &0.2670(1)          &0.2666(1)       &&Y1(4\emph{c}) \emph{z}       &0.48264(4)      &0.48299(3)       &0.48340(3)      \\
Y2(2\emph{a}) \emph{x}                        &&0.7331(1)       &0.7330(1)          &0.7325(1)       &&                             &                &                 &                \\
Y2(2\emph{a}) \emph{y}                        &&0.9327(1)       &0.9331(1)          &0.9340(1)       &&                             &                &                 &                \\
Y2(2\emph{a}) \emph{z}                        &&0.2326(1)       &0.2330(1)          &0.2334(1)       &&                             &                &                 &                \\
Cr1(2\emph{a}):                               &&                &                   &                &&Cr1(4\emph{b}):              &                &                 &                \\
$(x, y, z)$                                   &&(0, 0, 0.75)    &(0, 0, 0.75)       &(0, 0, 0.75)    &&$(x, y, z)$                  &(0, 0, 0.5)     &(0, 0, 0.5)      &(0, 0, 0.5)     \\
Cr2(2\emph{a}):                               &&                &                   &                &&                             &                &                 &                \\
$(x, y, z)$                                   &&(0.5, 0, 0.75)  &(0.5, 0, 0.75)     &(0.5, 0, 0.75)  &&                             &                &                 &                \\
O1(2\emph{a}) \emph{x}                        &&0.1427(1)       &0.1473(1)          &0.1505(1)       &&O1(4\emph{c}) \emph{x}       &0.25            &0.25             &0.25            \\
O1(2\emph{a}) \emph{y}                        &&0.4680(1)       &0.4663(1)          &0.4647(1)       &&O1(4\emph{c}) \emph{y}       &0.96643(27)     &0.96632(22)      &0.96519(22)     \\
O1(2\emph{a}) \emph{z}                        &&0.1470(1)       &0.1473(1)          &0.1476(1)       &&O1(4\emph{c}) \emph{z}       &0.60153(27)     &0.60268(22)      &0.60192(22)     \\
O2(2\emph{a}) \emph{x}                        &&0.8573(1)       &0.8527(1)          &0.8495(1)       &&                             &                &                 &                \\
O2(2\emph{a}) \emph{y}                        &&0.5320(1)       &0.5337(1)          &0.5353(1)       &&                             &                &                 &                \\
O2(2\emph{a}) \emph{z}                        &&0.3530(1)       &0.3527(1)          &0.3524(1)       &&                             &                &                 &                \\
O3(2\emph{a}) \emph{x}                        &&0.3664(1)       &0.3603(1)          &0.3537(1)       &&O2(8\emph{d}) \emph{x}       &0.05380(15)     &0.05378(11)      &0.05375(11)     \\
O3(2\emph{a}) \emph{y}                        &&0.3144(1)       &0.3030(1)          &0.2969(1)       &&O2(8\emph{d}) \emph{y}       &0.30362(20)     &0.30297(15)      &0.30204(16)     \\
O3(2\emph{a}) \emph{z}                        &&0.5689(1)       &0.5565(1)          &0.5474(1)       &&O2(8\emph{d}) \emph{z}       &0.30675(20)     &0.30654(16)      &0.30671(16)     \\
O4(2\emph{a}) \emph{x}                        &&0.6336(1)       &0.6397(1)          &0.6463(1)       &&                             &                &                 &                \\
O4(2\emph{a}) \emph{y}                        &&0.6856(1)       &0.6970(1)          &0.7031(1)       &&                             &                &                 &                \\
O4(2\emph{a}) \emph{z}                        &&0.9311(1)       &0.9435(1)          &0.9526(1)       &&                             &                &                 &                \\
O5(2\emph{a}) \emph{x}                        &&0.7359(1)       &0.7527(1)          &0.7646(1)       &&                             &                &                 &                \\
O5(2\emph{a}) \emph{y}                        &&0.2034(1)       &0.1970(1)          &0.1930(1)       &&                             &                &                 &                \\
O5(2\emph{a}) \emph{z}                        &&0.0457(1)       &0.0565(1)          &0.0658(1)       &&                             &                &                 &                \\
O6(2\emph{a}) \emph{x}                        &&0.2641(1)       &0.2473(1)          &0.2354(1)       &&                             &                &                 &                \\
O6(2\emph{a}) \emph{y}                        &&0.7966(1)       &0.8030(1)          &0.8070(1)       &&                             &                &                 &                \\
O6(2\emph{a}) \emph{z}                        &&0.4543(1)       &0.4435(1)          &0.4342(1)       &&                             &                &                 &                \\
\hline
$R_\textrm{p}$                                &&19.4            &16.2               &14.4            &&                             &12.5            &8.95             &7.93            \\
$R_\textrm{wp}$                               &&25.5            &21.6               &19.4            &&                             &15.3            &11.4             &11.4            \\
$R_\textrm{exp}$                              &&6.60            &6.57               &6.45            &&                             &6.87            &6.55             &6.40            \\
$\chi^2$                                      &&14.9            &10.9               &9.02            &&                             &4.96            &3.02             &3.16            \\
\hline
\end{tabular}
\end{table*}

\begin{figure*} [!t]
\centering \includegraphics[width=0.65\textwidth]{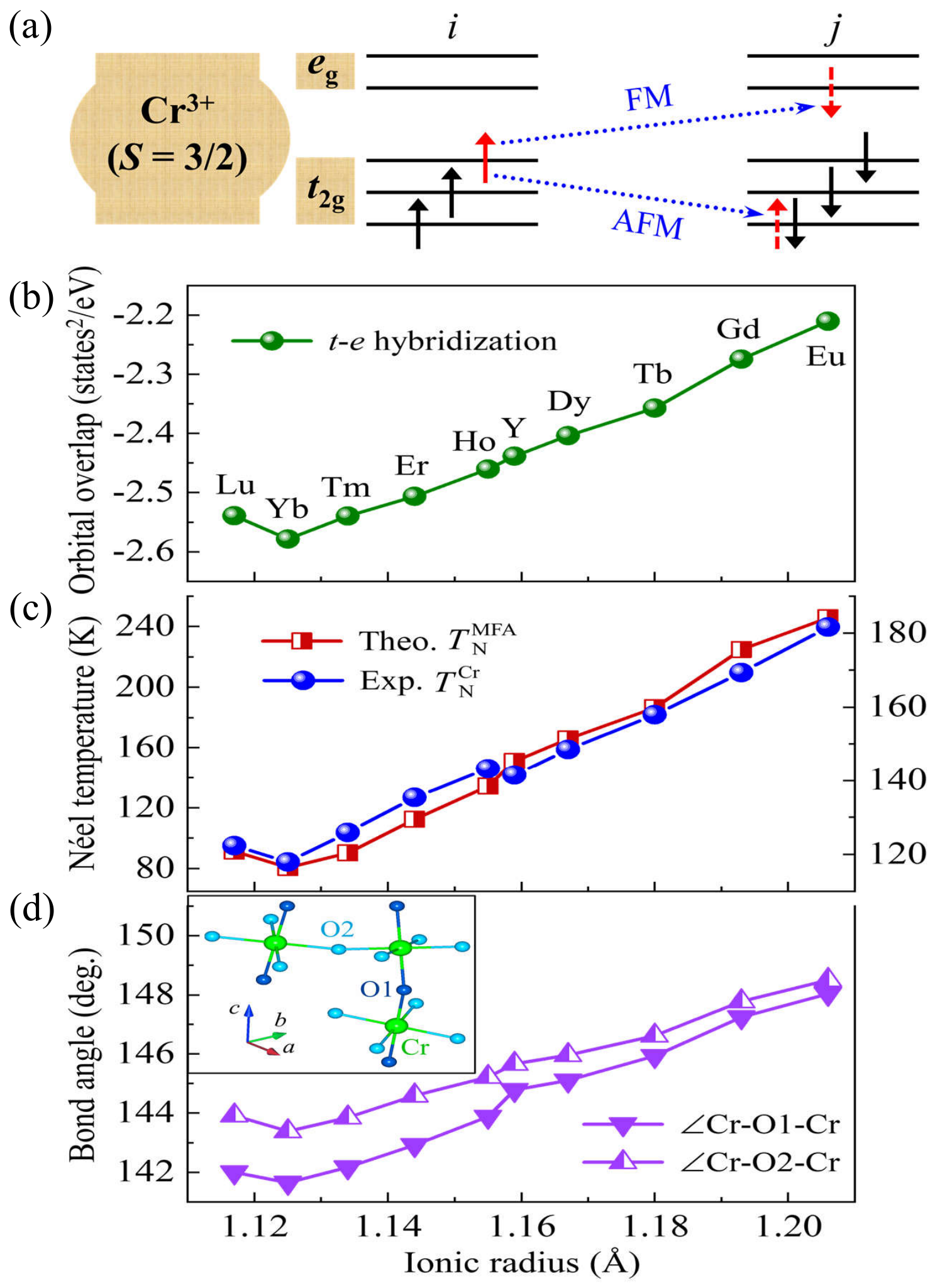}
\caption{
(a) Crystal-field splitting of the degenerate \emph{d} orbitals of Cr$^{3+}$ (3$d^3$) ions in a cubic structure. This crystalline environment splits the fivefold \emph{d}-level into $e_\texttt{g}$ (twofold degenerate) and $t_{\texttt{2g}}$ (threefold degenerate) levels. The arrows are Cr$^{3+}$ spins, schematically illustrating the virtual charge transfer resulting in AFM and FM states, respectively.
(b) Calculation of the overlapping degree of \emph{t}-\emph{e} orbitals (i.e., $I_{t_{\texttt{2g{$\downarrow$}}}-e_{\texttt{g{$\uparrow$}}}}$),
(c) Experimental ($T^{\texttt{Cr}}_\texttt{N}$) (right axis) and calculation ($T^\texttt{MFA}_\texttt{N}$) of the AFM transition temperatures (left axis), and
(d) the optimized $\angle$Cr-O1-Cr and $\angle$Cr-O2-Cr bond angles of RECrO$_3$ compounds (RE = Eu, Gd, Tb, Dy, Y, Ho, Er, Tm, Yb, and Lu). Inset of (d) displays the configuration of the bond angles. The horizontal axis stands for ionic radii of RE$^{3+}$ ions. Figures (a-d) were reproduced with permission from Ref. \cite{Zhu2021-1}. Copyright (2021) Yinghao Zhu.
}
\label{FM-AMF}
\end{figure*}

\section{Controversies and issues}

Even after nearly 69 years intensive studies on orthochromates since the first synthesis of these compounds \cite{Looby1954}, there are still some unsolved problems and controversies as summarized below. We systematically analyzed them.

\subsection{Cr oxidation state}

\subsubsection{YCrO$_4$}

As shown in \textcolor[rgb]{0.00,0.00,1.00}{Fig.}~\ref{Structure-1}(c) \cite{Duran2012-2}, upon heating of hydroxides from 473 to 1073 K, the stable crystalline phase was demonstrated to be YCrO$_4$, indicating the presence of Cr$^{5+}$ ions. With increasing temperature up to 1273--1373 K, the orthorhombic YCrO$_3$ phase is then formed. Based on the Debye-Scherrer equation \cite{Duran2012-2}, the grain sizes were calculated as 18 nm (YCrO$_4$, annealed between 873--1073 K) and 20 nm (YCrO$_3$, sintered at 1373 K). The nanocrystals are closely coalesced with each other, leading to agglomerates and thus low sinterability and high porosity in the powder samples \cite{Duran2012-2}. Therefore, heat treatments at different temperatures strongly influence the oxidation state of chromium ions.

\subsubsection{Cr$^{3+}$ and Cr$^{4+}$}

The X-ray photoelectron spectroscopy investigation demonstrates the existences of cation vacancies and a mixture of Cr$^{3+}$ and Cr$^{4+}$ ions in YCrO$_3$ compound \cite{Duran2012-2}, which may originate from the evaporation of Cr-based oxides during syntheses, or the oxygen over stoichiometry. In the YCrO$_3$ ceramics synthesized by a field-assisted sintering technique, the X-ray photoelectron spectroscopy study also reveals the existence of Cr$^{4+}$ ions (see \textcolor[rgb]{0.00,0.00,1.00}{Fig.}~\ref{Crstate-3}(b)) \cite{Zhang2014}.

\subsubsection{Cr$^{3+}$}

The X-ray absorption spectroscopy study of single-crystal YCrO$_3$ thin film (32 nm) demonstrates that the chromium oxidation state is approximately 3+ as displayed in \textcolor[rgb]{0.00,0.00,1.00}{Fig.}~\ref{Crstate-2}(b) (right panel) \cite{Sharma2020}, indicating both Cr and oxygen ions are stoichiometry in the grown thin film.

\subsection{Single crystal growth}

Currently, most chromium-based single crystals have been fabricated by the flux technique. There exist many problems: (i) The mass is only in µg or mg scale. (ii) The presence of RECrO$_4$ impurity phase. (iii) A residue of the Cr$_2$O$_3$ raw compound that also displays a weak ferromagnetism. The presence of Cr$_2$O$_3$ in RECrO$_3$ confuses the understanding of magnetism of RECrO$_3$. (iv) The flux produces some unwanted phases like Pb$_2$CrO$_5$. This leads to the main resultant RECrO$_3$ compound inhomogeneous.

The single-crystal RECrO$_3$ grown by the floating-zone technique displays some cracks inside \cite{Zhu2021}.

\subsection{Crystalline structure model}

So far, the determined structural models at low- and room-temperature were summarized in \textcolor[rgb]{0.00,0.00,1.00}{Table}~\ref{SM-parameters}. The theoretically proposed monoclinic model ($P12_11$, No. 4, noncentrosymmetric) is the only non-centrosymmetric one \cite{Serrao2005}, whereas all others are centrosymmetric. Time-of-flight NPD studies from 321--1200 K on a single-crystal YCrO$_3$ show that an orthorhombic (space group: $Pmnb$) structure model is insistent (see \textcolor[rgb]{0.00,0.00,1.00}{Fig.}~\ref{CrShift-1}(b)) \cite{Zhu2020}. The lattice parameters of \emph{a}, \emph{b}, \emph{c}, and \emph{V} agree well with theoretical calculations based on the Gr$\ddot{\textrm{u}}$neisen law with the presence of an anomaly around 900 K indicative of a possible isosymmetric structural phase transition as shown in \textcolor[rgb]{0.00,0.00,1.00}{Fig.}~\ref{LatticePara-2}(a) \cite{Zhu2020}. High-resolution NPD studies at 295 K (in the ferroelectric-like state) and 550 K (in the paraelectric state) show that the average crystallograhpic structure of YCrO$_3$ is orthorhombic ($Pnma$) and centrosymmetric above and below the dielectric transition \cite{Ramesha2007}. It is pointed out that the noncentrosymmetric monoclinic structure model ($P12_11$, No. 4) has not yet been confirmed experimentally, though Zhao \emph{et al}. \cite{zhao2021insights} reported the corresponding crystallographic information of lattice parameters and atomic positions (see \textcolor[rgb]{0.00,0.00,1.00}{Table}~\ref{RefinParas1}) of a YCrO$_3$ single crystal, taking advantages of the high-resolution synchrotron X-ray diffraction. The YCrO$_3$ compound presently possesses a centrosymmetric crystal structure that holds an inversion center, i.e., no net charge polar could be formed. However, the weak electric polarization in dielectric measurements was indeed observed around $\sim$ 473 K for YCrO$_3$, which requires atomic off centering.

\subsection{Magnetic structure}

The consensus is that YCrO$_3$ adopts a canted \emph{G}-type AFM structure with $t_{\textrm{2g}}$-$t_{\textrm{2g}}$ superexchange interactions within the \emph{ab}-plane in the $Pbnm$ space group; along the \emph{c} axis, weak ferromagnetism produced by the \emph{t}-\emph{e} orbital coupling was observed (\textcolor[rgb]{0.00,0.00,1.00}{Fig.}~\ref{FM-AMF}), which was attributed to the DM interactions theoretically. But whether DM interactions exist in YCrO$_3$ or not has not been confirmed experimentally. It is stressed that the magnetic structure model was reached based on in-house magnetization measurements with single crystals \cite{Judin1966, Ziel1969-1} and theoretical calculations \cite{Serrao2005, Zhu2021-1}. Zhao \emph{et al}. \cite{zhao2021insights} tentatively demonstrated that the local Cr2O$_6$ distortion modes give rise to the appearance of CAFM order by spin interactions of Cr1-Cr2 ions, mainly via the intermediate of O3 and O4 ions in the $P12_11$ space group (see \textcolor[rgb]{0.00,0.00,1.00}{Fig.}~\ref{Mon4}).

For future studies, some issues need to be addressed:

(i) As far as we know, the proposed magnetic structure model has not yet been confirmed by neutron scattering studies \cite{Zhu2020-2}, especially the canting angle has not been determined yet because with a few observed Bragg magnetic peaks, it is difficult to determine the expected small canting degree of the Cr$^{3+}$ AFM spins \cite{Ziel1969-1}.

(ii) Self-consistent theoretical calculations determine that the local magnetic moment of Cr$^{3+}$ ions is approximately 3 $\mu_\textrm{B}$, indicating fully occupied $t_{\textrm{2g}}$ states of Cr$^{3+}$ ions \cite{Serrao2005, Zhu2021-1}. In contrast, the NPD study of pulverized single-crystal YCrO$_3$ reveals a magnetic moment of 2.45(6) $\mu_\textrm{B}$ for Cr$^{3+}$ ions at 12 K.

(iii) SR was suggested at $\sim$ 60 K (see \textcolor[rgb]{0.00,0.00,1.00}{Table}~\ref{SM-parameters}) \cite{Duran2010}, and negative magnetization was observed \cite{Kumar2017, Jana2018-2}, but both were not displayed in the NPD study (see \textcolor[rgb]{0.00,0.00,1.00}{Fig.}~\ref{mag-3}(c)) \cite{Zhu2020-2}. As listed in \textcolor[rgb]{0.00,0.00,1.00}{Table}~\ref{SM-parameters}, some other magnetic anomalies were suggested at $\sim$ 10 and 230 K. These need to be carefully checked with modern neutron scattering technique that is powerful and unique for solving magnetic structures and determining spin excitations and fluctuations \cite{li2008synthesis, li2009crystal, Gupta2020, PhysRevB.82.140503}.

\subsection{Intrinsic \emph{P}-\emph{E} loop of YCrO$_3$}

Most of the reported \emph{P}-\emph{E} loop measurements focused on nano-crystalline, poly-crystalline, and thin-film samples. It is noticed that YCrO$_3$ nanoparticles show better ferroelectric properties probably because of the low leakage current. This was ascribed to a reduced particle size and increased grain boundaries. The YCrO$_3$ samples synthesized with the microwave-assisted method did not show evidence of ferroelectricity \cite{Prado-Gonjal2013}.

The beautiful \emph{P}-\emph{E} loops of single-crystal YCrO$_3$ fabricated by the flux technique were demonstrated to be from local polar domains \cite{Sanina2018}. Presently, we don't have satisfactory \emph{P}-\emph{E} loops given for single-crystal YCrO$_3$ grown by the floating-zone method mainly due to the high electrical leakage. Existence of proper ferroelectricity requires evidence of saturated ferroelectric hysteresis loops. Presently, no such evidence has been reported for single-crystalline YCrO$_3$ samples that provide more reliable electrical information than other types of matter do. Therefore, it is still a remaining question whether YCrO$_3$ has an intrinsic ferroelectricity or not. Thus in this review we call YCrO$_3$ as ferroelectric. Further improving the quality of single crystals and minimizing the leakage loss are in progress \cite{sotnikova2021radiometric}.

\subsection{Possible reasons for the high leakage current}

First of all, the charge-fluctuation-induced conductive process in polar site is crucial to the ferroelectric characteristic, and this intrinsic feature (i.e., semiconducting behavior) gives rise to the leakage current, which interferes with pooling process and polarization switching in biferroic compounds \cite{Duran2012-1}.

Current leakage makes a $P$-$E$ loop difficult to be measured and saturated, consequently, preventing us from acquiring decisive experimental evidence of whether the samples are ferroelectric or not. Mall \emph{et al}. summarized the reasons of electrical leakage in ferroelectric materials \cite{Mall2018-2}: (i) Various conduction mechanisms \cite{Qi2005-3}. (ii) Material defects, such as vacancies and interstitials. (iii) Planar defects like grain boundaries \cite{Qi2005-3} and Cr valence fluctuations.

Bahadur \emph{et al}. classifies the factors that affect dielectric constant and loss into two categories: (i) One is intrinsic in terms of the interaction of phonons with alternating electric field, depending on crystal structure. (ii) The other is extrinsic, depending on the heterogeneity of the medium such as porosity, impurities, grain boundaries, micro-cracks, random crystal orientation, vacancies, etc \cite{Bahadur2008}.

From the point of view of materials synthesis, porosity, connectivity, and particle size between YCrO$_3$ grains formed with thermal treatments at different temperatures have strong effects on the dielectric property \cite{Case1984}.

\subsection{Origin of the ferroelectricity}

Presently, the structural models determined in experiments are all centrosymmetric within accuracy, which does not qualify YCrO$_3$ as a ferroelectric compound.

\subsubsection{Cr ions}

By analyzing different length scales existing in YCrO$_3$ through neutron scattering studies, the development of weak electric polarization in poly-crystalline YCrO$_3$ was attributed to local noncentrosymmetry resulting from a small displacement of Cr ions in CrO$_6$ octahedra owning to off-centring distortion \cite{Ramesha2007}. Raman scattering studies on YCrO$_3$ show that the suppression of ferroelectricity was related to the Y-site cationic displacements \cite{Mall2019}. The observed weak ferroelectricity in permittivity was ascribed to local polarization field \cite{Duran2012-1}. Zhao \emph{et al}. \cite{zhao2021insights} tentatively demonstrated that the strain-balanced bonds of Cr1-O3(O4) and Cr2-O5(O5) and the local Cr1O$_6$ and Cr2O$_6$ octahedral distortion modes in $P12_11$ symmetry (see \textcolor[rgb]{0.00,0.00,1.00}{Fig.}~\ref{Mon4}) could be the origin of the dielectric anomaly reported previously.

\subsubsection{Local polar domains}

The observed local electric polarization in YCrO$_3$ single crystals was ascribed to two kinds of local polar domains: (i) One is with structure distortions locating near Pb$^{2+}$ ions. (ii) The other is the separation domains of magnetic phase near Pb$^{4+}$ ions. Thus, the inclusion of Pb$^{2+}$ and Pb$^{4+}$ impurity ions in the structural position of Y$^{3+}$ sites leads to the energetically beneficial processes of the formation of local ferroelectricity in separated domains \cite{Sanina2018}. The local domains also lead to a super-paraelectric state. In the frozen super-paraelectric state, a pyrocurrent maximum was observed.

\subsubsection{Y and O2 ions}

Time-of-flight NPD on single-crystal YCrO$_3$ \cite{Zhu2020} shows that the local crystalline distortions of O1, O2, and Y ions are significantly larger than that of Cr ions, approximately two orders of magnitude larger. The extracted modes of local distortion of Cr, O, and Y ions are displayed in \textcolor[rgb]{0.00,0.00,1.00}{Figs.}~\ref{LatticePara-3}(a-c), which strongly influences the properties of YCrO$_3$ compound. Moreover, both O2 and Y ions show an obvious displacement and a large charge deviation from 3+ and 2$-$, respectively, implying that local crystalline symmetry breaking of O2 and Y ions may be one of the reasons for the development of ferroelectric-like behavior in YCrO$_3$ \cite{Zhu2020}.

Theoretical calculations on YCrO$_3$ \cite{Serrao2005} show that: (i) The largest component of structural distortions originates from rotations of oxygen octahedral. (ii) Y and oxygen displacements produce a small component of distortion modes. (iii) No structure distortion occurs with Cr displacements. The modes in (i) keep the inversion symmetry, whereas, the relatively weak distortions in (ii) could lead to ferroelectricity.

\subsection{Size of the charge displacement}

YCrO$_3$ displays a low electric polarization, which could be due to a small displacement from the lattice centrality. Experimentally, it is not easy to determine the size of such a small charge displacement.

The neutron scattering study with PDF method collects information of short-range or local structure by diffuse scattering. While refining PDF data, one can obtain the information by limiting the crystallograhpic length scales. When the length scale (\emph{r}) was limited to 1.6--20 {\AA}, the PDF analysis implies that the $Pnma$ model fits better in contrast to $P12_11$ (No. 4, noncentrosymmetric), indicating an average centrosymmetric orthorhombic structure \cite{Ramesha2007}. As shown in \textcolor[rgb]{0.00,0.00,1.00}{Fig.}~\ref{CrShift-2}(a), when $r =$ 1.6--6 {\AA}, the high-temperature (at 550 K $> T_\textrm{C} =$ 473 K) PDF data agrees well with the $Pnma$ model, whereas in the low-temperature ferroelectric state (295 and 15 K), the PDF data was finally fit to the non-centrosymmetric $P12_11$ (No. 4, noncentrosymmetric) model by releasing local symmetry of Cr$^{3+}$ ions \cite{Ramesha2007}. This generates 0.01 {\AA} off centring (see \textcolor[rgb]{0.00,0.00,1.00}{Fig.}~\ref{CrShift-2}(b)) of Cr$^{3+}$ ions along the \emph{c}-axis with the $Pnma$ space group. In contrast, the Y and O ions almost hold the local symmetry, i.e., no displacements \cite{Ramesha2007}. As far as we know, this has been the only study extracting the size of charge displacement for the weak ferroelectricity of YCrO$_3$ compound.

\subsection{Magnetoelectric effect}

Theoretical studies suggested a close connection between magnetic and dielectric properties of YCrO$_3$ \cite{Serrao2005}. A clear magnetoelectric effect was observed in the single-crystal YCrO$_3$ thin film (32 nm) \cite{Sharma2020}. Temperature dependence of dielectric measurements of YCrO$_3$ show an anomaly at $\sim$ 230 K, which was attributed to a possible formation of short-range AFM correlations above $T_\textrm{N}$ rather than a structural phase transition \cite{Mall2017-1}. This indicates a presence of magnetodielectric coupling in YCrO$_3$. The dielectric constant of bulk YCrO$_3$ is smaller than that in nanoparticles. There exists a strong magnetoelectric coupling in YCrO$_3$ nanoparticles \cite{Apostolov2019}. In contrast, the microwave-synthesized RECrO$_3$ materials did not show a coupling between magnetic and dielectric properties \cite{Prado-Gonjal2013}.

\section{Conclusions and outlook}

First, we summarize general results and consensuses. The syntheses, physical properties, and potential applications were discussed for parent- and doped-YCrO$_3$ compounds in forms of poly-, nano-, and single-crystalline and thin-film materials. Most of the reported results show that YCrO$_3$ adopts the orthorhombic system with $Pnma$, $Pmnb$, or $Pbnm$ space group and approximate lattice parameters of $a$ = 5.2392(1) {\AA}, $b$ = 5.5198(1) {\AA}, $c$ = 7.5297(1) {\AA}, and $V$ = 217.75(1) {\AA}$^3$ ($Pbnm$). These orthorhombic structural models are centrosymmetric, making ferroelectric polarizations and DM interactions hardly form. Magnetization measurements as well as Raman scattering studies reveal that YCrO$_3$ forms AFM couplings along the shortest crystallographic axis and FM correlations along the axis with the longest constant, thus adopting a CAFM structure with weak ferromagnetism. This was attributed to possible DM interactions of 3$d^3$ Cr$^{3+}$ ions. Dielectric measurements with frequency and temperature display an anomaly at $\sim$ 473 K, showing a relaxor-ferroelectric behavior. \emph{P}-\emph{E} loop measurements, in most cases, exhibit non-saturated electric polarization. Possible reasons for the observed ferroelectric state were proposed, including local non-centrosymmetry of Cr ions, local polar domains (related to structural distortions, or magnetic phase separation), and a weak instability of Y displacements with regard to O$_6$ cage. These make orthochromates unique among multiferroic materials. So far, there is still subject of numerous studies.

Second, we propose future potential studies to address the long-standing issues.

(1) How to control the valence state of chromium ions and investigate its effect on the property of YCrO$_3$ compound remain problems. Building the correlations between chemical defects and macroscopic properties of YCrO$_3$ necessitates accurate control studies, by which one can introduce ionic defects by artificially removing Cr$^{3+}$ ions and/or changing oxygen stoichiometry and simultaneously monitor the corresponding modifications of macroscopic properties, e.g., saturation moment, dielectric constant, piezoelectric property, \emph{P}-\emph{E} loop, and ferroelasticity, thereby, determining their intrinsic connections.

(2) As to the single crystal growth, an alternative is that RECrO$_3$ crystals could be grown by the chemical hydrothermal method. Utilizing this method, one can start with a stoichiometric mixture of the RE$_2$O$_3$ and Cr$_2$O$_3$ raw materials, neglecting the flux and the loss of Cr$_2$O$_3$ during a reaction, and control the single crystal’s stoichiometry because the reaction temperature is largely decreased so that there is no reactant loss, avoiding a deviation from the ideal stoichiometry. This method is a particular competence in the exploitation of novel functional materials.

(3) So far, saturated ferroelectricity was reported in nanocrystals and thin films of YCrO$_3$. In most cases, the ferroelectricity was unsaturation in poly-crystalline YCrO$_3$ samples. In contrast, the intrinsic ferroelectricity has been never observed in YCrO$_3$ single crystals. This makes it unconfirmed whether YCrO$_3$ is ferroelectric or not. Solving this problem necessitates a greater improvement in the quality of grown single crystals, especially a larger decrease in leakage current.

(4) To solve magnetism-related problems, we suggest that: (i) Neutron diffraction study on single crystals on a basis of reactor or a time-of-flight neutron source are strongly urged. (ii) Inelastic neutron scattering study on single crystalline samples are suggested to get the superexchange parameters and reveal the origin of the weak ferromagnetism, shedding light on its forming mechanism.

(5) To reconcile the long-standing contradiction between structure and ferroelectricity, we propose high-resolution neutron/synchrotron diffraction studies on high-quality YCrO$_3$ single crystals to explore the actual crystallographic symmetry. This is not easy, but it could shed light on the dielectric anomaly of YCrO$_3$ compound.

Last but not least, the cleverest housewife cannot cook a meal without rice. High-quality materials are always priority first for reliable studies. We do hope that the existing knowledge, highlighted in the present review, can be of help in shedding light on the properties of parent- and doped-YCrO$_3$ compounds and providing a guidance for future experimental and theoretical studies and that other advanced applications will emerge in the future.

\emph{P.S. We present a comprehensive review, trying to include the related important works and patents as much as possible. We deeply apologize if we lost sight of the important works in this field. Last updated until September 28, 2022.}

\newpage

\section*{Declaration of Competing Interest}

The authors declare that they have no known competing financial interests or personal relationships that could have appeared to inﬂuence the work reported in this paper.

\section*{Acknowledgments}

Authors acknowledge financial support from the Science and Technology Development Fund, Macao SAR (File Nos. 0051/2019/AFJ, 0090/2021/A2, and 0049/2021/AGJ), Guangdong Basic and Applied Basic Research Foundation (Guangdong-Dongguan Joint Fund No. 2020B1515120025), University of Macau (MYRG2020-00278-IAPME and EF030/IAPME-LHF/2021/GDSTIC), Guangdong-Hong Kong-Macao Joint Laboratory for Neutron Scattering Science and Technology (Grant No. 2019B121205003), and the National Natural Science Foundation of China (51871232).

\appendix

\section*{Appendix A}

Here we show the relationship between orthorhombic $Pnma$ (setting = 1), $Pmnb$ (setting = 2), and $Pbnm$ (setting = 3) symmetries \cite{li2008synthesis}. These space groups are structurally centrosymmetric.

\begin{table*}[!h]
\small
\caption{\newline Lattice parameters and atomic positions of YCrO$_3$ compound with space groups of $Pnma$, $Pmnb$, and $Pbnm$. We listed the atomic number ($Z$) and Wyckoff site of each ion.}
\label{62_1-3}
\setlength{\tabcolsep}{3.2mm}{}
\renewcommand{\arraystretch}{1.1}
\begin{tabular}{lllllllllllll}
\hline
\hline
\multicolumn {13}{c}{Orthorhombic structural system}                                                                                                                \\
\hline
Space group                                      &&\multicolumn{3}{c}{$Pnma$}         &&\multicolumn{3}{c}{$Pmnb$}           &&\multicolumn{3}{c}{$Pbnm$}           \\
\hline
$a, b, c$ ({\AA})                                &&$a_0$    &$b_0$    &$c_0$          &&$b_0$   &$a_0$  &$c_0$               &&$c_0$     &$a_0$     &$b_0$          \\
$\alpha (= \gamma = \beta)$ $(^\circ)$           &&90       &90       &90             &&90      &90     &90                  &&90        &90        &90             \\
\hline
Y1(4\emph{c}, $Z =$ 39) (\emph{x, y, z})         &&$x_0$    &0.25     &$z_0$          &&0.25    &$x_0$  &$z_0$               &&$z_0$     &$x_0$     &0.25           \\
Cr1(4\emph{b}, $Z =$ 24) (\emph{x, y, z})        &&0.00     &0.00     &0.50           &&0.00    &0.00   &0.50                &&0.50      &0.00      &0.00           \\
O1(4\emph{c}, $Z =$ 8) (\emph{x, y, z})          &&$x_0$    &0.25     &$z_0$          &&0.25    &$x_0$  &$z_0$               &&$z_0$     &$x_0$     &0.25           \\
O2(8\emph{d}, $Z =$ 8) (\emph{x, y, z})          &&$x_0$    &$y_0$    &$z_0$          &&$y_0$   &$x_0$  &$z_0$               &&$z_0$     &$x_0$     &$y_0$          \\
\hline
\hline
\end{tabular}
\end{table*}

\bibliographystyle{elsarticle-num}
\bibliography{YCORV}

\begin{thebibliography}{100}
\expandafter\ifx\csname url\endcsname\relax
  \def\url#1{\texttt{#1}}\fi
\expandafter\ifx\csname urlprefix\endcsname\relax\def\urlprefix{URL }\fi
\expandafter\ifx\csname href\endcsname\relax
  \def\href#1#2{#2} \def\path#1{#1}\fi

\bibitem{matthias1963superconductivity}
B.~T. Matthias, T.~H. Geballe, V.~B. Compton, Superconductivity, Reviews of
  Modern Physics 35~(1) (1963) 1.

\bibitem{mcmillan1968transition}
W.~McMillan, Transition temperature of strong-coupled superconductors, Physical
  Review 167~(2) (1968) 331.

\bibitem{maple1976superconductivity}
M.~B. Maple, Superconductivity, Applied physics 9~(3) (1976) 179--204.

\bibitem{bednorz1986possible}
J.~G. Bednorz, K.~A. M{\"u}ller, Possible high ${T}_{C}$ superconductivity in
  the {Ba}-{La}-{Cu}-{O} system, Zeitschrift f{\"u}r Physik B Condensed Matter
  64~(2) (1986) 189--193.

\bibitem{nagamatsu2001superconductivity}
J.~Nagamatsu, N.~Nakagawa, T.~Muranaka, Y.~Zenitani, J.~Akimitsu,
  Superconductivity at 39 {K} in magnesium diboride, Nature 410~(6824) (2001)
  63--64.

\bibitem{tinkham2004introduction}
M.~Tinkham, Introduction to superconductivity, Courier Corporation, 2004.

\bibitem{plakida2010high}
N.~Plakida, High-temperature cuprate superconductors: Experiment, theory, and
  applications, Vol. 166, Springer Science \& Business Media, 2010.

\bibitem{zimmer1995rate}
B.~I. Zimmer, W.~Jeitschko, J.~H. Albering, R.~Glaum, M.~Reehuis, The rate
  earth transition metal phosphide oxides lnfepo, {Ln}{Ru}{P}{O} and
  {Ln}{Co}{P}{O} with {Zr}{Cu}{Si}{As} type structure, Journal of Alloys and
  Compounds 229~(2) (1995) 238--242.

\bibitem{quebe2000quaternary}
P.~Quebe, L.~Terb{\"u}chte, W.~Jeitschko, Quaternary rare earth transition
  metal arsenide oxides {R}{T}{As}{O} ({T} = {Fe}, {Ru}, {Co}) with zrcusias
  type structure, Journal of alloys and compounds 302~(1-2) (2000) 70--74.

\bibitem{kamihara2008iron}
Y.~Kamihara, T.~Watanabe, M.~Hirano, H.~Hosono, Iron-based layered
  superconductor {La}[{O}$_{1-x}${F}$_x$]{FeAs}$(x =$ {0.05}--{0.12}) with
  ${T}_{C}$ = 26 {K}, Journal of the American Chemical Society 130~(11) (2008)
  3296--3297.

\bibitem{chen2008superconductivity}
X.~Chen, T.~Wu, G.~Wu, R.~Liu, H.~Chen, D.~Fang, Superconductivity at 43 {K} in
  {Sm}{Fe}{As}{O}$_{1-x}${F}$_x$, Nature 453~(7196) (2008) 761--762.

\bibitem{kresin2021superconducting}
V.~Kresin, S.~Ovchinnikov, S.~Wolf, Superconducting State: Mechanisms and
  Materials, Vol. 170, Oxford University Press, 2021.

\bibitem{fiebig2002observation}
M.~Fiebig, T.~Lottermoser, D.~Fr{\"o}hlich, A.~V. Goltsev, R.~V. Pisarev,
  Observation of coupled magnetic and electric domains, Nature 419~(6909)
  (2002) 818--820.

\bibitem{kimura2003magnetic}
T.~Kimura, T.~Goto, H.~Shintani, K.~Ishizaka, T.-h. Arima, Y.~Tokura, Magnetic
  control of ferroelectric polarization, Nature 426~(6962) (2003) 55--58.

\bibitem{cheong2007multiferroics}
S.-W. Cheong, M.~Mostovoy, Multiferroics: a magnetic twist for
  ferroelectricity, Nature materials 6~(1) (2007) 13--20.

\bibitem{kagawa2010ferroelectricity}
F.~Kagawa, S.~Horiuchi, M.~Tokunaga, J.~Fujioka, Y.~Tokura, Ferroelectricity in
  a one-dimensional organic quantum magnet, Nature Physics 6~(3) (2010)
  169--172.

\bibitem{valencia2011interface}
S.~Valencia, A.~Crassous, L.~Bocher, V.~Garcia, X.~Moya, R.~Cherifi,
  C.~Deranlot, K.~Bouzehouane, S.~Fusil, A.~Zobelli, et~al., Interface-induced
  room-temperature multiferroicity in batio 3, Nature materials 10~(10) (2011)
  753--758.

\bibitem{tokunaga2012electric}
Y.~Tokunaga, Y.~Taguchi, T.-h. Arima, Y.~Tokura, Electric-field-induced
  generation and reversal of ferromagnetic moment in ferrites, Nature Physics
  8~(11) (2012) 838--844.

\bibitem{li2021understanding}
X.~Li, Z.~Cheng, X.~Wang, Understanding the mechanism of the oxygen evolution
  reaction with consideration of spin, Electrochemical Energy Reviews 4~(1)
  (2021) 136--145.

\bibitem{YANG201972}
L.~Yang, X.~Kong, F.~Li, H.~Hao, Z.~Cheng, H.~Liu, J.-F. Li, S.~Zhang,
  Perovskite lead-free dielectrics for energy storage applications, Progress in
  Materials Science 102 (2019) 72--108.

\bibitem{SUN2014124}
E.~Sun, W.~Cao, Relaxor-based ferroelectric single crystals: Growth, domain
  engineering, characterization and applications, Progress in Materials Science
  65 (2014) 124--210.

\bibitem{YOON2003275}
D.-S. Yoon, J.~S. Roh, S.-M. Lee, H.~K. Baik, Alteration for a diffusion
  barrier design concept in future high-density dynamic and ferroelectric
  random access memory devices, Progress in Materials Science 48~(4) (2003)
  275--371.

\bibitem{Hur2004}
N.~Hur, S.~Park, P.~Sharma, J.~Ahn, S.~Guha, S.-W. Cheong, Electric
  polarization reversal and memory in a multiferroic material induced by
  magnetic fields, Nature 429 (2004) 392--395.

\bibitem{Cheong2007}
S.-W. Cheong, M.~Mostovoy, Multiferroics: a magnetic twist for
  ferroelectricity, Nat. Mater. 6 (2007) 13--20.

\bibitem{Kagawa2010}
F.~Kagawa, S.~Horiuchi, M.~Tokunaga, J.~Fujioka, Y.~Tokura, Ferroelectricity in
  a one-dimensional organic quantum magnet, Nat. Phys. 6 (2010) 169--172.

\bibitem{Valencia2011}
S.~Valencia, A.~Crassous, L.~Bocher, V.~Garcia, X.~Moya, R.~Cherifi,
  C.~Deranlot, K.~Bouzehouane, S.~Fusil, A.~Zobelli, et~al., Interface-induced
  room-temperature multiferroicity in {BaTiO}$_{3}$, Nat. Mater. 10 (2011)
  753--758.

\bibitem{Tokunaga2012}
Y.~Tokunaga, Y.~Taguchi, T.-h. Arima, Y.~Tokura, Electric-field-induced
  generation and reversal of ferromagnetic moment in ferrites, Nat. Phys. 8
  (2012) 838--844.

\bibitem{Aken2004}
B.~B. Van~Aken, T.~T. Palstra, A.~Filippetti, N.~A. Spaldin, The origin of
  ferroelectricity in magnetoelectric {YMnO}$_{3}$, Nat. Mater. 3 (2004)
  164--170.

\bibitem{Kenzelmann2005}
M.~Kenzelmann, A.~B. Harris, S.~Jonas, C.~Broholm, J.~Schefer, S.~B. Kim, C.~L.
  Zhang, S.-W. Cheong, O.~P. Vajk, J.~W. Lynn, Magnetic inversion symmetry
  breaking and ferroelectricity in {TbMnO}$_{3}$, Phys. Rev. Lett. 95 (2005)
  087206.

\bibitem{evans2020domains}
D.~M. Evans, V.~Garcia, D.~Meier, M.~Bibes, Domains and domain walls in
  multiferroics, Physical Sciences Reviews 5~(9) (2020) 20190067.

\bibitem{lottermoser2020short}
T.~Lottermoser, D.~Meier, A short history of multiferroics, Physical Sciences
  Reviews 6~(2) (2021) 20200032.

\bibitem{davydova2020spin}
M.~D. Davydova, K.~A. Zvezdin, A.~A. Mukhin, A.~K. Zvezdin, Spin dynamics,
  antiferrodistortion and magnetoelectric interaction in multiferroics. the
  case of {BiFeO}$_3$, Physical Sciences Reviews 5~(12) (2020) 20190070.

\bibitem{huang2020manipulating}
Y.-L. Huang, D.~Nikonov, C.~Addiego, R.~V. Chopdekar, B.~Prasad, L.~Zhang,
  J.~Chatterjee, H.-J. Liu, A.~Farhan, Y.-H. Chu, et~al., Manipulating
  magnetoelectric energy landscape in multiferroics, Nature communications
  11~(1) (2020) 1--8.

\bibitem{cano2021multiferroics}
A.~Cano, D.~Meier, M.~Trassin, Multiferroics: Fundamentals and Applications,
  Walter de Gruyter GmbH \& Co KG, 2021.

\bibitem{wu2021100}
M.~Wu, 100 years of ferroelectricity, Nature Reviews Physics (2021) 1--1.

\bibitem{zhao2021dzyaloshinskii}
H.~J. Zhao, P.~Chen, S.~Prosandeev, S.~Artyukhin, L.~Bellaiche,
  Dzyaloshinskii--moriya-like interaction in ferroelectrics and
  antiferroelectrics, Nature Materials 20~(3) (2021) 341--345.

\bibitem{schierle2020promising}
E.~Schierle, A promising birthplace for skyrmions, Nature materials 19~(4)
  (2020) 369--370.

\bibitem{zhang2021purely}
L.~Zhang, C.~Tang, S.~Sanvito, A.~Du, Purely one-dimensional ferroelectricity
  and antiferroelectricity from van der waals niobium oxide trihalides, npj
  Computational Materials 7~(1) (2021) 1--7.

\bibitem{meier2021ferroelectric}
D.~Meier, S.~M. Selbach, Ferroelectric domain walls for nanotechnology, Nature
  Reviews Materials (2021) 1--17.

\bibitem{choi2020nanoengineering}
E.-M. Choi, T.~Maity, A.~Kursumovic, P.~Lu, Z.~Bi, S.~Yu, Y.~Park, B.~Zhu,
  R.~Wu, V.~Gopalan, et~al., Nanoengineering room temperature ferroelectricity
  into orthorhombic {SmMnO}$_3$ films, Nature communications 11~(1) (2020)
  1--9.

\bibitem{stein2021combined}
J.~Stein, S.~Biesenkamp, T.~Cronert, T.~Fr{\"o}hlich, J.~Leist, K.~Schmalzl,
  A.~Komarek, M.~Braden, Combined arrhenius-merz law describing domain
  relaxation in {Type}-{II} multiferroics, Physical review letters 127~(9)
  (2021) 097601.

\bibitem{zhou2021terahertz}
J.~Zhou, S.~Zhang, Terahertz optics-driven phase transition in two-dimensional
  multiferroics, npj 2D Materials and Applications 5~(1) (2021) 1--8.

\bibitem{morozovska2021combined}
A.~N. Morozovska, D.~V. Karpinsky, D.~O. Alikin, A.~Abramov, E.~A. Eliseev,
  M.~D. Glinchuk, A.~D. Yaremkevich, O.~M. Fesenko, T.~V. Tsebrienko,
  A.~Pakalni{\v{s}}kis, et~al., A combined theoretical and experimental study
  of the phase coexistence and morphotropic boundaries in
  ferroelectric-antiferroelectric-antiferrodistortive multiferroics, Acta
  Materialia 213 (2021) 116939.

\bibitem{kumar2021recent}
M.~Kumar, P.~C. Sati, A.~Kumar, M.~Sahni, P.~Negi, H.~Singh, S.~Chauhan, S.~K.
  Chaurasia, Recent advances on magnetoelectric coupling in {BiFeO}$_3$:
  Technological achievements and challenges, Materials Today: Proceedings.

\bibitem{wang2021competition}
H.~Wang, X.~Liu, K.~Sun, X.~Ma, H.~Guo, I.~Bobrikov, Y.~Sui, Q.~Liu, Y.~Xia,
  X.~Chen, et~al., Competition of ferromagnetism and antiferromagnetism in
  mn-doped orthorhombic {Y}{Cr}{O}$_3$, Journal of Magnetism and Magnetic
  Materials 535 (2021) 168022.

\bibitem{martin2016thin}
L.~W. Martin, A.~M. Rappe, Thin-film ferroelectric materials and their
  applications, Nature Reviews Materials 2~(2) (2016) 1--14.

\bibitem{uchino2018ferroelectric}
K.~Uchino, Ferroelectric devices, CRC press, 2018.

\bibitem{nicola2005}
N.~A. Spaldin, M.~Fiebig, The renaissance of magnetoelectric multiferroics,
  Science 309~(5733) (2005) 391--392.

\bibitem{mishra2021observation}
S.~Mishra, K.~Rudrapal, A.~Rahaman, P.~Pal, A.~Sagdeo, R.~Mishra, D.~Topwal,
  A.~R. Chaudhuri, V.~Adyam, D.~Choudhury, et~al., Observation of
  room-temperature ferroelectricity in spark-plasma sintered {Gd}{Cr}{O}$_3$,
  Physical Review B 104~(18) (2021) L180101.

\bibitem{fita2021pressure}
I.~Fita, R.~Puzniak, A.~Wisniewski, Pressure-tuned spin switching in
  compensated {Gd}{Cr}{O}$_3$ ferrimagnet, Physical Review B 103~(5) (2021)
  054423.

\bibitem{sharma2021tuning}
Y.~Sharma, B.~Paudel, J.~Lee, W.~S. Choi, Z.~Yang, H.~Wang, Y.~Du, K.~T. Kang,
  G.~Pilania, A.~Chen, Tuning magnetic and optical properties through strain in
  epitaxial {La}{Cr}{O}$_3$ thin films, Applied Physics Letters 119~(7) (2021)
  071902.

\bibitem{otsuka2021effect}
A.~Otsuka, R.~S. Silva~Jr, C.~dos Santos, N.~S. Ferreira, M.~d.~S. Rezende,
  M.~C. dos Santos, Effect of chemical and hydrostatic pressures on the
  structural and mechanical properties of orthorhombic rare-earth
  {R}{Ni}{O}$_3$, Computational Materials Science 197 (2021) 110691.

\bibitem{eerenstein2006multiferroic}
W.~Eerenstein, N.~Mathur, J.~F. Scott, Multiferroic and magnetoelectric
  materials, Nature 442~(7104) (2006) 759--765.

\bibitem{nan2008multiferroic}
C.-W. Nan, M.~Bichurin, S.~Dong, D.~Viehland, G.~Srinivasan, Multiferroic
  magnetoelectric composites: Historical perspective, status, and future
  directions, Journal of applied physics 103~(3) (2008) 1.

\bibitem{dong2015multiferroic}
S.~Dong, J.-M. Liu, S.-W. Cheong, Z.~Ren, Multiferroic materials and
  magnetoelectric physics: symmetry, entanglement, excitation, and topology,
  Advances in Physics 64~(5-6) (2015) 519--626.

\bibitem{spaldin2019advances}
N.~A. Spaldin, R.~Ramesh, Advances in magnetoelectric multiferroics, Nature
  materials 18~(3) (2019) 203--212.

\bibitem{fiebig2016evolution}
M.~Fiebig, T.~Lottermoser, D.~Meier, M.~Trassin, The evolution of
  multiferroics, Nature Reviews Materials 1~(8) (2016) 1--14.

\bibitem{zhao2014near}
H.~J. Zhao, W.~Ren, Y.~Yang, J.~{\'I}{\~n}iguez, X.~M. Chen, L.~Bellaiche, Near
  room-temperature multiferroic materials with tunable ferromagnetic and
  electrical properties, Nature communications 5~(1) (2014) 1--7.

\bibitem{SEKINE2022214663}
Y.~Sekine, R.~Akiyoshi, S.~Hayami, Recent advances in ferroelectric metal
  complexes, Coordination Chemistry Reviews 469 (2022) 214663.

\bibitem{hill2000there}
N.~A. Hill, Why are there so few magnetic ferroelectrics?, The journal of
  physical chemistry B 104~(29) (2000) 6694--6709.

\bibitem{ederer2011mechanism}
C.~Ederer, T.~Harris, R.~Kov{\'a}{\v{c}}ik, Mechanism of ferroelectric
  instabilities in {non}-${d}^0$ {perovskites}: {LaCrO}$_3$ versus {CaMnO}$_3$,
  Phys. Rev. B 83 (2011) 054110.

\bibitem{balke2012enhanced}
N.~Balke, B.~Winchester, W.~Ren, Y.~H. Chu, A.~N. Morozovska, E.~A. Eliseev,
  M.~Huijben, R.~K. Vasudevan, P.~Maksymovych, J.~Britson, et~al., Enhanced
  electric conductivity at ferroelectric vortex cores in {BiFeO}$_3$, Nature
  Physics 8~(1) (2012) 81--88.

\bibitem{SHI2019561}
C.~Shi, X.-B. Han, W.~Zhang, Structural phase {transition}-{associated}
  dielectric transition and ferroelectricity in coordination compounds,
  Coordination Chemistry Reviews 378 (2019) 561--576, special issue on the 8th
  Chinese Coordination Chemistry Conference.

\bibitem{gibbs2011high}
A.~S. Gibbs, K.~S. Knight, P.~Lightfoot, High-temperature phase transitions of
  hexagonal {YMnO}$_3$, Physical Review B 83~(9) (2011) 094111.

\bibitem{mostovoy2006ferroelectricity}
M.~Mostovoy, Ferroelectricity in spiral magnets, Physical review letters 96~(6)
  (2006) 067601.

\bibitem{choi2008ferroelectricity}
Y.~Choi, H.~Yi, S.~Lee, Q.~Huang, V.~Kiryukhin, S.-W. Cheong, Ferroelectricity
  in an ising chain magnet, Physical review letters 100~(4) (2008) 047601.

\bibitem{aschauer2013strain}
U.~Aschauer, R.~Pfenninger, S.~M. Selbach, T.~Grande, N.~A. Spaldin,
  Strain-controlled oxygen vacancy formation and ordering in {Ca}{Mn}{O}$_3$,
  Physical Review B 88~(5) (2013) 054111.

\bibitem{PhysRevMaterials.2.104414}
N.~S. Fedorova, Y.~W. Windsor, C.~Findler, M.~Ramakrishnan, A.~Bortis,
  L.~Rettig, K.~Shimamoto, E.~M. Bothschafter, M.~Porer, V.~Esposito, Y.~Hu,
  A.~Alberca, T.~Lippert, C.~W. Schneider, U.~Staub, N.~A. Spaldin,
  Relationship between crystal structure and multiferroic orders in
  orthorhombic perovskite manganites, Phys. Rev. Materials 2 (2018) 104414.

\bibitem{wang2003epitaxial}
J.~Wang, J.~Neaton, H.~Zheng, V.~Nagarajan, S.~Ogale, B.~Liu, D.~Viehland,
  V.~Vaithyanathan, D.~Schlom, U.~Waghmare, et~al., Epitaxial {BiFeO}$_3$
  multiferroic thin film heterostructures, science 299~(5613) (2003)
  1719--1722.

\bibitem{katsufuji2001dielectric}
T.~Katsufuji, S.~Mori, M.~Masaki, Y.~Moritomo, N.~Yamamoto, H.~Takagi,
  Dielectric and magnetic anomalies and spin frustration in hexagonal
  {RMnO}$_3$ {(R} = {Y, Yb, and Lu}), Physical review B 64~(10) (2001) 104419.

\bibitem{fennie2005ferroelectric}
C.~J. Fennie, K.~M. Rabe, Ferroelectric transition in {YMnO}$_3$ from first
  principles, Physical Review B 72~(10) (2005) 100103.

\bibitem{Serrao2005}
C.~R. Serrao, A.~K. Kundu, S.~B. Krupanidhi, U.~V. Waghmare, C.~N.~R. Rao,
  Biferroic {YCrO}$_{3}$, Phys. Rev. B 72 (2005) 220101.

\bibitem{Prado2013}
J.~Prado-Gonjal, R.~Schmidt, J.-J. Romero, D.~\'{A}vila, U.~Amador,
  E.~Mor\'{a}n, Microwave-assisted synthesis, microstructure, and physical
  properties of rare-earth chromites, Inorg. Chem. 52 (2013) 313--320.

\bibitem{Sardar2011}
K.~Sardar, M.~R. Lees, R.~J. Kashtiban, J.~Sloan, R.~I. Walton, Direct
  hydrothermal synthesis and physical properties of rare-earth and yttrium
  orthochromite perovskites, Chem. Mater. 23 (2011) 48--56.

\bibitem{Zhu2020}
Y.~Zhu, S.~Wu, B.~Tu, S.~Jin, A.~Huq, J.~Persson, H.~Gao, D.~Ouyang, Z.~He,
  D.-X. Yao, Z.~Tang, H.-F. Li, High-temperature magnetism and crystallography
  of a {YCrO}$_{3}$ single crystal, Phys. Rev. B 101 (2020) 014114.

\bibitem{Zhu2020-2}
Y.~Zhu, Y.~Fu, B.~Tu, T.~Li, J.~Miao, Q.~Zhao, S.~Wu, J.~Xia, P.~Zhou, A.~Huq,
  W.~Schmidt, D.~Ouyang, Z.~Tang, Z.~He, H.-F. Li, Crystalline and magnetic
  structures, magnetization, heat capacity, and anisotropic magnetostriction
  effect in a yttrium-chromium oxide, Phys. Rev. Materials 4 (2020) 094409.

\bibitem{Taran2020-1}
S.~Taran, B.~Biswas, S.~Pal, H.~D. Yang, Improvement of the biferroic property
  by {Ca}-doping in bulk perovskite {YCrO}$_{3}$ system, AIP Conf. Proc. 2220
  (2020) 110002.

\bibitem{Mall2020}
A.~K. Mall, A.~K. Pramanik, Impedance spectroscopy study on {Ca}$^{2+}$ doped
  {YCrO}$_{3}$ ceramics, AIP Conf. Proc. 2220 (2020) 110005.

\bibitem{Shi2020}
C.~Shi, Y.~Su, J.~Guo, G.~Gong, H.~Hu, Y.~Gao, Y.~Wang, Influence of
  {Ga}$^{3+}$ ion dopant on the structure and magnetic properties of
  {YCrO}$_{3}$, Ceram. Int. 46 (2020) 27457--27462.

\bibitem{Seybolt1966}
A.~U. Seybolt, High temperature oxidation of chromium containing
  {Y}$_{2}${O}$_{3}$, Corros. Sci. 6 (1966) 263--269.

\bibitem{McCarthy1982}
G.~J. McCarthy, H.~B. SilberJames, J.~RhyneFaye, M.~Kalina, The Rare Earths in
  Modern Science and Technology (1982) New York, Plenum Press.

\bibitem{Poplawski2000}
K.~Poplawski, J.~Lichtenberger, F.~J. Keil, K.~Schnitzlein, M.~D. Amiridis,
  Catalytic oxidation of 1,2-dichlorobenzene over {ABO}$_{3}$-type perovskites,
  Catal. Today 62 (2000) 329--336.

\bibitem{Kim2003}
J.~H. Kim, H.-S. Shin, S.-H. Kim, J.-H. Moon, B.-T. Lee, Formation of
  {YCrO}$_{3}$ thin films using radio-frequency magnetron sputtering method for
  a wide range thermistor application, Jpn. J. Appl. Phys. 42 (2003) 575.

\bibitem{Zhang2014}
B.~Zhang, Q.~Zhao, A.~Chang, Y.~Liu, Y.~Li, Y.~Wu, Synthesis of {YCrO}$_{3}$
  ceramics through a field-assisted sintering technique, J. Mater. Sci.: Mater.
  Electron 25 (2014) 1400--1403.

\bibitem{Sharma2014-2}
Y.~Sharma, P.~Misra, R.~S. Katiyar, Unipolar resistive switching behavior of
  amorphous {YCrO}$_{3}$ films for nonvolatile memory applications, J. Appl.
  Phys. 116 (2014) 084505.

\bibitem{Sharma2014-3}
Y.~Sharma, P.~Misra, R.~K. Katiyar, R.~S. Katiyar, Photovoltaic effect and
  enhanced magnetization in 0.9({BiFeO}$_{3}$)--0.1({YCrO}$_{3}$) composite
  thin film fabricated using sequential pulsed laser deposition, J. Phys. D:
  Appl. Phys. 47 (2014) 425303.

\bibitem{Tiwari2015}
B.~Tiwari, A.~Dixit, R.~Naik, G.~Lawes, M.~S.~R. Rao, Magnetostructural and
  magnetocaloric properties of bulk {LaCrO}$_{3}$ system, Mater. Res. Express 2
  (2015) 026103.

\bibitem{Oliveira2016}
G.~N.~P. Oliveira, P.~Machado, A.~L. Pires, A.~M. Pereira, J.~P. Ara{\'u}jo,
  A.~M.~L. Lopes, Magnetocaloric effect and refrigerant capacity in
  polycrystalline {YCrO}$_{3}$, J. Phys. Chem. Solids 91 (2016) 182--188.

\bibitem{Goncharov2016}
V.~S. Goncharov, M.~V. Goncharov, E.~V. Vasil’ev, Diffusion permeability of
  yttrium-based heat-resistant ion-plasma coatings, Phys. Met. Metallogr. 117
  (2016) 896--900.

\bibitem{Hussain2017}
M.~H.~A. Jabbar, I.~Robinson, K.-J. Pan, B.~M. Blackburn, C.~Pellegrinelli,
  Y.-L. Huang, E.~D. Wachsman, Chromate-based oxide anodes for low-temperature
  operating solid oxide fuel cells, ECS Trans. 78 (2017) 1319.

\bibitem{Malcev2020}
D.~D. Malcev, O.~Y. Sinelshchikova, V.~I. Popkov, On the electrical
  conductivity of {YCrO}$_{3}$ porous ceramics, J. Phys.: Conf. Ser. 1697
  (2020) 012196.

\bibitem{Mall2020-1}
A.~K. Mall, A.~K. Pramanik, Structural and magnetic properties of {Gd} doped
  {YCrO}$_{3}$, AIP Conf. Proc. 2220 (2020) 110026.

\bibitem{Tiwari2020}
B.~Tiwari, A.~Dixit, M.~R. Rao, Magnetic entropy change in a non-collinear weak
  ferromagnetic {YCrO}$_{3}$, Vacuum 179 (2020) 109519.

\bibitem{Bhowmik2021}
T.~K. Bhowmik, T.~P. Sinha, Al-dependent electronic and magnetic properties of
  {YCrO}$_{3}$ with magnetocaloric application: {An} ab-initio and {Monte}
  {Carlo} approach, Physica B 606 (2021) 412659.

\bibitem{Chakraborty2021}
P.~Chakraborty, S.~Dey, S.~Basu, Structural, electrical and magnetic properties
  of {Eu} doped {YCrO}$_{3}$ nanoparticles, Physica B 601 (2021) 412677.

\bibitem{Chakraborty2021-1}
P.~Chakraborty, S.~Basu, Structural, electrical and magnetic properties of {Er}
  doped {YCrO}$_{3}$ nanoparticles, Mater. Chem. Phys. 259 (2021) 124053.

\bibitem{Mao2021}
Z.~Mao, L.~Xiong, S.~Liu, The formation of the complex oxide in {Ni}-based
  alloy powder during mechanical milling and heat treatment, J. Alloys Compd.
  879 (2021) 160333.

\bibitem{Looby1954}
J.~T. Looby, L.~Katz, Yttrium chromium oxide, a new compound of the perowskite
  type, J. Am. Chem. Soc. 76 (1954) 6029--6030.

\bibitem{Bertaut1966}
E.~F. Bertaut, G.~Bassi, G.~Buisson, P.~Burlet, J.~Chappert, A.~Delapalme,
  J.~Mareschal, G.~Roult, R.~Aleonard, R.~Pauthenet, J.~P. Rebouillat, Some
  neutron-diffraction investigations at the nuclear center of grenoble, J.
  Appl. Phys. 37 (1966) 1038--1039.

\bibitem{Case1984}
E.~D. Case, C.~J. Glinka, Characterization of microcracks in {YCrO}$_{3}$ using
  small-angle neutron scattering and elasticity measurements, J. Mater. Sci. 19
  (1984) 2962--2968.

\bibitem{Kuznetsov1998}
M.~V. Kuznetsov, I.~P. Parkin, Convenient, rapid synthesis of rare earth
  orthochromites {LnCrO}$_{3}$ by self-propagating high-temperature synthesis,
  Polyhedron 17 (1998) 4443--4450.

\bibitem{Xing2010}
L.~Xing, J.~Lu, Q.~Bi, Z.~Pu, M.~Guo, Y.~Wang, M.~Luo, In situ raman
  spectroscopy of phase transformation in {CrO$_{x}$-Y$_{2}$O$_{3}$} system at
  elevated temperatures, Appl. Surf. Sci. 256 (2010) 3586--3591.

\bibitem{Holder1980}
J.~D. Holder, R.~A. Hartzell, C.~G. W, Eutectics of {LaCrO}$_{3}$ and
  {YCrO}$_{3}$ with {W}, {Mo}, and {Cr}, J. Am. Ceram. Soc. 63 (1980) 344--345.

\bibitem{Duran2010}
A.~Dur{\'a}n, A.~M. Ar{\'e}valo-L{\'o}pez, E.~Castillo-Mart{\'\i}nez,
  M.~Garc{\'\i}a-Guaderrama, E.~Moran, M.~P. Cruz, F.~Fern{\'a}ndez, M.~A.
  Alario-Franco, Magneto-thermal and dielectric properties of biferroic
  {YCrO}$_{3}$ prepared by combustion synthesis, J. Solid State Chem. 183
  (2010) 1863--1871.

\bibitem{Geller1957-2}
S.~Geller, Crystallographic studies of perovskite-like compounds. {V}.
  {Relative} ionic sizes, Acta Cryst. 10 (1957) 248--251.

\bibitem{Prado-Gonjal2013}
J.~Prado-Gonjal, R.~Schmidt, J.-J. Romero, D.~\'{A}vila, U.~Amador,
  E.~Mor\'{a}n, Microwave-assisted synthesis, microstructure, and physical
  properties of rare-earth chromites, Inorg. Chem. 52 (2013) 313--320.

\bibitem{Belik2012}
A.~A. Belik, Y.~Matsushita, M.~Tanaka, E.~Takayama-Muromachi, Crystal
  structures and properties of perovskites {ScCrO}$_{3}$ and {InCrO}$_{3}$ with
  small ions at the {A} site, Chem. Mater. 24 (2012) 2197--2203.

\bibitem{Zhang2015}
X.~Zhang, X.~Zeng, J.~Dou, X.~Pu, R.~Xie, Multiferroic and magnetoelectric
  properties of {BiFeO}$_{3}$--{YCrO}$_{3}$ ceramics at the
  rhombohedral--orthorhombic phase boundary, Mater. Lett. 141 (2015) 168--171.

\bibitem{Bedekar2007}
V.~Bedekar, R.~Shukla, A.~K. Tyagi, Nanocrystalline {YCrO}$_{3}$ with
  onion-like structure and unusual magnetic behaviour, Nanotechnology 18 (2007)
  155706.

\bibitem{Duran2012-2}
A.~Dur{\'a}n, G.~G.~C. Arizaga, et~al., Hydroxide precursors to produce
  nanometric {YCrO}$_{3}$: Characterization and conductivity analysis, Mater.
  Res. Bull. 47 (2012) 1442--1447.

\bibitem{Park2012}
J.~J. Park, S.~M. Hong, E.~K. Park, M.~K. Lee, C.~K. Rhee, Synthesis of {Fe}
  based {ODS} alloys by a very high speed planetary milling process, J. Nucl.
  Mater. 428 (2012) 35--39.

\bibitem{Ahmad2016}
T.~Ahmad, I.~H. Lone, Citrate precursor synthesis and multifunctional
  properties of {YCrO}$_{3}$ nanoparticles, New J. Chem. 40 (2016) 3216--3224.

\bibitem{Apostolov2019}
A.~T. Apostolov, I.~N. Apostolova, J.~M. Wesselinowa, Magnetic and dielectric
  properties of pure and ion doped {RCrO}$_{3}$ nanoparticles, Eur. Phys. J. B
  92 (2019) 1--7.

\bibitem{Sinha2021}
R.~Sinha, M.~Halder, S.~Basu, A.~K. Meikap, Dielectric relaxation and
  electrical conduction mechanism of {Neodymium} doped {Yttrium} {Chromite},
  Physica B 615 (2021) 413035.

\bibitem{Chakraborty2021-3}
P.~Chakraborty, D.~K. Rana, S.~Basu, Enhanced electrical and magnetic
  properties of {Sm}-doped {YCrO}$_{3}$ nanoparticles, Bull. Mater. Sci. 44
  (2021) 1--11.

\bibitem{Kagawa1997}
M.~Kagawa, Y.~Kato, Y.~Syono, Ultrafine particles and thin films of
  {YCrO}$_{3}$ synthesized by the spray-icp technique, J. Aerosol. Sci. 1001
  (1997) S475--S476.

\bibitem{Cheng2010}
Z.~X. Cheng, X.~L. Wang, S.~X. Dou, H.~Kimura, K.~Ozawa, A novel multiferroic
  system: rare earth chromates, J. Appl. Phys. 107 (2010) 09D905.

\bibitem{Seo2013}
J.-D. Seo, J.~Y. Son, Room temperature ferroelectricity of {YCrO}$_{3}$ thin
  films on {Rh} single crystals, J. Cryst. Growth 375 (2013) 53--56.

\bibitem{Seo2015}
J.~Seo, Y.~Ahn, J.~Y. Son, Multiferroic properties of {YCrO}$_{3}$ thin films
  on glass substrate, Ceram. Int. 41 (2015) 12471--12474.

\bibitem{Seo2016}
J.~Seo, Y.~Ahn, J.~Y. Son, Multiferroic {YCrO}$_{3}$ thin films grown on glass
  substrate: Resistive switching characteristics, Electron. Mater. Lett. 12
  (2016) 87--90.

\bibitem{Araujo2014}
C.~M. Araujo, S.~Nagar, M.~Ramzan, R.~Shukla, O.~Jayakumar, A.~Tyagi, Y.-S.
  Liu, J.-L. Chen, P.-A. Glans, C.~Chang, et~al., Disorder-induced room
  temperature ferromagnetism in glassy chromites, Sci. Rep. 4 (2014) 1--6.

\bibitem{Duran2014}
A.~Dur{\'a}n, H.~Tiznado, J.~Romo-Herrera, D.~Dom{\'\i}nguez, R.~Escudero,
  J.~Siqueiros, Nanocomposite {YCrO}$_{3}$/{Al}$_{2}${O}$_{3}$:
  characterization of the core--shell, magnetic properties, and enhancement of
  dielectric properties, Inorg. Chem. 53 (2014) 4872--4880.

\bibitem{Tiznado2014}
H.~Tiznado, D.~Dom{\'\i}nguez, F.~Mu{\~n}oz-Mu{\~n}oz, J.~Romo-Herrera,
  R.~Machorro, O.~E. Contreras, G.~Soto, Pulsed-bed atomic layer deposition
  setup for powder coating, Powder Technol. 267 (2014) 201--207.

\bibitem{Duran2016}
A.~Dur{\'a}n, L.~Moxca, H.~Tiznado, J.~M. Romo-Herrera, M.~Herrera, J.~M.
  Siqueiros, {YCrO}$_{3}$/{Al}$_{2}${O}$_{3}$ core-shell design: The effect of
  the nanometric {Al}$_{2}${O}$_{3}$-shell on dielectric properties, J. Am.
  Ceram. Soc. 99 (2016) 3382--3388.

\bibitem{Arciniega2016}
J.~J. Gervacio-Arciniega, F.~J. Flores-Ruiz, C.~J. Diliegros-Godines,
  E.~Broitman, C.~I. Enriquez-Flores, F.~J. Espinoza-Beltr{\'a}n, J.~Siqueiros,
  M.~P. Cruz, Nanofrictional behavior of amorphous, polycrystalline and
  textured {Y-Cr-O} films, Appl. Surf. Sci. 378 (2016) 157--162.

\bibitem{Pal2018}
B.~Pal, X.~Liu, F.~Wen, M.~Kareev, A.~T. N'Diaye, P.~Shafer, E.~Arenholz,
  J.~Chakhalian, Electronic properties of ultra-thin {YCrO}$_{3}$ films, Appl.
  Phys. Lett. 112 (2018) 252901.

\bibitem{Arciniega2018}
J.~J. Gervacio-Arciniega, E.~Murillo-Bracamontes, O.~Contreras, J.~M.
  Siqueiros, O.~Raymond, A.~Dur{\'a}n, D.~Bueno-Baques, D.~Valdespino,
  E.~Cruz-Valeriano, C.~I. Enr{\'\i}quez-Flores, et~al., Multiferroic
  {YCrO}$_{3}$ thin films: Structural, ferroelectric and magnetic properties,
  Appl. Surf. Sci. 427 (2018) 635--639.

\bibitem{Kuang2018}
D.~Kuang, F.~Yang, W.~Jing, Z.~Yang, Multiferroic properties of a
  {YCrO}$_{3}$/{BiFeO}$_{3}$ bilayered thin film prepared by a sol-gel method,
  Physica B 530 (2018) 295--299.

\bibitem{Sharma2020}
Y.~Sharma, E.~Skoropata, B.~Paudel, K.~T. Kang, D.~Yarotski, T.~Z. Ward,
  A.~Chen, Epitaxial stabilization of single-crystal multiferroic {YCrO}$_{3}$
  thin films, Nanomaterials 10 (2020) 2085.

\bibitem{Remeika1956}
J.~P. Remeika, Growth of single crystal rare earth orthoferrites and related
  compounds, J. Am. Chem. Soc. 78 (1956) 4259--4260.

\bibitem{Razdan1992}
A.~K. Razdan, P.~N. Kotru, B.~M. Wanklyn, Elemental compositional changes in
  surface microstructures of flux-grown {YCrO}$_{3}$ single crystals, Mater.
  Sci. Eng. B 15 (1992) 199--202.

\bibitem{Todorov2011}
N.~D. Todorov, M.~V. Abrashev, V.~G. Ivanov, G.~G. Tsutsumanova, V.~Marinova,
  Y.-Q. Wang, M.~N. Iliev, Comparative {Raman} study of isostructural
  {YCrO}$_{3}$ and {YMnO}$_{3}$: {Effects} of structural distortions and
  twinning, Phys. Rev. B 83 (2011) 224303.

\bibitem{Yin2017}
L.~H. Yin, J.~Yang, P.~Tong, W.~H. Song, J.~M. Dai, X.~B. Zhu, Y.~P. Sun,
  Temperature and field induced spin reorientation and dielectric properties in
  {YCr}$_{0. 88}${Fe}$_{0.12}${O}$_{3}$ single crystal, Appl. Phys. Lett. 111
  (2017) 072402.

\bibitem{Sanina2018}
V.~A. Sanina, B.~K. Khannanov, E.~I. Golovenchits, M.~P. Shcheglov, Electric
  polarization in {YCrO}$_{3}$ induced by restricted polar domains of magnetic
  and structural natures, Phys. Solid State 60 (2018) 2532--2540.

\bibitem{Grodkiewics1966}
W.~H. Grodkiewicz, D.~J. Nitti, Oxide crystal growth by flux evaporation, J.
  Am. Ceram. Soc. 49 (1966) 576.

\bibitem{Aoyagi1969}
K.~Aoyagi, K.~Tsushima, S.~Sugano, Direct observation of davydov splitting in
  antiferromagnetic {YCrO}$_{3}$, Solid State Commun. 7 (1969) 229--232.

\bibitem{Zhou2002-1}
J.-S. Zhou, J.~B. Goodenough, Probing structural inhomogeneities induced by
  exchange striction above ${T}_{N}$ in antiferromagnetic perovskites, Phys.
  Rev. B 66 (2002) 052401.

\bibitem{zvezdin2021multiferroic}
A.~Zvezdin, Z.~Gareeva, X.~Chen, Multiferroic order parameters in rhombic
  antiferromagnets {RCrO}$_3$, Journal of Physics: Condensed Matter 33~(38)
  (2021) 385801.

\bibitem{bajaj2021magnetoelastic}
N.~Bajaj, A.~P. Roy, A.~Khandelwal, M.~K. Chattopadhyay, V.~Sathe, S.~K.
  Mishra, R.~Mittal, P.~D. Babu, M.~D. Le, J.~L. Niedziela, et~al.,
  Magnetoelastic coupling and spin contributions to entropy and thermal
  transport in biferroic yttrium orthochromite, Journal of Physics: Condensed
  Matter 33~(12) (2021) 125702.

\bibitem{chakraborty2021structural}
P.~Chakraborty, S.~Dey, S.~Basu, Structural, electrical and magnetic properties
  of eu doped {YCrO}$_3$ nanoparticles, Physica B: Condensed Matter 601 (2021)
  412677.

\bibitem{taheri2019structural}
M.~Taheri, F.~Razavi, Z.~Yamani, R.~Flacau, C.~Ritter, S.~Bette, R.~Kremer,
  Structural, magnetic, and thermal properties of
  {Ce}$_{1-x}${Eu}$_x${Cr}{O}$_3$ orthochromite solid solutions, Physical
  Review B 99~(5) (2019) 054411.

\bibitem{wang2019structure}
H.~Wang, X.~Liu, L.~Hao, X.~Ma, W.~Han, K.~Sun, D.~Chen, H.~Guo, Z.~Fu, C.-W.
  Wang, et~al., The structure and magnetism of orthochromites
  {Ho}$_{1-x}${Y}$_x${Cr}{O}$_3$, Journal of Magnetism and Magnetic Materials
  473 (2019) 428--434.

\bibitem{ye2020temperature}
Y.~Ye, A.~Cui, M.~Bian, K.~Jiang, L.~Zhu, J.~Zhang, L.~Shang, Y.~Li, Z.~Hu,
  J.~Chu, et~al., Temperature and pressure manipulation of magnetic ordering
  and phonon dynamics with phase transition in multiferroic {Gd}{Fe}{O}$_3$:
  Evidence from raman scattering, Physical Review B 102~(2) (2020) 024103.

\bibitem{zhou2020structural}
J.-S. Zhou, Structural distortions in rare-earth transition-metal oxide
  perovskites under high pressure, Physical Review B 101~(22) (2020) 224104.

\bibitem{tan2020charge}
Z.~Tan, F.~D. Romero, T.~Saito, M.~Goto, M.~A. Patino, A.~Koedtruad, Y.~Kosugi,
  W.-T. Chen, Y.-C. Chuang, H.-S. Sheu, et~al., Charge disproportionation and
  interchange transitions in twelve-layer {Ba}{Fe}{O}$_3$, Physical Review B
  102~(5) (2020) 054404.

\bibitem{fita2019spin}
I.~Fita, R.~Puzniak, A.~Wisniewski, V.~Markovich, Spin switching and unusual
  exchange bias in the single-crystalline {Gd}{Cr}{O}$_3$ compensated
  ferrimagnet, Physical Review B 100~(14) (2019) 144426.

\bibitem{yoshii2019dielectric}
K.~Yoshii, N.~Ikeda, Dielectric and magnetocaloric study of {Tm}{Cr}{O}$_3$,
  Journal of Alloys and Compounds 804 (2019) 364--369.

\bibitem{ahmed2019ab}
S.~S. Ahmed, G.~D. Ngantso, M.~Boujnah, A.~Benyoussef, A.~El~Kenz, Ab initio
  and monte carlo studies of phase transitions and magnetic properties of
  {Y}{Cr}{O}$_3$: Heisenberg model, Physics Letters A 383~(2-3) (2019)
  121--126.

\bibitem{li2008synthesis}
H.~Li, Synthesis of {CMR} manganites and ordering phenomena in complex
  transition metal oxides, Forschungszentrum J$\ddot{\rm u}$lich GmbH Press,
  2008.

\bibitem{li2006correlation}
H.~Li, Y.~Su, J.~Persson, P.~Meuffels, J.~Walter, R.~Skowronek, T.~Br{\"u}ckel,
  Correlation between structural and magnetic properties of
  {La}$_{7/8}${Sr}$_{1/8}${Mn}$_{1-\gamma}${O}$_{3+\delta}$ with controlled
  nonstoichiometry, Journal of Physics: Condensed Matter 19~(1) (2006) 016003.

\bibitem{li2007neutron}
H.~Li, Y.~Su, J.~Persson, P.~Meuffels, J.~Walter, R.~Skowronek, T.~Br{\"u}ckel,
  Neutron-diffraction study of structural transition and magnetic order in
  orthorhombic and rhombohedral
  {La}$_{7/8}${Sr}$_{1/8}${Mn}$_{1-\gamma}${O}$_{3+\delta}$, Journal of
  Physics: Condensed Matter 19~(17) (2007) 176226.

\bibitem{thompson2008scattering}
R.~B. Thompson, F.~Margetan, P.~Haldipur, L.~Yu, A.~Li, P.~Panetta, H.~Wasan,
  Scattering of elastic waves in simple and complex polycrystals, Wave Motion
  45~(5) (2008) 655--674.

\bibitem{CAROLAN2017128}
D.~Carolan, Recent advances in germanium nanocrystals: Synthesis, optical
  properties and applications, Progress in Materials Science 90 (2017)
  128--158.

\bibitem{HANUS20131056}
M.~J. Hanus, A.~T. Harris, Nanotechnology innovations for the construction
  industry, Progress in Materials Science 58~(7) (2013) 1056--1102.

\bibitem{FU201731}
Y.~Fu, J.~Luo, N.~Nguyen, A.~Walton, A.~Flewitt, X.~Zu, Y.~Li, G.~McHale,
  A.~Matthews, E.~Iborra, H.~Du, W.~Milne, Advances in piezoelectric thin films
  for acoustic biosensors, acoustofluidics and lab-on-chip applications,
  Progress in Materials Science 89 (2017) 31--91.

\bibitem{RAMANUJAM2020100619}
J.~Ramanujam, D.~M. Bishop, T.~K. Todorov, O.~Gunawan, J.~Rath, R.~Nekovei,
  E.~Artegiani, A.~Romeo, Flexible cigs, cdte and a-si:h based thin film solar
  cells: A review, Progress in Materials Science 110 (2020) 100619.

\bibitem{ZHOU201487}
Q.~Zhou, K.~H. Lam, H.~Zheng, W.~Qiu, K.~K. Shung, Piezoelectric single crystal
  ultrasonic transducers for biomedical applications, Progress in Materials
  Science 66 (2014) 87--111.

\bibitem{REVCOLEVSCHI1997321}
A.~Revcolevschi, J.~Jegoudez, Growth of large {high}-${T}_{c}$ single crystals
  by the floating zone {method}: {A} review, Progress in Materials Science
  42~(1) (1997) 321--339.

\bibitem{STEMPER2022100873}
L.~Stemper, M.~A. Tunes, R.~Tosone, P.~J. Uggowitzer, S.~Pogatscher, On the
  potential of aluminum crossover alloys, Progress in Materials Science 124
  (2022) 100873.

\bibitem{GALANO2022100831}
M.~Galano, F.~Audebert, Novel al based nanoquasicrystalline alloys, Progress in
  Materials Science 123 (2022) 100831, a Festschrift in Honor of Brian Cantor.

\bibitem{GAO2021100813}
F.~Gao, K.~Zhang, Y.~Guo, J.~Xu, M.~Szafran, {(Ba, Sr)TiO}$_3$/polymer
  dielectric composites–progress and perspective, Progress in Materials
  Science 121 (2021) 100813.

\bibitem{YANG2021100710}
Y.~Yang, Y.~Xu, Y.~Ji, Y.~Wei, Functional epoxy vitrimers and composites,
  Progress in Materials Science 120 (2021) 100710.

\bibitem{li2009crystal}
H.-F. Li, Y.~Su, Y.~Xiao, J.~Persson, P.~Meuffels, T.~Br{\"u}ckel, Crystal and
  magnetic structure of single-crystal {La}$_{1-{x}}${Sr}$_{x}${Mn}{O}$_3$ ({x}
  $\approx$ 1{/}8), The European Physical Journal B 67~(2) (2009) 149--157.

\bibitem{li2019superconductivity}
D.~Li, K.~Lee, B.~Y. Wang, M.~Osada, S.~Crossley, H.~R. Lee, Y.~Cui, Y.~Hikita,
  H.~Y. Hwang, Superconductivity in an infinite-layer nickelate, Nature
  572~(7771) (2019) 624--627.

\bibitem{li2020absence}
Q.~Li, C.~He, J.~Si, X.~Zhu, Y.~Zhang, H.-H. Wen, Absence of superconductivity
  in bulk {Nd}$_{1-x}${Sr}$_{x}${Ni}{O}$_2$, Communications Materials 1~(1)
  (2020) 1--8.

\bibitem{valenzuela2005magnetic}
R.~Valenzuela, Magnetic ceramics, no.~4, Cambridge university press, 2005.

\bibitem{Ikeda2015}
S.~Ikeda, S.~Hara, T.~Sakurai, S.~Okubo, H.~Ohta, H.~Sakurai, High-field {ESR}
  measurements of {YCrO}$_{3}$, Appl. Magn. Reson. 46 (2015) 1053--1058.

\bibitem{zhao2021insights}
Q.~Zhao, S.~Wu, Y.~Zhu, J.~Xia, H.-F. Li, Insights into the structural symmetry
  of single-crystal {YCrO}$_3$ from synchrotron {X}-{ray} diffraction, arXiv
  preprint arXiv:2112.02495.

\bibitem{kroger1977defect}
F.~Kroger, Defect chemistry in crystalline solids, Annual Review of Materials
  Science 7~(1) (1977) 449--475.

\bibitem{maier1993defect}
J.~Maier, Defect chemistry: composition, transport, and reactions in the solid
  state; part i: thermodynamics, Angewandte Chemie International Edition in
  English 32~(3) (1993) 313--335.

\bibitem{Cruz2014}
M.~P. Cruz, D.~Valdespino, J.~J. Gervacio, M.~Herrera, D.~Bueno-Baques,
  A.~Dur{\'a}n, J.~Munoz, A.~C. Garc{\'\i}a-Castro, F.~J. Espinoza-Beltr{\'a}n,
  M.~Curiel, J.~M. Siqueirosa, Piezoelectric and ferroelectric response
  enhancement in multiferroic {YCrO}$_{3}$ films by reduction in thickness,
  Mater. Lett. 114 (2014) 148--151.

\bibitem{Mall2018-2}
A.~K. Mall, A.~Garg, R.~Gupta, Dielectric relaxation and ac conductivity in
  magnetoelectric {YCrO}$_{3}$ ceramics: {A} temperature dependent impedance
  spectroscopy analysis, J. Eur. Ceram. Soc. 38 (2018) 5359--5366.

\bibitem{wang2005general}
X.~Wang, J.~Zhuang, Q.~Peng, Y.~Li, A general strategy for nanocrystal
  synthesis, Nature 437~(7055) (2005) 121--124.

\bibitem{COROT20061471}
C.~Corot, P.~Robert, J.-M. Idée, M.~Port, Recent advances in iron oxide
  nanocrystal technology for medical imaging, Advanced Drug Delivery Reviews
  58~(14) (2006) 1471--1504.

\bibitem{STOLLE2013160}
C.~J. Stolle, T.~B. Harvey, B.~A. Korgel, Nanocrystal photovoltaics: a review
  of recent progress, Current Opinion in Chemical Engineering 2~(2) (2013)
  160--167.

\bibitem{cozzoli2006synthesis}
P.~D. Cozzoli, T.~Pellegrino, L.~Manna, Synthesis, properties and perspectives
  of hybrid nanocrystal structures, Chemical Society Reviews 35~(11) (2006)
  1195--1208.

\bibitem{CHANG2011608}
T.-C. Chang, F.-Y. Jian, S.-C. Chen, Y.-T. Tsai, Developments in nanocrystal
  memory, Materials Today 14~(12) (2011) 608--615.

\bibitem{Ahmad2018}
T.~Ahmad, I.~H. Lone, Structural characterization and properties of
  {YCrO}$_{3}$ nanoparticles prepared by reverse micellar method, Bulletin of
  Materials Science 41 (2018) 1--5.

\bibitem{Zhu2021}
H.~Li, Y.~Zhu, S.~Wu, Z.~Tang, A method of centimeter-sized single crystal
  growth of chromate compounds and related storage device, China Patent
  CN110904497B, 2021.

\bibitem{Zhu2021-1}
Y.~Zhu, J.~Xia, S.~Wu, K.~Sun, Y.~Yang, Y.~Zhao, H.~W. Kan, Y.~Zhang, L.~Wang,
  H.~Wang, J.~Fang, C.~Wang, T.~Wu, Y.~Shi, J.~Yu, R.~Zhang, H.-F. Li, Crystal
  growth engineering and origin of the weak ferromagnetism in antiferromagnetic
  matrix of orthochromates from {\emph{t}}-{\emph{e}} orbital hybridization,
  iScience 25~(4) (2022) 104111.

\bibitem{Kim2007}
J.~W. Kim, Y.~S. Oh, K.~S. Suh, Y.~D. Park, K.~H. Kim, Specific heat of a
  {YCrO}$_{3}$ single crystal as investigated by a {Si}--{N} membrane based
  microcalorimeter, Thermochim. Acta 455 (2007) 2--6.

\bibitem{Wu2020}
S.~Wu, Y.~Zhu, H.~Gao, Y.~Xiao, J.~Xia, P.~Zhou, D.~Ouyang, Z.~Li, Z.~Chen,
  Z.~Tang, H.-F. Li, Super-necking crystal growth and structural and magnetic
  properties of {SrTb}$_{2}${O}$_{4}$ single crystals, ACS omega 5 (2020)
  16584--16594.

\bibitem{Zhu2020-3}
Y.~Zhu, P.~Zhou, T.~Li, J.~Xia, S.~Wu, Y.~Fu, K.~Sun, Q.~Zhao, Z.~Li, Z.~Tang,
  Y.~Xiao, Z.~Chen, H.-F. Li, Enhanced magnetocaloric effect and magnetic phase
  diagrams of single-crystal {GdCrO}$_{3}$, Phys. Rev. B 102 (2020) 144425.

\bibitem{QIAO2021100622}
H.~Qiao, C.~Wang, W.~S. Choi, M.~H. Park, Y.~Kim, Ultra-thin ferroelectrics,
  Materials Science and Engineering: R: Reports 145 (2021) 100622.

\bibitem{Mall2017-1}
A.~K. Mall, A.~Dixit, A.~Garg, R.~Gupta, Temperature dependent electron
  paramagnetic resonance study on magnetoelectric {YCrO}$_{3}$, J. Phys.:
  Condens. Matter 29 (2017) 495805.

\bibitem{Kumar2008}
A.~Kumar, A.~S. Verma, S.~R. Bhardwaj, Prediction of formability in
  perovskite-type oxides, The Open Applied Physics Journal 1 (2008) 11--19.

\bibitem{webionicsize}
M.~Winter, Source: Webelements [http://www.webelements.com/],
  WebElements$^{\textrm{TM}}$.

\bibitem{Platonov2011}
A.~N. Platonov, V.~S. Urusov, K.~Langer, A remark on correlation between
  relaxation parameter and covalence of {Cr}$^{3+}$--{O} bonds in some mineral
  structures, Mineral. Journ. (Ukraine). 33 (2011) 49--52.

\bibitem{Katz1955}
L.~Katz, On the unit cell of {YCrO}$_{3}$, Acta Cryst. 8 (1955) 121--122.

\bibitem{Geller1956}
S.~Geller, E.~A. Wood, Crystallographic studies of perovskite-like compounds.
  {I}. {Rare} earth orthoferrites and {YFeO}$_{3}$, {YCrO}$_{3}$, {YAlO}$_{3}$,
  Acta Cryst. 9 (1956) 563--568.

\bibitem{Fabian2017}
F.~A. Fabian, K.~O. Moura, C.~C.~S. Barbosa, E.~B. Peixoto, F.~Garcia, J.~G.~S.
  Duque, C.~T. Meneses, Structural and magnetic phase transition observed in
  the {YCrO}$_{3+{\gamma}}$ compound, J. Alloys Compd. 702 (2017) 244--248.

\bibitem{Mall2019}
A.~K. Mall, B.~Paul, A.~Garg, R.~Gupta, Temperature dependent {X}-ray
  diffraction and raman spectroscopy studies of polycrystalline {YCrO}$_{3}$
  ceramics across the ${T}_{c}$ $\sim$ 460 {K}, J. Raman Spectrosc. 51 (2020)
  537--545.

\bibitem{Ardit2010}
M.~Ardit, G.~Cruciani, M.~Dondi, M.~Merlini, P.~Bouvier, Elastic properties of
  perovskite {YCrO}$_{3}$ up to 60 {GPa}, Phys. Rev. B 82 (2010) 064109.

\bibitem{Ramesha2007}
K.~Ramesha, A.~Llobet, T.~Proffen, C.~R. Serrao, C.~N.~R. Rao, Observation of
  local non-centrosymmetry in weakly biferroic {YCrO}$_{3}$, J. Phys.: Condens.
  Matter 19 (2007) 102202.

\bibitem{Jana2018-1}
R.~Jana, A.~Chandra, G.~D. Mukherjee, High pressure studies on nanocrystalline
  {YCrO}$_{3}$, AIP Conf. Proc. 1953 (2018) 030081.

\bibitem{Jana2018-2}
R.~Jana, V.~Pareek, P.~Khatua, P.~Saha, A.~Chandra, G.~D. Mukherjee, Pressure
  induced anomalous magnetic behaviour in nanocrystalline {YCrO}$_{3}$ at room
  temperature, J. Phys.: Condens. Matter 30 (2018) 335401.

\bibitem{Geller1957-1}
S.~Geller, Crystallographic studies of perovskite-like compounds. {IV}. {Rare}
  earth scandates, vanadites, galliates, orthochromites, Acta Cryst. 10 (1957)
  243--248.

\bibitem{Weber2019}
S.~F. Weber, S.~M. Griffin, J.~B. Neaton, Topological semimetal features in the
  multiferroic hexagonal manganites, Phys. Rev. Materials 3 (2019) 064206.

\bibitem{fu2021gapless}
H.~Fu, K.~Huang, K.~Watanabe, T.~Taniguchi, J.~Zhu, Gapless spin wave transport
  through a quantum canted antiferromagnet, Physical Review X 11~(2) (2021)
  021012.

\bibitem{pan2021giant}
Y.~Pan, C.~Le, B.~He, S.~J. Watzman, M.~Yao, J.~Gooth, J.~P. Heremans, Y.~Sun,
  C.~Felser, Giant anomalous nernst signal in the antiferromagnet ybmnbi2,
  Nature materials (2021) 1--7.

\bibitem{kim2021spin}
T.~Kim, S.~Hwang, S.~Hamh, S.~Yoon, S.~Han, B.~Cho, Spin orbit torque switching
  in an antiferromagnet through n{\'e}el reorientation in a rare-earth ferrite,
  Physical Review B 104~(5) (2021) 054406.

\bibitem{davidson2021pressure}
R.~A. Davidson, J.~S. Miller, Pressure and temperature dependences of the
  canting angle and increase in the magnetic ordering temperature,
  ${T}_c$({P}), for the weak ferromagnet {Li}$^{+}$[{TCNE}]-({TCNE} =
  tetracyanoethylene), Dalton Transactions 50~(39) (2021) 13859--13865.

\bibitem{ning2021antisite}
S.~Ning, A.~Kumar, K.~Klyukin, E.~Cho, J.~H. Kim, T.~Su, H.-S. Kim, J.~M.
  LeBeau, B.~Yildiz, C.~A. Ross, An antisite defect mechanism for room
  temperature ferroelectricity in orthoferrites, Nature Communications 12~(1)
  (2021) 1--7.

\bibitem{flynn2021two}
M.~O. Flynn, T.~E. Baker, S.~Jindal, R.~R. Singh, Two phases inside the bose
  condensation dome of {Yb}$_2${Si}$_2${O}$_7$, Physical Review Letters 126~(6)
  (2021) 067201.

\bibitem{li2021antiferromagnetic}
N.~Li, J.~Tang, L.~Su, Y.-J. Ke, W.~Zhang, Z.-K. Xie, R.~Sun, X.-Q. Zhang,
  W.~He, Z.-H. Cheng, Antiferromagnetic spin dynamics in exchanged-coupled
  {Fe}/{Gd}{Fe}{O}$_3$ heterostructure, Chinese Physics B 30~(11) (2021)
  117502.

\bibitem{zeng2021smoothing}
X.~Zeng, Y.~Liang, H.~Zhang, X.~Xi, J.~Cao, B.~Li, J.~Zhou, Smoothing method to
  directly denoise terahertz signals in rare-earth orthoferrite
  antiferromagnets, Journal of the American Ceramic Society 104~(7) (2021)
  3325--3333.

\bibitem{das2021anisotropic}
S.~Das, A.~Ross, X.~Ma, S.~Becker, C.~Schmitt, F.~van Duijn, F.~Fuhrmann, M.-A.
  Syskaki, U.~Ebels, V.~Baltz, et~al., Anisotropic long-range spin transport in
  canted antiferromagnetic orthoferrite {Y}{Fe}{O}$_3$, arXiv preprint
  arXiv:2112.05947.

\bibitem{zhou2020weak}
J.-S. Zhou, L.~Marshall, Z.-Y. Li, X.~Li, J.-M. He, Weak ferromagnetism in
  perovskite oxides, Physical Review B 102~(10) (2020) 104420.

\bibitem{Judin1966}
V.~M. J{\"u}din, A.~B. Sherman, Weak ferromagnetism of {YCrO}$_{3}$, Solid
  State Commun. 4 (1966) 661--663.

\bibitem{HFLi2016}
H.-F. Li, Possible ground states and parallel magnetic-field-driven phase
  transitions of collinear antiferromagnets, npj Comput. Mater. 2 (2016) 16032.

\bibitem{Tsushima1970}
K.~Tsushima, K.~Aoyagi, S.~Sugano, Magnetic and magneto-optical properties of
  some rare-earth and {Yttrium} orthochromites, J. Appl. Phys. 41 (1970)
  1238--1240.

\bibitem{Krishnan2011}
S.~Krishnan, C.~S. Suchand~Sandeep, R.~Philip, N.~Kalarikkal, The open aperture
  z-scan studies on biferroic {YCrO}$_{3}$, AIP Conf. Proc. 1349~(1) (2011)
  1277--1278.

\bibitem{Carini1991-1}
G.~F. Carini~II, H.~U. Anderson, M.~M. Nasrallah, D.~M. Sparlin, Defect
  structure, nonstoichiometry, and phase stability of {Ca}-doped {YCrO}$_{3}$,
  J. Solid State Chem. 94 (1991) 329--336.

\bibitem{Tachiwaki2001}
T.~Tachiwaki, Y.~Kunifusa, M.~Yoshinaka, K.~Hirota, O.~Yamaguchi, Formation,
  densification, and electrical conductivity of air-sinterable
  {Y}({Cr}$_{1-x}${Mg}$_{x}$){O}$_{3}$ prepared by the hydrazine method, Mater.
  Sci. Eng. B 86 (2001) 255--259.

\bibitem{Sahu2008}
J.~R. Sahu, C.~R. Serrao, C.~N.~R. Rao, Modification of the multiferroic
  properties of {YCrO}$_{3}$ and {LuCrO}$_{3}$ by mn substitution, Solid State
  Commun. 145 (2008) 52--55.

\bibitem{Singh2013}
I.~Singh, A.~K. Nigam, K.~Landfester, R.~Mu{\~n}oz-Esp{\'\i}, A.~Chandra,
  Anomalous magnetic behavior below 10 {K} in {YCrO}$_{3}$ nanoparticles
  obtained under droplet confinement, Appl. Phys. Lett. 103 (2013) 182902.

\bibitem{Sharma2014-1}
Y.~Sharma, S.~Sahoo, W.~Perez, S.~Mukherjee, R.~Gupta, A.~Garg, R.~Chatterjee,
  R.~S. Katiyar, Phonons and magnetic excitation correlations in weak
  ferromagnetic {YCrO}$_{3}$, J. Appl. Phys. 115 (2014) 183907.

\bibitem{Ziel1969-3}
J.~P. Van Der~Ziel, L.~G. Van~Uitert, Magnon-assisted optical emission in
  {YCrO}$_{3}$ and {LuCrO}$_{3}$, Phys. Rev. 179 (1969) 343--351.

\bibitem{Jacobs1971}
I.~S. Jacobs, H.~F. Burne, L.~M. Levinson, Field-induced spin reorientation in
  {YFeO}$_{3}$ and {YCrO}$_{3}$, J. Appl. Phys. 42 (1971) 1631--1632.

\bibitem{Sugano1971-1}
S.~Sugano, Y.~Uesaka, I.~Tsujikawa, K.~Aoyagi, K.~Tsushima, Exciton lines in
  rare-earth orthochromites, J. Phys. Colloques 32 (1971) C1--798--C1--800.

\bibitem{Sugano1971-2}
S.~Sugano, K.~Aoyagi, K.~Tsushima, Exciton absorption lines in
  antiferromagnetic rare-earth orthochromites--with particular reference to
  {YCrO}$_{3}$, J. Phys. Soc. Jpn. 31 (1971) 706--722.

\bibitem{Zhou2002-2}
J.-S. Zhou, J.~B. Goodenough, Pressure-induced transition from localized
  electron toward band antiferromagnetism in {LaCrO}$_{3}$, Phys. Rev. Lett. 89
  (2002) 087201.

\bibitem{Ray2008}
N.~Ray, U.~V. Waghmare, Coupling between magnetic ordering and structural
  instabilities in perovskite biferroics: {A} first-principles study, Phys.
  Rev. B 77 (2008) 134112.

\bibitem{Nair2013}
V.~G. Nair, C.~Ganeshraj, P.~N. Santhosh, V.~Subramanian, Ab initio
  calculations of yttrium chromite 1512 (2013) 846--847.

\bibitem{Apostolov2017}
A.~T. Apostolov, I.~N. Apostolova, J.~M. Wesselinowa, Influence of spin--phonon
  interactions and spin-reorientation transitions on the phonon properties of
  {RCrO}$_{3}$, Mod. Phys. Lett. B 31 (2017) 1750009.

\bibitem{Kumar2017}
S.~Kumar, I.~Coondoo, A.~Rao, B.-H. Lu, Y.-K. Kuo, A.~L. Kholkin, N.~Panwar,
  Impact of low level praseodymium substitution on the magnetic properties of
  {YCrO}$_{3}$ orthochromites, Physica B 510 (2017) 104--108.

\bibitem{Ahmed2019}
S.~S. Ahmed, G.~D. Ngantso, M.~Boujnah, A.~Benyoussef, A.~El~Kenz, Ab initio
  and monte carlo studies of phase transitions and magnetic properties of
  {YCrO}$_{3}$: Heisenberg model, Phys. Lett. A 383 (2019) 121--126.

\bibitem{Ziel1969-1}
J.~P. van~der Ziel, L.~G. Van~Uitert, Magnon-assisted chromium emission in
  {YCrO}$_{3}$, {LuCrO}$_{3}$, and {GdCrO}$_{3}$, J. Appl. Phys. 40 (1969)
  997--998.

\bibitem{Sanina1970}
V.~A. Sanina, E.~I. Golovenchits, T.~A. Fomina, A.~G. Gurevich,
  Antiferromagnetic resonance in orthorhombic weak ferromagnet {YCrO}$_{3}$,
  Phys. Lett. A 33 (1970) 291--292.

\bibitem{Alvarez2010}
G.~Alvarez, M.~P. Cruz, A.~C. Dur{\'a}n, H.~Montiel, R.~Zamorano, Weak
  ferromagnetism in the magnetoelectric {YCrO}$_{3}$ detected by microwave
  power absorption measurements, Solid State Commun. 150 (2010) 1597--1600.

\bibitem{Jara2018}
A.~N.~L. Jara, J.~F. Carvalho, A.~F. J{\'u}nior, L.~J.~Q. Maia, R.~C. Santana,
  On the optical and magnetic studies of {YCrO}$_{3}$ perovskites, Physica B
  546 (2018) 67--72.

\bibitem{Saha2014}
R.~Saha, A.~Sundaresan, C.~N.~R. Rao, Novel features of multiferroic and
  magnetoelectric ferrites and chromites exhibiting magnetically driven
  ferroelectricity, Mater. Horiz. 1 (2014) 20--31.

\bibitem{Rao1968}
G.~V.~S. Rao, G.~V. Chandrashekhar, C.~N.~R. Rao, Are rare earth orthochromites
  ferroelectric?, Solid State Communications 6 (1968) 177--179.

\bibitem{Mall2016}
A.~K. Mall, A.~Garg, R.~Gupta, High temperature x-ray diffraction, {Raman}
  spectroscopy and dielectric studies on yttrium orthochromites, AIP Conf.
  Proc. 1728 (2016) 020239.

\bibitem{Lizarraga2012}
R.~Liz{\'a}rraga, M.~Ramzan, C.~M. Araujo, A.~Blomqvist, R.~Ahuja,
  E.~Holmstr{\"o}m, Structural characterization of amorphous {YCrO}$_{3}$ from
  first principles, Europhys. Lett. 99 (2012) 57010.

\bibitem{Patil1988}
D.~S. Patil, N.~Venkatramani, V.~K. Rohatgi, Infrared spectra of the ceramic
  {YCrO}$_{3}$ doped with strontium, J. Mater. Sci. Lett. 7 (1988) 413--414.

\bibitem{Eguchi2005-1}
K.~Eguchi, Y.~Kawabe, E.~Hanamura, Second harmonic generation of yttrium
  orthochromite with magnetic origin, J. Phys. Soc. Japan 74 (2005) 1075--1076.

\bibitem{Krishnan2013}
S.~Krishnan, N.~Kalarikkal, Synthesis of {YCrO}$_{3}$ nanoparticles through paa
  assisted sol--gel route, J. Sol.-Gel. Sci. Technol. 66 (2013) 6--14.

\bibitem{Krishnan2014}
S.~Krishnan, K.~Shafakath, R.~Philip, N.~Kalarikkal, Size dependent nonlinear
  optical properties of {YCrO}$_{3}$ nanosystems, AIP Conf. Proc. 1576 (2014)
  138--140.

\bibitem{Krishnan2012}
S.~Krishnan, C.~S.~S. Sandeep, R.~Philip, N.~Kalarikkal, Two-photon assisted
  excited state absorption in multiferroic {YCrO}$_{3}$ nanoparticles, Chem.
  Phys. Lett. 529 (2012) 59--63.

\bibitem{Weber2012}
M.~C. Weber, J.~Kreisel, P.~A. Thomas, M.~Newton, K.~Sardar, R.~I. Walton,
  Phonon raman scattering of {RCrO}$_{3}$ perovskites {(R = Y, La, Pr, Sm, Gd,
  Dy, Ho, Yb, Lu)}, Phys. Rev. B 85 (2012) 054303.

\bibitem{Tiwari2013}
B.~Tiwari, M.~K. Surendra, M.~S.~R. Rao, {HoCrO}$_{3}$ and {YCrO}$_{3}$: a
  comparative study, J. Phys.: Condens. Matter 25~(21) (2013) 216004.

\bibitem{Mall2014}
A.~K. Mall, S.~Mukherjee, Y.~Sharma, A.~Garg, R.~Gupta, Temperature dependent
  raman scattering in {YCrO}$_{3}$ 1591 (2014) 1753--1754.

\bibitem{Mannepalli2016}
V.~rao Mannepalli, R.~Ramadurai, Studies on local structural inhomogeneity and
  origin of ferroelectricity in yttrium chromite ceramics, MRS Advances 1
  (2016) 609--614.

\bibitem{Mannepalli2017-3}
V.~R. Mannepalli, R.~Raghunathan, R.~Ramadurai, A.~David, W.~Prellier, Local
  structural distortion and interrelated phonon mode studies in yttrium
  chromite, J. Mater. Res. 32 (2017) 1541--1547.

\bibitem{Mall2018-1}
A.~K. Mall, B.~Paul, A.~Garg, R.~Gupta, Evidence for incipient ferroelectricity
  in {YCrO}$_{3}$ (2018) arXiv 1805.10077.

\bibitem{Weber1986}
W.~J. Weber, C.~W. Griffin, J.~L. Bates, Electrical and thermal transport
  properties of the {Y}$_{1-x}${M}$_{x}${CrO}$_{3}$ system, J. Mater. Res. 1
  (1986) 675--684.

\bibitem{Tiwari2019}
S.~Tiwari, M.~Saleem, A.~Mishra, D.~Varshney, Structural, optical, and
  dielectric studies on {Sr}-doped biferroic {YCrO}$_{3}$, J. Supercond. Nov.
  Magn. 32 (2019) 2521--2531.

\bibitem{Carini1991-2}
G.~F. Carini~II, H.~U. Anderson, D.~M. Sparlin, M.~M. Nasrallah, Electrical
  conductivity, seebeck coefficient and defect chemistry of {Ca}-doped
  {YCrO}$_{3}$, Solid State lonics 49 (1991) 233--243.

\bibitem{Duran2012-1}
A.~Dur{\'a}n, E.~Verdin, R.~Escamilla, F.~Morales, R.~Escudero, Mechanism of
  small-polaron formation in the biferroic {YCrO}$_{3}$ doped with calcium,
  Mater. Chem. Phys. 133 (2012) 1011--1017.

\bibitem{li2012-1}
H.~Li, Y.~Xiao, B.~Schmitz, J.~Persson, W.~Schmidt, P.~Meuffels, G.~Roth,
  T.~Br{\"u}ckel, Possible magnetic-polaron-switched positive and negative
  magnetoresistance in the {GdSi} single crystals, Sci. Rep. 2 (2012) 750.

\bibitem{Escamilla2017}
R.~Escamilla, L.~Huerta, M.~Romero, E.~Verdin, A.~Dur{\'a}n, Evidence of mixed
  valence {Cr}$^{+3}$/{Cr}$^{+4}$ in {Y}$_{1-x}${Ca}$_x${CrO}$_3$
  polycrystalline ceramics by {X-ray} photoelectron spectroscopy, J. Mater.
  Sci. 52 (2017) 2889--2894.

\bibitem{Westphal2000}
D.~Westphal, A.~Laske, S.~Jakobs, U.~Guth, Structural and electrical
  investigations of ({Y,Gd})$_{1-x}${Cr}$_{0.8}${Ga}$_{0.2}${O}$_{3}$, Ionics 6
  (2000) 346--350.

\bibitem{Ardit2009}
M.~Ardit, M.~Dondi, G.~Cruciani, F.~Matteucci, {Ti--Ca--Al}-doped {YCrO}$_{3}$
  pigments: {XRD} and {UV}--vis investigation, Mater. Res. Bull. 44 (2009)
  666--673.

\bibitem{Mall2015}
A.~K. Mall, A.~Garg, R.~Gupta, Composition-dependent structural and {Raman}
  spectroscopic studies on {Y}$_{1-x}${Ho}$_{x}${CrO}$_{3}$
  (0$\leq$x$\leq$0.1), AIP Conf. Proc. 1665 (2015) 140052.

\bibitem{Mannepalli2017-1}
V.~R. Mannepalli, R.~Ramadurai, {AC}-conductivity studies on
  {Y}$_{1-x}${Bi}$_{x}${CrO}$_{3}$ solid solution, AIP Conf. Proc. 1832 (2017)
  110017.

\bibitem{Mannepalli2017-2}
V.~R. Mannepalli, R.~Ramadurai, Structural and electrical transport studies in
  {Bi}-substituted yttrium chromite, J. Mater. Sci.: Mater. Electron 28 (2017)
  8087--8092.

\bibitem{Sinha2016}
R.~Sinha, S.~Kundu, S.~Basu, A.~K. Meikap, Effect of {La} doping on optical and
  electrical transport properties of nanocrystalline {YCrO}$_{3}$, Solid State
  Sci. 60 (2016) 75--84.

\bibitem{Kumar2010}
N.~Kumar, A.~Sundaresan, On the observation of negative magnetization under
  zero-field-cooled process, Solid State Commun. 150 (2010) 1162--1164.

\bibitem{Duran2018-3}
A.~Dur\'an, R.~Escamilla, R.~Escudero, F.~Morales, E.~Verd\'{\i}n, Reversal
  magnetization, spin reorientation, and exchange bias in {YCrO}$_{3}$ doped
  with praseodymium, Phys. Rev. Materials 2 (2018) 014409.

\bibitem{Deng2015}
D.~Deng, J.~Zheng, D.~Yu, B.~Wang, D.~Sun, M.~Avdeev, Z.~Feng, C.~Jing, B.~Lu,
  S.~Cao, J.~Zhang, Cooling field tuned magnetic phase transition and exchange
  bias-like effect in {Y}$_{0.9}${Pr}$_{0.1}${CrO}$_{3}$, Appl. Phys. Lett. 107
  (2015) 102404.

\bibitem{Sinha2018}
R.~Sinha, S.~Basu, A.~K. Meikap, The investigation of the electrical transport
  properties of {Gd} doped {YCrO}$_{3}$ nanoparticles, Mater. Res. Bull. 97
  (2018) 578--587.

\bibitem{Sinha2019}
R.~Sinha, S.~Basu, A.~K. Meikap, Effect of neodymium doping on electrical
  properties and relaxor ferroelectric behavior of {YCrO}$_{3}$ nanoparticles,
  Physica E 113 (2019) 194--201.

\bibitem{Mall2017-3}
A.~K. Mall, A.~Garg, G.~Rajeev, Modifications of the structure and magnetic
  properties of ceramic {YCrO}$_{3}$ with {Fe}/{Ni} doping, Mater. Res. Express
  4 (2017) 076104.

\bibitem{Duran2018-2}
A.~Dur\'{a}n, J.~M. Jim\'{e}nez, M.~Sol\'{o}rzano, R.~Falconi, Improvement of
  the ferroelectric properties of {Ti}-doped {YCrO}$_{3}$ ceramic, J. Phys.
  Chem. Solids. 123 (2018) 228--234.

\bibitem{Kamlo2011}
A.~N. Kamlo, J.~Bernard, C.~Lelievre, D.~Houivet, Synthesis and {NTC}
  properties of {YCr}$_{1-x}${Mn}$_{x}${O}$_{3}$ ceramics sintered under
  nitrogen atmosphere, J. Eur. Ceram. Soc. 31 (2011) 1457--1463.

\bibitem{Sinha2015}
R.~Sinha, S.~Basu, A.~K. Meikap, Investigation of dielectric and electrical
  behavior of {Mn} doped {YCrO}$_{3}$ nanoparticles synthesized by the sol gel
  method, Physica E 69 (2015) 47--55.

\bibitem{hachimi2017}
A.~G. El~hachimi, M.~L. OuldNE, H.~Zaari, A.~Benyoussef, A.~El~Kenz,
  Ferromagnetism and electronic structure in manganese doped {YCrO}$_{3}$
  perovskite oxide: Ab initio study, J. Supercond. Nov. Magn. 30 (2017)
  483--488.

\bibitem{Li2017}
C.~L. Li, S.~Huang, X.~X. Li, C.~M. Zhu, G.~Zerihun, C.~Y. Yin, C.~L. Lu, S.~L.
  Yuan, Negative magnetization induced by {Mn} doping in {YCrO}$_{3}$, J. Magn.
  Magn. Mater. 432 (2017) 77--81.

\bibitem{Panwar2018}
N.~Panwar, S.~Kumar, I.~Coondoo, M.~Vasundhara, N.~Kumar, Low temperature
  magnetic and magnetocaloric studies in {YCr}$_{0.85}${Mn}$_{0.15}${O}$_3$
  ceramic, Physica B 545 (2018) 352--357.

\bibitem{Taran2020}
S.~Taran, B.~Biswas, H.~D. Yang, Structural, magnetic, and ferroelectric
  properties of {Zr}-doped {Y}$_{1-x}${Zr}$_{x}${CrO}$_{3}$ bulk
  polycrystalline system, J. Supercond. Nov. Magn. 33 (2020) 2483--2491.

\bibitem{Saxena2020}
P.~Saxena, P.~Choudhary, A.~Yadav, V.~N. Rai, A.~Mishra, Effect of strontium
  doping on the structural and dielectric properties of {YCrO}$_{3}$, J. Mater.
  Sci.: Mater. Electron. 31 (2020) 12444--12454.

\bibitem{Mall2021-2}
A.~K. Mall, A.~K. Pramanik, Temperature dependent $^{89}${Y} {NMR} study on
  multiferroic {YCrO}$_{3}$, J. Phys.: Condens. Matter 33 (2021) 125803.

\bibitem{Cruciani2009}
G.~Cruciani, M.~Ardit, M.~Dondi, F.~Matteucci, M.~Blosi, M.~C. Dalconi,
  S.~Albonetti, Structural relaxation around {Cr}$^{3+}$ in
  {YAlO}$_{3}$-{YCrO}$_{3}$ perovskites from electron absorption spectra, J.
  Phys. Chem. A 113 (2009) 13772--13778.

\bibitem{Duran2018-1}
A.~Dur\'{a}n, E.~Verd\'{\i}n, A.~Conde, R.~Escamilla, Effect of {Al}-doped
  {YCrO}$_{3}$ on structural, electronic and magnetic properties, J. Magn.
  Magn. Mater. 453 (2018) 36--43.

\bibitem{Wold1959}
A.~Wold, W.~Croft, Preparation and properties of the systems
  {LnFe}$_{x}${Cr}$_{1-x}${O}$_{3}$ and {LaFe}$_{x}${Co}$_{1-x}${O}$_{3}$, J.
  Phys. Chem. 63 (1959) 447--448.

\bibitem{Kovachev2010}
S.~Kovachev, D.~Kovacheva, S.~Aleksovska, E.~Svab, K.~Krezhov, Structure and
  properties investigation of mixed oxides {YCr}$_{1-x}${Fe}$_x${O}$_3$
  (0$\leq$x$\leq$1), AIP Conf. Proc. 1203 (2010) 199.

\bibitem{Tachiwaki2001-1}
T.~Tachiwaki, Y.~Kunifusa, M.~Yoshinaka, K.~Hirota, O.~Yamaguchi, Formation,
  powder characterization and sinering of {YCrO}$_{3}$ prepared by a sol-gel
  technique using hydrazine, Int. J. Inorg. Mater. 3 (2001) 107--111.

\bibitem{MalagaReddy2012}
V.~MalagaReddy, B.~Rai, S.~Mishra, C.~Rong, J.~Liu, Facile mechanochemical
  synthesis and magnetic properties of pervoskite
  {YCr}$_{x}${Fe}$_{1-x}${O}$_{3}$, (0$\leq$x$\leq$1), MRS Online Proceedings
  Library 1397 (2012) 71--77.

\bibitem{Fabian2016}
A.~Dur\'{a}n, E.~Verd\'{\i}n, A.~Conde, R.~Escamilla, Reversal magnetization
  dependence with the {Cr} and {Fe} oxidation states in
  {YFe}$_{1-x}${Cr}$_x${O}$_3$ (0$\leq$x$\leq$1) perovskites, J. Magn. Magn.
  Mater. 408 (2016) 94--98.

\bibitem{schiffer1995low}
P.~Schiffer, A.~Ramirez, W.~Bao, S.-W. Cheong, Low temperature
  magnetoresistance and the magnetic phase diagram of
  {La}$_{1-{x}}${Ca}$_{x}${MnO}$_{3}$, Physical Review Letters 75~(18) (1995)
  3336.

\bibitem{hemberger2002structural}
J.~Hemberger, A.~Krimmel, T.~Kurz, H.-A.~K. Von~Nidda, V.~Y. Ivanov, A.~Mukhin,
  A.~Balbashov, A.~Loidl, Structural, magnetic, and electrical properties of
  single-crystalline {La}$_{1-x}${Sr}$_x${MnO}$_{3}$ {(0.4} $<$ {x} $<$
  {0.85}), Physical Review B 66~(9) (2002) 094410.

\bibitem{Wang2021}
H.~Wang, X.~Liu, K.~Sun, X.~Ma, H.~Guo, I.~Bobrikov, Y.~Sui, Q.~Liu, Y.~Xia,
  X.~Chen, Z.-Y. Li, L.~Hao, Y.~Liu, D.~Chen, Competition of ferromagnetism and
  antiferromagnetism in {Mn}-doped orthorhombic {YCrO}$_{3}$, J. Magn. Magn.
  Mater. 535 (2021) 168022.

\bibitem{Jin2004}
H.~K. Jin, J.-H. Moon, B.-T. Lee, K.-S. Hur, S.-H. Kim, Method for
  manufacturing a {NTC} thermistor of {Y}{Cr}{O}$_3$ thin film type, South
  Korea Patent {KR}100436980{B}1, 2004.

\bibitem{Zenzo2009}
I.~Zenzo, O.~Masahiro, High temperature oxidation resistant iron-base alloy
  composite member, method for producing the same, and fuel cell separator
  using the same, Japan Patent JP4381158B2, 2009.

\bibitem{Tatsuaki2013}
Y.~Tatsuaki, C.~Katsunao, Sintered component made of stainless steel with high
  corrosion resistance and production method therefor, Europan Patent
  EP1522601B1, 2013.

\bibitem{Mori1996}
K.~Mori, H.~Miyamoto, T.~Matsudaira, Interconnector material, Japan Patent
  JPH0883620A, 1996.

\bibitem{Kuzuoka2005}
K.~Kuzuoka, High temperature thermistor material manufacturing method and high
  temperature thermistor, Japan Patent JP3721701B2, 2005.

\bibitem{Takeuchi2021}
A.~Takeuchi, A.~Takeuchi, N.~Shinseki, N.~Shinseki, Thermistor sintered body
  and temperature sensor element, Japan Patent JPWO2020090489A1, 2021.

\bibitem{iwaya1997ceramic}
M.~Iwaya, K.~Hayashi, Ceramic composition for thermistor, {US} {Patent}
  5,610,111A (1997).

\bibitem{sorg2001yttrium}
D.~J. Sorg, Yttrium chromite chromia thermistors, {US} {Patent} 6,204,748
  (2001).

\bibitem{taira2020temperature}
M.~Taira, Y.~Sato, K.~Yano, Temperature sensor and manufacturing method of
  temperature sensor, {US} {Patent} {App}. 16/588,248 (2020).

\bibitem{iwaki2011temperature}
T.~Iwaki, H.~Wado, Y.~Takeuchi, Temperature sensor and manufacturing method of
  temperature sensor, {US} {Patent} {App}. 13/043,842 (2011).

\bibitem{Taira2020}
M.~Taira, Y.~Sato, K.~Yano, Temperature sensor and method for manufacturing
  temperature sensor, {Japan} {Patent} {JP}6785925B2 (2020).

\bibitem{Takeuchi2021-1}
A.~Takeuchi, N.~Shinseki, Temperature sensor element and method for
  manufacturing temperature sensor element, WIPO (PCT) Patent WO2021065541A1,
  2021.

\bibitem{Murata1998}
T.~Murata, I.~Murata, N.~Murata, N.~Murata, K.~Murata, T.~Murata,
  Semiconductive ceramics having negative temperature coefficient of
  resistance, European Patent EP0609888B1, 1998.

\bibitem{Yoshihara2021}
T.~Yoshihara, M.~Kirihara, Temperature sensor, China Patent CN113167659A, 2021.

\bibitem{Kuzuoka1999}
K.~Kuzuoka, A method for manufacturing thermistor materials and thermistors,
  {Europan} {Patent} {EP}0798275B1, 1999.

\bibitem{Hardwicke2006}
C.~Hardwicke, S.~Rutkowski, Sensor and method for making same, {US} {Patent}
  US20060020415A1, 2006.

\bibitem{Hendrickson2011}
P.~W. Hendrickson, L.~Mickelson, P.~H. Larson, S.~Lindroth, M.~Mogensen, A
  multi-layer coating, {China} Patent {CN}101438439{B}, 2011.

\bibitem{jabbar2019chromate}
M.~H.~A. JABBAR, E.~D. Wachsman, P.~Ke-Ji, Chromate based ceramic anode
  materials for solid oxide fuel cells, {US} {Patent} {App}. 16/141,051 (2019).

\bibitem{barnett2003fuel}
S.~Barnett, J.~Liu, B.~Madsen, Z.~Ji, Fuel-flexible anodes for solid oxide fuel
  cells, {US} {Patent} {App}. 10/291,875 (2003).

\bibitem{Chu2013}
X.~Chu, W.~Zhu, L.~Shen, Y.~Zhang, Cathode material of {Y}{Cr}{O}$_3$-{base}
  sulfur-oxygen fuel cell, {China} {Patent} {CN}102142565{B}, 2013.

\bibitem{Chen2017}
Y.~Chen, Q.~Gu, X.~Lu, Y.~Ding, D.~Tian, B.~Lin, Preparation method for
  connection material membrane and electrolyte membrane for ceramic membrane
  fuel cell, {China} {Patent} CN103985888B, 2017.

\bibitem{Gupta2020}
M.~K. Gupta, R.~Mittal, S.~K. Mishra, P.~Goel, B.~Singh, Spin–phonon coupling
  and thermodynamic behaviour in {YCrO}$_{3}$ and {LaCrO}$_{3}$: inelastic
  neutron scattering and lattice dynamics, J. Phys.: Condens. Matter 32~(50)
  (2020) 505402.

\bibitem{PhysRevB.82.140503}
H.-F. Li, C.~Broholm, D.~Vaknin, R.~M. Fernandes, D.~L. Abernathy, M.~B. Stone,
  D.~K. Pratt, W.~Tian, Y.~Qiu, N.~Ni, S.~O. Diallo, J.~L. Zarestky, S.~L.
  Bud'ko, P.~C. Canfield, R.~J. McQueeney, Anisotropic and quasipropagating
  spin excitations in superconducting
  {Ba}({Fe}$_{0.926}${Co}$_{0.074})_{2}${As}$_{2}$, Phys. Rev. B 82 (2010)
  140503.

\bibitem{sotnikova2021radiometric}
G.~Y. Sotnikova, G.~Gavrilov, K.~Muratikov, R.~Passet, E.~Smirnova, Radiometric
  studies of leakage currents in dielectrics, Physics of the Solid State (2021)
  1--5.

\bibitem{Qi2005-3}
X.~Qi, J.~Dho, R.~Tomov, M.~G. Blamire, J.~L. MacManus-Driscoll, Greatly
  reduced leakage current and conduction mechanism in aliovalent-ion-doped
  {BiFeO}$_{3}$, Appl. Phys. Lett. 86 (2005) 062903.

\bibitem{Bahadur2008}
J.~Bahadur, D.~Sen, S.~Mazumder, R.~Shukla, A.~K. Tyagi, Non-{Debye} to {Debye}
  transition of ac dielectric response in {YCrO}$_{3}$ nanoceramic under
  sintering: effect of pore structure, J. Phys.: Condens. Matter 20 (2008)
  345201.

\end{thebibliography}

\end{document}